\documentclass[traditabstract]{aa}
\usepackage{graphicx}
\usepackage[varg]{txfonts}
\usepackage{longtable}
\usepackage{lscape}
\usepackage{natbib}
\usepackage{url}
\usepackage{color}
\usepackage{multirow,bigdelim}
\usepackage{cases}
\usepackage[colorinlistoftodos]{todonotes}
\bibpunct{(}{)}{;}{a}{}{,} 
\newcommand\kms{{\rm\,km\,s^{-1}}}

\newcommand\rgal{R_{\rm gal}}

\newcommand\teff{T_{\rm eff}}

\usepackage{hyperref}
\hypersetup{
     breaklinks,
     colorlinks   = true,
     citecolor    = blue,
     urlcolor     = blue,
     linkcolor    = blue
}

\begin{document}

\title{
Chemical evolution of the Galactic bulge as traced by\\ microlensed dwarf and subgiant stars\thanks{Based on data obtained with the European Southern Observatory telescopes (Proposal ID:s 87.B-0600, 88.B-0349, 89.B-0047, 90.B-0204, 91.B-0289, 92.B-0626, 93.B-0700, 94.B-0282), the Magellan Clay telescope at the Las Campanas observatory, and the Keck~I telescope at the W.M. Keck Observatory, which is operated as a scientific partnership among the California Institute of Technology, the University of California and the National Aeronautics and Space Administration.}
\fnmsep
\thanks{Tables~\ref{tab:parameters} and \ref{tab:ews} are only available in electronic form at the CDS via anonymous ftp to \url{cdsarc.u-strasbg.fr (130.79.128.5)} or via \url{http://cdsweb.u-strasbg.fr/cgi-bin/qcat?J/A+A/XXX/AXX}.
}}
\subtitle{
VI. Age and abundance structure of the stellar populations\\ in the central sub-kpc of the Milky Way}
\titlerunning{Chemical evolution of the Galactic bulge as traced by microlensed dwarf and subgiant stars. VI.}

\author{
T.~Bensby\inst{1}
\and
S.~Feltzing\inst{1}
\and
A.~Gould\inst{2,3,4}
\and
J.C.~Yee\inst{5}
\and
J.A.~Johnson\inst{4}
\and
M.~Asplund\inst{6}
\and
J.~Mel\'endez\inst{7}
\and
S.~Lucatello\inst{8}
\and
L.M.~Howes\inst{1}
\and
A.~McWilliam\inst{9}
\and
A.~Udalski\inst{10,18} 
\and
M.K.~Szyma\'nski\inst{10,18}
\and
I.~Soszy\'nski\inst{10,18}
\and
R.~Poleski\inst{10,4,18}
\and
{\L}.~Wyrzykowski\inst{10,18}
\and
K.~Ulaczyk\inst{10,11,18}
\and
S.~Koz{\l}owski\inst{10,18}
\and
P.~Pietrukowicz\inst{10,18}
\and
J.~Skowron\inst{10,18}       
\and\\
P.~Mr{\'o}z\inst{10,18}
\and
M.~Pawlak\inst{10,18} 
\and
F.~Abe\inst{12,19}
\and
Y.~Asakura\inst{12,19}
\and
A.~Bhattacharya\inst{13,19}
\and
I.A.~Bond\inst{14,19}
\and\\
D.P.~Bennett\inst{15,19}
\and
Y.~Hirao\inst{16,19}
\and
M.~Nagakane\inst{16,19}
\and
N.~Koshimoto\inst{16,19}
\and\\
T.~Sumi\inst{16,19}
\and
D.~Suzuki\inst{15,19}
\and
P.J.Tristram\inst{17,19}
 }

\institute{
Lund Observatory, Department of Astronomy and Theoretical Physics, Box 43, SE-221\,00 Lund, Sweden
\and
Max Planck Institute for Astronomy, K\"onigstuhl 17, D-69117 Heidelberg, Germany
\and
Korea Astronomy and Space Science Institute Institute, Daejon 305-348, Republic of Korea
\and
Department of Astronomy, Ohio State University, 140 W. 18th Avenue, Columbus, OH 43210, USA
\and
Smithsonian Astrophysical Observatory, 60 Garden St., Cambridge, MA 02138, USA
\and
Research School of Astronomy \& Astrophysics, Australian National University, Cotter Road, Canberra, ACT 2611, Australia
\and
Departamento de Astronomia do IAG/USP, Universidade de S\~ao Paulo, Rua do Mat\~ao 1226, S\~ao Paulo, 05508-900, SP, Brasil
\and
INAF-Astronomical Observatory of Padova, Vicolo dell'Osservatorio 5, 35122 Padova, Italy
\and
Carnegie Observatories, 813 Santa Barbara Street, Pasadena, CA 91101, USA
\and
Warsaw University Observatory, Al.~Ujazdowskie~4, 00-478~Warszawa,
Poland
\and
Department of Physics, University of Warwick, Gibbet Hill Road,
Coventry, CV4~7AL,~UK
\and
Institute for Space-Earth Environmental Research, Nagoya University, Nagoya 464-8601, Japan
\and
Department of Physics, University of Notre Dame, Notre Dame, IN 46556, USA
\and
Institute for Information and Mathematical Sciences, Massey University, 1330 Auckland, New Zealand
\and
Laboratory for Exoplanets and Stellar Astrophysics, NASA/Goddard Space Flight Center, Greenbelt, MD 20771, USA
\and
Department of Earth and Space Science, Graduate School of Science, Osaka University, Toyonaka, Osaka 560-0043, Japan
\and
Mt. John University Observatory, P.O. Box 56, Lake Tekapo 8770, New Zealand
\and
The OGLE Collaboration
\and
The MOA Collaboration
}


\date{Received 4 February 2017 / Accepted 7 July 2017}
 \abstract{
We present a detailed elemental abundance study of 90 F and G dwarf, turn-off, and subgiant stars in the Galactic bulge. Based on high-resolution spectra acquired during gravitational microlensing events, stellar ages and abundances for 11 elements (Na, Mg, Al, Si, Ca, Ti, Cr, Fe, Zn, Y and Ba) have been determined.
Four main findings are presented: (1) a wide metallicity distribution with distinct peaks at $\rm [Fe/H]=-1.09,\,-0.63,\,-0.20,\,+0.12,\,+0.41$; (2) a high fraction of intermediate-age to young stars where at $\rm [Fe/H]>0$ more than 35\,\% are younger than 8\,Gyr, and for $\rm [Fe/H]\lesssim-0.5$ most stars are 10\,Gyr or older; (3) several episodes of significant star formation in the bulge has been identified: 3,\,6,\,8, and 11\,Gyr ago; (4) tentatively the ``knee'' in the $\alpha$-element abundance trends of the sub-solar metallicity bulge is located at a slightly higher [Fe/H] than in the local thick disk. 
These findings show that the Galactic bulge has complex age and abundance properties that appear to be tightly connected to the main Galactic stellar populations. In particular, the peaks in the metallicity distribution, the star formation episodes, and the abundance trends, show similarities with the properties of the Galactic thin and thick disks. At the same time, the star formation rate appears to have been slightly faster in the bulge than in the local thick disk, which most likely is an indication of the denser stellar environment closer to the Galactic centre. There are also additional components not seen outside the bulge region, and that most likely can be associated with the Galactic bar.  Our results strengthen the observational evidence that support the idea of a secular origin for the Galactic bulge, formed out of the other main Galactic stellar populations present in the central regions of our Galaxy. Additionally, our analysis of this enlarged sample suggests that the $(V-I)_0$ colour of the bulge red clump should be revised to 1.09.
}
   \maketitle

\section{Introduction}
\label{sec:introduction}

How spiral galaxies like our own Milky Way form and evolve into the complex spiral structures that we see today is a very active research field in contemporary astrophysics.  Bulges are massive major components of many spiral galaxies \citep[e.g.][]{gadotti2009} and have for a long time been regarded as the oldest components of spiral galaxies \citep[see review by e.g.][]{wyse1997}, formed during the initial monolithic collapse era of galaxy formation \citep{eggen1962}, or merging of clumps within the disk at high red-shift \citep[e.g.][]{noguchi1999,bournaud2009}. However, the picture of how our Milky Way bulge formed and evolved has in recent years undergone a dramatic change. It is now widely believed that the bulge is a boxy peanut-shaped \citep[e.g.][]{dwek1995,wegg2013} pseudo-bulge of a secular origin \citep{kormendy2004} rather than being a classical spheroid. This means that the formation of the bulge occurred later and most likely out of disk material. Observations of high-redshift galaxies have revealed similar indications that also their bulges formed at later times. Using data from the 3D-HST and CANDELS Treasury surveys, \cite{vandokkum2013} found that galaxies with Milky Way-like masses built most of their stellar masses at redshifts $z\lesssim2.5$ and that the overall star formation had almost ceased by $z\approx 1$. During the same redshift interval, between $1<z<2.5$, the star forming activity in the inner 2\,kpc of central regions of these galaxies increased by a factor of about 2.5, indicating that the bulges were not fully assembled at high redshifts. An interpretation of these results is that the bulges and disks in these galaxies formed and evolved in parallel, and possibly that the bulges originated out of disk material. These discoveries and theoretical interpretation render the previous general consensus that bulges are the oldest components of spiral galaxies uncertain. However, as very distant galaxies cannot be resolved to reveal all the intricate details that a spiral galaxy possesses, such as barred bulges, spiral arms, and different intertwined stellar populations, one must rely on the overall integrated properties, making it difficult to tell how the bulges and disks that form in parallel are connected. As the Milky Way is the only spiral galaxy for which large numbers of individual stars can be resolved and studied in great detail, and may serve as a benchmark galaxy when constraining theoretical models of galaxy formation and evolution, it is therefore important to have as a complete picture as possible of the Milky Way \citep{freeman2002,blandhawthorn2016}.

\begin{figure*}
\centering
\resizebox{0.75\hsize}{!}{
\includegraphics[viewport= 0 0 648 500,clip]{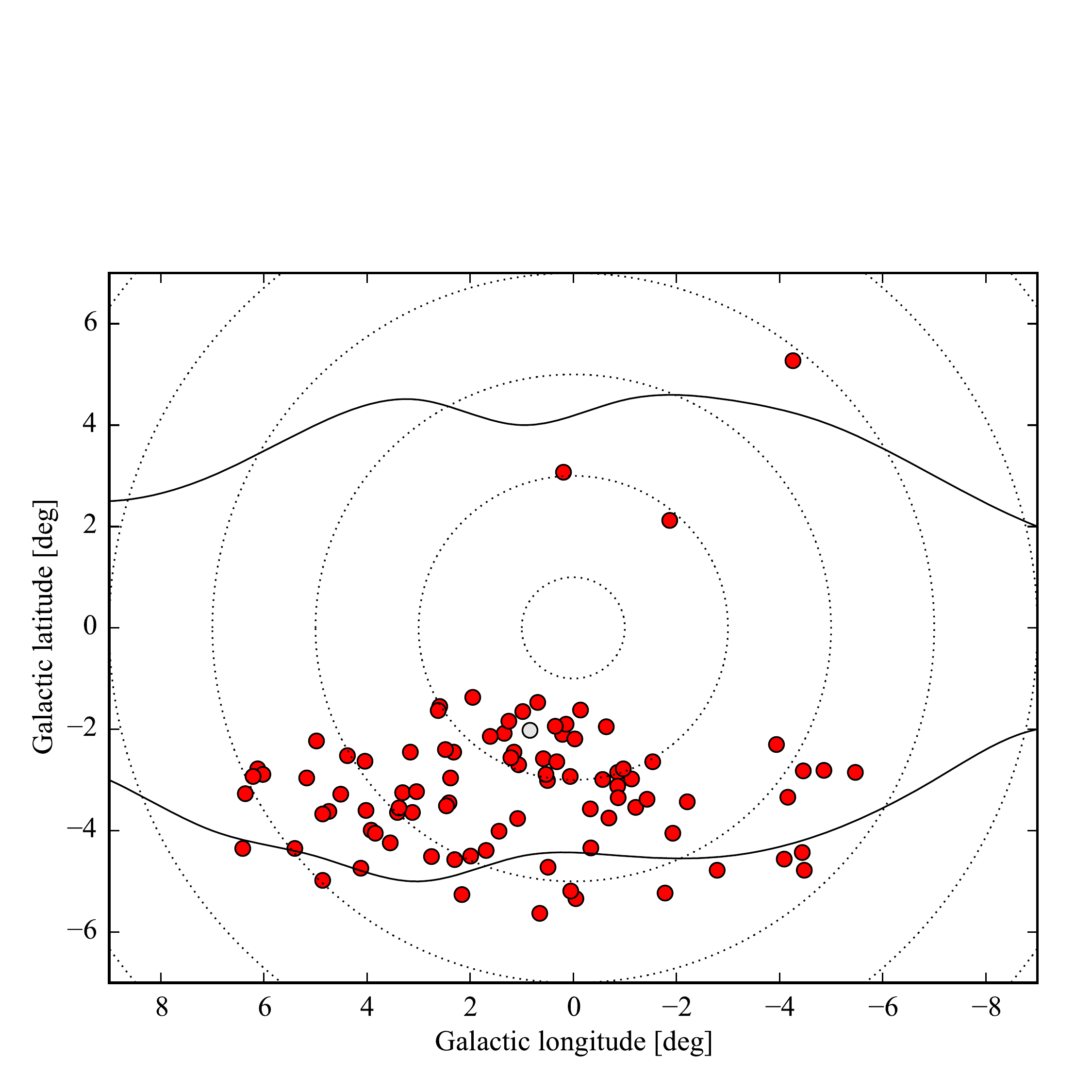}}
\caption{
Positions on the sky for the microlensed dwarf sample. The bulge contour lines based on observations with the COBE satellite are shown as solid lines \citep{weiland1994}. The dotted lines are concentric circles in steps of 2{\degr}. OGLE-2013-BLG-0911S that is excluded from the final bulge sample (see Sect.~\ref{sec:ob130911}) is marked by a lightgrey circle.
\label{fig:glonglat}
}
\end{figure*}

The first detailed abundance studies of red giants in the Milky Way bulge showed very high $\alpha$-enhancement even at super-solar metallicities, indicating a very fast chemical enrichment in the bulge \citep{mcwilliam1994,zoccali2006,fulbright2007,lecureur2007}. At the same time photometric studies seemed to show that the bulge had an old turn-off \citep{zoccali2003,clarkson2008}. Together with other spectroscopic data showing that it contained a vertical metallicity gradient \citep{minniti1995,zoccali2008}, all these results fitted very well into the picture of the Galactic bulge being a very old population originating from the initial collapse of the proto-galaxy and subsequent mergers, i.e. the scenario for a so called ``classical bulge'' \citep[e.g.][]{white1978,matteucci1990,ferreras2003,rahimi2010}.

This picture started to change when the \cite{melendez2008} study of red giants in the bulge found $\alpha$-enhancements that were lower at solar metallicities, and that the abundance trends more followed what was observed in the nearby thick disk. \cite{alvesbrito2010} then re-analysed the stars from \cite{fulbright2007}, using their actual equivalent width measurements, and found the $\alpha$-enhancement at solar metallicities to be at solar levels, and again that the abundance trends indeed seem to agree quite well with the thick disk. Solar $\rm [\alpha/Fe]$ ratios were also seen in the first very metal-rich ($\rm [Fe/H]>+0.3$) microlensed dwarf stars in the bulge \citep{johnson2007,johnson2008,cohen2008,cohen2009}. Subsequent studies of microlensed bulge dwarf stars have further strengthened the similarities between the (sub-solar metallicity) bulge and the thick disk \citep{bensby2010,bensby2011}, although \cite{bensby2013} saw a tendency of the bulge $\alpha$-element trends to be located closer to the upper envelope of the thick disk trends, possible hinting towards a slightly faster chemical enrichment in the bulge. Later studies of red giants in the bulge have generally also found the abundance trends to be similar to the local thick disk \citep{ryde2010,ryde2015,gonzalez2011,gonzalez2015,ryde2016,jonsson2017}, and some of them also tend to place the bulge abundance trends closer to the upper envelope of the thick disk trends \citep{johnson2014,jonsson2017}.

The median of the metallicity distribution of the bulge has, from high-resolution spectroscopic studies, been found to be metal-rich, peaking at or slightly above solar metallicities \citep{rich1988,fulbright2006,zoccali2008,hill2011,johnson2013,johnson2014}. Several studies have also found vertical metallicity gradients in the bulge \citep[e.g.][]{zoccali2008,sansfuentes2014}. This has often been interpreted as a signature of a classical bulge formed through dissipative processes (but see \citealt{martinezvalpuesta2013} who show that this is also possible through disk instabilities). However, later studies have shown that the metallicity distribution (MDF) of the bulge is a composite of two MDFs and that the relative strengths of the different components are dependent on Galactic latitude, i.e. the distance from the Galactic plane \citep{babusiaux2010,gonzalez2013,rojasarriagada2014,massari2014,babusiaux2014,zoccali2017,rojasarriagada2017,schultheis2017}, meaning that the observed gradient is just the manifestation of different stellar populations with different scale-heights. This has been taken one step further by \cite{ness2013} who identified up to five gaussian peaks in the bulge MDF, representing all Galactic stellar populations and a separate bar population, and showing that the strengths of the peaks vary with Galactic latitude. The up to five components were also seen in the \cite{bensby2013} sample of 58 microlensed dwarfs, with very good alignment with the \cite{ness2013} peaks.

The age of the Galactic bulge has been widely debated in the last few years. While photometric studies have claimed that the bulge should be a genuinely old stellar population \citep[e.g.][]{zoccali2003,clarkson2008},
studies based on ages on individual stars,  have claimed that the bulge at super-solar metallicities may contain as many as 50\,\% stars of young or intermediate ages, less than about 7\,Gyr \citep{bensby2011,bensby2013}. The main reason these young stars are not seen in the colour-magnitude diagrams could be because the photometric studies have no metallicity information on the stars, and intermediate age metal-rich isochrones and old metal-poor isochrones are essentially indistinguishable from each other \citep{bensby2013,haywood2016}, resulting in an apparent old turn-off. It is clear that not many stars have very low ages (around or less than about 1\,Gyr), but how large fraction of young to intermediate-age stars (4-8\,Gyr) can hide in the turn-off region?

Another recent major finding is that the Galactic bulge rotates as a cylinder \citep{howard2009,kunder2012,zoccali2014,ness2016}. This has been interpreted as the Milky Way being a pure-disk galaxy with no or a very small classical component \citep{shen2010}. Small classical bulges might however be lost in the secular evolution \citep{saha2015}.

As seen, our picture of the Galactic bulge has changed rather dramatically over the last decade. Many of the new results come from studies of evolved red giant stars. However, compared to the many detailed studies of the Solar neighbourhood that utilise the intrinsically much fainter dwarf and turn-off stars, the analysis of the rich giant spectra is more challenging, and is usually associated with larger uncertainties in stellar parameters and elemental abundance ratios. In addition, while stellar ages can easily be estimated from isochrones for dwarf and turn-off stars, it is very difficult for giant stars due to the crowding and overlap of the isochrones on the red giant branch. Hence, the studies of microlensed dwarf, turn-off, and subgiant stars play an important role in the mapping and understanding of the age and abundance structure of the enigmatic inner region of our Galaxy. 

The current study presents new high-resolution spectroscopic observations of 33 microlensed dwarf and subgiant stars. However, one of these is excluded from the final bulge sample as it was deemed likely to be located outside the bulge region (see Sect.~\ref{sec:ob130911}). Adding the first 58 microlensed dwarfs that were originally published in \cite{cavallo2003,johnson2007,johnson2008,cohen2008,cohen2009,bensby2009,bensby2010,epstein2010,bensby2011,bensby2013}, the sample now consists of 90 dwarf and subgiant stars in the bulge that have been homogeneously analysed in a consistent manner.

\section{Stellar sample and data analysis}

\begin{table*}
\centering
\caption{
Summary of the 33 microlensed microlensing events towards the Galactic bulge that are new in this study (sorted by observation date)\tablefootmark{$\dagger$}. Note that OGLE-2013-BLG-0911S is excluded from the final bulge sample (see Sect.~\ref{sec:ob130911}). 
\label{tab:events}
}
\setlength{\tabcolsep}{1.4mm}
\tiny
\begin{tabular}{lccrrrrrrrcrr}
\hline\hline
\noalign{\smallskip}
\multicolumn{1}{c}{Object}     &
RAJ2000                        &
DEJ2000                        &
 \multicolumn{1}{c}{$l$}       &
 \multicolumn{1}{c}{$b$}                           &
 \multicolumn{1}{c}{$T_{E}$}   &
 \multicolumn{1}{c}{$T_{max}$}                     &  
 \multicolumn{1}{c}{$A_{max}$}                     &
 \multicolumn{1}{c}{$T_{obs}$} &  
 Exp.                          &
 $S/N$                         &
 Spec.                         &
 \multicolumn{1}{c}{$R$}       \\
                               &
[hh:mm:ss]                     &
[dd:mm:ss]                     &
 [deg]                         &    
 [deg]                         &  
 [days]                        &
 \multicolumn{1}{c}{[HJD]}                         &
                               &   
 \multicolumn{1}{c}{[MJD]}     &  
 \multicolumn{1}{c}{[s]}       &
                               &
                               &
                               \\  
\noalign{\smallskip}
\hline
\noalign{\smallskip}
MOA-2013-BLG-063S              & 17:45:13.33 & $-$33:29:50.26 & $-3.94$ & $-2.30$ &  70 & 6353.00 &   48 & 6353.298 & 7200 & 20,\,30 & UVES  & 42\,000  \\
MOA-2013-BLG-068S              & 17:54:21.86 & $-$31:11:39.95 & $-0.97$ & $-2.78$ &  52 & 6361.97 &   41 & 6361.312 & 7200 & 30,\,50 & UVES  & 42\,000  \\
MOA-2013-BLG-299S              & 18:07:43.23 & $-$27:48:50.10 & $ 3.41$ & $-3.64$ &  38 & 6421.09 &  290 & 6420.372 & 4200 & 35,\,55 & MIKE  & 40\,000  \\
OGLE-2013-BLG-0692S            & 18:16:06.66 & $-$27:11:06.00 & $ 4.86$ & $-4.98$ &  20 & 6428.57 &   21 & 6428.199 & 7200 & 40,\,60 & UVES  & 42\,000  \\
OGLE-2013-BLG-0446S            & 18:06:56.18 & $-$31:39:27.20 & $-0.05$ & $-5.34$ &  78 & 6446.05 & 3300 & 6445.248 & 7200 &160,\,220& UVES  & 42\,000  \\
OGLE-2013-BLG-0835S            & 17:52:59.39 & $-$29:05:57.10 & $ 0.69$ & $-1.47$ &  10 & 6449.97 &   18 & 6450.258 & 7200 & 18,\,55 & UVES  & 42\,000  \\
MOA-2013-BLG-402S              & 18:03:00.14 & $-$29:54:24.42 & $ 1.08$ & $-3.76$ &  36 & 6463.23 &   11 & 6462.312 & 7200 & 25,\,60 & UVES  & 42\,000  \\
OGLE-2013-BLG-1114S            & 17:54:24.40 & $-$28:56:32.10 & $ 0.98$ & $-1.65$ &  71 & 6474.95 &  135 & 6474.244 & 7200 & 20,\,40 & UVES  & 42\,000  \\
MOA-2013-BLG-524S              & 18:02:29.65 & $-$33:06:32.20 & $-1.78$ & $-5.23$ &  10 & 6501.59 &   30 & 6501.009 & 7200 & 25,\,45 & UVES  & 42\,000  \\
OGLE-2013-BLG-1147S            & 18:08:39.19 & $-$26:40:44.80 & $ 4.51$ & $-3.28$ &  53 & 6506.75 &   37 & 6506.165 & 7200 & 30,\,40 & UVES  & 42\,000  \\
MOA-2013-BLG-517S              & 18:13:36.42 & $-$27:43:22.83 & $ 4.12$ & $-4.74$ &  39 & 6508.21 &   21 & 6507.087 & 7200 & 40,\,45 & UVES  & 42\,000  \\
OGLE-2013-BLG-1259S            & 18:10:23.34 & $-$27:58:45.60 & $ 3.55$ & $-4.24$ &  25 & 6511.42 &   74 & 6510.970 & 7200 & 90,\,120& UVES  & 42\,000  \\
OGLE-2013-BLG-1015S            & 17:52:48.18 & $-$35:00:51.40 & $-4.44$ & $-4.43$ &  60 & 6533.37 &   48 & 6532.026 & 7200 & 80,\,100& UVES  & 42\,000  \\
OGLE-2013-BLG-0911S            & 17:55:31.98 & $-$29:15:13.80 & $ 0.84$ & $-2.02$ & 101 & 6537.32 &  207 & 6536.985 & 7200 & 70,\,110& UVES  & 42\,000  \\
OGLE-2013-BLG-1793S            & 17:54:04.69 & $-$29:38:04.10 & $ 0.35$ & $-1.94$ &  20 & 6548.67 &   21 & 6548.038 & 7200 & 75,\,125& UVES  & 42\,000  \\
OGLE-2013-BLG-1768S            & 17:52:26.58 & $-$31:36:43.80 & $-1.54$ & $-2.64$ &  21 & 6553.64 &   24 & 6553.000 & 7200 & 20,\,45 & UVES  & 42\,000  \\
OGLE-2013-BLG-1125S            & 17:53:27.01 & $-$29:47:34.10 & $ 0.14$ & $-1.90$ &  52 & 6563.56 &   10 & 6560.996 & 7200 & 45,\,70 & UVES  & 42\,000  \\
MOA-2013-BLG-605S              & 17:58:42.88 & $-$29:23:53.60 & $ 1.06$ & $-2.70$ &  21 & 6573.06 &   13 & 6571.994 & 7200 & 42,\,65 & UVES  & 42\,000  \\
OGLE-2013-BLG-1868S            & 18:05:35.56 & $-$30:53:06.50 & $ 0.49$ & $-4.72$ &  27 & 6578.89 &   37 & 6577.989 & 7200 & 50,\,85 & UVES  & 42\,000  \\
OGLE-2013-BLG-1938S            & 17:46:01.89 & $-$34:12:31.50 & $-4.46$ & $-2.82$ &   8 & 6581.02 &   77 & 6579.985 & 7200 & 45,\,85 & UVES  & 42\,000  \\
OGLE-2014-BLG-0157S            & 17:58:16.85 & $-$33:46:18.80 & $-2.79$ & $-4.78$ &  35 & 6730.83 &   95 & 6733.320 & 7200 & 40,\,60 & UVES  & 42\,000  \\
MOA-2014-BLG-131S              & 17:59:02.50 & $-$31:01:54.70 & $-0.33$ & $-3.57$ &  36 & 6750.88 &  500 & 6750.313 & 7200 & 35,\,55 & UVES  & 42\,000  \\
OGLE-2014-BLG-0801S            & 17:54:02.42 & $-$32:35:31.60 & $-2.21$ & $-3.43$ &  13 & 6806.55 &  102 & 6806.197 & 7200 & 70,\,120& UVES  & 42\,000  \\
OGLE-2014-BLG-0953S            & 18:08:52.98 & $-$27:16:04.10 & $ 4.02$ & $-3.60$ &  18 & 6814.44 &   27 & 6814.209 & 7200 & 25,\,35 & UVES  & 42\,000  \\
OGLE-2014-BLG-0987S            & 18:16:53.61 & $-$25:31:31.00 & $ 6.41$ & $-4.35$ &  32 & 6829.10 &   35 & 6828.253 & 7200 & 35,\,50 & UVES  & 42\,000  \\
OGLE-2014-BLG-1122S            & 18:05:55.57 & $-$29:40:56.30 & $ 1.69$ & $-4.39$ &  22 & 6838.30 &   23 & 6838.112 & 7200 & 30,\,45 & UVES  & 42\,000  \\
OGLE-2014-BLG-1469S            & 17:56:07.54 & $-$30:57:13.00 & $-0.57$ & $-2.99$ &   9 & 6864.55 &   15 & 6864.005 & 7200 & 40,\,55 & UVES  & 42\,000  \\
OGLE-2014-BLG-1370S            & 17:55:42.33 & $-$31:53:34.30 & $-1.43$ & $-3.38$ &  57 & 6866.73 &  670 & 6866.127 & 7200 &110,\,150& UVES  & 42\,000  \\
OGLE-2014-BLG-1418S            & 18:14:47.62 & $-$26:24:38.90 & $ 5.40$ & $-4.35$ &  69 & 6901.49 &   88 & 6901.023 & 7200 & 95,\,130& UVES  & 42\,000  \\
OGLE-2014-BLG-2040S            & 17:55:59.48 & $-$31:16:35.80 & $-0.86$ & $-3.12$ &  30 & 6959.14 &   36 & 6958.986 & 6600 & 15,\,35 & UVES  & 42\,000  \\
OGLE-2015-BLG-0078S            & 17:55:30.69 & $-$27:57:36.50 & $ 1.95$ & $-1.37$ &  27 & 7089.88 &   55 & 7089.281 & 7200 & 15,\,60 & UVES  & 42\,000  \\
OGLE-2015-BLG-0159S            & 17:43:38.53 & $-$35:05:12.10 & $-5.47$ & $-2.85$ &  41 & 7093.99 &   19 & 7092.311 & 9000 & 25,\,50 & UVES  & 42\,000  \\
MOA-2015-BLG-111S              & 18:10:56.89 & $-$25:01:37.88 & $ 6.21$ & $-2.93$ &  27 & 7112.41 &  300 & 7112.271 & 7200 & 25,\,40 & UVES  & 42\,000  \\
\noalign{\smallskip}
\hline
\end{tabular}
\tablefoot{
\tablefoottext{$\dagger$}{Given for each microlensing event is: RA and DEC coordinates (J2000) read from the
fits headers of the spectra (the direction towards which the telescope pointed during observation); 
Galactic coordinates ($l$ and $b$); 
duration of the event in days ($T_{E}$); time when maximum magnification occured ($T_{max}$); 
maximum magnification ($A_{max}$); time when the event was observed
with high-resolution spectrograph ($T_{obs}$);
the exposure time (Exp.),
the measured signal-to-noise ratios per pixel at $\sim$6400\,{\AA} (on the UVES REDL spectrum) and 
at $\sim$8000\,{\AA} (on the UVES REDU spectrum) ($S/N$); the
spectrograph that was used (Spec); the spectral resolution ($R$).
}
}
\end{table*}

\subsection{Observations}

Main-sequence/turn-off/subgiant stars at a distance 8\,kpc from the Sun in the direction of the Galactic bulge have apparent $V$ magnitudes in the range 18-20 \citep[e.g.][]{feltzing2000b} and are too faint to observe with high-resolution spectrographs under normal conditions. Obtaining a spectrum with $S/N\approx 50$ would require more than 50 hours of integration time even with an instrument such as UVES on the VLT. However, during gravitational microlensing events the brightnesses of the background source stars can increase by factors of several hundred. It is then possible to achieve high-resolution and high $S/N$ spectra during 2 hour (or even shorter) exposures. However, microlensing events are random events and can happen essentially anywhere and anytime. Due to the unpredictability of microlensing events we have therefore since ESO observing period P82 (starting in October 2008) been running Target-of-Opportunity (ToO) programs with UVES \citep{dekker2000} on the ESO Very Large Telescope on Cerro Paranal, allowing us to trigger observations with only a few hours notice. 

\begin{figure}
\centering
\resizebox{0.9\hsize}{!}{
\includegraphics[viewport= 10 30 525 535,clip]{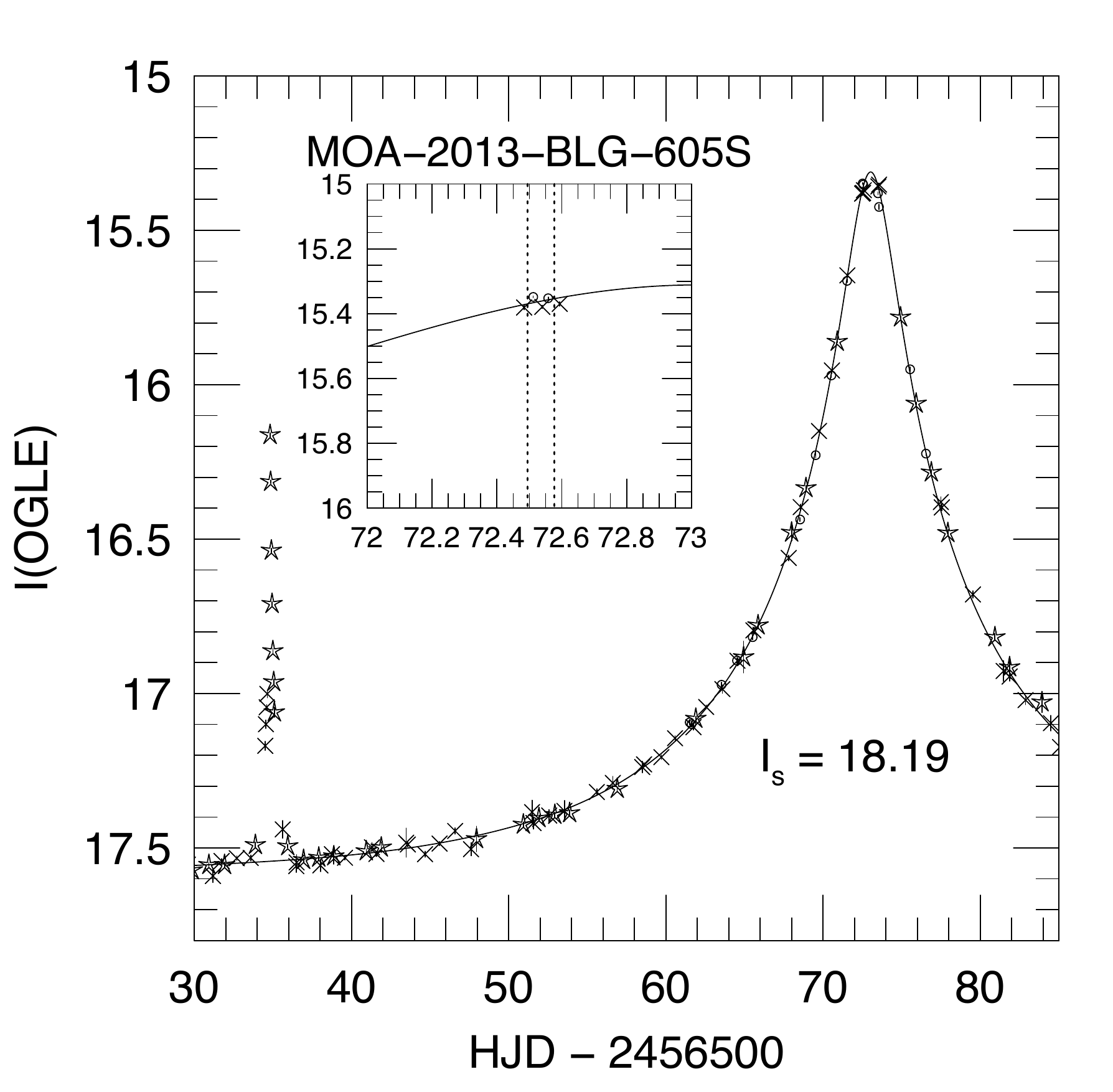}
}
\caption{
An example showing the light curve for OGLE-2013-BLG0605S, one of the new targets (light curves for the other new targets can be found in Fig.~\ref{fig:lightcurves} in the Appendix). The plot has a zoom-in window, showing the time interval when the source star was observed with UVES. The un-lensed magnitude of the source star is also given ($I_{\rm S}$).
\label{fig:events1}
}
\end{figure}

To identify potential targets we utilised the early warning systems from the MOA \citep{bond2001} and OGLE \citep{udalski2015} projects that survey the Galactic bulge every night. Figure~\ref{fig:glonglat} shows the positions on the sky for the 91 microlensing events towards the Galactic bulge that were observed (including OGLE-2013-BLG-0911S that is excluded from the final bulge sample, see Sect.~\ref{sec:ob130911}). All but three stars are located at negative Galactic latitudes, approximately 2 to 5 degrees from the plane, and within 6 degrees of the Galactic centre in longitude. As discussed in our earlier papers, the locations of the microlensing events is due to the observing strategy of the bulge areas covered by MOA and OGLE surveys.

Our goal has been to observe and obtain high-resolution spectra of the targets as close to peak brightness as possible, and to observe targets that reach at least $I\approx 15$ during the observations. Figure~\ref{fig:events1} shows the light curve for one of the targets (the light curves for all new events are shown in Fig.~\ref{fig:lightcurves} in the appendix). In the plot the time interval when the spectroscopic observation was carried out has been marked  (see inset in the plot). Generally, we were able to catch many objects very close to peak brightness. Note that for one of the events, MOA-2013-BLG-605S shown in Fig.~\ref{fig:events1}, the light curve marks the discovery of the first Neptune analog exoplanet or super-Earth with a Neptune-like orbit \citep{sumi2016}. The observations and microlensing light curves for the previous 58 stars are shown in our previous papers \citep{bensby2009,bensby2010,bensby2011,bensby2013}. 

The UVES ToO observations were carried out with a $1"$ wide slit which means that the spectra have a resolving power of $R\approx 42\,000$ and should have a minimum signal-to-noise ratio of around $S/N\approx 50-60$ and in the UVES 760\,nm setting (according to the UVES exposure time calculator) for a star that brightens to at least $I\approx15$. However, sometimes the observed microlensing light curve does not follow the predicted one, meaning that the obtained spectra could be better or worse than anticipated. Also, observing conditions (poor seeing and clouds) have affected the quality of some of the observations. Hence, the signal-to-noise ratios vary between $S/N\approx 15$ for the poorest ones and up to $S/N\approx 200$ for the best ones (see Table~\ref{tab:events}). 

The ToO observations with VLT have been augmented with a few observations obtained with the MIKE spectrograph \citep{bernstein2003} on the 6.5-meter Magellan Clay telescope on Las Campanas, or the HIRES spectrograph \citep{vogt1994} on the Keck telescope on Hawaii. Of the 33 new stars presented in this study, 32 stars were observed with UVES and one star with MIKE. More information about the observations and data reduction of the previous 58 targets, and with those instruments, can be found in our previous papers \citep{johnson2007,johnson2008,cohen2008,bensby2009,cohen2009,bensby2010,bensby2011,bensby2013}

The overwhelming majority of the real-time alerts that triggered the spectra analysed here were a by-product of the search for extra-solar planets by the Microlensing Follow Up Network ($\mu$FUN, \citealt{gould2010b}), which focused on high-magnification events because these are exceptionally sensitive to planets \citep{griest1998} and are also accessible to amateur-class telescopes \citep[e.g.][]{udalski2005}.  The timely identification of these events required about 700 hours of highly specialised labour per year, which could not have been justified for this spectroscopic program alone.  Hence, as microlensing planet searches have evolved toward ``pure survey mode'', especially with the advent of the OGLE-IV survey and the Korea Microlensing Telescope Network \citep{kim2016}, with $\mu$FUN essentially discontinuing operations, the source of spectroscopic alerts has likewise disappeared.  For this reason, this paper presents the final results of our project. That is, the project was made possible by the relatively brief overlap, when VLT ToO capability had already been achieved and before microlensing alerts were superseded by new microlensing technology.

\subsection{Data reduction}

Data reductions of the 32 UVES spectra were carried out with the UVES pipeline \citep{ballester2000} versions 4.9.0 through 5.0.7, depending on when the stars were observed. The MIKE spectrum was reduced with the Carnegie Observatories python pipeline\footnote{Available at \url{http://obs.carnegiescience.edu/Code/mike}}. All spectra have resolving powers between $R\approx 40\,000-90\,000$ (see Table~\ref{tab:events}). Information about the data reductions for the first 58 microlensing events can be found in our previous papers \citep{bensby2009,bensby2010,bensby2011,bensby2013}.   

\begin{figure}
\resizebox{\hsize}{!}{
\includegraphics[viewport=0 10 520 648,clip]{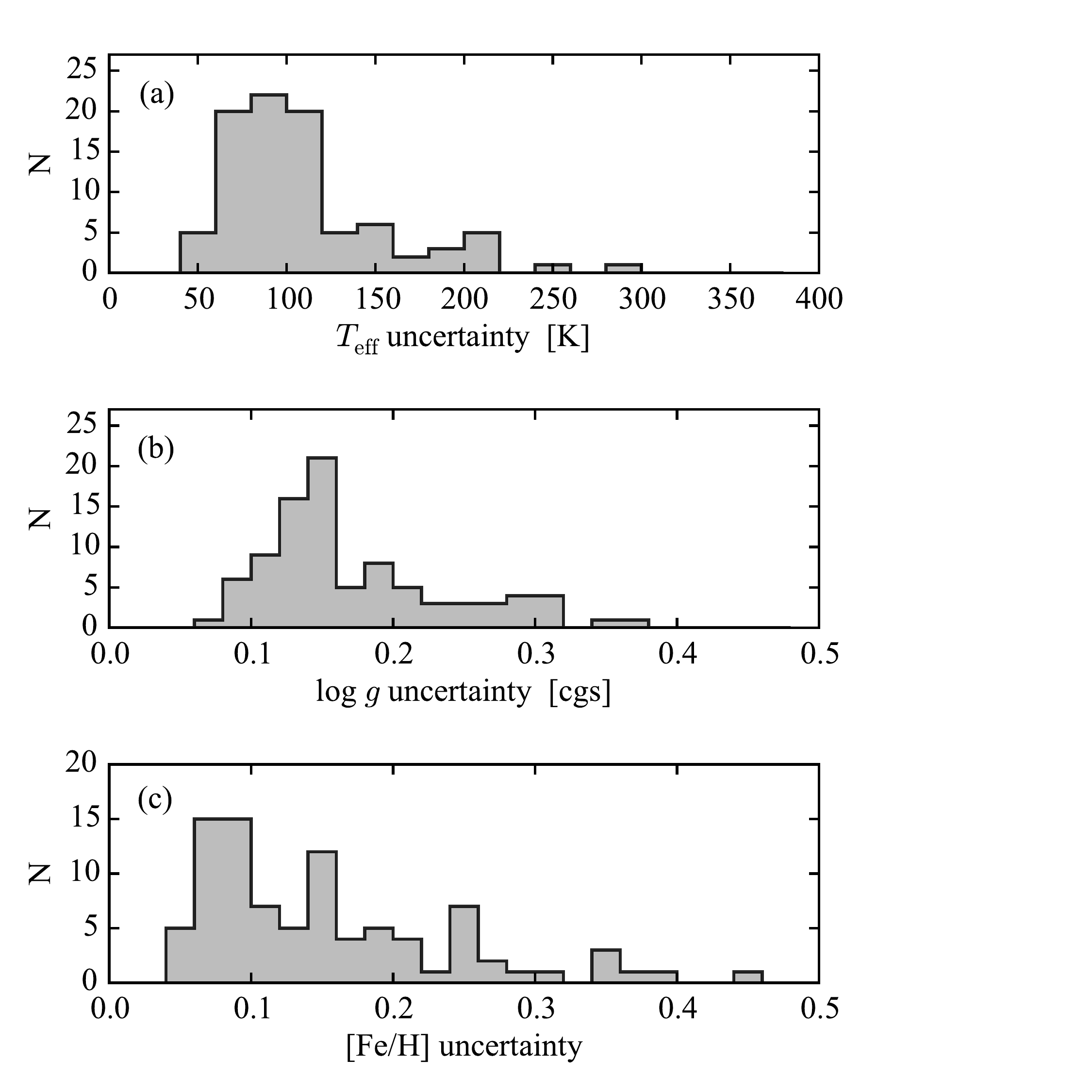}}
\caption{
The distribution of the uncertainties in $\teff$, $\log g$, and [Fe/H] for the microlensed dwarf sample.
\label{fig:feherrors}
}
\end{figure}

\subsection{Stellar parameters and elemental abundances}
\label{sec:parameters}

The methods to determine fundamental stellar parameters, detailed elemental abundances, uncertainties and errors, are fully described in \cite{bensby2013} and \cite{bensby2014}. Briefly, we use standard 1-D plane-parallel model stellar atmospheres calculated with the Uppsala MARCS code  \citep{gustafsson1975,edvardsson1993,asplund1997}. (For consistency with our previous analyses, we continue to use the old MARCS models. As shown in \citealt{gustafsson2008}, the differences between the new and old MARCS models are very small for our types of stars, i.e. mainly F and G dwarf and subgiant stars). Elemental abundances are calculated with the Uppsala EQWIDTH program using equivalent widths that were measured by hand using the IRAF\footnote{IRAF is distributed by the National Optical Astronomy Observatories, which are operated by the Association of Universities for Research in Astronomy, Inc., under cooperative agreement with the National Science Foundation \citep{tody1986,tody1993}.} task SPLOT. Effective temperatures were determined from excitation balance of abundances from \ion{Fe}{i} lines, surface gravities from ionisation balance between abundances from \ion{Fe}{i} and \ion{Fe}{ii} lines, and the microturbulence parameters by requiring that the abundances from \ion{Fe}{i} lines are independent of line strength. In all steps, line-by-line NLTE corrections from \cite{lind2012} are applied to all \ion{Fe}{i} lines.

\begin{figure}
\resizebox{\hsize}{!}{
\includegraphics{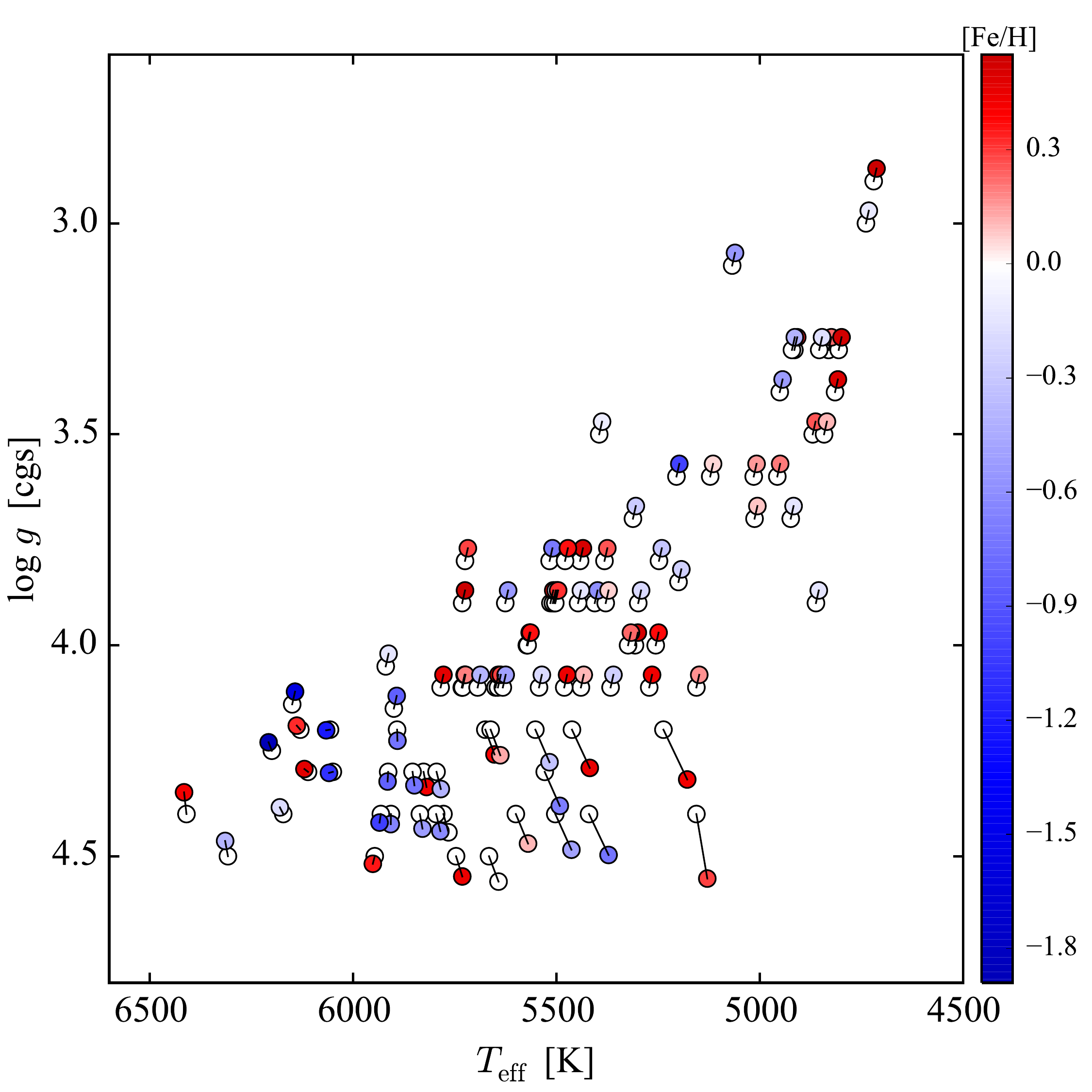}}
\caption{
The empty circles show the stellar parameters before applying the empirical corrections from \cite{bensby2014} and the coloured circles the parameters after the corrections. The values before and after the corrections are connected with solid lines. On average the effective temperatures change by $-9\pm 12$\,K and the surface gravities by $-0.01\pm0.04$. The colours represent [Fe/H], as shown by the colour bar on the right-hand side.
\label{fig:newparameters}
}
\end{figure}

The error analysis follows the method outlined in \cite{epstein2010}, and is  summarised in \cite{bensby2013}. This method takes into account the uncertainties in the four observables that were used to find the stellar parameters: the uncertainty of the slope in the graph of \ion{Fe}{i} abundances versus lower excitation potential; the uncertainty of the slope in the graph of \ion{Fe}{i} abundances versus line strength; the uncertainty in the difference between \ion{Fe}{i} and \ion{Fe}{ii} abundances; and the uncertainty in the difference between input and output metallicities. The method also accounts for abundance spreads (line-to-line scatter) as well as how the average abundances for each element reacts to changes in the stellar parameters.
Figure~\ref{fig:feherrors} shows the distributions of the uncertainties in $\teff$, $\log g$, and [Fe/H]. Typical errors are around or below 100\,K in $\teff$, $0.1-0.2$\,dex in $\log g$, and $0.05-0.15$\,dex in [Fe/H]. The median values are 94\,K, 0.15\,dex, and 0.13\,dex, respectively.

In the \cite{bensby2014} study that contains 714 nearby F and G dwarf and subgiant stars, analysed in the exact same way as the current sample of microlensed bulge dwarf stars, it was discovered that our methodology that utilises Fe ionisation and excitation balances to determine $\teff$ and $\log g$ results in an apparently flat lower main sequence. The reason for this behaviour is currently unclear, and \cite{bensby2014} applied small empirical corrections to the temperatures and surface gravities. These corrections have now been applied to all stars in the current sample and we have re-calculated abundances and stellar ages based on the corrected parameters. Figure~\ref{fig:newparameters} contains a Hertzsprung-Russell diagram with both the corrected and un-corrected effective temperatures and surface gravities shown. On average the effective temperatures change by $-9\pm12$\,K and the surface gravities by $-0.01\pm0.04$\,dex after applying the corrections. These corrections are very small but have been applied to all stars for consistency with the 714 F and G dwarf stars analysed in \cite{bensby2014}. Table~\ref{tab:parameters} contains updated parameters, abundances, ages, etc. for all 91 stars (including OGLE-2013-BLG-0911S that is excluded from the final bulge sample, see Sect.~\ref{sec:ob130911}), and supersedes the tables from previous papers.

\begin{figure}
\resizebox{\hsize}{!}{
\includegraphics{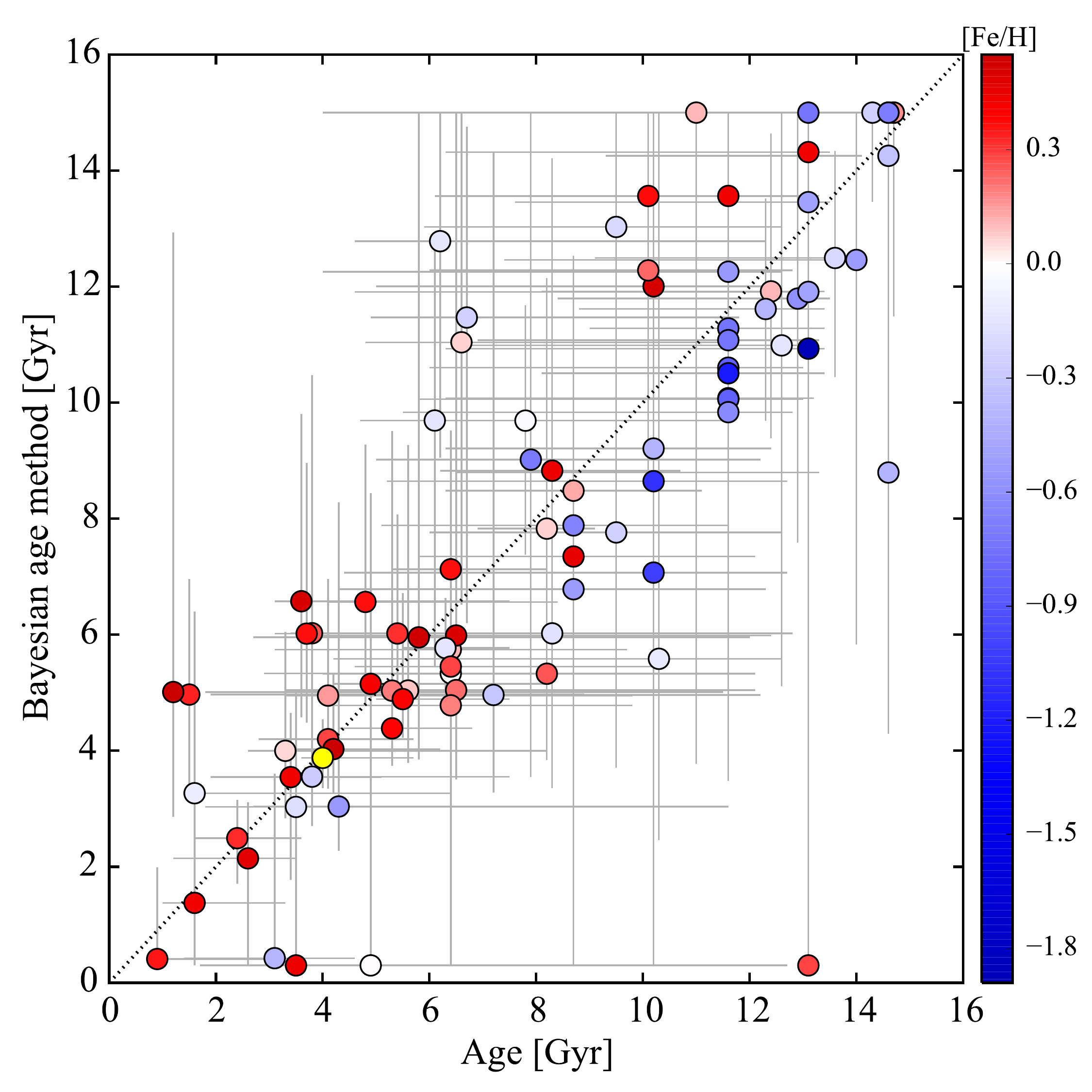}}
\caption{
Ages estimated from Bayesian method versus ages as determined from our age probability distribution methods. The stars have been colour-coded according to their effective temperatures (as shown by the colour bar on the right-hand side). The error bars indicate the upper and lower age estimates from the two methods. OGLE-2013-BLG-0911S that is excluded from the final bulge sample (see Sect.~\ref{sec:ob130911}) is marked by a yellow circle.
\label{fig:newages}
}
\end{figure}

\begin{figure*}
\centering
\resizebox{0.9\hsize}{!}{
\includegraphics[viewport= 0 0 648 420,clip]{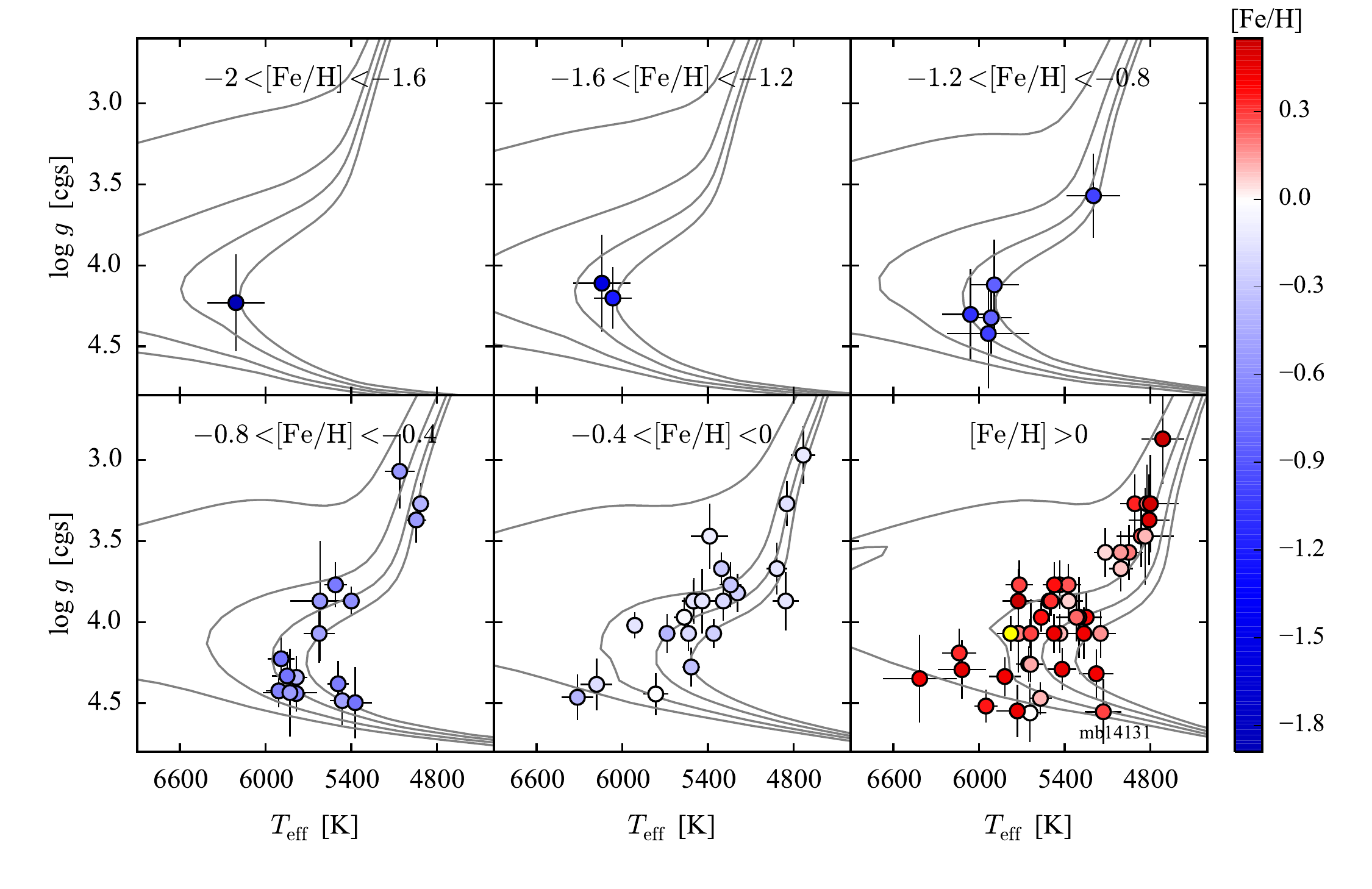}}
\caption{
Hertzsprung-Russel diagrams for the microlensed dwarf sample, split into six metallicity bins (as indicated in the plots). In each subplot one set of isochrones is shown (with ages 1, 5, 10, and 15 Gyr delineated), representing the mean metallicity in the range indicated. The isochrones are calculated with the $\alpha$-enhanced Yonsei-Yale models from \cite{demarque2004}. The stars have been colour-coded according to their metallicity (as shown by the colour bar on the right-hand side). MOA-2014-BLG-131S (mb14131) has been identified in the $\rm [Fe/H]>0$ subplot on the lower right-hand side. OGLE-2013-BLG-0911 that is excluded from the final bulge sample (see Sect.~\ref{sec:ob130911}) is marked by a yellow circle).
\label{fig:hr}
}
\end{figure*}

\subsection{Age determinations}
\label{sec:newages}

As in our previous studies stellar ages, masses, luminosities, absolute $I$ magnitudes ($M_I$), and colours ($V-I$) have been estimated from $Y^2$ isochrones \citep{demarque2004} by maximising probability distribution functions as described in \cite{bensby2011}. The estimations have been updated for all stars using the corrected stellar parameters (as described above and illustrated in Fig.~\ref{fig:newparameters}). The individual age probability distributions are shown in Fig.~\ref{fig:agefunctions} in the appendix for all stars.

For the stellar ages we have also used another method, the Bayesian estimation from isochrones as described in \cite{jorgensen2005}. The probability distribution function of an individual star's age, or so-called G function, is calculated by combining the likelihood function of the associated observed data ($\teff$, $\log g$, and $\rm [Fe/H]$) with a prior probability density of the model parameters (initial stellar mass, chemical composition Z, and age). This combination is then integrated with respect to the initial mass and Z, resulting in the G function. As a prior, we assumed no knowledge of the metallicity distribution or star formation rate, but we did assume a Salpeter IMF for masses greater than 1\,$M_{\odot}$. The isochrones used were from the $\alpha$-enhanced Yonsei-Yale models \citep{demarque2004}, and we made use of the interpolator provided with the grid\footnote{\url{http://www.astro.yale.edu/demarque/yyiso.html}}. The estimated ages are the most probable ages from the G functions, and the confidence intervals quoted are the range of values for which the normalised G function is above a value of 0.6. This corresponds to 1$\sigma$ for Gaussian errors. The individual G functions are shown in Fig.~\ref{fig:agefunctions} in the appendix.

Figure~\ref{fig:newages} shows a comparison between the Bayesian ages as described above and the ages as determined from probability distribution functions. The agreement is overall good, with some scatter, but no clear systematic offsets or trends. A striking example is MOA-2014-131S that with the age probability distribution method was old, around 13\,Gyr, but that with the Bayesian method is below 1\,Gyr. Looking at the G function and the age probability distribution function for this star (first plot on the top left-hand side in Fig.~\ref{fig:agefunctions} in the appendix) it is clear that the age for this star is hard to estimate and that it could have essentially any age. This could be due to that it is located below the turn-off, on the lower main-sequence, where the isochrones are very close and ages are difficult to estimate (see Fig.~\ref{fig:hr}, where MOA-2014-131S has been marked out on the lower right-hand side). From Fig.~\ref{fig:agefunctions} (top left panel) it appears however more likely that it is older, more in line with the age estimated from the age probability distribution method. 

In summary we find that the age probability distribution functions are very similar to the G functions (Fig.~\ref{fig:agefunctions} in the appendix) and they provide very similar ages (Fig.~\ref{fig:newages}). As the Bayesian method currently does not provide stellar masses, absolute magnitudes, and colours for the stars, we will keep using the ages from the age probability distribution function method to be internally consistent. Our ages have also been verified to be in good agreement with the age estimation method by \cite{valle2015}.

\subsection{Sample age estimation}
\label{sec:sampleage}

The individual ages from the Bayesian method will be used to estimate the sample age distribution, using the code and techniques of \cite{jorgensen2005conf,jorgensen2005phdt}. This code takes the individual G functions of all 90 stars (as described above), and uses that information to predict a total distribution of their ages combined. The only new prior introduced at this stage is the assumption that the star formation history of the sample is a non-negative and smooth function; aside from that it is left free to adapt to the data. A range of possible distributions are created by finding the most-likely distribution statistically, and then varying this within some small allowed degree of statistical error. Within this variation a minimum is found that corresponds to the least complicated form of the distribution. In practice, a function H is added to the original distribution which penalises more complicated distributions, and then the total function is minimised. The end product is the most likely age distribution of the sample, given a smooth star formation history. The results of this method are further examined in Sect.~\ref{sec:popages}.

\begin{figure}
\resizebox{\hsize}{!}{
\includegraphics{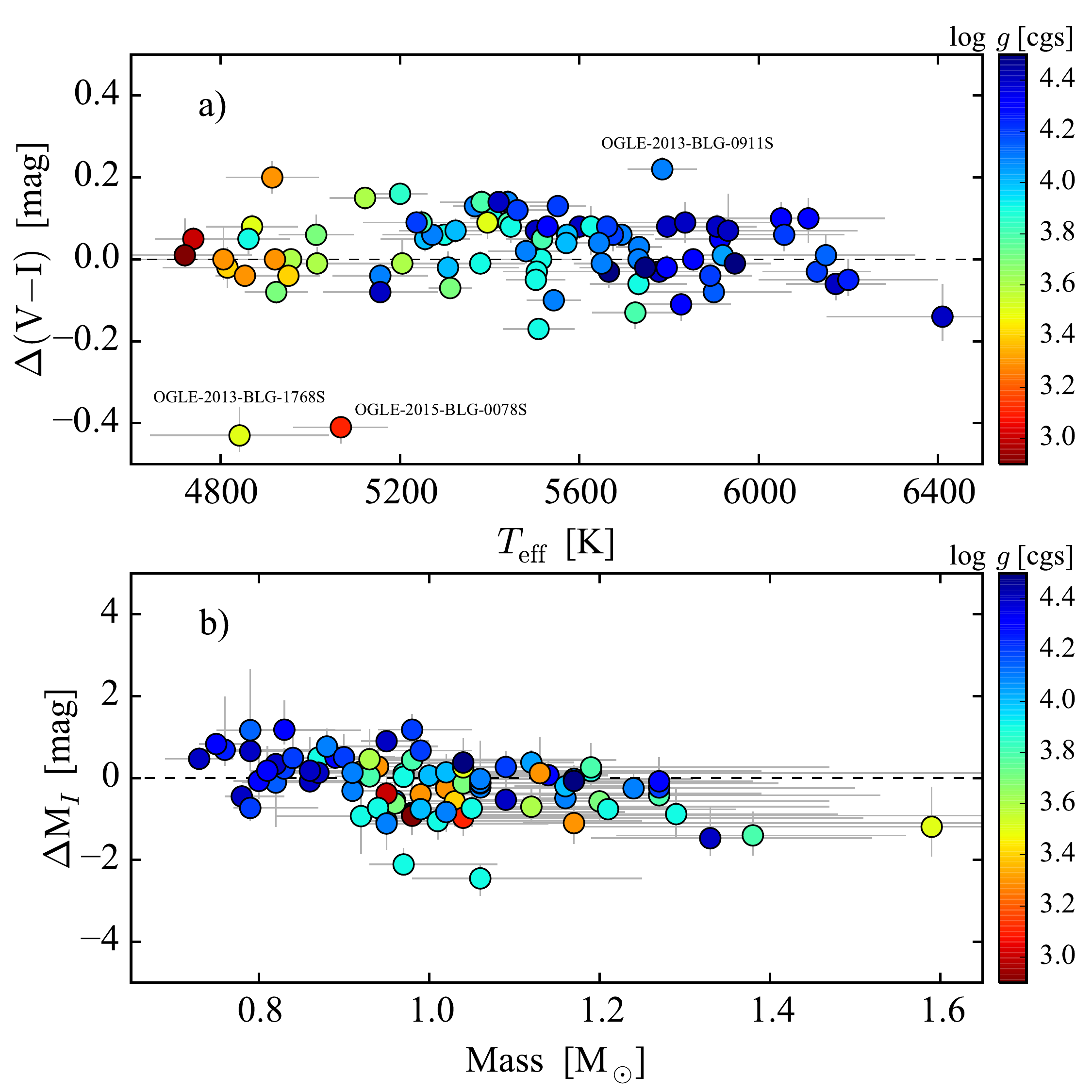}}
\resizebox{\hsize}{!}{
\includegraphics[viewport=0 0 648 325,clip]{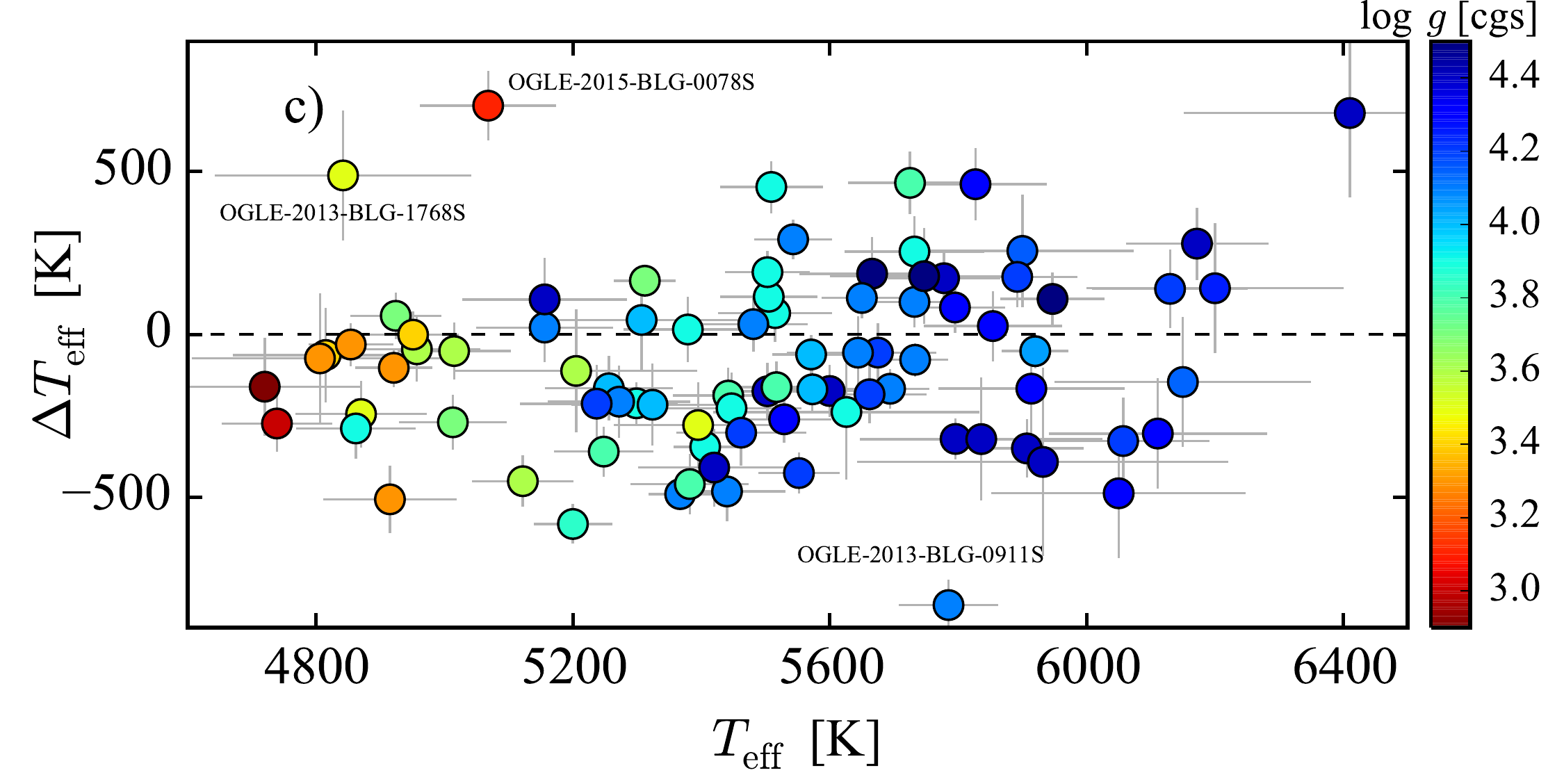}}
\caption{
(a) Difference between microlensing colours and the spectroscopic colours versus spectroscopic $\teff$. (b) Difference between the absolute $I$ magnitudes from microlensing techniques and from spectroscopy versus stellar mass (derived from spectroscopy). Error bars represent the uncertainties in the spectroscopic values. (c) The difference between the spectroscopic $\teff$ and the temperature based on the microlensing colours and the \cite{casagrande2010} $\teff$--colour calibration. The microlensing values used in the plots are based on the assumption that the bulge red clump has $(V-I)_0 = 1.06$ and $M_I=-0.12$. The stars have been colour coded according to their surface gravities (as shown by the colour bar on the right-hand side).
\label{fig:delta_vi}
}
\end{figure}

\section{Consistency checks}
\label{sec:consistency}

\subsection{Hertzsprung-Russel diagrams}

Figure~\ref{fig:hr} shows Hertzsprung-Russel diagrams in six metallicity bins for the microlensed bulge dwarf sample.  The spectroscopic effective temperatures and surface gravities categorise them mainly as turn-off and subgiant stars, and a few have started to ascend the giant branch. As these stars are low-luminosity giants that have just left the subgiant branch they should still have retained their natal chemical compositions as the major internal mixing processes occur further up on the giant branch \citep[e.g.][]{pinsonneault1997}. It should also be mentioned that age estimates are still possible for the low-luminosity giants, since the isochrones have not yet fully converged.

As our method to determine stellar parameters is purely spectroscopic and independent of the distance to the stars this shows that the underlying assumptions on how to select and observe un-evolved low-mass stars in the Galactic bulge is sound and works well. 

\subsection{Balmer wing line profiles}

The wings of the H$\alpha$ Balmer line are sensitive to the effective temperature and provides an independent way of checking the determined effective temperature of for instance a reddened star \citep[e.g.][]{cayrel2011}. Figure~\ref{fig:balmer} in the appendix shows comparisons between synthetic H$\alpha$ line profiles to observed H$\alpha$ line profiles based on the spectroscopic temperatures (blue lines) for the 33 new stars (similar plots are shown for the other stars in previous papers).  There is generally very good agreement between synthetic and observed spectra for the spectroscopic effective temperatures. Hence, we believe that the effective temperatures we have determined should be good. The synthetic spectra were calculated with MARCS model atmospheres \citep{gustafsson2008} using the SME package (Spectroscopy Made Easy, v. 2011-12-05, \citealt{valenti1996}).

\subsection{Microlensing parameters}
\label{sec:mulens}

The $(V-I)_0$ colour and the $M_I$ magnitude can be estimated from microlensing techniques assuming that the reddening towards the microlensed source is the same as towards the red clump in the same field, that $(V-I)_0$ and $M_I$ of the bulge red clump is known, and (for $M_I$) that the distance to the source and the red clump is the same \citep{yoo2004}. The de-reddened magnitude and colour of the source can then be derived from the offsets between the microlensing source and the red clump in the instrumental colour-magnitude diagram. The microlensing values for $M_I$ and $(V-I)$ given in Table~\ref{tab:parameters} are based on the assumption that the bulge red clump has $(V-I)_{0} = 1.06$ (as determined in \citealt{bensby2011}) and $M_{I} = -0.12$ \citep{nataf2013}. 

For 5 of the events we could not estimate microlensing colours and magnitudes due to lacking photometric observations, but for the other 86 events, Figs.~\ref{fig:delta_vi}a and b show comparisons between the spectroscopic $(V-I)_{0}$ colours and $M_I$ magnitudes to those determined from microlensing techniques. In \cite{bensby2013} wherein we analysed 58 targets, there was a significant slope present between the colour difference, $\Delta(V-I)$, and the spectroscopic $\teff$. With larger numbers, this is no longer the case. There are, however, two stars with larger discrepancies in the $(V-I)_{0}$ colours than the rest, about $-0.4$\,mag. These are OGLE-2013-BLG-1768S (with $\teff\approx4850$\,K and $\log g =3.5$) and OGLE-2015-BLG-0078S (with $\teff=5068$\,K and $\log g = 3.1$) and they have been marked in Fig.~\ref{fig:delta_vi}a.

For the 86 stars that have $(V - I)_0$ colours from microlensing techniques  we calculate effective temperatures using the \cite{casagrande2010} $\teff$--colour calibration. Figure~\ref{fig:delta_vi}c shows the difference between the two temperatures versus spectroscopic $\teff$. The microlensing temperatures are on average 85\,K higher than the spectroscopic ones. If we restrict ourselves to stars with (spectroscopic) $\teff > 5500$\,K the microlensing temperatures are on average still higher, but only by 20\,K. The dispersion in both cases is around 280\,K. Figure~\ref{fig:balmer} also shows synthetic spectra of the H$\alpha$ Balmer line based on temperatures from the microlensing $(V - I)_0$ colours (red dashed lines). It is clear that the microlensing temperatures produce synthetic spectra that overall also match the observed spectra well, but in a few cases they clearly do not. As for example in the case of OGLE-2013-BLG-1768S, whose $(V-I)_{0}$ colour which is almost 500\,K below the spectroscopic determination, the synthetic line profile is too narrow compared to the observed spectrum (fifth plot from the top in the third column of Fig.~\ref{fig:balmer}).

\begin{figure}
\resizebox{\hsize}{!}{
\includegraphics[viewport=0 0 648 375,clip]{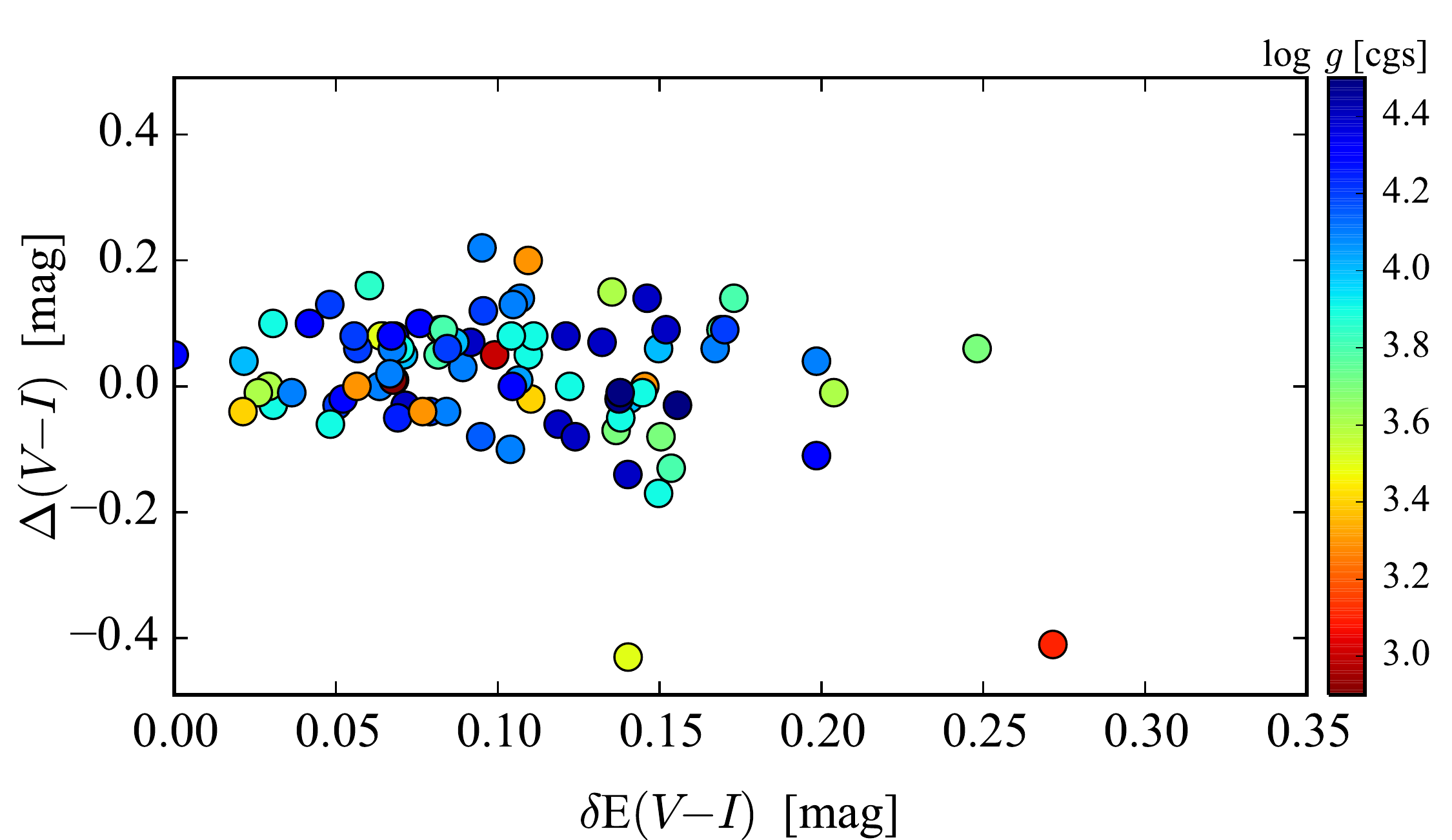}
}
\caption{
Difference between microlensing colours and the spectroscopic colours versus the amount of differential reddening based on the maps by \cite{nataf2013}. The microlensing values used in the plots are based on the assumption that the bulge red clump has $(V-I)_0 = 1.06$ and $M_I=-0.12$. The stars have been colour coded according to their surface gravities (as shown by the colour bar on the right-hand side).
\label{fig:delta_vi_red}
}
\end{figure}

The bulge is known to have patchy and irregular reddening. \cite{nataf2013} measured reddening and differential reddening for more than 9000 sight lines towards the bulge. Figure~\ref{fig:delta_vi_red} shows the difference in the $(V-I)_{0}$ colours versus differential reddening and there appears to be no preference for large differential reddening values with the colour difference. It is however clear that OGLE-2015-BLG-0078S which is one of the two stars with the largest deviations in $(V-I)_{0}$ also is located along a sight line that has the largest differential reddening among the stars in the sample, and for OGLE-2013-BLG-1768S the differential reddening is relatively large, around 0.14\,mag. This indicates that differential reddening can to some extent be the cause for the discrepancies between the microlensing and spectroscopic $(V-I)_{0}$ colours.

At the time of observations it was difficult to obtain good $V$ band observations of OGLE-2013-BLG-1768S, for three reasons: it is intrinsically red, it is highly reddened (see above), and it was not very bright at peak brightness (see Fig.~\ref{fig:lightcurves}). Hence, for OGLE-2013-BLG-1768S the microlensing colour is deemed uncertain, and it will not be used below when re-determining the colour of the bulge red clump. For OGLE-2015-BLG-0078S the colour is also deemed uncertain, due to the very high differential reddening (see above), and it will also be left out the re-determination of the red clump colour. In addition, OGLE-2013-BLG-0911S is left out, but for very different reasons (see Sect.~\ref{sec:ob130911}). This leaves us with 83 stars that have good colour measurements.

The difference between the spectroscopic and the microlensing colours for these 83 stars is $+0.03\pm0.07$\,mag (1-$\sigma$ standard deviation), and for the absolute magnitudes it is $-0.17\pm0.69$\,mag (1-$\sigma$ standard deviation). As the microlensing colours have been estimated assuming that the colour of the bulge red clump is 1.06, this means that the colour of the bulge red clump could be slightly adjusted. By bootstrapping the sample (83 stars) 10\,000 times we get a distribution of differences in the $(V-I)_{0}$ colour with a mean value of $+0.03$\,mag, with a standard error of the mean of 0.01\,mag. Hence we suggest that the $(V-I)_{0}$ colour of the bulge red clump should be revised to 1.09.  

\section{Are the microlensed dwarfs in the bulge?}
\label{sec:inthebulge}

A question that has often been brought up is whether the microlensed dwarfs that have been observed are located in the bulge or not. If the stars had a distance of about 8\,kpc from the Sun, this would mean that they would lie between 280-700\,pc below the plane, and up to 800\,pc on either side of the Galactic centre. Of course, the stars do not all have a distance of exactly 8\,kpc from the Sun, but there are a series of different indicators that support the conclusion of them being located in the bulge, and not in the disk in front of the bulge, or in the disk far behind the bulge. We have addressed these questions in our previous papers and summarise them again:
\begin{itemize}
\item \cite{nair1999} estimate that about 15\% of the microlensing events toward the Bulge could have source stars belonging to the disk on far side of the bulge, more than 3 kpc away from the Galactic centre. On the other hand, more recent theoretical calculations of the distance to microlensed sources, assuming a constant disk density and an exponential bulge, show that the distance to the sources is strongly peaked in the bulge, with the probability of having $D < 7$\,kpc very small \citep{kane2006}.
\item Stars in the bulge have been observed to have higher radial velocities and velocity dispersions than those in the Galactic disk. The radial velocities measured for the microlensed dwarfs (see Fig.~\ref{fig:rvel}) are fully consistent with other studies of large samples of bulge stars \citep[e.g.][]{kunder2012,ness2013b}. 
\item Microlensing analysis yields the baseline colour and magnitude completely disentangled from any blended light.  These quantities identify  the stars as turn-off and subgiant stars at the approximate distance of the bulge. The spectroscopic parameters that are determined in a distance independent way are consistent with this picture. The alternative would be if some stars were actually giant stars in the disk on the other side of the bulge, that could have the same un-lensed magnitudes as intrinsically fainter dwarf stars in the bulge.  We investigated this in detail in \cite{bensby2011}, when the sample consisted of 26 stars, and found strong support for a great majority of the sample being located within the bulge.  In \cite{bensby2011} the spectroscopic values of the absolute $I$ magnitudes were on average $-0.13$\,mag higher (with a dispersion of 0.56\,mag) than the microlensing ones. Based on the current sample, we now find that the spectroscopic ones are higher by $-0.17$\,mag (with a dispersion of 0.68\,mag), see Fig.~\ref{fig:delta_vi}b, which is quite similar as before. The reasoning in Sect~4.3 in \cite{bensby2011} should hold also for the present sample, i.e., that a great majority are located within the bulge.
\end{itemize}

\paragraph{\sl \bfseries The case of OGLE-2013-BLG-0911S:}
\label{sec:ob130911}

For one star, OGLE-2013-BLG-0911S, we have indications that it is part of the disk population. First, the microlensing event OGLE-2013-BLG-0911 shows strong finite source effects, and from these we can estimate the source-lens relative proper motion to be about 0.3\,mas/yr. This is about 10 times slower than typical.  Since the probability of measuring $\mu<\mu_{0}$ scales as $\mu_{0}^3$ (for a given population), this is extremely small.  However, the proper motions of disk-source events are expected to be much slower (none have ever been measured to our knowledge) because the observer, lens, and source are all in the disk, hence all moving approximately the same speed. Second, OGLE-2013-BLG-0911S got a lot of observations in the $V$ band and has a very secure colour measurement of $(V-I)_{0} = 0.49$ (assuming that the source is in the bulge, and that it is behind as much dust as the red clump). The colour based on the spectroscopic analysis gives $(V-I)_{0} = 0.71$ (OGLE-2013-BLG-0911S is the star that shows the largest difference at $\teff = 5785$\,K in Fig.~\ref{fig:delta_vi}). Hence, the microlensing parameters appear to be inconsistent with this star being a solar-type star with $\rm [Fe/H] = +0.47$. Third, OGLE-2013-BLG-0911S shows a galactocentric radial velocity of $-35\kms$ (heliocentric radial velocity of $-46\kms$), which is consistent with a disk star. It does however not rule out the possibility of it being a bulge star as the bulge velocity distribution does have a large dispersion. 

These observations could be indirect evidence for a disk-source/disk-lens microlensing event. As it is likely that the source is located outside the bulge region we will leave it out of the sample from now on (as it was also left out of the sample when re-determining the colour of the bulge red clump in Sect.~\ref{sec:mulens}). An important consequence of this microlensing event is that it shows that we are able to identify a disk source when we see one.

\begin{figure}
\resizebox{\hsize}{!}{
\includegraphics{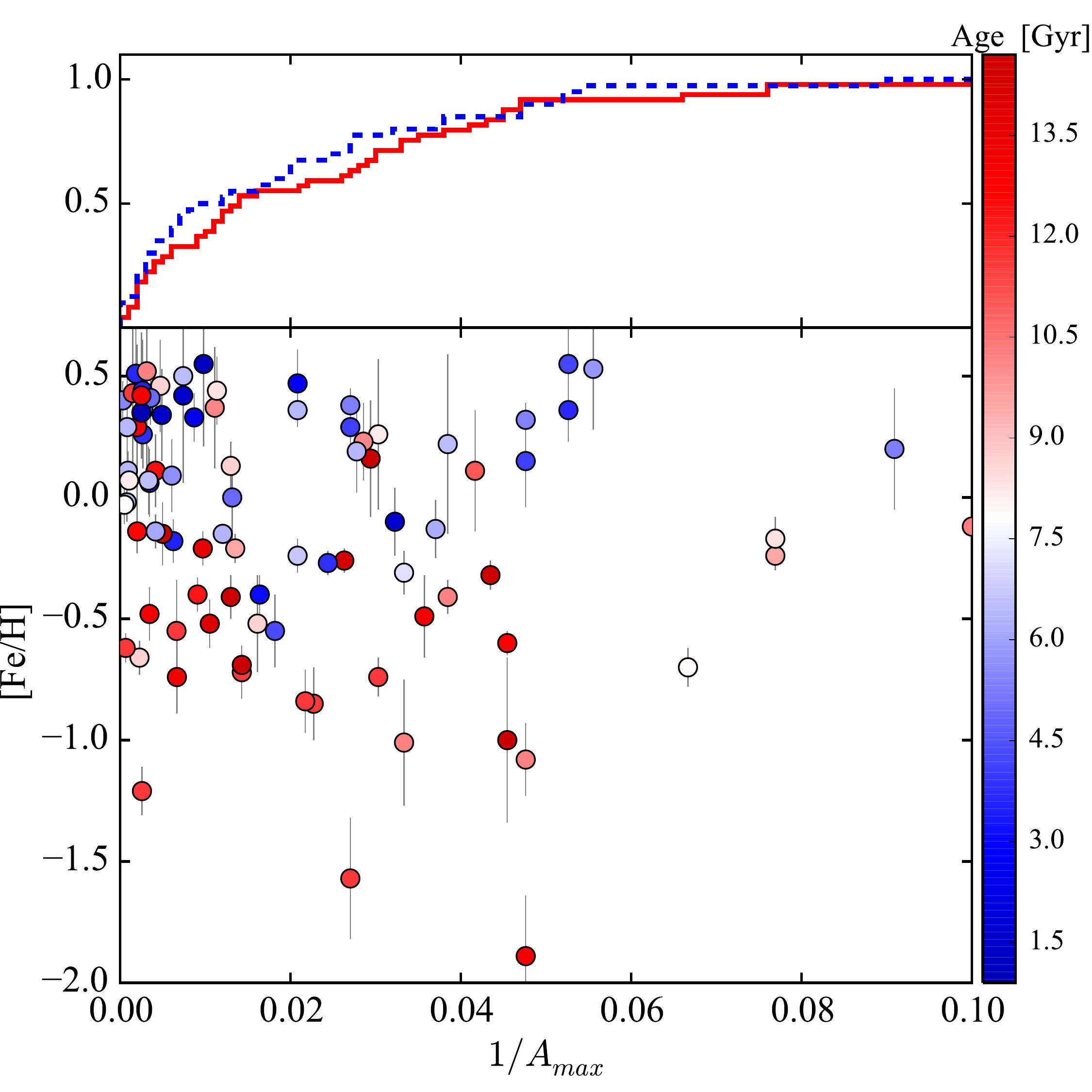}}
\caption{
[Fe/H] versus $1 / A_{\rm max}$ for the microlensed dwarf sample. The stars have been colour coded according to their ages (as shown by the colour bar on the right-hand side). The top panel shows the cumulative distributions for stars younger than 7\,Gyr (dashed blue line) and older than 7\,Gyr (red solid line). 
\label{fig:fehamax}
}
\end{figure}

\section{The magnification-metallicity relation}

When the microlensed bulge dwarf sample contained about 16 stars, \cite{cohen2010puzzle} discovered that there was a strong correlation between the brightness magnification and the spectroscopic metallicity. At that time no resolution to the mystery could be found. In \cite{bensby2013} with a sample of 58 microlensed bulge dwarfs we investigated this further and found that the microlensed dwarf sample probably was biased toward metal-rich and younger targets. The level of that bias was however not enough to resolve the issue of the apparent $A_{\rm max}-{\rm [Fe/H]}$ relation.

Now with 90 microlensed dwarf stars in hand we investigate the existence of this relation again. The bottom panel in Fig.~\ref{fig:fehamax} shows the $1/A_{\rm max}-{\rm [Fe/H]}$ plot. The underlying distribution of microlensing events should be roughly uniform in this quantity, meaning that any deviations from a straight line will show our intrinsic bias toward high magnification. As can be seen this is not the case, the stars appear to be uniformly distributed. However, the metal-rich stars are on average younger, as can be seen from the colour-coding. The upper panel shows the cumulative $1/A_{\rm max}$ distributions for stars older than 7\,Gyr and for stars younger than 7\,Gyr. A two-sided Kolmogorov-Smirnov test gives a KS-statistic of $D=0.16$ (i.e. the maximum vertical distance between the two distributions) and a p-value of 0.56, hence we cannot reject the hypothesis that they have been drawn from the same underlying population.

Doing the two-sided KS-test for only the 58 stars that were published in \cite{bensby2013} gives $D=0.31$ and a p-value of 0.11. This means that the differences have become significantly smaller now with the larger sample of 90 stars, and it might well be possible that they would disappear completely if we had an even larger sample.

\section{Metallicity distribution}
\label{sec:mdf}

\begin{figure}
\resizebox{\hsize}{!}{
\includegraphics[viewport=0 0 648 648,clip]{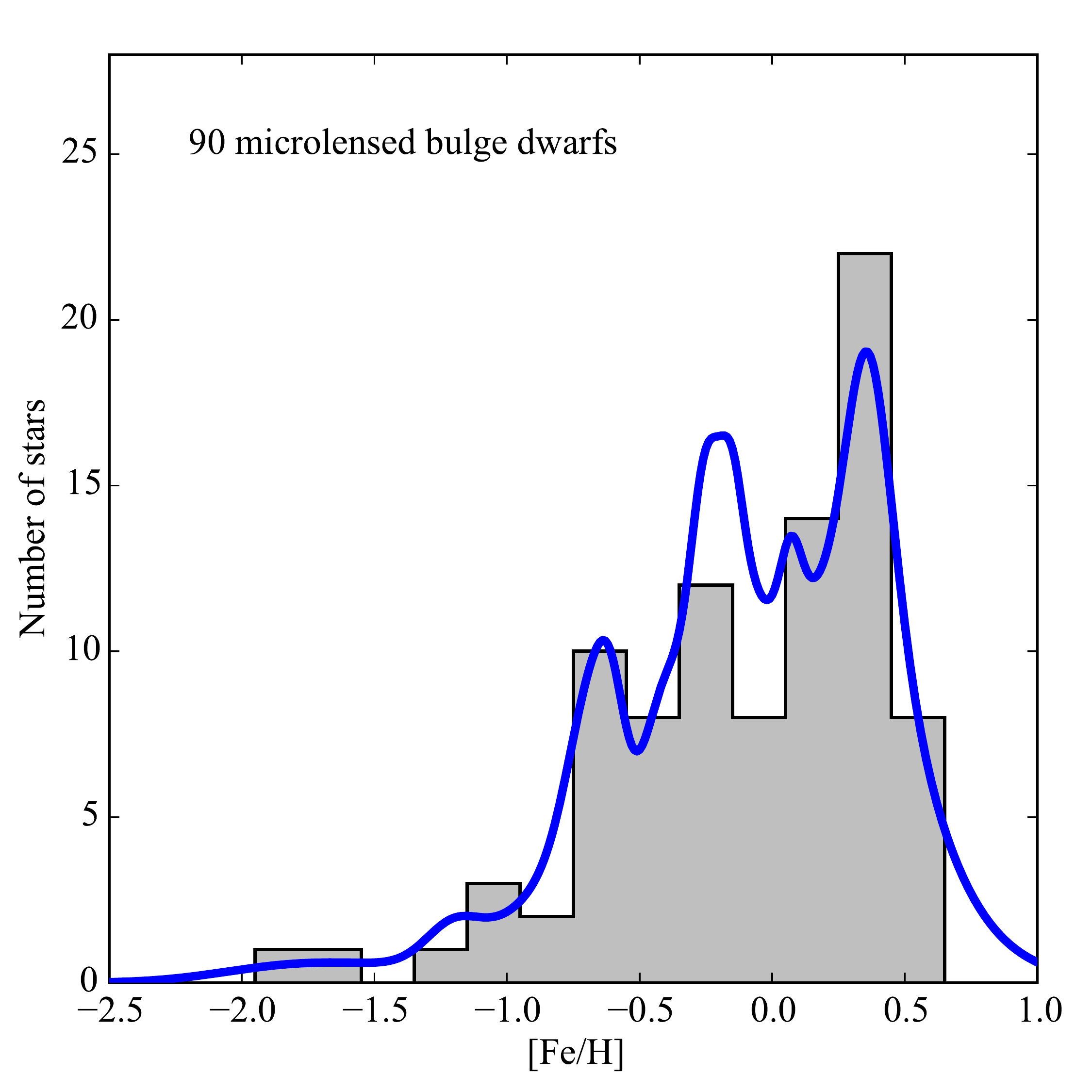}}
\caption{
The MDF for the microlensed dwarf sample, now in total 90 stars, both as a regular histogram (grey shaded) as well as a generalised histogram (curved blue line).
\label{fig:mdf}
}
\end{figure}

Figure~\ref{fig:mdf} shows the metallicity distribution of the 90 microlensed bulge dwarf and subgiant stars, both as a regular histogram and as a generalised one representing the sum of the individual Gaussians for all stars (where the central locations of the Gaussians are the measured metallicities, and the widths of the Gaussians are the estimated uncertainties). The regular histogram shows a wide distribution, ranging from $\rm [Fe/H]\approx -2$ to $+0.5$\,dex, with a prominent peak at super-solar metallicities, and then a tail towards lower metallicities. There appears to be structure in the regular histogram. As the microlensed stars have a range of uncertainties in [Fe/H] (from about 0.05\,dex to about 0.2\,dex, and a few up to 0.4\,dex, see Fig.~\ref{fig:feherrors}c), which is dependent on the quality of the spectra, it is difficult to choose a good representative value for the bin size. The appearance of the regular histogram is therefore very sensitive to the binning. By showing the MDF as a generalised histogram, where each star is represented by a Gaussian, making a MDF that is independent of the bin size, this problem is circumvented. The generalised histogram is shown as the solid blue line in Figure~\ref{fig:mdf} and a lot of structure is revealed, with several peaks in the metallicity distribution.

\begin{figure}
\centering
\resizebox{\hsize}{!}{
\includegraphics[viewport=0 0 648 648,clip]{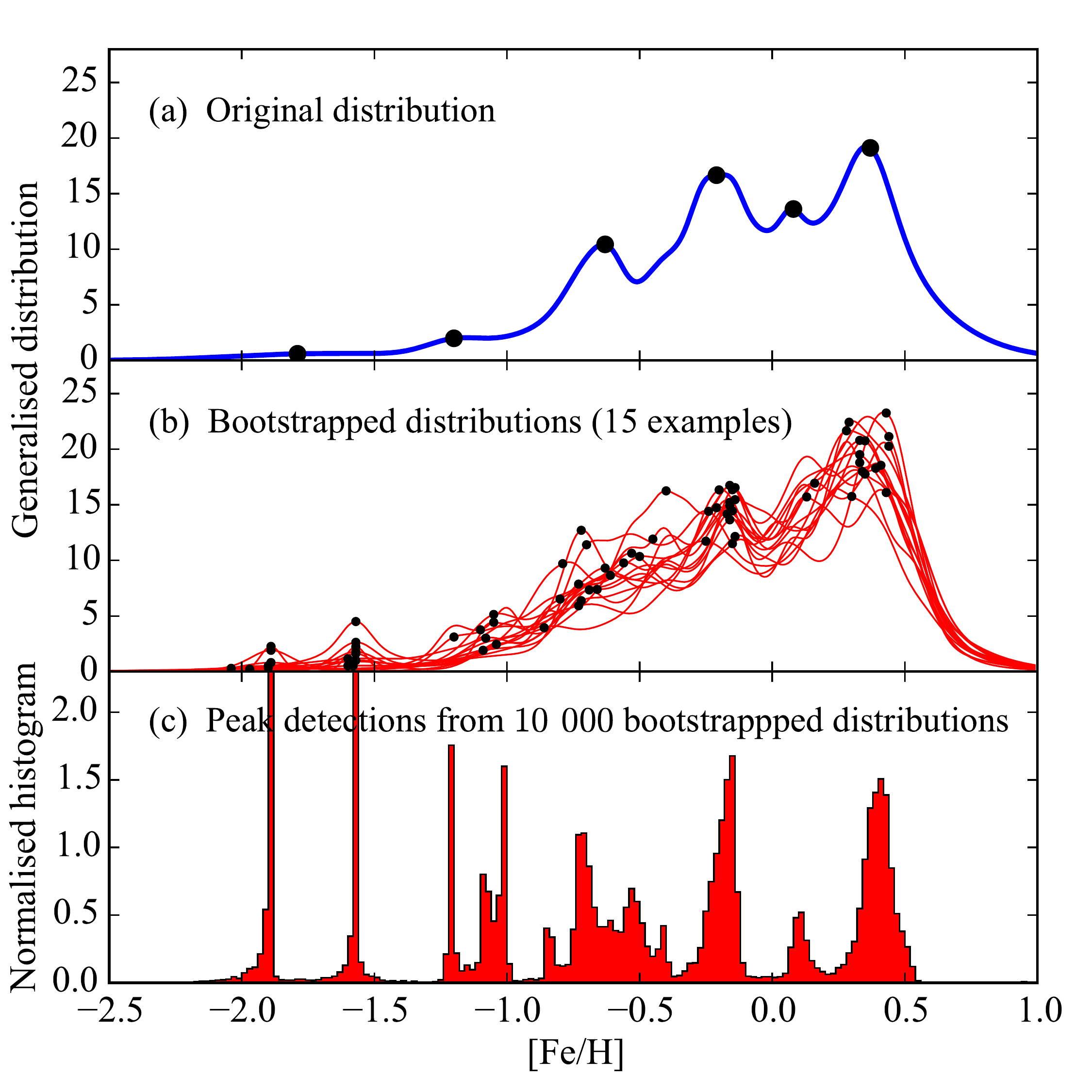}
}
\caption{
(a) The observed metallicity distribution with the positions of the detected peaks marked out. (b)  Examples for 15 bootstrapped metallicity distributions. In each of the distributions the positions of the detected peaks are marked by circles. (c) The distribution of all detected peak locations in 10\,000 bootstrapped samples.
\label{fig:mdfboot2}
}
\end{figure}

Testing whether the peaky nature of the MDF seen in Fig.~\ref{fig:mdf} is real and whether the peaks can be claimed to be statistically significant are very difficult tasks. We will start by estimating the locations of the peaks, and their uncertainties. To do that we re-sample the observed metallicity distribution 10\,000 times. In each re-sampling both the [Fe/H] values and their associated errors (Fig.~\ref{fig:feherrors}c) are re-sampled individually, so that new combinations, and duplications, are created. New generalised histograms are then created that are used to identify peaks using the Python algorithm {\tt scipy.signal.find\underline{{ }}peaks\underline{{ }}cwt} (SciPy version 0.14.0) in each of the 10\,000 bootstrapped distributions. Figure~\ref{fig:mdfboot2}b shows a few examples. For each bootstrapped distribution between 3 and 7 peaks are identified by the Python routine. Their locations are shown in the histogram in Fig.~\ref{fig:mdfboot2}c.

The distributions of the detected peak positions are used to estimate the locations of the peaks in the MDF and their uncertainties. The five metal-rich peaks are located at the following metallicities: 
\begin{align}
[{\rm Fe/H}]_{1} &= +0.41\pm0.06 \label{eq:1}\\
[{\rm Fe/H}]_{2} &= +0.12\pm0.04 \label{eq:2}\\
[{\rm Fe/H}]_{3} &= -0.20\pm0.06 \label{eq:3}\\
[{\rm Fe/H}]_{4} &= -0.63\pm0.11 \label{eq:4}\\
[{\rm Fe/H}]_{5} &= -1.09\pm0.08 \label{eq:5}
\end{align}
For each metallicity peak we give the mean value and the one-sigma dispersion around the mean. Even though we have two weak peaks at the lowest metallicities between $\rm -2<[Fe/H]<-1.5$ (see Fig.~\ref{fig:mdfboot2}c) we refrain from giving locations for them as they are based on very few stars in that metallicity range of the microlensed dwarf sample.

\begin{figure}
\centering
\resizebox{\hsize}{!}{
\includegraphics[viewport=0 0 648 325,clip]{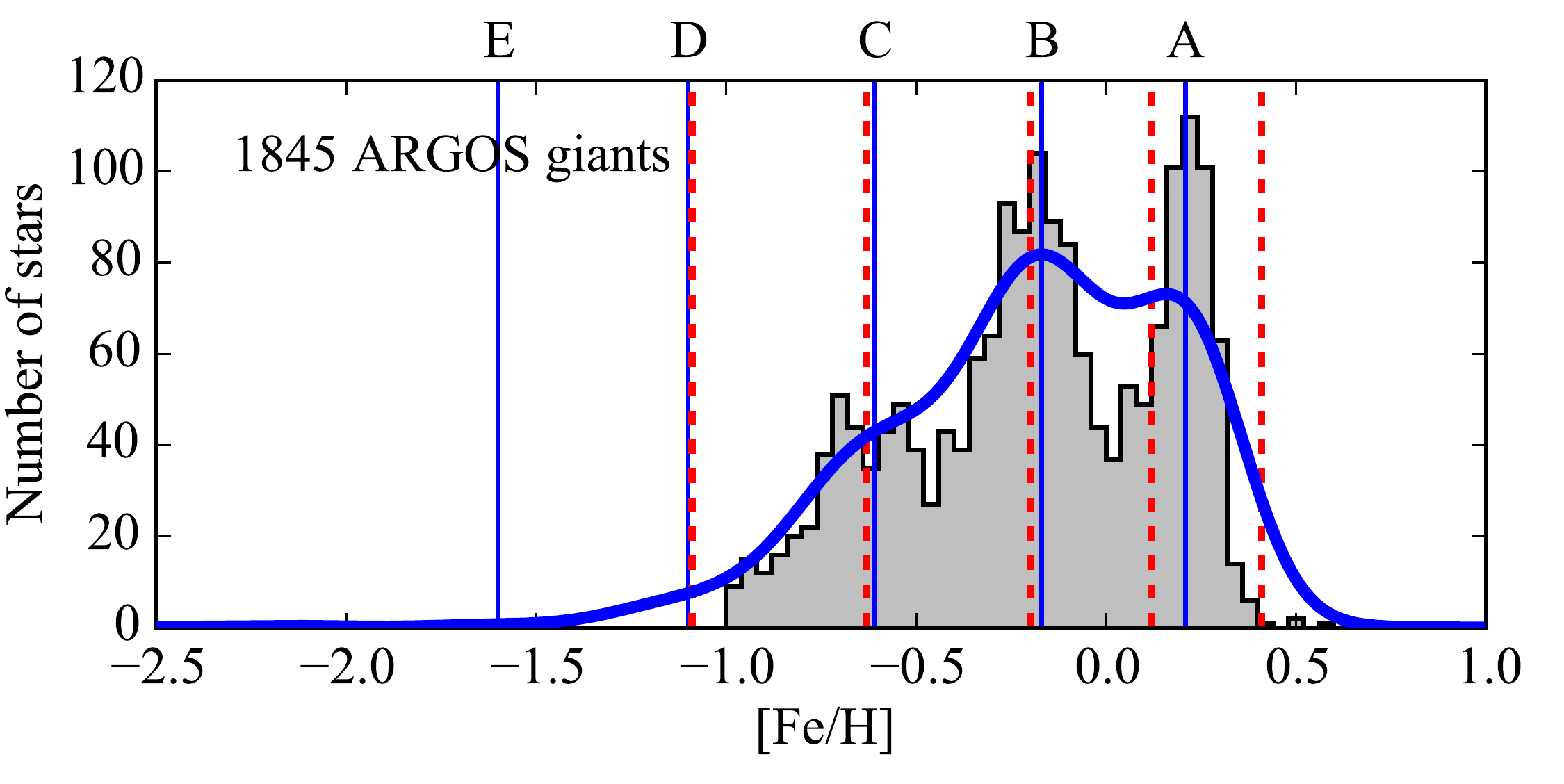}
}
\caption{
Metallicity distribution based on the 1845 giant stars in the $b=-5^{\circ}$ fields from the ARGOS survey \citep{ness2013}. The metallicity distribution is shown as both a regular histogram as well as a generalised histogram (assuming an uncertainty of 0.13\,dex for all stars, Melissa Ness privat communication). The vertical blue lines indicate the locations of the Gaussian peaks (A to E) claimed by \cite{ness2013}, and the vertical red dashed lines indicate the locations of the peaks in the MDF based on the microlensed dwarfs (Eqs.~\ref{eq:1}-\ref{eq:5}).  
\label{fig:mdfargos}
}
\end{figure}

Based  on intermediate resolution spectroscopy of red giant stars from the ARGOS survey, \cite{ness2013} identified five Gaussian components in the bulge MDF. The metallicity distribution based on the ARGOS $b=-5^{\circ}$ is reproduced in Fig.~\ref{fig:mdfargos}. The ARGOS peaks are located at $\rm [Fe/H] = +0.21\,(A),\,-0.17\,(B),\,-0.61\,(C),\,-1.10\,(D)$ and $\rm -1.60\,(E)$, as indicated by vertical blue lines in Fig.~\ref{fig:mdfargos}. \cite{ness2013} associated the peaks different stellar populations: A - a thin boxy/peanut bulge; B - a thicker boxy/peanut bulge; C - the pre-instability thick disk; D - the metal-weak thick disk; and E - the stellar halo. Three of the peaks in the microlensed dwarf MDF, given by Eqs.~(\ref{eq:3})-(\ref{eq:5}), align very well with ARGOS peaks (B, C, and D) while there appears to be an offset for the most metal-rich peak ($+0.21$\,dex for ARGOS peak A versus $+0.41$\,dex for the microlensed peak, Eq.~\ref{eq:1}). This offset is most likely because the ARGOS metallicity distribution is compressed at the metal-rich end due to the lack of calibration stars at these high metallicities (Melissa Ness, private communication). Regarding the metal-poor ARGOS peak E at $-1.60$\,dex, we do have indications of a peak in that metallicity range as well, but as mentioned above, we refrain from claiming a peak here due to the low number statistics in the microlensed sample at these metallicities. In addition to the ARGOS peaks we have one peak at $+0.12$\,dex (Eq.~\ref{eq:2}), that might be less significant as it is not as clearly detected in all cases as the other peaks seen in Fig.~\ref{fig:mdfboot2}c.

Given the agreement between our peaks and the peaks claimed by \cite{ness2013}, whose study is completely different from ours, and that is based on a twenty times larger sample, we find that it is unlikely that the detected peaks are just Poisson noise from an underlying smooth distribution.  We are somewhat uncertain about the peak at $\rm [Fe/H]=+0.12$ as it was not as clearly detected in our bootstrapped distributions as the other peaks. It was also not detected by \cite{ness2013}, but that might be due to issues with the metallicity calibration at the high metallicity end in the ARGOS data. ARGOS did also detect a peak at $-1.6$, which we actually also see in the microlensed distribution, so that one might also be real. However, due to the limited sample size of the microlensed sample, we only have two stars with metallicities below $\rm [Fe/H]=-1.5$, and we cannot claim a significant peak there.

\section{Ages}

\subsection{Age-metallicity relation}
\label{sec:agefe}

\begin{figure}
\centering
\resizebox{\hsize}{!}{
\includegraphics{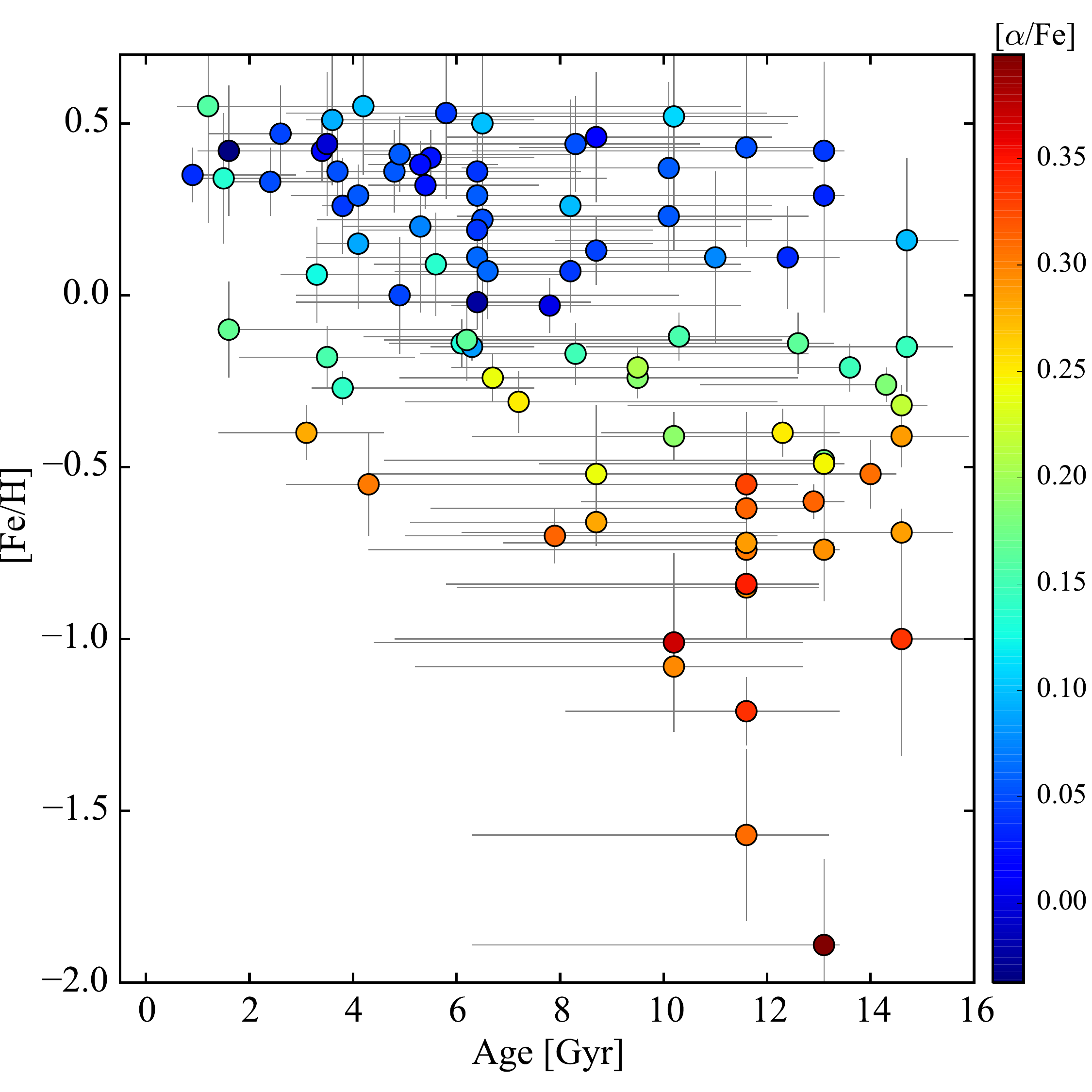}}
\caption{
Age versus [Fe/H] for the microlensed dwarf sample. The stars have been colour-coded according to their level of $\alpha$-enhancement (as shown by the colour bar on the right-hand side).
\label{fig:agefe}
}
\end{figure}

\begin{figure}
\resizebox{\hsize}{!}{
\includegraphics[viewport=0 0 642 642,clip]{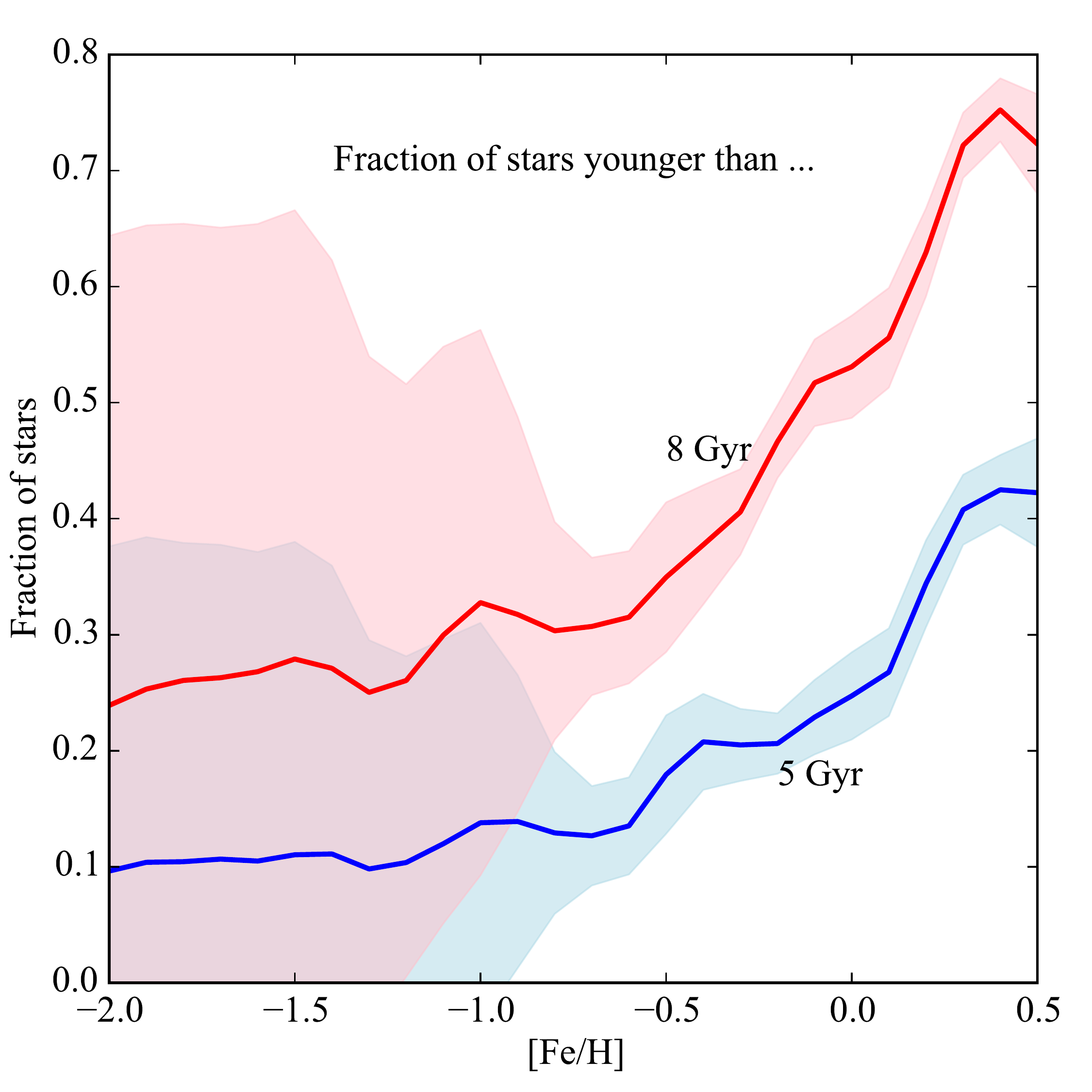}}
\caption{
The fraction of stars younger than 5\,Gyr and 8\,Gyr, respectively, at different metallicities. The thick blue and red lines represent the mean values from 10\,000 age-metallicity distributions where individual ages and metallicities have been re-sampled using the [Fe/H] uncertainties and the individual age probability distributions (see text for more details). The shaded areas show the the formal errors of the mean (1-sigma dispersion divided by $\sqrt{N}$).
\label{fig:agefrac}
}
\end{figure}

Figure~\ref{fig:agefe} shows the age-metallicity diagram for the current sample of 90 microlensed bulge dwarfs. A great majority of the metal-poor stars below $\rm [Fe/H]\approx -0.5$ have old ages around or greater than 10\,Gyr. At higher metallicities the stars span a wide range of ages, the youngest being around 1\,Gyr, and the oldest ones as old as the metal-poor stars in the sample. Especially at high metallicities there appears to be a large fraction of young to intermediate-age stars. Figure~\ref{fig:agefrac} shows that fraction of stars younger than 5\,Gyr and younger than 8\,Gyr, respectively, as a function of metallicity. The relationships are based on 10\,000 samples of 90 stars, where in each sample each star was given a metallicity randomly drawn from a normal distribution defined by the metallicity and uncertainty of that star, and an age randomly drawn from that star's age probability distribution function (see Appendix~\ref{sec:agefunctions}). For each sample of 90 stars the fraction of young stars in 0.2\,dex wide metallicity bins were calculated and the relationships shown in Fig.~\ref{fig:agefrac} represents the average values (solid lines), and formal errors of the means (shaded regions), from the 10\,000 samples. It is clear that the fractions of younger stars increase drastically with metallicity. At metallicities lower than about $\rm [Fe/H]\approx-0.7$ the formal errors of the means increase dramatically as well. This is due to the low number of stars in our sample in this regime. Hence, the fractions here should be taken lightly, they are most likely overestimated.

The high-metallicity regime ($\rm [Fe/H]>0$) shows a clear dominance of young and intermediate-age stars, the low-metallicity bin ($\rm [Fe/H]\leq-0.5$) is dominated by old stars, while in the intermediate-metallicity bin ($\rm -0.5<[Fe/H]\leq 0$) shows a wider and flatter distribution with no preferred age range. For these three metallicity bins the fraction of stars younger than 8\,Gyr are about 65\,\%, 45\,\%, and 25\,\% respectively.

Based on the un-predictability and random nature of microlensing events, one naively believes that the microlensed bulge dwarf sample should be a completely un-biased sample. However, as demonstrated in \cite{bensby2013}, the adopted observing strategy seems to favour young and metal-rich stars. This has to do with the fact that in order to trigger spectroscopic follow-up observations, we required that the estimated peak magnitude of the source star should be at least $I\approx 15$, which lead to a slight bias towards metal-rich and young stars as these are intrinsically brighter than old and metal-poor stars. In \cite{bensby2013} it was concluded that the fraction of young and metal-rich could be over-estimated by up to 50\,\%. Conservative estimates of the fraction of young stars in the three metallicity bins are therefore 35\,\%, 20\,\%, and 9\,\%, respectively. Still, even when including this bias, more than one third of the metal-rich bulge population is younger than about 8\,Gyr. 

\cite{clarkson2011} found that at most 3.4\,\% of the bulge population could be genuinely young, i.e. less than 5\,Gyr. For the microlensed sample we find that 26\,\% of the total sample is younger than 5\,Gyr. Reducing this percentage due to the observational bias that favours young and metal-rich stars, we estimate that the fraction of stars younger than 5\,Gyr in the microlensed sample could be around 15\,\%. Hence, there is still a factor of 4 to 5 discrepancy between our estimate and the \cite{clarkson2011} estimate of so called genuinely young stars in the bulge.

\begin{figure}
\resizebox{\hsize}{!}{
\includegraphics[viewport=0 0 520 634,clip]{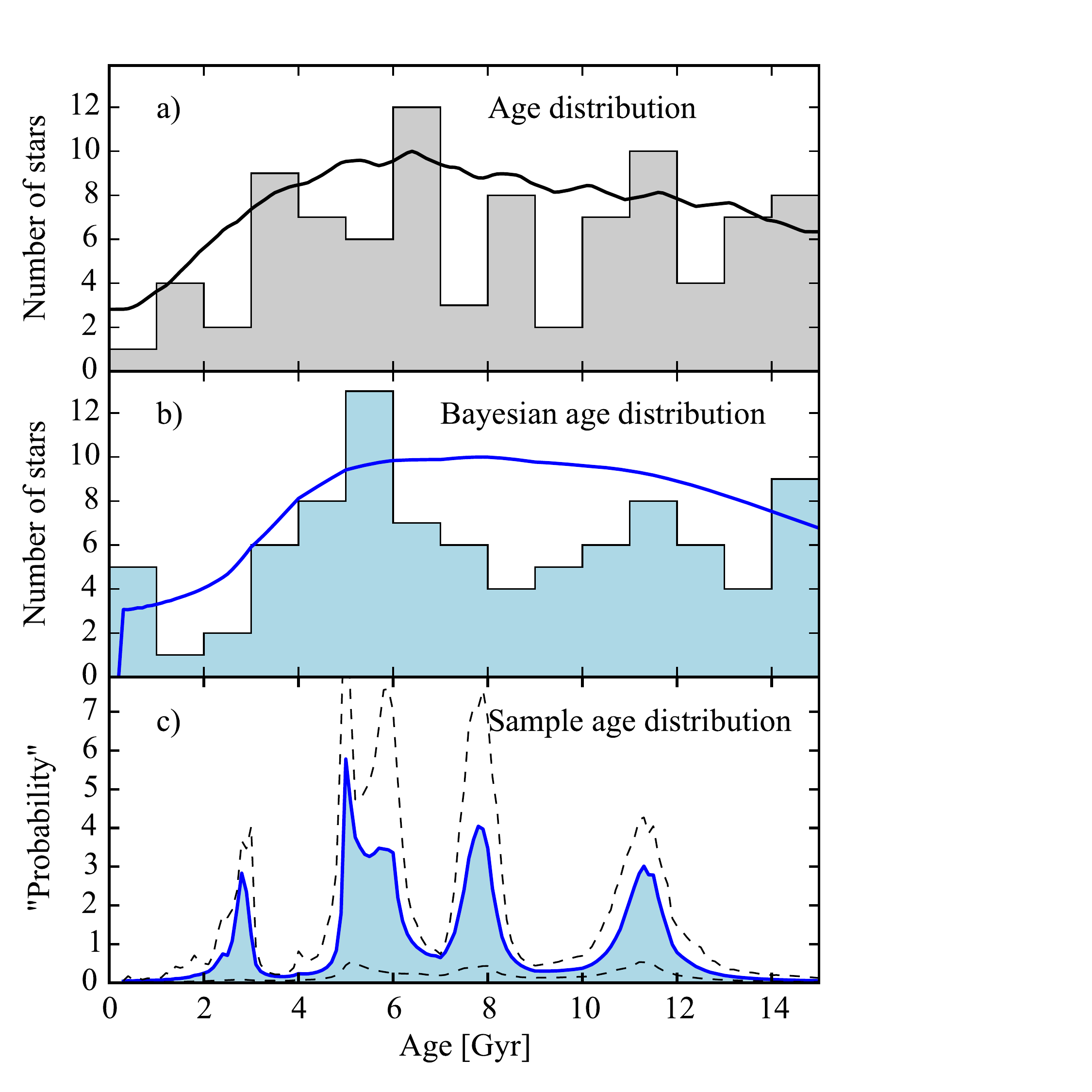}}
\caption{
a) A histogram of the estimated ages. The solid line shows the sum of all individual age probability distribution functions.
b) A histogram based on the ages from the Bayesian method, and the lighter shaded under the solid blue line shows the sum of individual age G functions.
c) The sample age distribution determined using the method described in Sect.~\ref{sec:sampleage}. The peaks show episodes of significant star formation.
\label{fig:ages}
}
\end{figure}
\begin{figure}
\centering
\resizebox{\hsize}{!}{
\includegraphics[viewport= 0 7 520 642,clip]{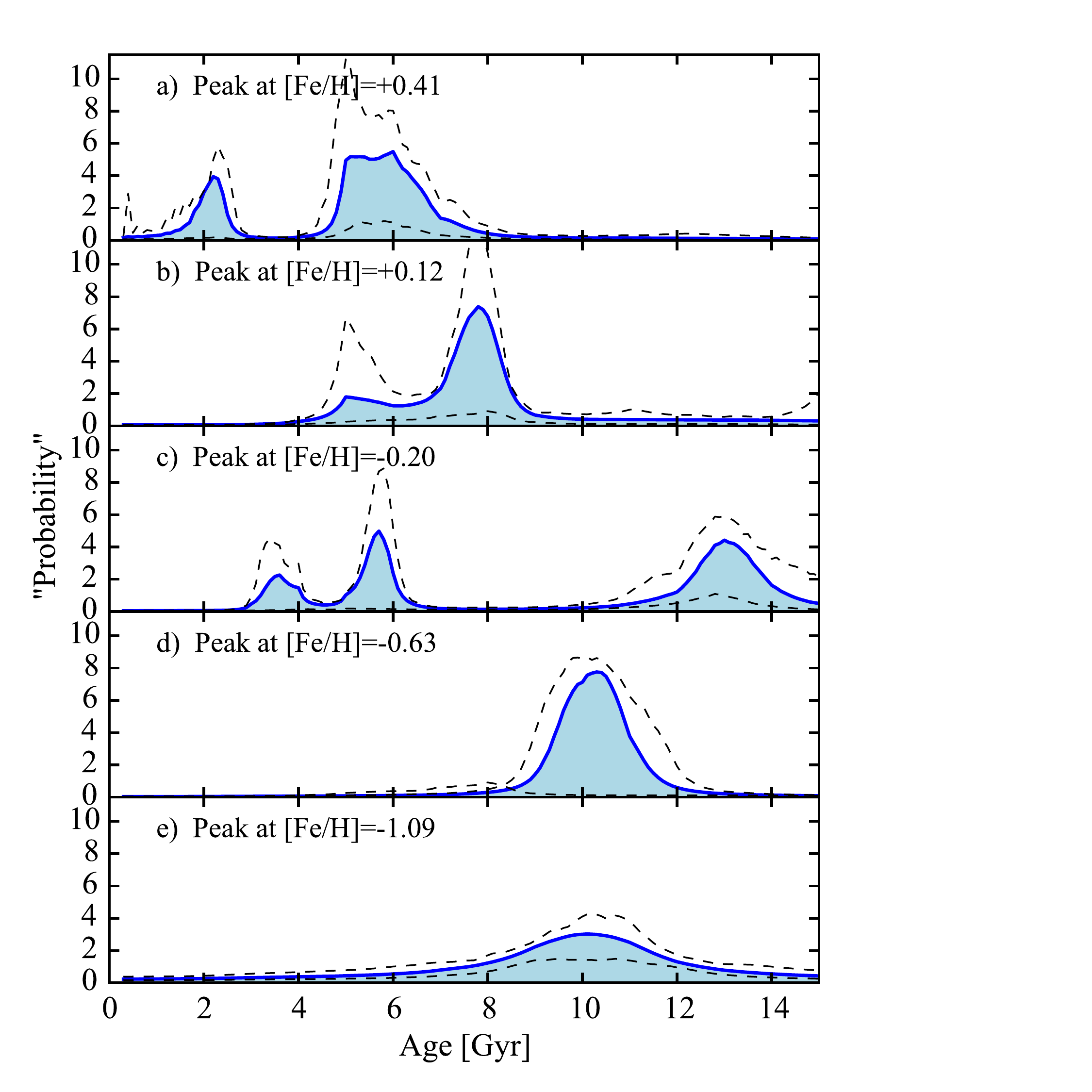}}
\caption{
Star formation episodes for the five peaks in the MDF. For each metallicity peak, stars within 2\,$\sigma$ of the mean metallicity given by Eqs.~(1)-(5) were selected. The metallicities are indicated in the top left corner of each subplot.
\label{fig:ages2}
}
\end{figure}

\subsection{Sample age distribution}
\label{sec:popages}

Figure~\ref{fig:ages}a shows the age distribution for the full sample of 90 microlensed bulge dwarf stars. The sample spans a wide range of ages with no clear dominant age. The youngest stars are around 1-2 Gyr and then there is flat distribution up to the oldest stars around 12 Gyr. There might be some structure in the age histogram, with two or maybe three groups of stars with different ages, but, again, as any structure in a regular histogram is highly dependent on the binning of the data no structure can be claimed. Also shown in Fig.~\ref{fig:ages}a is the sum of the individual age probability distribution functions (see Appendix~\ref{sec:agefunctions}) as a solid line, but this reveals no further information, except a very wide age distribution.

Figure~\ref{fig:ages}b shows again the age distribution, but using the Bayesian ages determined in Sect.~\ref{sec:newages}, with the sum of the individual G functions (see Appendix~\ref{sec:agefunctions}) shown as the solid blue line. Again, this reveals no further information except that the age distribution is very wide and spans essentially all possible ages. The main point here is to show that the distribution of the Bayesian ages is very similar to the distribution of our adopted ages, as the Bayesian ages will be used below in an attempt to reveal if there are any major episodes of star formation in the bulge.

The star formation history of the bulge is investigated using the Bayesian method described in \cite{jorgensen2005phdt,jorgensen2005conf}. Instead of individual ages, this code uses the available information from all individual G functions in order to estimate the likely periods of star formation in the history of the Galactic bulge. Figure~\ref{fig:ages}c shows the results, and several peaks can be seen; at about 11, 8, 6, and 3\,Gyr ago. These peaks, through estimating the likely ages of the distribution as a whole, can be interpreted as representative of when there were significant episodes of star formation. The sharp spike at about 5\,Gyr is most likely due to a one star with narrow age uncertainties, and cannot be claimed as significant.

Figure~\ref{fig:ages2} shows the same analysis but for different sub-sets of stars that have metallicities within 2\,$\sigma$ of the metallicity peaks given by Eqs.~(\ref{eq:1})-(\ref{eq:5}). The two metal-poor peaks show age peaks around 11\,Gyr (Figs.~\ref{fig:ages2}d and e), the metallicity peak at $+0.41$\,dex shows no clear age peak but rather an extended bi-modal distribution in the range 2 to 8\,Gyr (Fig.~\ref{fig:ages2}a).  The metallicity peak at $+0.12$\,dex is interesting as it shows a clear age peak at 8\,Gyr (Fig.~\ref{fig:ages2}b). The ages seen for the metallicity peak at $-0.20$\,dex are hard to interpret, one very old and some much younger (Fig.~\ref{fig:ages2}c). An interpretation of these peaks will be discussed in Sect.~\ref{sec:whatisthebulge}.

\subsection{Metallicity distributions for different age bins}

\begin{figure}
\resizebox{\hsize}{!}{
\includegraphics[viewport= 0 0 648 648,clip]{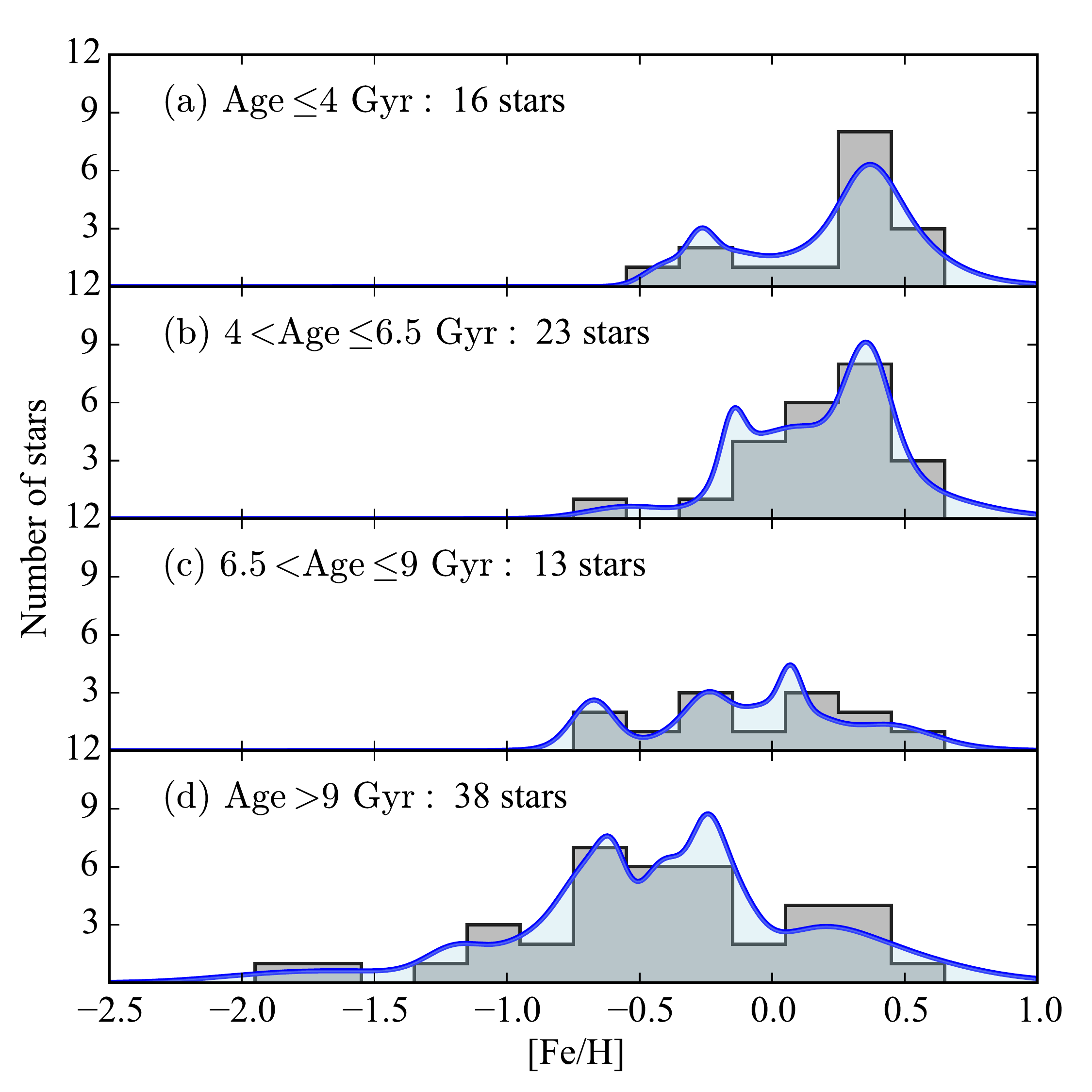}}
\caption{
Metallicity distribution for different age bins (as indicated in the plots).
\label{fig:mdfage}
}
\end{figure}

The previous sections have shown that the bulge MDF has structure and that it most likely is composed of several metallicity peaks (Fig.~\ref{fig:mdfboot2} and Eqs.~\ref{eq:1}-\ref{eq:5}), that appear to have different age distributions (Fig.~\ref{fig:ages2}). At the same time, when considering the whole sample, four different age peaks are revealed (bottom panel of Fig.~\ref{fig:ages}). Are the metallicity distributions for these age peaks consistent with what we have seen so far?
Figure~\ref{fig:mdfage} shows the metallicity distributions for stars that have individual ages centred around the four age peaks in Fig.~\ref{fig:ages}. A metal-rich distribution is dominating for the two youngest age bins, while the two oldest age bins show a wide range of metallicities. The same information could have been read from the age-metallicity diagram in Fig.~\ref{fig:agefe}, but now the age bins are centred on the peaks found in Fig.~\ref{fig:ages}.

\begin{figure*}
\resizebox{\hsize}{!}{
\includegraphics[viewport= 0 0 642 642,clip]{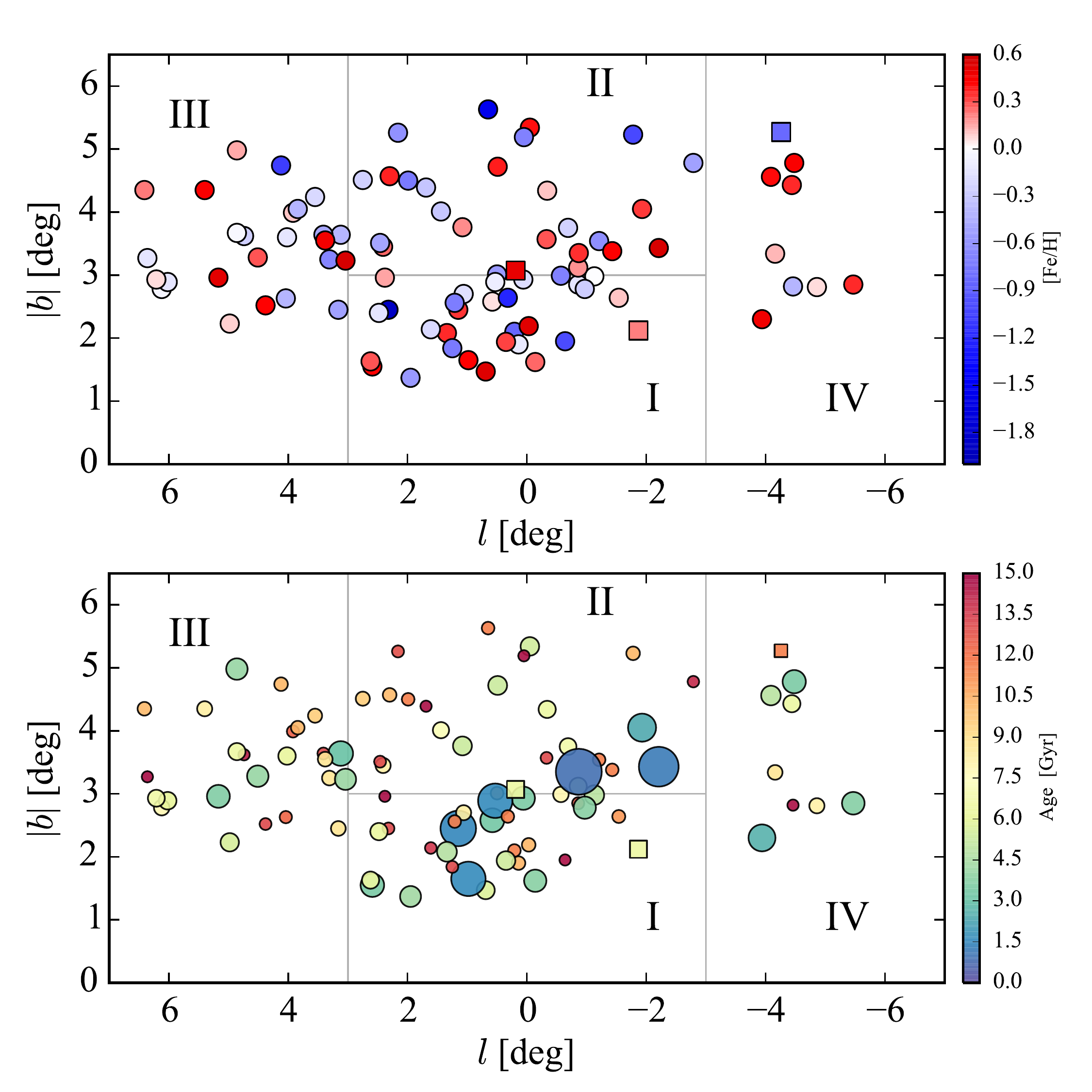}
\includegraphics[viewport= -30 0 560 642,clip]{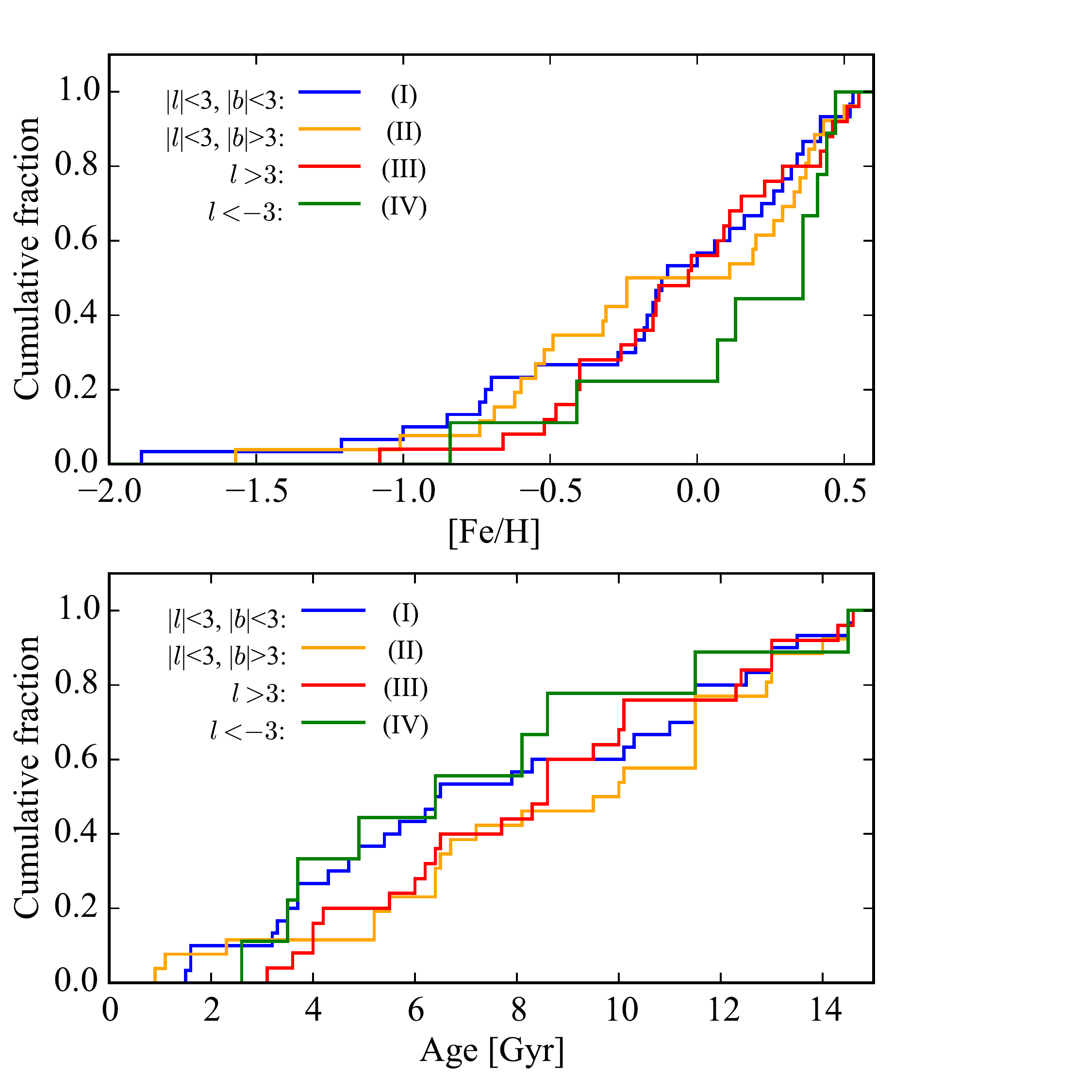}}
\caption{
{\sl Left-hand side:} Two plots of Galactic latitude versus Galactic longitude. In the top plot the markers have been colour-coded based on the metallicities of the stars, and in the bottom plot based on the ages of the stars. Stars at positive latitudes are marked by squares and stars at negative latitudes by circles. In each plot four regions have been marked out. The age and metallicity distributions for these regions are shown in the plots on the right-hand side.
{\sl Right-hand side:} The cumulative metallicity and age distributions for the stars located within regions I-IV as indicated in the plots and illustrated in the plots on the left-hand side.
\label{fig:mdflat}
}
\end{figure*}

\subsection{Summary of age results}

In summary, the microlensed dwarf and subgiant sample shows that the Galactic bulge contains a significant fraction of young and intermediate age stars. In general, the stars at low metallicities are old, while the metal-rich stars show a wide range of ages. At least 18\,\% of the stars, at all metallicities, appear to be younger than 5\,Gyr. At super-solar metallicities more than 35\,\% are younger than 8\,Gyr. In addition we find that the bulge most likely has had four episodes of significant star formation. A first estimation is that they happened 3, 6, 8, and 11 Gyr ago. Whether these episodes can be associated with major episodes in the other main Galactic stellar populations needs to be investigated, and is further discussed in Sect.~\ref{sec:whatisthebulge}.

\section{Variation of ages and metallicities with position}

Based on N-body and star-forming simulations \cite{debattista2016astroph} found that the fraction of young stars were different at positive and negative longitudes, depending on whether the far- or near-side of the Bulge was considered. The near-side showed an excess of old stars at negative longitudes, while the far-side showed an excess of old stars at positive longitudes. This difference is attributed to the population of the X-shaped bar, that is believed to contain a higher fraction of young- to intermediate-age stars. From our vantage point the near-side bar is located at positive longitudes, and the far-side bar at negative longitudes. The question is if this longitudal age-asymmetry is present in the microlensed sample?

The probabilities of microlensing events towards the Galactic bulge favours events where the source stars are on the far-side of the Galactic centre (see Sect.~\ref{sec:inthebulge}. Hence, our sample should contain a higher relative fraction of stars in Galactic bar at positive longitudes (where the bar is mainly on this side of the Galactic centre), than at negative longitudes. If the bar has a higher fraction of younger stars, we should then see a higher fraction of younger stars at negative longitudes than at positive longitudes. Then it might also be likely to expect the metallicities to be on average higher at negative longitudes?

The panels on the left-hand side of Fig.~\ref{fig:mdflat} show the locations of the sample in Galactic coordinates, with the stars being colour-coded based on either metallicity (top panel) or age (lower panel). The plots also indicate four different regions (I-IV) for which the cumulative metallicity and age distributions are shown in the panels on the right-hand side of Fig.~\ref{fig:mdflat}.

The cumulative age and metallicity distributions shown in Fig.~\ref{fig:mdflat} are different at negative and positive longitudes. The stars at positive longitude stars (region III, red lines) are metal-poorer and older than the stars at negative longitudes (region IV, green lines). As most stars in the sample are likely to be located on the far-side of the Galactic centre, this means that the ones at negative longitudes (that are then in the bar) do have higher metallicities and younger ages than those at positive longitudes (that are beyond the bar). This means that the apparent longitudal age- and metallicity asymmetry seen in the microlensed sample can be attributed to the Galactic bar.

The range in latitude is too short to reveal if there are any vertical abundance gradients, but we do see that the sub-sample closest to the plane (region I, blue line) is somewhat younger that the sample farthest from the plane (region II, orange line). This is in agreement with the finding by \cite{ness2014} who showed that the existence of young stars in the bulge is a natural outcome of its internal evolution, and the most likely place to find them is closer to the plane.

\begin{figure}
\centering
\resizebox{\hsize}{!}{
\includegraphics[viewport= 0 0 648 540,clip]{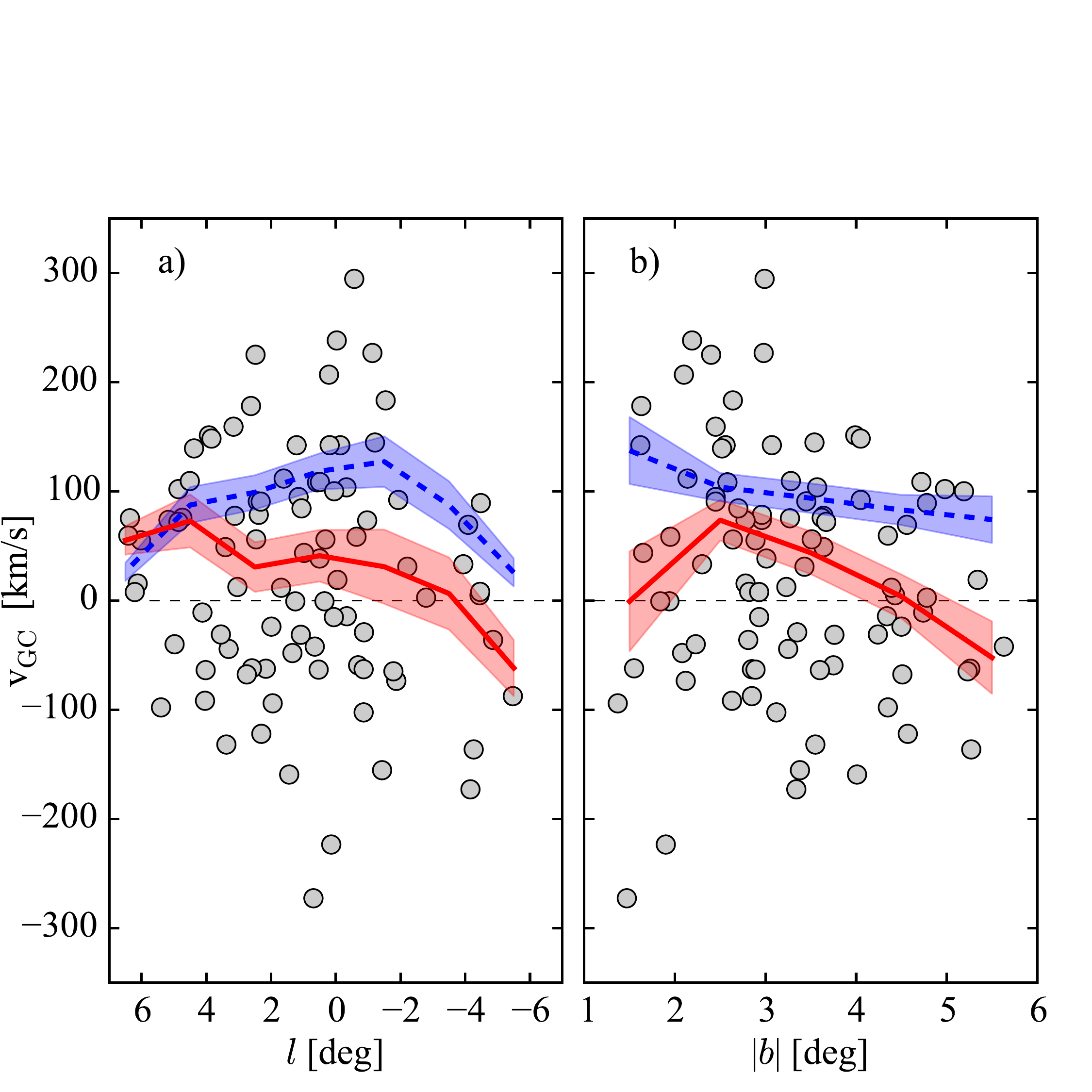}}
\caption{
Galactocentric radial velocity versus Galactic longitude and absolute latitude. The dashed blue lines represent the running velocity dispersion and the solid red lines the running median radial velocity (calculated in bin widths of 2$^{\circ}$ in $l$ and 1$^{\circ}$ in $b$). The shaded regions represent the errors in velocity (calculated as $\sigma/\sqrt{N-1}$) and errors in dispersion (calculated as $\sigma/\sqrt{2\,N}$).
\label{fig:rvel}
}
\end{figure}

\begin{figure*}
\centering
\resizebox{0.7\hsize}{!}{
\includegraphics[viewport= 0 57 648 310,clip]{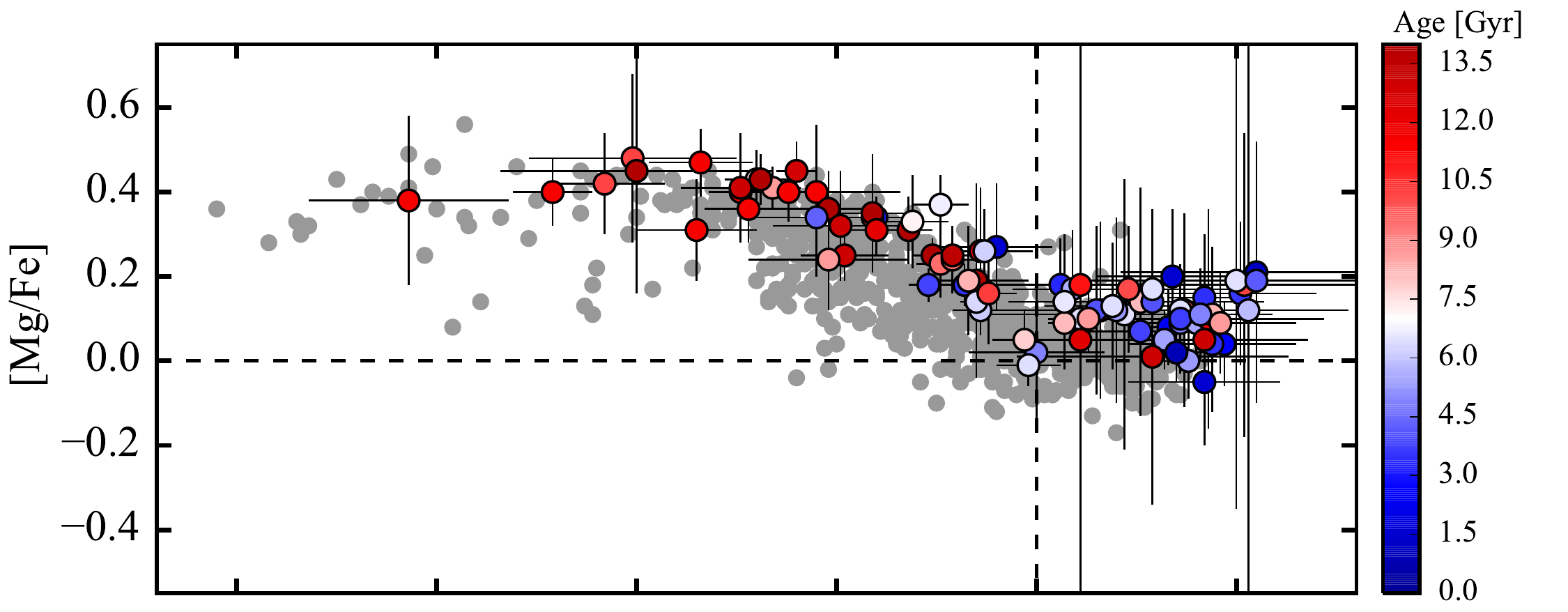}}
\resizebox{0.7\hsize}{!}{
\includegraphics[viewport= 0 57 648 293,clip]{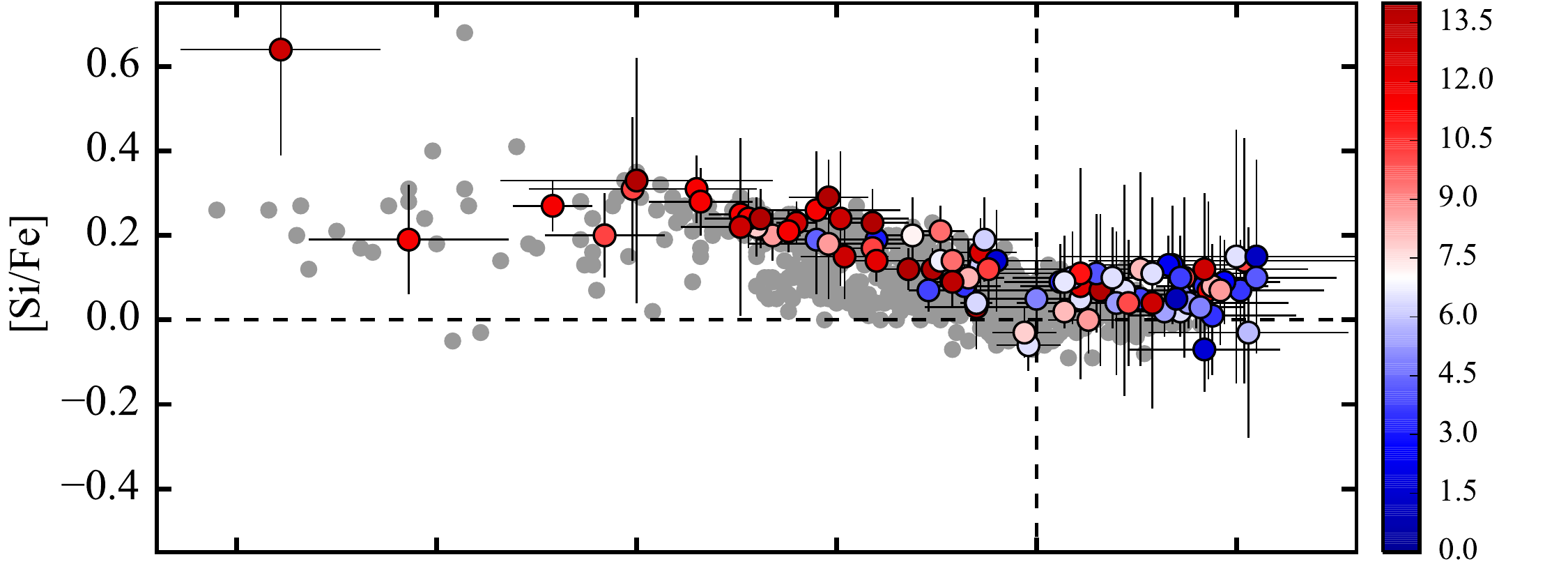}}
\resizebox{0.7\hsize}{!}{
\includegraphics[viewport= 0 57 648 293,clip]{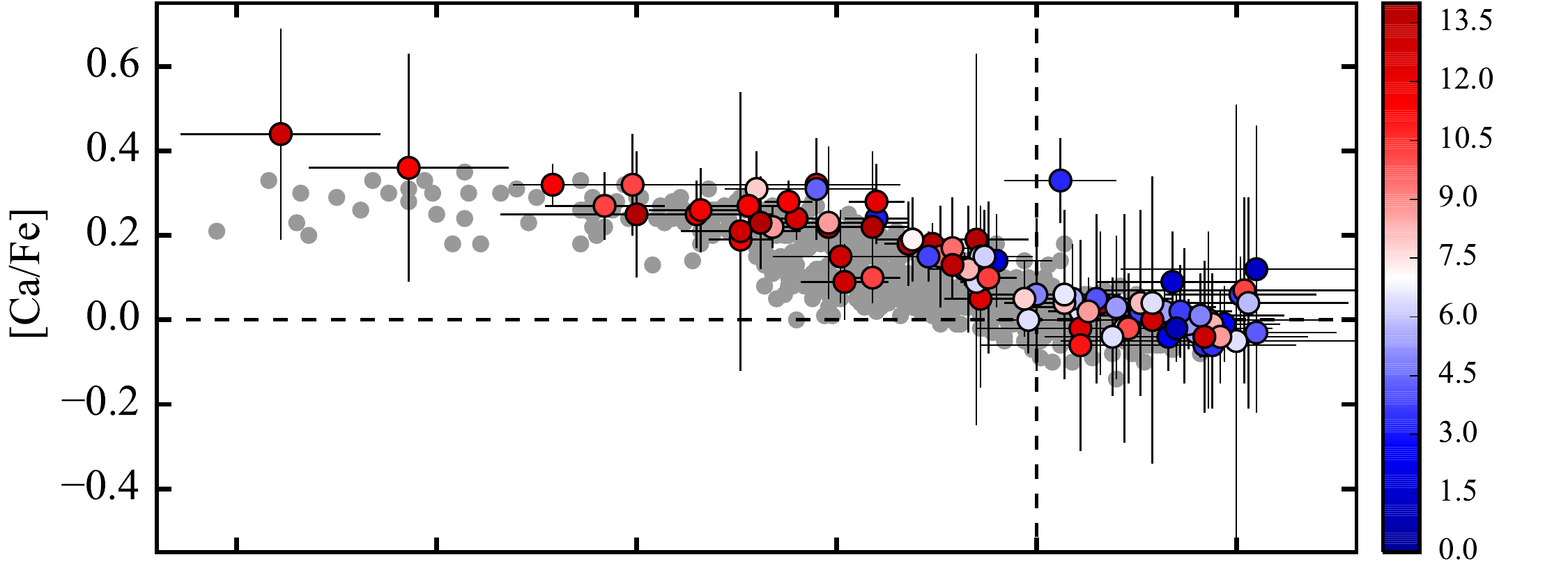}}
\resizebox{0.7\hsize}{!}{
\includegraphics[viewport= 0 0 648 293,clip]{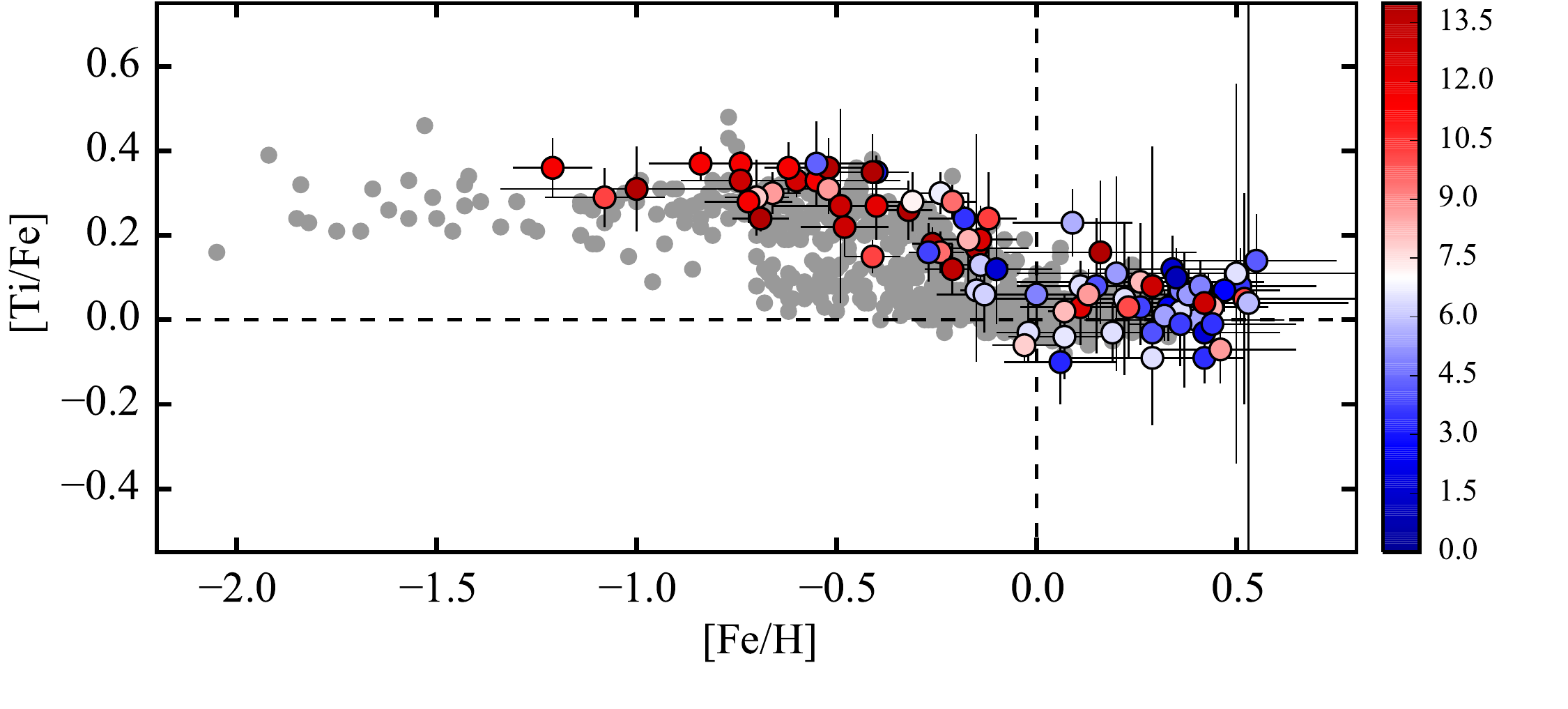}}
\caption{
Abundance trends with Fe as reference element for the microlensed dwarf sample.
The stars have been colour coded according to their ages (colour bar on the right-hand side). Grey circles in the background are the Solar neighbourhood dwarf stars from \cite{bensby2014}. 
\label{fig:abundances}
}
\end{figure*}

\section{Radial velocities}
\label{sec:rv}

Heliocentric radial velocities were determined from measurements of the doppler shift of 10-15 \ion{Fe}{i} lines that have accurate laboratory wavelengths from \cite{nave1994}. The lines are distributed over the whole wavelength range of the spectra, and the precisions of the estimated radial velocities are $0.1-0.2\,\kms$, which for the purposes of this study is negligible.   
Figures~\ref{fig:rvel}a and b shows the galactocentric radial velocities\footnote{We correct our measured heliocentric radial velocities to the Galactic centre using the relation $v_{\rm GC} = v_{\rm r, HC} + 232 \sin(l) \cos(b) + 9 \cos(l) \cos(b) + 7 \sin(b)$ (obtained from \url{http://leda.univ-lyon1.fr}) which takes the motion of the relative to the local standard of rest (LSR), and the motion of the LSR relative to the Galactic centre, into account.} as a function of Galactic longitude and latitude, respectively. Each plot contains lines that represent the running median values (red lines) and the running velocity dispersions (dashed blue lines). 

The velocity dispersion (blue dashed line) appears to decrease slowly with increasing distance from the plane (|$b$|), which is not very clear from the figure. However, by splitting the sample in two latitude bins (closer or farther than 3 degrees from the plane) the velocity dispersion is $120\kms$ for the low-latitude sample and $89\kms$ for the high-latitude sample, i.e. significantly higher closer to the plane. The median velocity (red line) tend to show a slight decrease, $45\kms$ and $9\kms$ for the low- and high latitude samples, respectively. There are essentially no differences in these trends or values whether we consider the whole sample, or if the sample is split into a low-metallicity and a high-metallicity sample (not shown in the figure).

The velocity dispersion with Galactic longitude appears to show its largest values closer to the Galactic centre ($l=0$), with slightly decreasing values for greater $l$. Again, this does not come out very clearly  in the plots, but by plotting in longitude bins (closer or farther than 2 degrees from the Galactic centre direction) we find that the velocity dispersion is $120\kms$ for $-2<l<2$, $90\kms$ for $l>2$, and $80\kms$ for $l<-2$, i.e. significantly lower in the outer samples than for the sample closer to the Galactic centre. The median velocity is positive at positive longitudes and negative at negative longitudes ($+36\kms$ for $l>2$, $+28\kms$ for $-2<l<2$, and $-18\kms$ for $l>2$). Again, there appear to be no significant differences whether we split the sample based on the metallicities, or not (not shown in figure). 

\cite{williams2016} found that the velocity dispersion for stars with super-solar metallicities drops steeply with latitude, from about 100\,$\kms$ at $b=-4^{\circ}$ to about 50\,$\kms$ at $-10^{\circ}$, while the stars with sub-solar metallicities only show a shallow decline. Our data span a shorter range in latitude and are located slightly closer to the plane. In Fig.~\ref{fig:rvel} the velocity dispersion has been indicated, and we see indeed that it is lower for the part of the sample that is farther from the plane. However, we do see about the same decrease for metal-rich as well as for metal-poor stars. Our sample is significantly smaller than the more than 2000 giants in \cite{williams2016}.

We have also checked the velocities and velocity dispersions versus metallicity and age, respectively. We see no difference in velocity or velocity dispersion between young and old stars. Also, there are no trends of the median velocity nor velocity dispersion with age. The stars younger than 7\,Gyr have a median velocity of 20\,$\kms$ with a dispersion of 107\,$\kms$, while the stars older than 7\,Gyr have a median velocity of 32\,$\kms $ with a dispersion of 104\,$\kms$. 

In all, and even though the sample is much smaller, the kinematics of the microlensed bulge dwarfs are consistent with what has been found from other surveys such as BRAVA \citep{kunder2012} and ARGOS \citep{ness2013b} that each contain several thousands of stars.

\begin{figure*}
\centering
\resizebox{0.7\hsize}{!}{
\includegraphics[viewport= 0 57 648 310,clip]{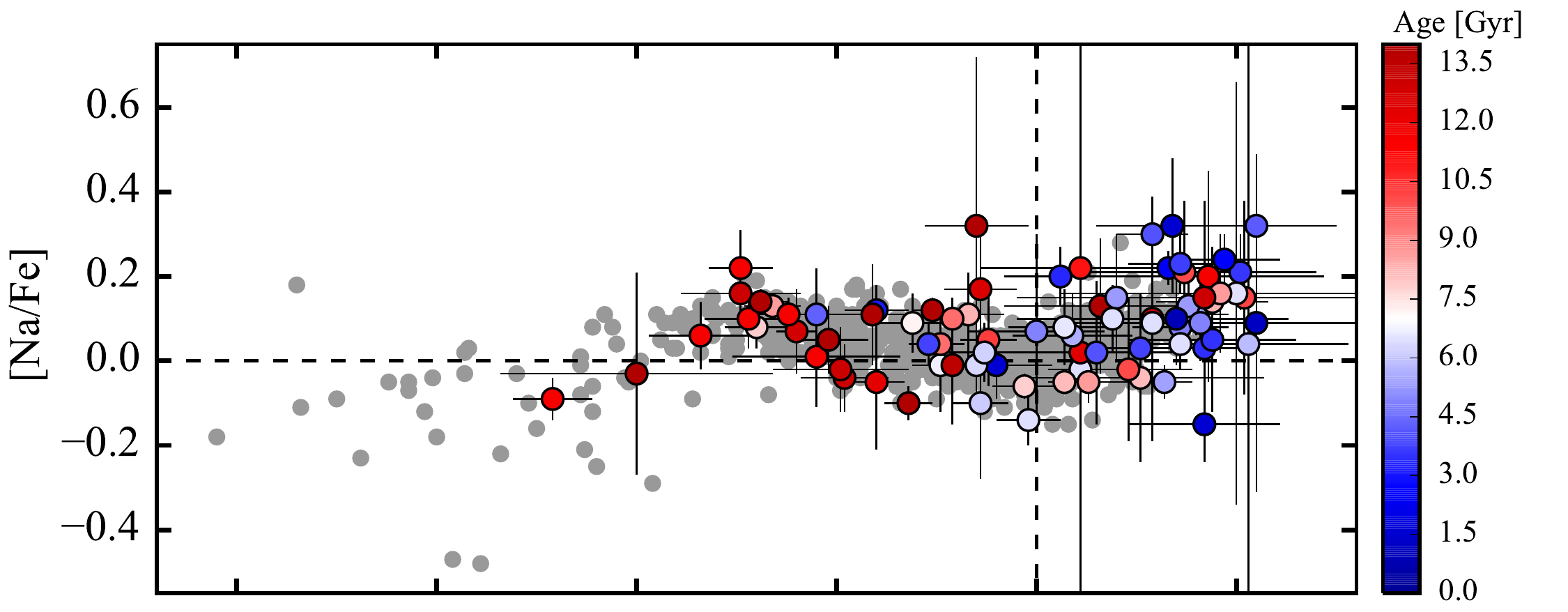}}
\resizebox{0.7\hsize}{!}{
\includegraphics[viewport= 0 0 648 293,clip]{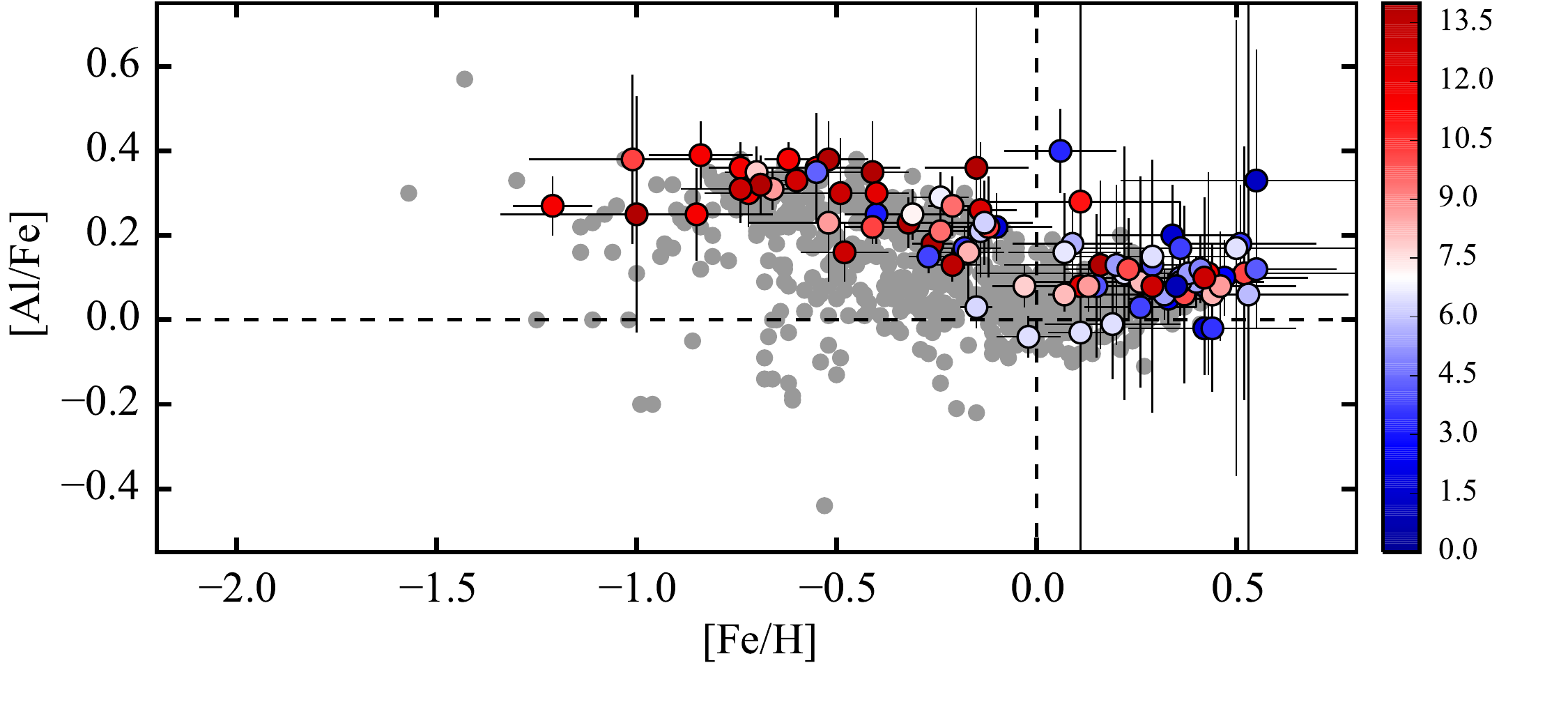}}
\caption{
Abundance trends with Fe as reference element for the microlensed dwarf sample.
The stars have been colour coded according to their ages (colour bar on the right-hand side). Grey circles in the background are the Solar neighbourhood dwarf stars from \cite{bensby2014}. 
\label{fig:abundances2}
}
\end{figure*}
\begin{figure*}
\centering
\resizebox{0.7\hsize}{!}{
\includegraphics[viewport= 0 57 648 310,clip]{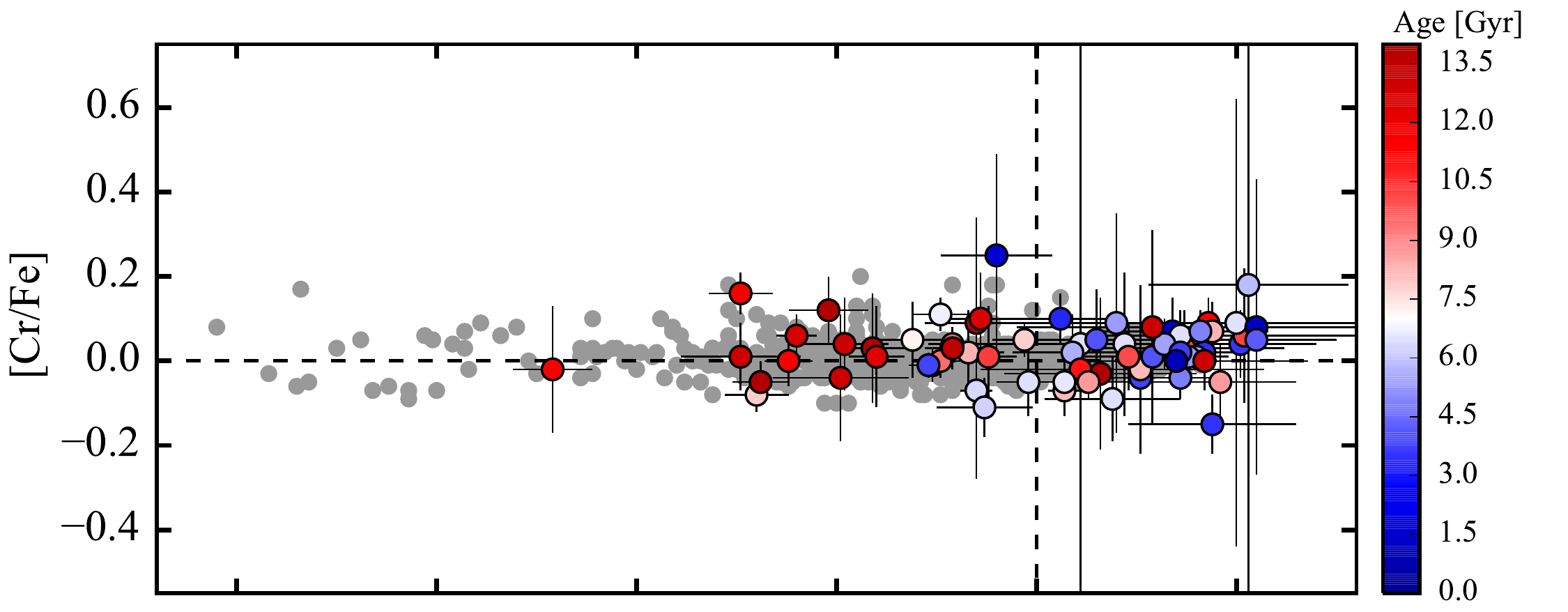}}
\resizebox{0.7\hsize}{!}{
\includegraphics[viewport= 0 57 648 293,clip]{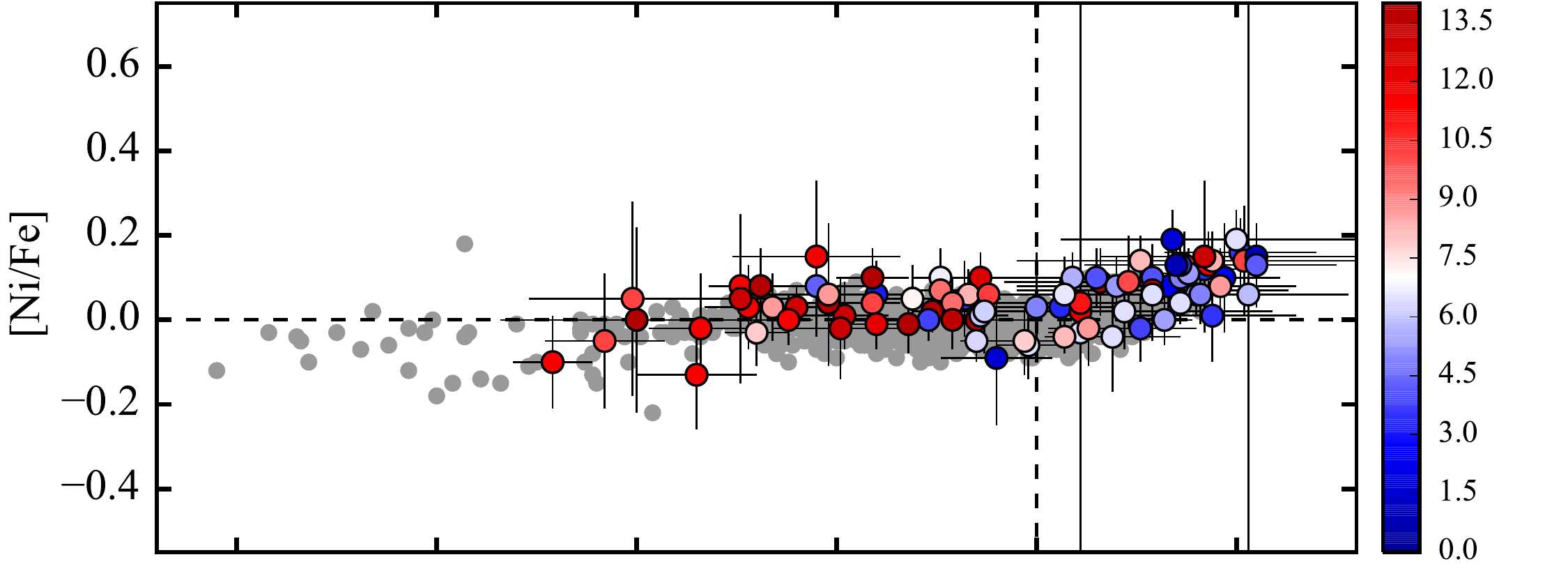}}
\resizebox{0.7\hsize}{!}{
\includegraphics[viewport= 0 0 648 293,clip]{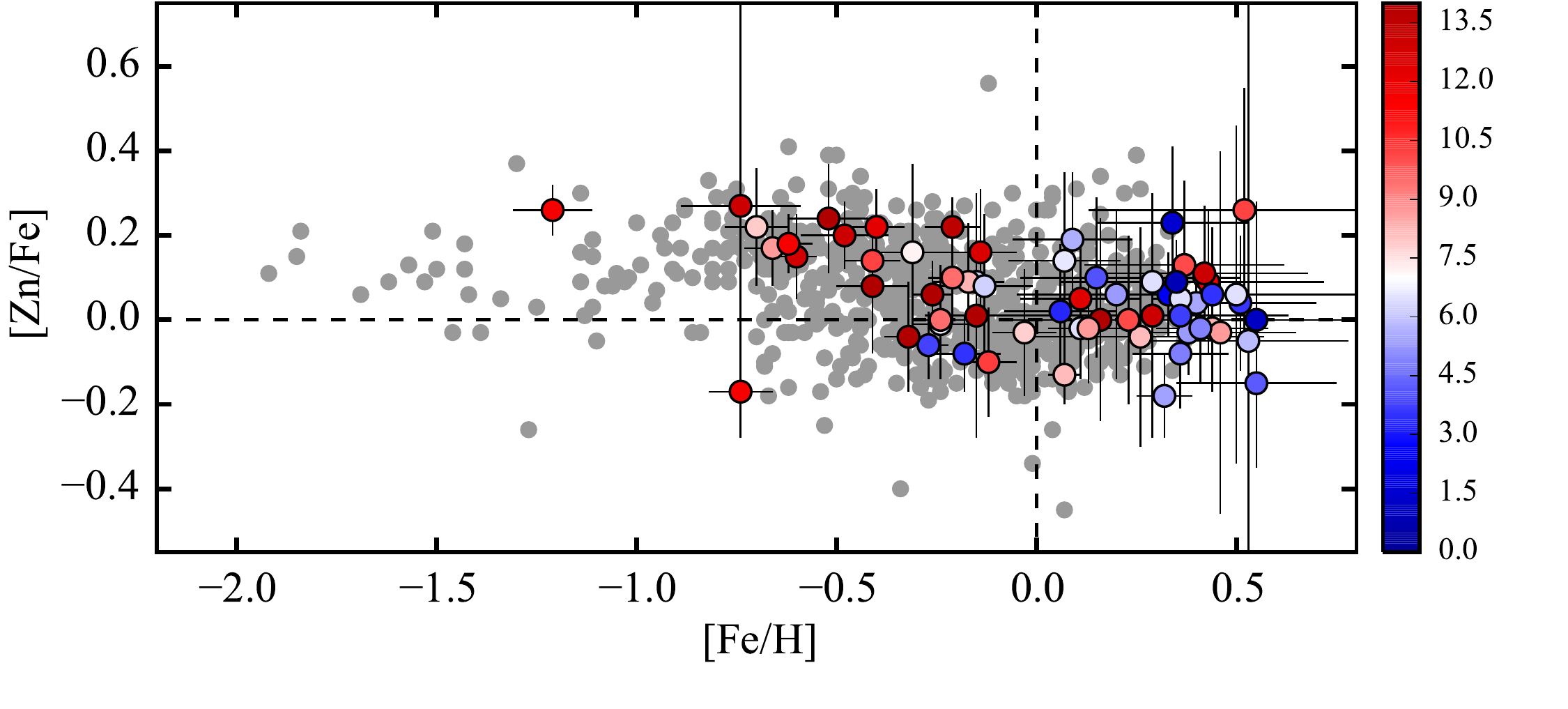}}
\caption{
Abundance trends with Fe as reference element for the microlensed dwarf sample.
The stars have been colour coded according to their ages (colour bar on the right-hand side). Grey circles in the background are the Solar neighbourhood dwarf stars from \cite{bensby2014}.
\label{fig:abundances3}
}
\end{figure*}
\begin{figure*}
\centering
\resizebox{0.7\hsize}{!}{
\includegraphics[viewport= 0 57 648 310,clip]{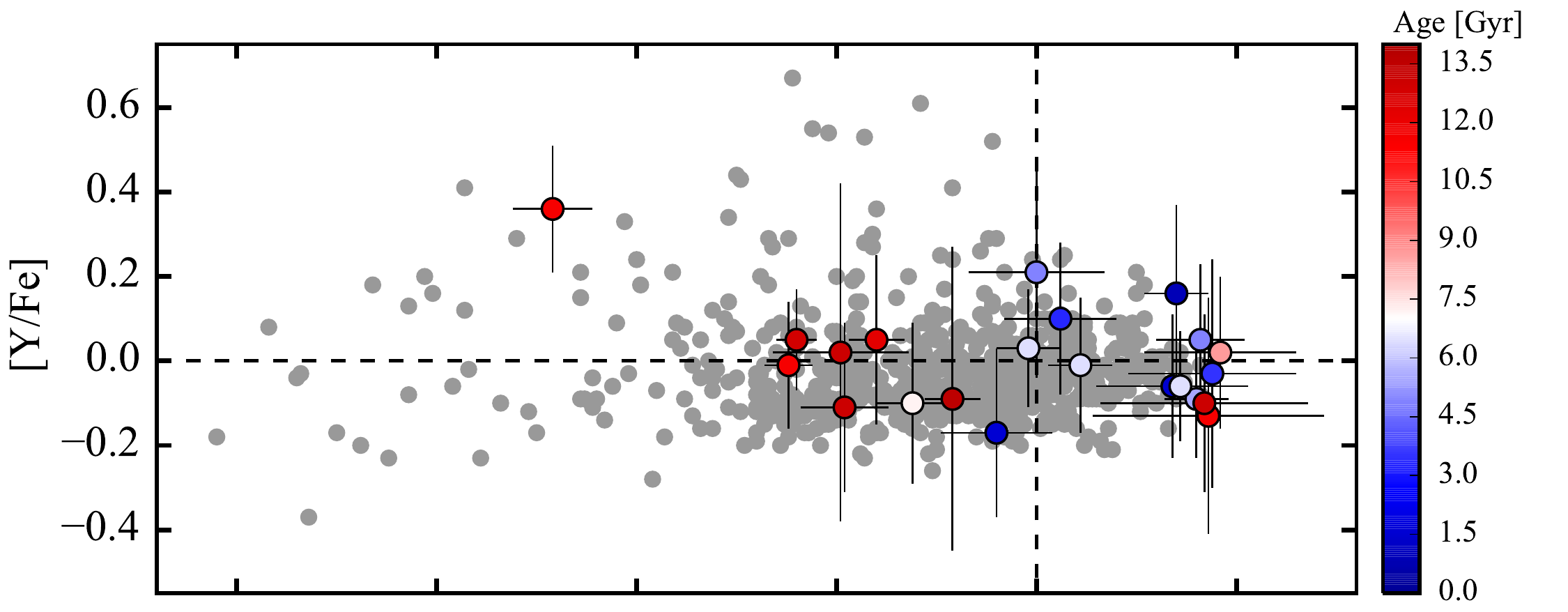}}
\resizebox{0.7\hsize}{!}{
\includegraphics[viewport= 0 0 648 293,clip]{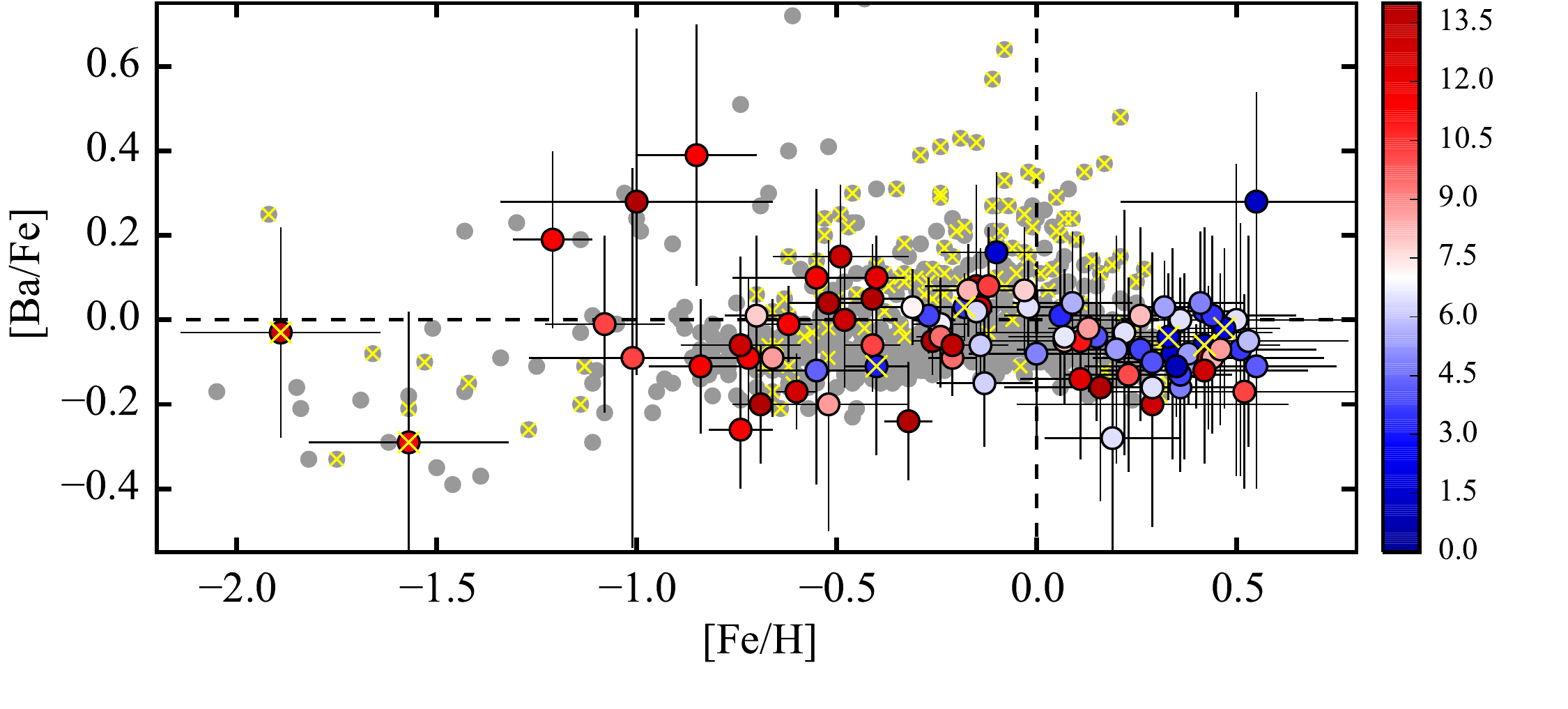}}
\caption{
Abundance trends with Fe as reference element for the microlensed dwarf sample.
The stars have been colour coded according to their ages (colour bar on the right-hand side). Grey circles in the background are the Solar neighbourhood dwarf stars from \cite{bensby2014}. In the Ba plot stars with $\teff>6100$\,K has been specially marked with crosses as those are likely to show significant NLTE effects at solar metallicities.
\label{fig:abundances4}
}
\end{figure*}

\section{Elemental abundances}

\subsection{Abundance trends}

Figures~\ref{fig:abundances}-\ref{fig:abundances4} show the abundance trends with Fe as the reference element for the 90 bulge dwarfs. The stars have been colour-coded based on their estimated ages (redder is older, bluer is younger), and the background stars are Solar neighbourhood thin and thick disk dwarf stars from \cite{bensby2014}. Again we point out that these stars have been analysed using the exact same methods as used for the microlensed bulge dwarfs. In general we find that the microlensed bulge dwarfs follow well-defined abundance trends.

\paragraph{\sl \bfseries $\alpha$-elements Mg, Si, Ca, and Ti:}
\label{sec:alpha}

The $\alpha$-elements are useful tracers of Galactic chemical evolution as they are believed to mainly come from one type of source, core-collapse supernovae \citep[e.g.][]{arnett1996,woosley1995}, even though some of them might have significant contributions from low-mass stars as well \citep[e.g.][]{thielemann2002}. Characteristic features in the $\rm [\alpha/Fe]-[Fe/H]$ diagram are the $\alpha$-enhanced plateau at low metallicities, due to fast enrichment from massive stars, that starts to decline once the enrichment from low-mass stars becomes significant, creating the ``knee'' in the $\rm [\alpha/Fe]-[Fe/H]$ diagram. The location of this ``knee'' depends on the relative roles of SNII and SNIa, and will shift to higher metallicities for higher star formation rates \citep{matteucci1986}.

The abundance trends for the $\alpha$-elements (Mg, Si, Ca, and Ti) show the typical signature of fast chemical enrichment, i.e. starting out at an enhanced $\rm [\alpha/Fe]$ representing the enrichment from short-lived massive stars. At $\rm [Fe/H]\approx-0.5$ there is a downturn in the $\rm [\alpha/Fe]$ ratio due to the enrichment from low- and intermediate mass stars that produce none or very little of the $\alpha$-elements. At sub-solar [Fe/H] the bulge dwarfs tend to follow the same trends as can be seen for nearby thick disk stars. However, at the same time it appears as if the bulge abundance ratios are somewhat closer to the upper envelope of the abundance trends defined by the local thick disk. This will be further investigated in Sect.~\ref{sec:kneeshift}.

\paragraph{\sl \bfseries Light odd-Z elements Na and Al:}

Sodium and aluminium are light odd-Z elements that appear to come from several sources. Na is partly made in massive stars, in the C-burning phase \citep[e.g.][]{woosley1995}, and partly from low-mass stars through the Ne-Na and Mg-Al cycles and mixed to the surface when ascending the red giant branch \citep[e.g.][]{karakas2010}. Al is made in C and Ne burning in massive stars \citep[e.g.][]{arnett1985}, but can also come from the Mg-Al cycle in massive AGB stars \citep[e.g.][]{doherty2014}.

Even though aluminium is a light odd-Z element it has been found to more or less behave as the $\alpha$-elements, i.e. with a flat $\alpha$-enhanced plateau at lower metallicities and then a ``knee'' at which the enhancement declines towards solar values. This was first noted by \cite{mcwilliam1997} and has since then been observed in several studies of the Galactic disk \citep[e.g.][]{edvardsson1993,bensby2014}. As can be seen in Fig.~\ref{fig:abundances2} this also holds for the bulge.  

Sodium has a different behaviour, being slightly enhanced at sub-solar metallicities, after which it rises rather strongly at super-solar metallicities. The bulge data shows the same trends as seen for the Solar neighbourhood stars. As discussed by \cite{mcwilliam2016} there appears to be a zig-zag trend of [Na/Fe], which suggests Na made by massive stars with a metal-dependent yield.  The trend of [Na/Fe] first goes up with [Fe/H], since only massive stars are making Na and Fe, but then when SNIa start to make a significant impact, the [Na/Fe] starts to decline with [Fe/H].  Finally, when the SNII/SNIa ratio has reach quasi-equilibrium, then [Na/Fe] begins to rise again.  This qualitative picture is consistent with the [$\alpha$/Fe] decline driven by time delay of SNIa with respect to SNII and the SFR.

In general the observed trends for Na and Al are poorly matched by chemical evolution models, and the outcomes are very sensitive to the adopted stellar yields. Especially the strong rise in [Na/Fe] at super-solar metallicities cannot be reproduced by the models \citep[e.g.][]{romano2010}. \cite{smiljanic2016} used data from the Gaia-ESO survey to investigate the behaviour of Na and Al and found that the chemical evolution models of the two elements could not be simultaneously matched to the observations.

The outcome from the microlensed bulge dwarfs does not add further information to the origin and evolution of Na and Al, besides that the bulge shares the abundance trends observed in the local thin and thick disks. Also, similar as for the $\alpha$-elements, the [Al/Fe] trend appears to be located at the upper envelope of the thick disk trend (see Sect.~\ref{sec:kneeshift} for further discussion). 

\paragraph{\sl \bfseries Iron-peak elements Cr, Ni, and Zn}

Chromium and nickel show extremely well-defined abundance patterns, scaling more or less perfectly with Fe (although with a slight upturn at super-solar [Fe/H] for Ni), and in full agreement with the trends seen in the Solar neighbourhood. In the Solar neighbourhood sample it is seen that the [Ni/Fe] is separated for the thin and thick disks at sub-solar metallicities with the thick disk being slightly more enhanced. Even though the separation between the thin and thick disks is less than about 0.05\,dex we see that the bulge dwarfs essentially fall on top of the thick disk sequence, again highlighting the similarities between the thick disk and the metal-poor bulge.

Zn is one of the most important elements for supernova physics \citep{nomoto2013} and is also an important element for studies of distant damped Lyman $\alpha$ systems as Zn is not depleted onto dust \citep[e.g.][]{kobayashi2006}. It is believed to be produced in both massive stars as well as low mass stars \citep[e.g.][]{nomoto1997a,nomoto1997b,mishenina2002}. The bulge data does not appear to reveal any new insights into the origin of Zn. It is, however, consistent with the results seen for the $\alpha$-elements in the sense that the metal-poor bulge stars follow the local thick disk trends. An important observation is that Zn does not scale directly with Fe, and can therefore not be used as a direct proxy for Fe in studies of the distant Universe. The somewhat large scatter in the [Zn/Fe]-[Fe/H] plane is most likely due to that the Zn abundances for the bulge stars are based on the analysis of a single \ion{Zn}{i} line.

\paragraph{\sl \bfseries Neutron-capture elements Y and Ba:}

Y and Ba are slow neutron-capture (s-process) elements that mainly are made in AGB stars \citep[e.g.][]{karakas2014}. 
The [Y/Fe] and [Ba/Fe] trends in Fig.~\ref{fig:abundances4} show good agreement between the bulge and the local disks. Both elements more or less appear to follow Fe, although with a larger scatter than in the disk. It should be noted that Ba shows large NLTE effects for stars with effective temperatures larger than about 6100\,K \citep[e.g.][]{korotin2011}. These have been marked with crosses in both the disk sample and in the bulge sample in Fig.~\ref{fig:abundances4}. It is specially around solar metallicities that the Ba abundances are enhanced due to this.

\begin{figure*}
\centering
\resizebox{\hsize}{!}{
\includegraphics[viewport= 0 0 500 648,clip]{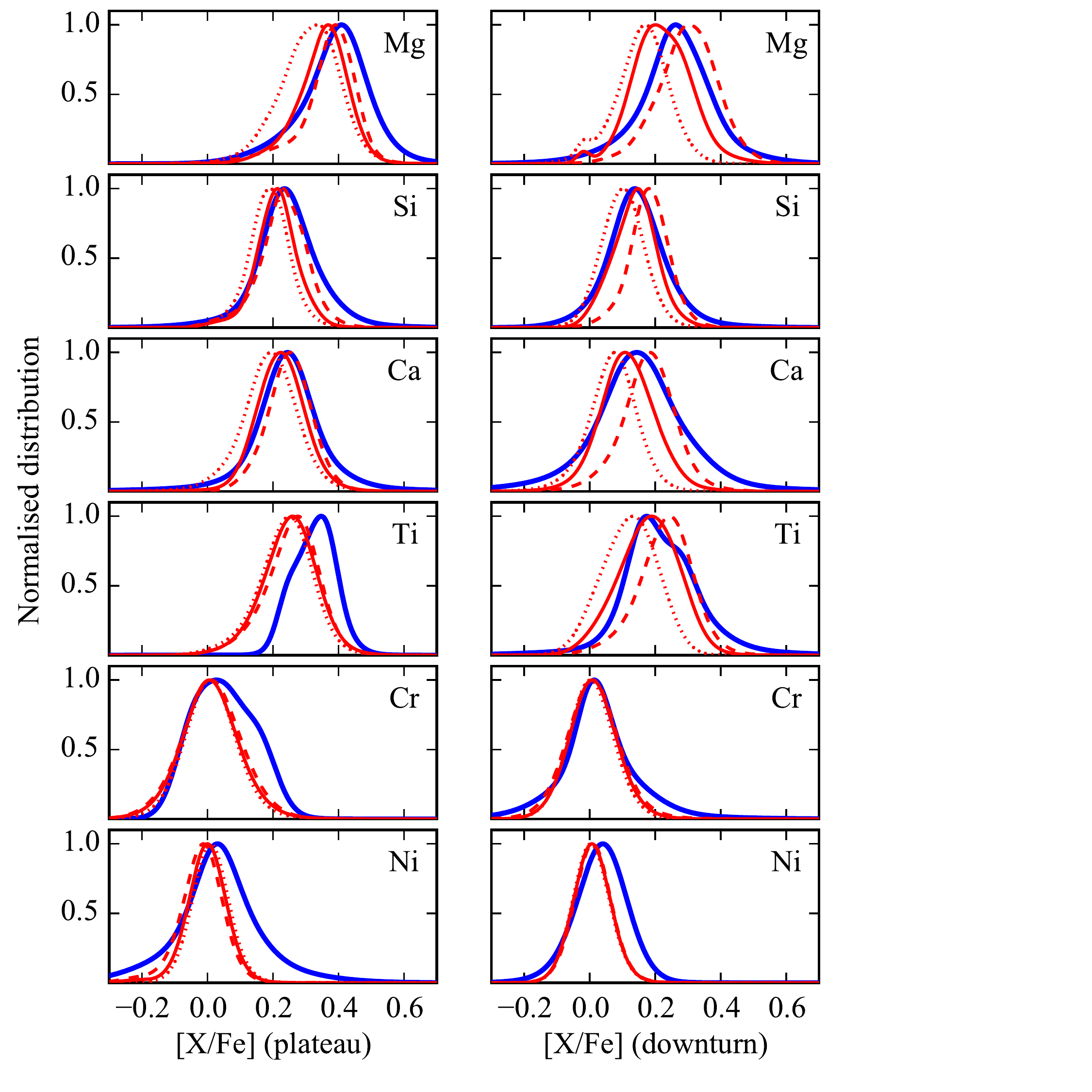}
\includegraphics[viewport= 0 0 500 648,clip]{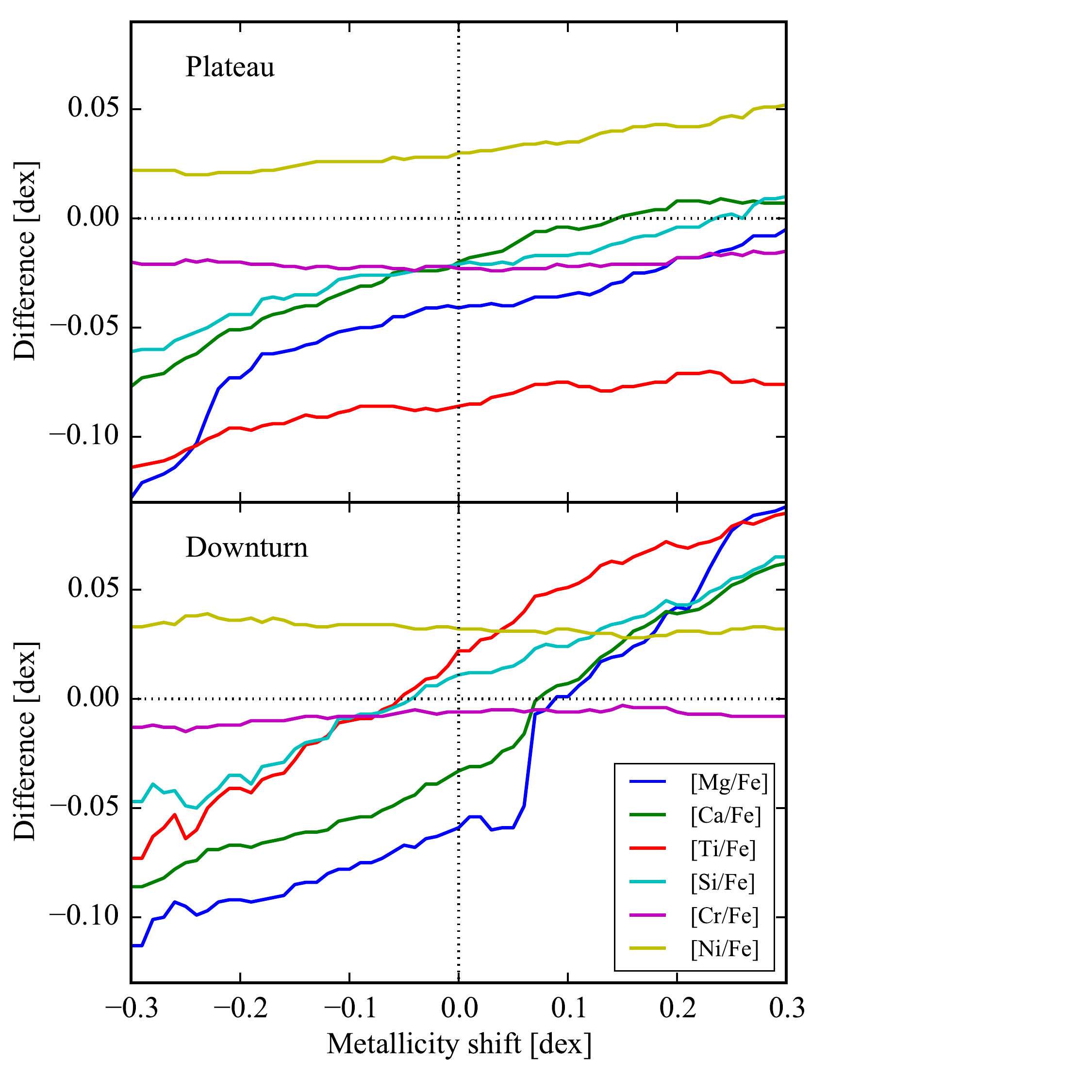}}
\caption{
{\sl Left-hand side:} Generalised distributions for $\rm [\alpha/Fe]$ in two metallicity intervals, $\rm -1<[Fe/H]<-0.5$ (the ``plateau regime'') and $\rm -0.5<[Fe/H]<-0.1$ (the ``downturn regime''). The microlensed dwarf sample is shown by blue lines, and local thick disk sample by red lines (dotted and dashed lines show the distribution if [Fe/H] shifted by minus or plus 0.20\,dex, respectively). Also included are the distributions for two iron-peak elements, Cr and Ni.
{\sl Right-hand side:} The difference between the bulge and thick disk [$X$/Fe] distributions, for metallicity shifts in the range $\pm 0.30$\,dex. Agreement is reached when the difference is zero (on the y-axis).
\label{fig:kneeshift}
}
\end{figure*}

\subsection{The location of the $\alpha$-knee}
\label{sec:kneeshift}

The vertical location of the $\rm [\alpha/Fe]$-plateau is an indicator of the initial mass function (IMF) and the horizontal location of the down-turn (the ``knee'') an indicator of the star formation rate (SFR) \citep[e.g.][]{mcwilliam1997}. Hence, if real, the slight vertical offset in $\rm [\alpha/Fe]$ between the bulge and the local thick disk could mean that the bulge IMF is different from that of the local disk and most other Galactic populations. However, observations of resolved stellar populations and integrated properties of distant galaxies have shown that there are no significant variations in the IMF, and that the IMF most likely can be considered to be universal \citep[e.g.][]{bastian2010}. Hence it is unlikely that the IMF would be the cause for the apparent offset seen between the bulge and nearby thick disk. Another explanation could therefore be that the SFR has been different in the bulge compared to the disk. This would manifest itself as a horizontal shift of the $\rm [\alpha/Fe]$-knee towards higher [Fe/H] for the bulge abundance trends.

To investigate if the $\rm [\alpha/Fe]$-knee for the metal-poor bulge stars is located at a higher metallicity than for the (local) thick disk or not, Fig.~\ref{fig:kneeshift} shows the generalised $\rm [\alpha/Fe]$ distributions for Mg, Si, Ca, and Ti, in two different [Fe/H] intervals; $\rm -1<[Fe/H]<-0.5$ and $\rm -0.5<[Fe/H]<-0.1$, referred to as the ``plateau'' and ``downturn'' regimes, respectively. The blue lines represent the microlensed bulge dwarf sample, and the red lines the local thick disk stars from \cite{bensby2014}, here selected as those stars that have lower age estimates greater than 8\,Gyr. To simulate horizontal shifts of the $\rm [\alpha/Fe]$-knee, the metallicities of the local disk sample have been shifted by $\pm 0.20$\,dex, shown by red dotted and dashed lines, respectively, in Fig.~\ref{fig:kneeshift}. The right-hand side of the figure shows how the $\alpha$-enhancement differences between the microlensed dwarf stars and the local thick disk vary for metallicity shifts between $-0.3$ to $+0.3$\,dex. The difference is defined as the distance between the peak values of the blue and read distributions (as illustrated for a few examples in the plots on the left-hand side). 

In the ``plateau regime'' the bulge sample shows slight $\alpha$-enhancement offsets of 0.03-0.05\,dex in Mg, Si, and Ca, and of about 0.09\,dex in Ti. To put the thick disk and bulge plateaus at the same level, shifts of about 0.1-0.3\,dex are needed in the Mg, Si, and Ca cases. For Ti we cannot put the two populations on the same $\alpha$-enhancement level on the plateau, no matter how much the metallicity is shifted.
In the case of the ``downturn regime'', positive metallicity shifts of about 0.08\,dex are needed for Mg and Ca, while negative shifts of about $-0.05$\,dex are needed for Si and Ti. These inconsistent results shows that it will be difficult to put strong claims that the bulge has experienced a much faster enrichment history than that of the thick disk. Mg and Si appears to indicate that it is the case, while Ca and Ti give indecisive results.  

Another possibility could be that the [Fe/H] scale of the bulge dwarfs is offset compared to the Solar neighbourhood stars, even though they have been analysed in the same way. If that were the case there would most likely also be an offset in the [Ni/Fe] and [Cr/Fe] trends between the bulge and the local disk. Actually, there appears to be a slight negative offset of around $0.01-0.02$\,dex for Cr and a positive offset of about 0.03\,dex for Ni. Both are more or less constant for different metallicity shift. As they do not both have either a positive or negative offset it is unlikely that there is an overall offset in [Fe/H]. The reason the offsets we see are hence unclear, but could be due to the fact that the microlensed bulge dwarf spectra in general have lower $S/N$ than the thick disk dwarfs analysed in \cite{bensby2014}, and that the analysis is based on a lower number spectral lines. This could be investigated further, but is outside the scope of the present paper. 

We conclude that the $\rm [\alpha/Fe]$-knee appears to be located at a slightly higher metallicity than in the local thick disk. This result is however inconclusive and needs further confirmation with larger samples.  

\begin{figure}
\centering
\resizebox{\hsize}{!}{
\includegraphics[viewport= 0 50 648 620,clip]{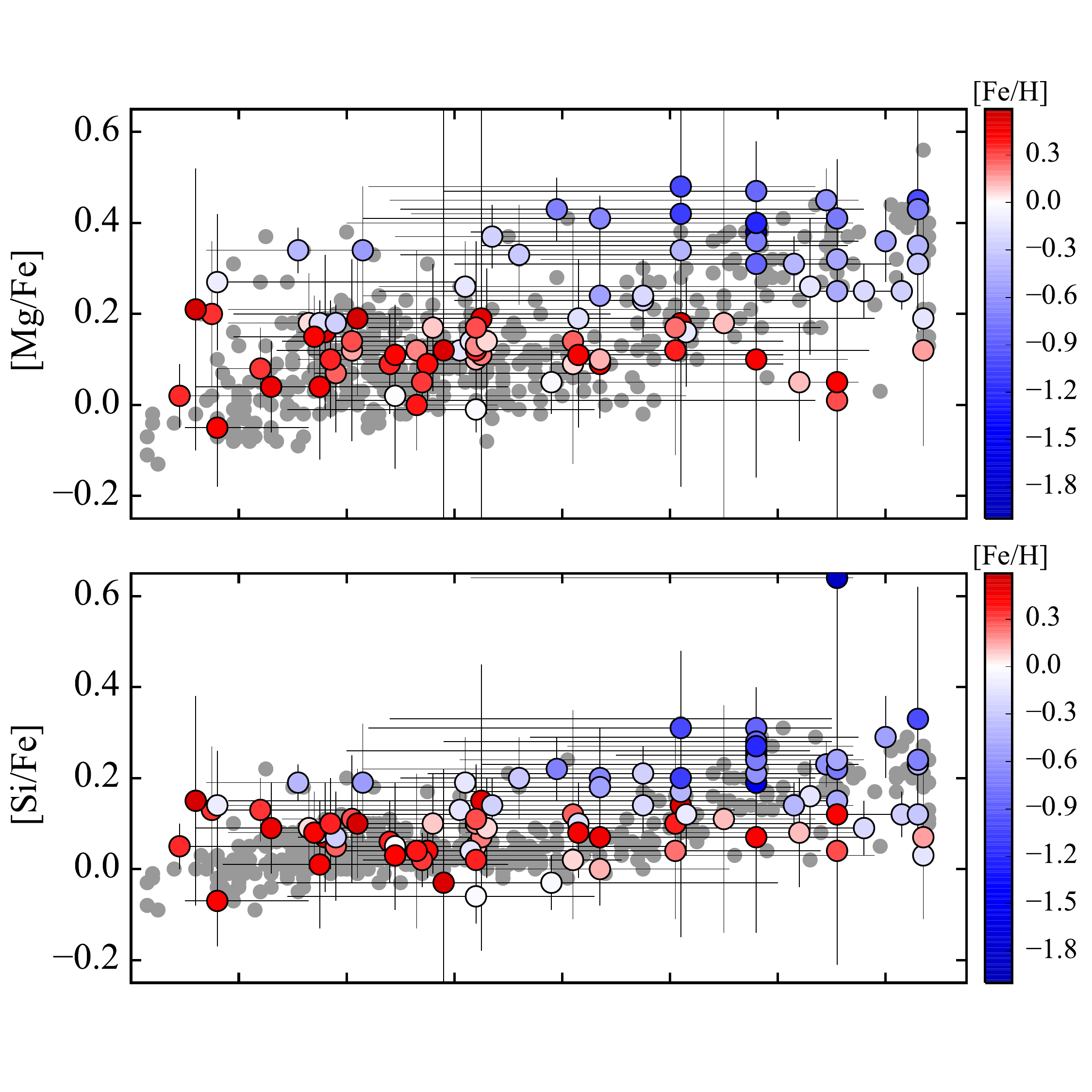}}
\resizebox{\hsize}{!}{
\includegraphics[viewport= 0 0 648 600,clip]{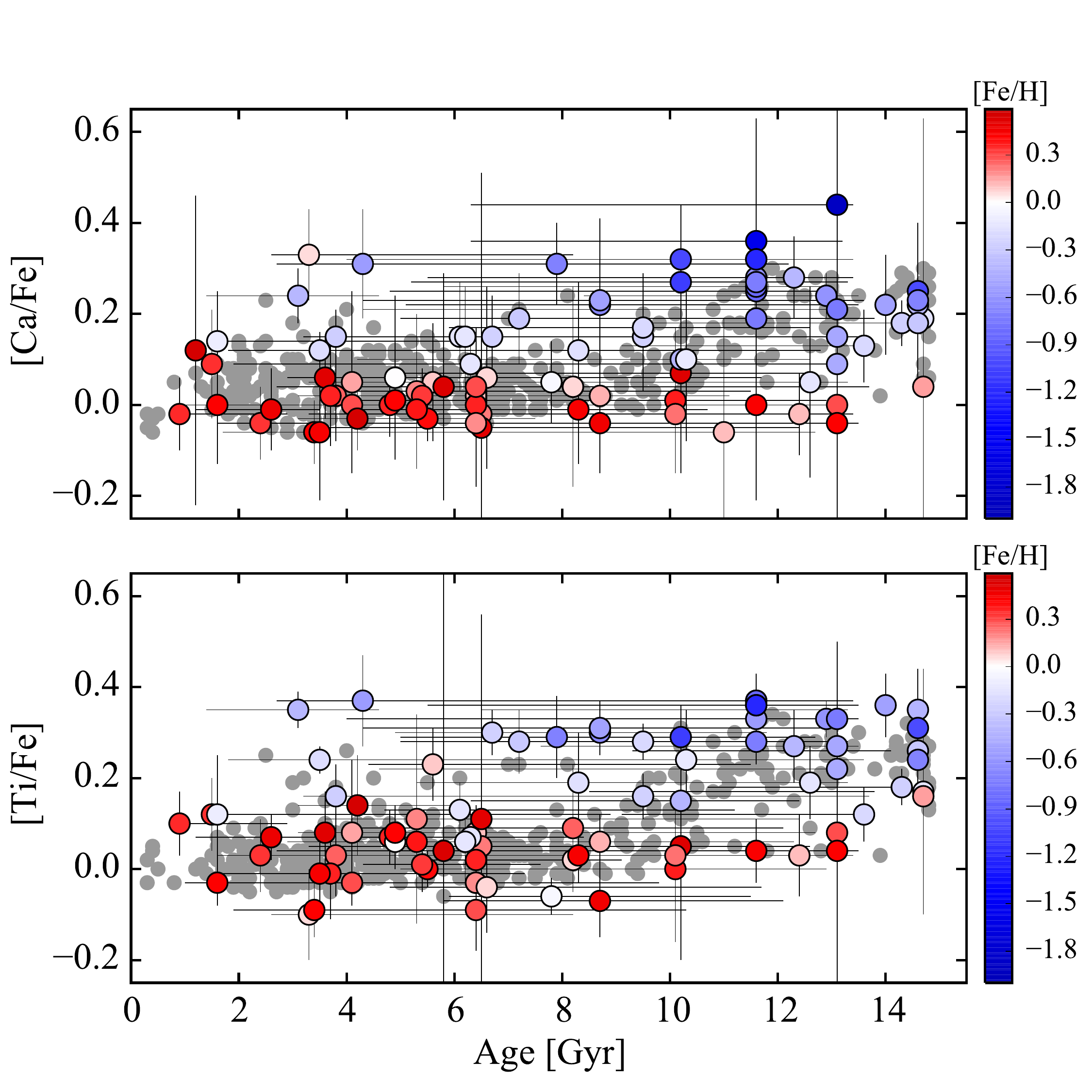}}
\caption{
$\alpha$-abundances versus age for the microlensed dwarf sample. The stars have been colour coded according to their metallicities (as shown by the colour bar on the right-hand side). The grey circles in the background are the Solar neighbourhood F and G dwarf stars from \cite{bensby2014}.
\label{fig:alphaage}
}
\end{figure}

\begin{figure}
\resizebox{\hsize}{!}{
\includegraphics[viewport= 0 0 648 390,clip]{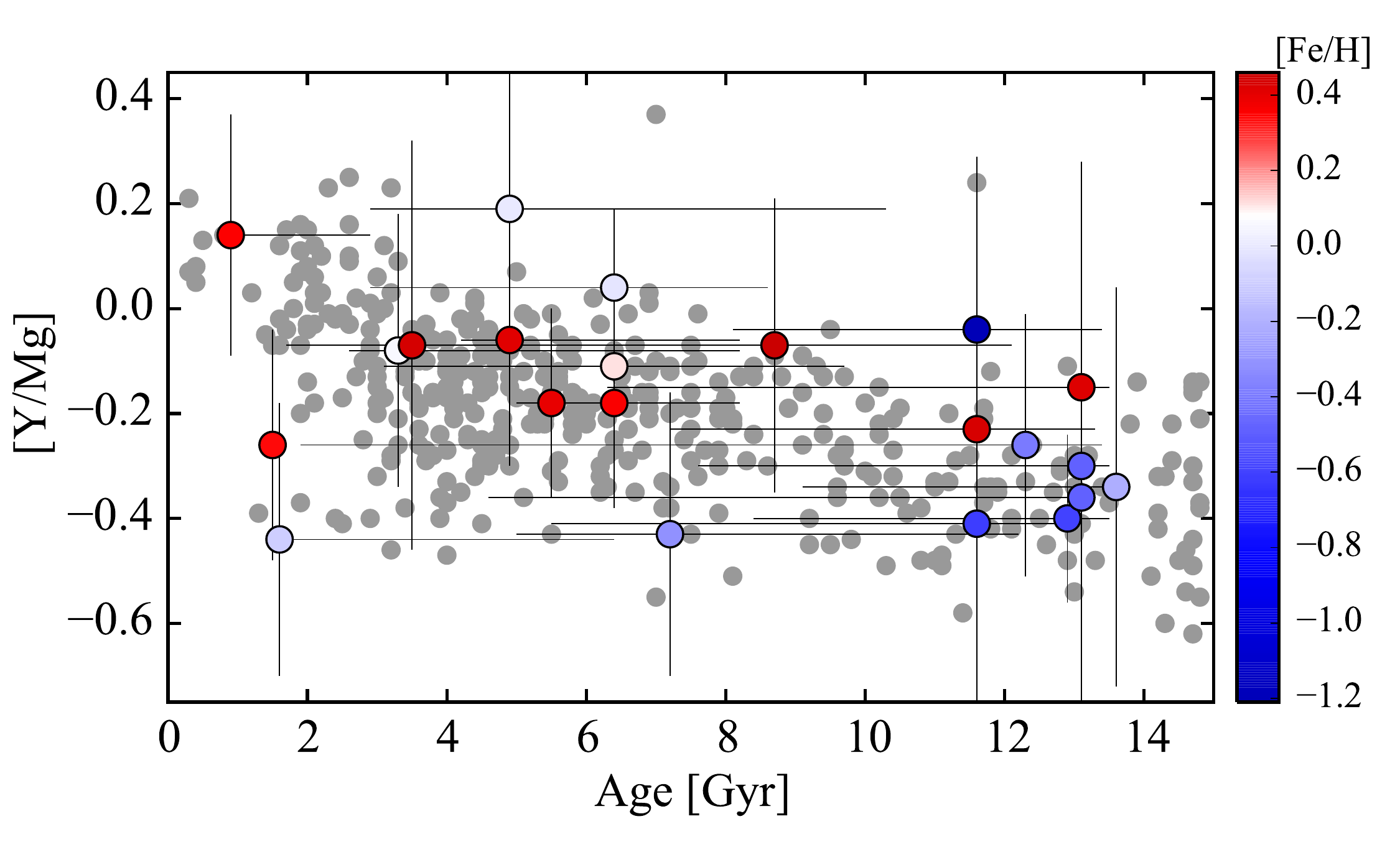}}
\caption{
[Y/Mg] versus age for the microlensed dwarf sample. The stars have been colour coded according to their metallicities (as shown by the colour bar on the right-hand side). The grey circles in the background are the Solar neighbourhood F and G dwarf stars from \cite{bensby2014}.
\label{fig:ymgage}
}
\end{figure}

\subsection{$\rm [\alpha/Fe]$ as a proxy for age?}

In the Solar neighbourhood F and G dwarf stars show a relatively tight relation between the enhancement of $\alpha$-elements, e.g. [Ti/Fe] or [Mg/Fe], and age \citep{haywood2013,bensby2014}. This fact has been used to assign an old age to stars with a high $\rm [\alpha/Fe]$ ratio and a younger age to stars with lower $\rm [\alpha/Fe]$ ratios. However, in the plots of $\rm [\alpha/Fe]$ versus age there are always a few stars that do not follow the trend, notably there are some stars with young age but high $\rm [\alpha/Fe]$ and some stars with high age and low $\rm [\alpha/Fe]$. A recent example of this are the high-$\alpha$, but young stars found in two studies using masses from asteroseismic studies to obtain ages for the stars \citep{chiappini2015,martig2015}. Recent radial velocity studies by \cite{yong2016} and \cite{jofre2016} argue that these stars are instead blue stragglers, i.e. they are binary stars that have gained mass via mass transfer in a binary. The higher mass make them look younger than they are. If this is the cause of the young ages seen in the local studies of stars that are clearly not in binaries remains to be verified.

Figure~\ref{fig:alphaage} shows for the microlensed bulge stars the four $\rm alpha$-elements Mg, Si, Ti, and Ca relative to Fe as a function of age. We see essentially the same type of structures in this figure as seen in local samples; however, the number of stars in different regions of the plot might differ. Overall the microlensed stars show that stars with iron abundances at solar or higher values have low [Ti/Fe] at ages 0-7\,Gyr, stars with low [Fe/H] in the same age range instead have high [Ti/Fe]. This compares well with what is seen in local sample (which is shown in the plot). At higher ages we find that more than 2/3 of our stars have high $\rm [\alpha/Fe]$ and low [Fe/H]; but there are also stars with high age and low [Ti/Fe] and super-solar metallicities.

The microlensed bulge dwarf sample and its similarity to that of larger samples in the Solar neighbourhood appears to point to; a) overall there is a trend such that older stars have higher $\rm [\alpha/Fe]$ abundances and lower [Fe/H]; b) these properties are universal; c) there are stars at all ages of all metallicities and $\alpha$-enhancements, in fact at high ages there can be a non-negligible fraction of super-solar metallicity stars. These last points weaken the usage of $\rm [\alpha/Fe]$ as a proxy for age.

\subsection{[Y/Mg] as proxy for age?}

Recently a number of studies have looked at the relation between various elemental abundances and stellar age for turn-off stars and subgiant stars \citep[e.g.][]{nissen2015,tuccimaia2016,spina2016}. These studies of Solar neighbourhood stars have used spectra with very high signal-to-noise ratios, and by selecting only stars with very similar stellar parameters (essentially solar-like) they have achieved remarkable precision in both elemental abundances and age determinations. A major finding from these studies is the very tight relation between [Y/Mg] and age for stars with thin disk kinematics and solar temperature and surface gravity. If the relation is universal, then this provides an interesting way to obtain stellar ages without having to compare the stellar parameters to isochrones or evolutionary tracks.

In a recent paper, \cite{feltzing2017} challenged the universality of this relation. Using a much larger sample of stars, albeit with somewhat lower signal-to-noise ratios in the stellar spectra but covering a much wider range in [Fe/H] they showed that the [Y/Mg] with age trend is a function of [Fe/H]. This can be understood from the nucleosynthesis of the three elements involved. Magnesium is a so called $\alpha$-element and is mainly produced in core collapse supernovae, while iron is produced also in the SNIa. This means that once the SNIa start to contribute the ratio of [Mg/Fe] will go down, this is the classical ``knee'' seen in all plots of $\alpha$-elements \citep[e.g.][]{bensby2004,hayden2015}. Yttrium, on the other hand, is produced through the s-process, which occurs in asymptotic giant branch stars. These are stars in the mass range of 1 to 8 solar masses. This means that the release of yttrium will increase with time as the lower mass stars will contribute at a later time. Thus younger stars have higher yttrium content than older stars and as the magnesium content does not increase as much, the [Y/Mg] ratio would become a clock.  However, the production of yttrium depends on the amount of seeds available, which in this case is iron. This means that when the iron content in the stars go up, the production of yttrium is much enhanced.

Figure~\ref{fig:ymgage} shows a comparison of the data from \cite{bensby2014} used by \cite{feltzing2017} and the data from the micro-lensed dwarf stars. Overall the bulge stars follow the trend found in the Solar neighbourhood, that on average the older stars have lower [Y/Mg]. The spread in [Y/Mg] in the Solar neighbourhood at a given age (especially between about 4-6\,Gyr) is due to the range of [Fe/H] at those ages (compare Fig.~1 in \citealt{feltzing2017}). On average the more metal-rich and young bulge dwarf stars also have the larger [Y/Mg] values and vice versa for the metal-poorer, young stars. We note that three solar metallicity stars in the bulge sample appears to have anomalously high [Y/Mg] for their age. If this is a true feature of the bulge or just statistical errors remains to be investigated. One way to investigate this could be to obtain ages from asteroseismic data for red giant branch stars in the bulge and compare their [Y/Mg] in same way. This would then also need to be normalised to studies in the Solar neighbourhood where the stars can be observed with high signal-to-noise ratio spectra.

To conclude, we find that on average the bulge stars appear to show the same trends as observed for Solar neighbourhood stars and also indicating, as outlined in \cite{feltzing2017}, that although the general trend of [Y/Mg] is a declining function of age, there is a second parameter (Fe) that removes this as a useful age indicator for dwarf stars.

\section{Discussion}

\subsection{The metallicity distribution of the Milky Way bulge}
\label{sec:mdfevolve}

\begin{figure}
\resizebox{\hsize}{!}{
\includegraphics[viewport= 0 0 648 648,clip]{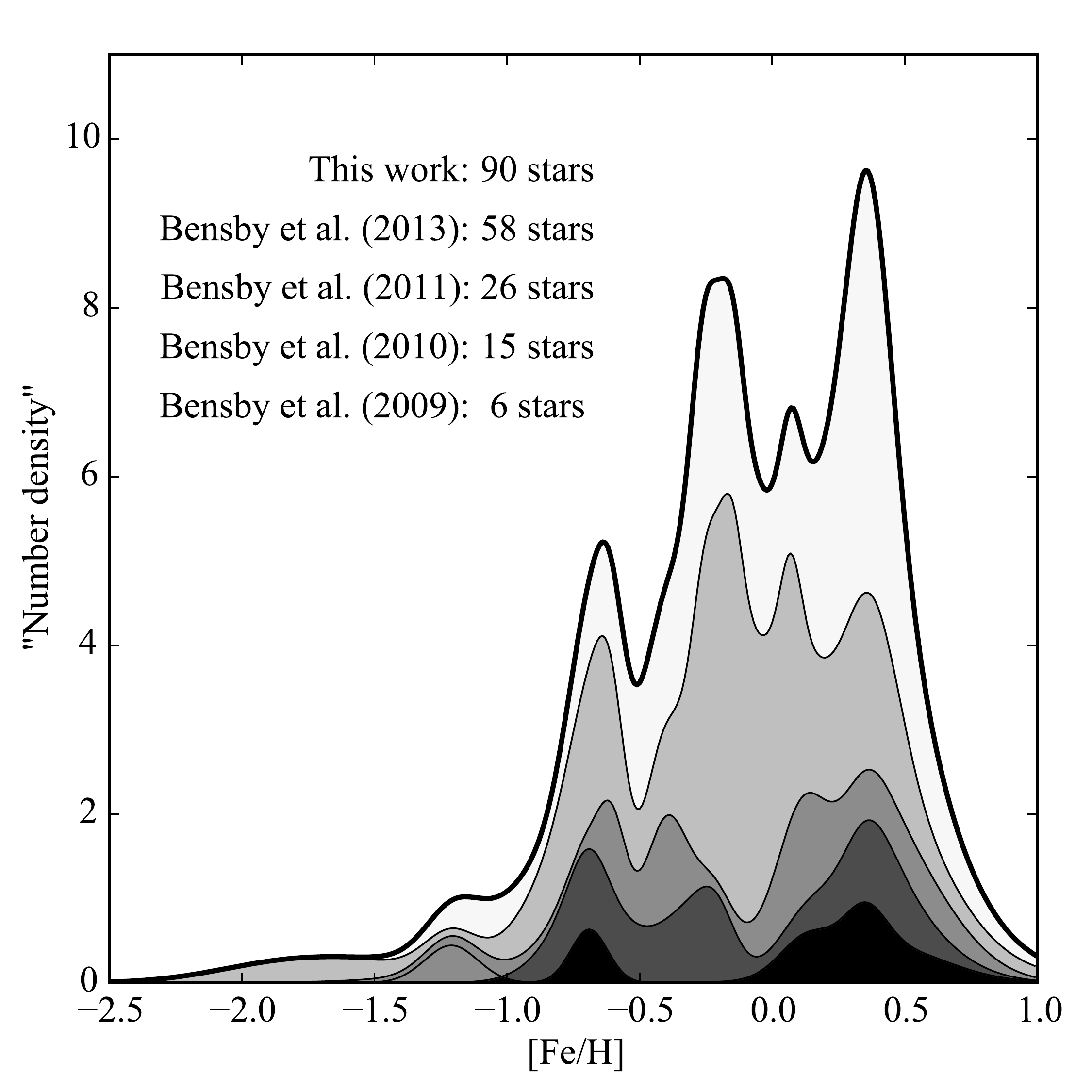}}
\caption{
Illustration on how the microlensed bulge dwarf MDF has evolved between the papers published so far. The lowest black histogram includes the first 6 stars from \cite{johnson2007,johnson2008,cohen2008,cohen2009,bensby2009}, and the in decreasing shades of grey the next histograms include, 9 stars from \cite{bensby2010}, 11 stars from \cite{bensby2011}, 32 stars from \cite{bensby2013}, and the final almost white histogram the 32 new stars from this work, now in total 90 stars. 
\label{fig:mdfevolve}
}
\end{figure}
The first spectroscopic observations of the source star during a microlensing event was carried out by \cite{lennon1996} who used the EMMI spectrograph at the ESO NTT telescope. The spectrum of 96-BLG-3 had a resolving power of about 1100 and they derived a very high metallicity in the range $+0.3$ to $+0.6$\,dex. Next \cite{minniti1998} observed a main sequence star in the Galactic bulge, using Keck as a ``15 meter telescope'' while the target was magnified approximately 2.5 times. These studies opened up new possibilities to study the chemical composition of the Galactic bulge, and was soon followed by \cite{cavallo2003} who presented the first detailed analysis of six microlensed stars, a mix of dwarf stars and giants. The spectra had a resolving power that is somewhat lower than what is considered as useful for detailed high-resolution spectroscopic abundance analysis, and the signal-to-noise ratios were also on the low side. Hence they could only make some overall estimates on the metallicities and deduce only a few individual abundances. The first detailed abundance analysis, using spectra of high-resolution and high signal-to-noise ($R\approx 40\,000$, $S/N>50$) was presented by \cite{johnson2007}. They found a very high metallicity at $\rm [Fe/H]=+0.56$, at that time the most metal-rich star ever observed. The subsequent two events by \cite{johnson2008} and \cite{cohen2008} also turned out to be very metal-rich, with both being more metal-rich than $\rm [Fe/H]>+0.30$. These results were in stark contrast with the results from the giant star studies, which showed a metallicity distribution that peaked around solar values, although with long tails \citep[e.g.][]{rich1988,zoccali2008}. The giant studies generally contained several hundreds of stars, but as the analysis of dwarf stars are considered to be easier due to less crowded spectra, and with smaller uncertainties, it was suggested that perhaps the metal-rich giants in the bulge for some unknown reason were missing, and that the bulge is much more metal-rich than previously thought. \cite{bensby2009} analysed the first microlensed dwarf at sub-solar metallicities, a subgiant at $\rm [Fe/H]=-0.33$, showing that the bulge actually does contain stars of lower metallicities.  The same year \cite{cohen2009} added another two events to the collection of now six microlensed bulge dwarfs and both of them were, again, extremely metal-rich. By randomly drawing six stars from the sample of more than 200 red giants from \cite{zoccali2008}, \cite{cohen2008} showed that it is extremely unlikely that the MDFs from microlensed bulge dwarf stars and bulge red giants represent the same underlying bulge MDF. An explanation that was brought forward was that the metal-rich red giant stars were missing due to strong stellar winds. The MDF at this time, consisting of five stars at super-solar [Fe/H] and one a sub-solar, is shown as the lowest (black coloured) generalised histogram in Fig.~\ref{fig:mdfevolve}. Shortly thereafter \cite{bensby2009letter} presented the analysis of a second metal-poor bulge dwarf, now at $\rm [Fe/H]\approx -0.86$, again showing that there truly exist metal-poor dwarf stars  in the bulge. \cite{bensby2010} then added a few events, several of them at sub-solar [Fe/H]  and re-analysed all previous events in order to have them analysed in a consistent manner. The sample now consisted of in total 16 events and the MDF they presented was clearly bi-modal with a lack of stars at solar metallicities (the next generalised histogram in Fig.~\ref{fig:mdfevolve}).  \cite{bensby2011} added another 10 events, now in total 26 events, and found indications of a bi-modal MDF, with no stars at solar metallicities. This was changed with the next paper, \cite{bensby2013}, that added another 32 events for a total 58 events, and where a substantial fraction actually had solar or around solar [Fe/H]. The MDF was wide, and indications of multiple components, similar to the ones claimed by \cite{ness2013} started to emerge.  With this last addition of 32 more events, to a total of 90 microlensed dwarf and subgiant stars in the bulge, the peaks in the MDF are even more pronounced, and by looking at the generalised histograms in Fig.~\ref{fig:mdfevolve}, one can see that they appear to have been present more or less from the start, just becoming better defined as the data sample has grown in size. The first few microlensing events all turned out to be very metal-rich, and the first 26 events had no stars with solar metallicities, most likely due to low number statistics, and demonstrates that it is dangerous to draw conclusions based on small samples.

\subsection{The age distribution of the Milky Way bulge}

The Galactic bulge has for a long time been viewed as a genuinely old, if not the oldest, stellar population of the Milky Way. The main piece of observational evidence has been the observed red colours of the turn-off in the colour-magnitude diagrams \citep[e.g.][]{terndrup1988,renzini1994,kuijken2002,zoccali2003,clarkson2008,valenti2013,gennaro2015}. As demonstrated in \cite{bensby2013} an old turn-off will be apparent if metallicity information for the stars are lacking; old and metal-poor isochrones (10-12\,Gyr and $\rm [Fe/H] \approx -1$) and intermediate-age and metal-rich isochrones (4-5\,Gyr and $\rm [Fe/H]>0$) are essentially indistinguishable from each other (see also \citealt{haywood2016}), and therefore the whole population will be estimated to be old. With the advent of the possibility of determining ages of individual stars in the bulge from the spectroscopic observations of microlensed dwarf, turn-off, and subgiant stars, we have seen that the bulge is not a genuinely old stellar population. While it appears to be so at metallicities below $\rm [Fe/H]\leq-0.5$, at higher metallicities there is a wide range of ages, and at super-solar metallicities the young to intermediate age stars actually seem to be in majority (see Fig.~\ref{fig:agefe}).  The age determinations of the microlensed dwarf stars in \cite{bensby2010,bensby2011,bensby2013} have been re-confirmed by \cite{valle2015}, and in this study we have further shown that our age determination method and the Bayesian method by \cite{jorgensen2005} give very similar results.

Other reasons for the discrepancies between the ages of the microlensed stars and the photometric colour-magnitude diagram studies have been proposed. One is that the bulge should be significantly more enriched in He than the other Galactic stellar populations \citep{nataf2012}. We investigated this in \cite{bensby2013} and found that there were small effects on the derived ages, but far from being sufficient to make any significant changes.

Additional observational indications of young components in the bulge are shown by the excess of very bright stars in the inner bulge, interpreted as coming from young star-forming regions near the Galactic centre \citep{lopezcorredoira2001}; AGB stars \citep{vanloon2003,cole2002,uttenthaler2007}; long-period Mira variables with ages around 5 Gyr that follow the bar structure in the bulge \citep{groenewegen2005,catchpole2016}; around Terzan 5 two turn-off ages, one metal-poor major population at 11 and one metal-rich minor at 4.5 Gyr \citep{ferraro2016}. Young stars in a predominantly old bulge tend to lie closer to the plane \citep{ness2014}. Additional arguments for and against an old Galactic bulge can be found in the review by \cite{nataf2016b}.

\subsection{Elemental abundance trends of the Milky Way bulge}

The elemental abundance trends of the Galactic bulge have in many studies been shown to be similar to those of the local thick disk for sub-solar metallicities. This was first recognised by \cite{melendez2008}, and has since then been further verified in studies of both red giants \citep[e.g.][]{alvesbrito2010,ryde2010,gonzalez2011,jonsson2017} as well as dwarf stars (see Figs.~\ref{fig:abundances}-\ref{fig:abundances4}, and \citealt{bensby2009,bensby2011,bensby2013}). In \cite{bensby2013,johnson2014} and \cite{jonsson2017} there were indications that the match between the bulge and the local thick disk $\rm [\alpha/Fe]-[Fe/H]$ trends were not perfect, but that the bulge trends tended to placed closer to the upper envelope of the thick disk trends. 
With the enlarged sample of microlensed dwarf stars in this study it appears as if the level of $\alpha$-enhancement is slightly higher in the bulge than in the thick disk. If it is real, the offset would indicate that the star formation rate has been slightly faster in the bulge than in the thick disk. This would not surprising since the stellar environment is much denser in the inner parts of the Milky Way. However, as demonstrated in Fig.~\ref{fig:kneeshift} and discussed in Sect.~\ref{sec:kneeshift} the evidence for this signature are inconclusive.  But also other studies have found indications of an offset between the bulge and thick disk $\rm [\alpha/Fe]$ abundance trends \citep{johnson2014,jonsson2017}, so we do not reject the hypothesis.  

\cite{bensby2010letter} found that the abundance trends of the inner disk were similar to the ones in the Solar neighbourhood (apart from a lack of low-$\alpha$ stars for $\rm [Fe/H]\lesssim-0.1$ in the inner disk, but seen in the local thin disk). \cite{bensby2011letter} further showed that the outer disk seemed to be void of stars with a thick disk abundance pattern. This was interpreted as being due to the thick disk having a very short scale-length compared to the thin disk. A short scale-length of the thick disk relative to the thin disk means that the fraction of thick disk stars should increase strongly as one gets closer to the Galactic centre, at the expense of the fraction of thin disk stars that should decrease for shorter galactocentric radii. This could explain why there are essentially no stars at sub-solar [Fe/H] in the bulge with low $\alpha$-enhancements (compare the thin disk stars in Figs~\ref{fig:abundances} shown by small grey circles). 

The abundance trends for the inner and outer disks seen by \cite{bensby2010letter,bensby2011letter} have later been confirmed by other larger surveys such as for example, \cite{hayden2015} who used APOGEE data to trace the abundance pattern throughout the Galactic disk at different Galactocentric radii and different heights above/below the Galactic plane. Looking at their $\rm [\alpha/Fe]$ abundance trend for the stars located $0-0.5$\,kpc from the plane, it appears as though the location of the $\rm [\alpha/Fe]$-knee is at approximately the same position at the different Galactocentric radii spanning the inner disk (4-8\,kpc). Hence, there appears to be no gradual change in the knee location with galactocentric radius that potentially could match a possible slight shift (of about 0.1\,dex) in the $\rm [\alpha/Fe]$-knee that tentatively is suggested by the microlensed dwarf stars. It should be noted that the abundance structure close to the plane as a function of Galactocentric radius is poorly constrained, and the APOGEE data might not have the accuracy to reveal a gradual change in the location of the $\rm [\alpha/Fe]$-knee of the order of 0.01\,dex/kpc. More data is clearly needed to probe the high-extinction inner regions of the Galactic plane.

The metal-rich parts of the bulge abundance trends more or less agree with what is observed in the solar neighbourhood. Whether this means that the metal-rich parts of the bulge can be directly associated with the thin disk is unsaid. If there is a thin disk component in the bulge it is clearly lacking the metal-poor parts down to about $\rm [Fe/H]\approx-0.8$ that have been observed in the thin disk in the Solar neighbourhood \citep[e.g.][]{adibekyan2011,bensby2014}. This could be due to strong radial metallicity gradient that has been observed for the $\alpha$-poor disk population, located at $\rm [Fe/H]\approx -0.4$ for $13<\rgal<15$\,kpc, at $\rm [Fe/H]\approx0$ at $\rgal=8$\,kpc, and at $\rm [Fe/H]\approx +0.4$ for $3<\rgal<5$\,kpc (see lower panels of Fig.~4 in \citealt{hayden2015}).

\subsection{What is the Milky Way bulge?}
\label{sec:whatisthebulge}

As discussed above, during the last decade our view of the Galactic bulge has changed dramatically: the metallicity distribution appears to be the composite of two or even more peaks; it is not uniquely old - at super-solar metallicities maybe up to about 50\,\% of the stars could be younger than 7-8\,Gyr; the $\alpha$-enhancements at high metallicities have been shown to be at solar-like levels, something that now also is supported by studies of red giant stars. Also, star counts have shown that there is a bar \citep{stanek1994} that now is known to be x-shaped \citep[e.g.][]{nataf2010,mcwilliam2010,saito2011,saito2012}. Furthermore, the bulge is where the highest density of the oldest stars are expected to be found \citep{tumlinson2010}. These are halo stars that are in the bulge, but not of the bulge. As the density of stars in the bulge is very high, the relative fraction of halo stars is very low and they are difficult to find \citep{howes2014,howes2015}. A major finding is that the bulge has cylindrical rotation \citep{howard2009}, which leaves very little room for a classical bulge component ($\lesssim8$\,\%, \citealt{shen2010}).

The above observations points to that the bulge should not be regarded as a unique stellar population, but rather a region of the Milky Way that contains a mixture of the other Galactic stellar populations (after all the central region of the Galaxy is where all the major stellar populations overlap). This question of overlapping disk populations in the bulge has recently been addressed by \cite{dimatteo2014} who associated various parts/populations of the Galactic disk with the bulge metallicity components found by \cite{ness2013} using the ARGOS survey data. Components A ($\rm [Fe/H]=+0.21$) and B ($\rm [Fe/H]=-0.17$) originate from the (thin) disk, with the A population being formed on average closer to the Galactic centre than the B population. Component C ($\rm [Fe/H]=-0.61$) is associated with the old thick disk, and the minor components D ($\rm [Fe/H]=-1.10)$ and E ($\rm [Fe/H]=-1.60$) perhaps to the halo and/or a small classical component.

The current study confirms the existence of these metallicity peaks, and tentatively adds another peak at $\rm [Fe/H]=-0.17$. The two metal-poor peaks show age peaks around 11\,Gyr (Figs.~\ref{fig:ages2}d and e), and it is tempting to associate these age and metallicity peaks with the stellar halo and thick disk. The metallicity peak at $+0.41$\,dex shows no clear age peak but rather an extended irregular distribution in the range 2 to 8\,Gyr (Fig.~\ref{fig:ages2}a). Possibly this can be associated with the thin disk, that could have undergone variations in star formation history. The metallicity peak at $+0.12$\,dex is interesting as it shows a clear age peak at 8\,Gyr (Fig.~\ref{fig:ages2}b). This is the time in the history of the Milky Way that has been associated with a transition between the thick and thin disk eras. Could this peak be a sign of a significant episode of star formation caused by a major event about this time, such as a major merger? The ages seen for the metallicity peak at $-0.20$\,dex are hard to interpret, one very old and some much younger (Fig.~\ref{fig:ages2}c). This is also a region of the metallicity regime where thin and thick disk stars largely overlap, and hence that might be why there are duality in the ages here as well. 

In summary, all of these observations shows that the Milky Way bulge has a very complicated structure and composition but taken together they clearly point to a formation scenario for the Galactic bulge through secular evolution in which a bulge and bar forms from buckling instabilities in the disk \citep[e.g.][]{combes1990,kormendy2004,athanassoula2005,ness2012}. However, maybe a more realistic possibility includes several scenarios. For instance, a low-mass classical bulge could have been lost in subsequent secular evolution \citep{saha2015}, and mixed scenarios has also been observed in other galaxies \citep[e.g.][]{prugniel2001}, and is predicted from theoretical models \citep[e.g.][]{tsujimoto2012,grieco2012}. An interesting observation is that of Milky Way-like galaxies in the high-redshift Universe, that has shown that the central regions (the bulges) form and evolve in lockstep with the disks \citep{vandokkum2013}, which matches very well the current understanding for a secular origin of the Milky Way bulge.

\section{Summary}

Using gravitational microlensing as Nature's magnifying glass we have obtained high-resolution spectra for a total of 91 dwarf, turn-off, and subgiant stars in the Galactic bulge. This has allowed us to study the Galactic bulge at a level of detail that has previously not been possible. In addition to high-precision abundances we have been able to estimate individual ages for all 91 stars, which currently form the only bulge sample with ages for individual stars. One of the targets, OGLE-2013-BLG-0911S is most likely not located within the bulge region and has been excluded from our final bulge sample (see Sect.~\ref{sec:ob130911}). This object is important as it demonstrates that we are able to identify a disk target located outside the bulge region when we see one. Hence, our findings based on the final sample of 90 microlensed bulge dwarfs are: 

\begin{enumerate}
\item We have shown that the bulge metallicity distribution is very wide and that the underlying population does not have a smooth distribution, but is dominated by several peaks. We identify five peaks at $\rm [Fe/H] = +0.41$, $+0.12$, $-0.20$, $-0.63$, and $-1.09$. Four of these peaks align with the peaks found in the ARGOS survey \citep{ness2013}. The peak at $\rm [Fe/H] = +0.12$ was not seen in the ARGOS data. 
\item We find that the bulge has a very wide age distribution. At metallicities below $\rm [Fe/H]\approx-0.5$ essentially all stars are older than 10\,Gyr. At higher metallicities the stars span all possible ages, from the youngest at around 1\,Gyr, to the oldest around 12-13\,Gyr. The fraction of young stars ($<8$\,Gyr) increases with metallicity. Below $\rm [Fe/H]\approx -0.5$ essentially all stars are old, below solar but more metal-rich than $-0.5$\,dex, the fraction is around 20\,\%, and above solar metallicity more than one third of the stars are younger than 8\,Gyr. 
\item We have found first indications that the star formation history of the bulge shows several peaks, with major episodes about 11, 8, 6, and 3\,Gyr ago. The two oldest ones could be associated with the two major disk populations, the onset of the thick and thin disks, while the younger ones could possibly be associated with the younger parts of the thin disk in connection with the Galactic bar.
\item The knee in the bulge $\rm [\alpha/Fe]-[Fe/H]$ abundance trends  appears to be located at slightly higher [Fe/H], about 0.1\,dex, than what is observed in the local thick disk. If the metal-poor bulge is connected to the thick disk, this means that the star formation rate was faster in the inner parts of the thick disk. This finding is tentative and needs further confirmation.
\item We suggest that the $(V-I)_{0}$ colour of the bulge red clump should be revised to 1.09.
\end{enumerate}

These findings, together with other findings such as the cylindrical rotation \citep{kunder2012}, points to a secular origin for the Galactic bulge. It cannot completely rule out a small contribution of a classical bulge, which maybe that the stellar halo is present also in the bulge region. The bulge is quite likely to be a conglomerate of stellar populations, i.e. it is not a unique stellar population on its own, but the central region of the Milky Way where all the other populations also reside and widely overlap. Due to buckling instabilities in the early phases of the history of the Milky Way, the Galactic bar was formed out of disk material.

We now believe that our sample is a statistically significant sample. In order to truly resolve new features in the age, metallicity, and abundance trends and distributions, we think that the sample needs to be increased by a factor of at least 5 to 10. Although the microlensing surveys themselves have substantially improved the last years and if significant efforts were put in, high-mag events could be alerted at perhaps double the rate that we found them. However, these efforts are currently unavailable. Hence, with these 90 microlensed dwarf and subgiant stars in the bulge we have ended our observing campaign in its current form.

\begin{acknowledgement}

We would like to thank Bengt Gustafsson, Bengt Edvardsson, and Kjell Eriksson for usage of the MARCS model atmosphere program and their suite of stellar abundance programs.  Lennart Lindegren and Paul McMillan are thanked for giving valuable advice on statistical methods. Ortwin Gerhard and Ken Freeman are thanked for interesting discussions about the microlensed dwarf sample during different conferences the last few years. T.B., S.F, and L.M.H. were supported by the project grant ``The New Milky Way'' from Knut and Alice Wallenberg Foundation. Work by J.C.Y. was supported by an SNSF Graduate Research Fellowship under Grant No. 2009068160. A.G. acknowledges support from AST-1516842, and J.C.Y. acknowledges support from NSF AST-1103471. M.A. gratefully acknowledges funding from the Australian Research Council (FL110100012). J.L.C. is grateful to NSF award AST-0908139 for partial support. S.L. research was partially supported by the DFG cluster of excellence `Origin and Structure of the Universe'.  J.M. thanks FAPESP (2014/18100-4) and CNPq (Bolsa de Produtividade). The OGLE Team thanks Profs. M. Kubiak and G. Pietrzy{\'n}ski, former members of the OGLE team, for their contribution to the collection of the OGLE photometric data over the past years. The OGLE project has received funding from the National Science Centre, Poland, grant MAESTRO 2014/14/A/ST9/00121 to AU. The MOA project is supported by JSPS KAKENHI Grant Number 23103002, 24253004, 26247023, 25103508 and 23340064. 
 
\end{acknowledgement}
\bibliographystyle{aa}
\bibliography{referenser}

\begin{appendix}

\section{Online tables}

We are providing two online tables. The first table (Table~\ref{tab:ews}) gives the equivalent widths and elemental abundances for individual lines in all stars as well as in the Sun. The second table (Table~\ref{tab:parameters}) gives the results, radial velocities, ages, abundance ratios, uncertainties, and microlensing parameters for the full sample of 91 stars (including OGLE-2013-BLG-0911S that is excluded from the final bulge sample, see Sect.~\ref{sec:ob130911}). Details on both tables are given below.

\begin{table}[h]
\centering
\caption{
Description of the online data table for the full sample of 91 stars. 
The table is only available in electronic form at the CDS via anonymous ftp to
\url{cdsarc.u-strasbg.fr (130.79.128.5)} or via
\url{http://cdsweb.u-strasbg.fr/cgi-bin/qcat?J/A+A/XXX/AXX}.
\label{tab:parameters}
}
\setlength{\tabcolsep}{1mm}
\footnotesize
\begin{tabular}{clll}
\hline\hline\noalign{\smallskip}
Column & Parameter & Unit & Description \\
\noalign{\smallskip}
\hline
\noalign{\smallskip}
 (1) & Object                         &                       &  \\
 (2) & RAJ2000                        & $\rm [hhmmss]$      &  \\
 (3) & DEJ2000                        & $\rm [ddmmss]$      &  \\
 (4) & $l$                            & $\rm [deg]$           &  Galactic longitude \\
 (5) & $b$                            & $\rm [deg]$           &  Galactic latitude \\
 (6) & $T_{E}$                        & $\rm [days]$          &  Duration of the event \\
 (7) & $T_{max}$                      & $\rm [HJD]$           &  Time of maximum magnification\\
 (8) & $A_{max}$                      &                       &  Maximum magnification \\
 (9) & $T_{obs}$                      & $\rm [MJD]$           &  Time when observed \\
(10) & Exp.                           & $\rm [s]$             &  Exposure time \\
(11) & $S/N$                          &                       &  $S/N$ per pixel at $\sim 6400$\,{\AA}\\
(11) & $S/N$                          &                       &  $S/N$ per pixel at $\sim 8000$\,{\AA}\\
(12) & Spec.                          &                       &  Spectrograph that was used\\
(13) & $R$                            &                       &  Spectral resolving power\\
(14) & $T_{\rm eff}$                  & $\rm [K]$             &  Effective temperature \\
(15) & $\log g$                       & $\rm [cgs]$           &  Surface gravity \\
(16) & $\xi_{\rm t}$                  & $\rm [\kms]$          &  Microturbulence parameter \\
(17) & $\rm [Fe/H]$                   &                       &  Iron abundance normalised to Sun\\
(18) & N(\ion{Fe}{i})                 &                       &  Number of \ion{Fe}{i} lines used \\
(19) & N(\ion{Fe}{ii})                &                       &  Number of \ion{Fe}{ii} lines used\\
(20) & $v_{\rm r}$                    & $\rm [km/s]$          &  Heliocentric radial velocity \\
     &                                &                       &  Galactocentric radial velocity \\
(22) & $\mathcal{M}$                  & $M_{\odot}$           &  Stellar mass \\
(23) & $\log L$                       & $L_{\odot}$           &  Luminosity \\
(24) & Age                            & $\rm [Gyr]$           &  Stellar age\\
(25) & $-$1$\sigma$                   & $\rm [Gyr]$           &  Lower limit on stellar age\\
(26) & $+$1$\sigma$                   & $\rm [Gyr]$           &  Upper limit on stellar age\\
(27) & $(V$--$I)_{0}$                 & $\rm [mag]$           &  From microlensing techniques\\
(28) & $M_{\rm V}$                    & $\rm [mag]$           &  From microlensing techniques\\
(29) & $-$1$\sigma$                   & $\rm [mag]$           &  \\
(30) & $+$1$\sigma$                   & $\rm [mag]$           &  \\
(31) & $M_{\rm I}$                    & $\rm [mag]$           &  \\
(32) & $M_{\rm I,\,\mu}$              & $\rm [mag]$           &  \\
(33) & $(V$--$I)_{\rm 0,\,\mu}$       & $\rm [mag]$           &  \\
(34) & $\teff^{\mu}$                  & $\rm [K]$             &  From microlensing techniques \\
(35) & $\rm [Na/Fe]$                  &                       &  Abundance ratio \\
(36) & $\rm \epsilon [Na/Fe] $        &                       &  Uncertainty in abundance ratio\\
(37) & N(Na)                          &                       &  Number of lines\\
(38) & $\rm \sigma (Na)$              &                       &  Line-to-line scatter \\
(39) - (42) &         Mg              &                       &  Same as cols 35-39 but for Mg\\
(43) - (46) &         Al              &                       &  Same as cols 35-39 but for Al\\
(47) - (50) &         Si              &                       &  Same as cols 35-39 but for Si\\
(51) - (54) &         Ca              &                       &  Same as cols 35-39 but for Ca\\
(55) - (58) &         Ti              &                       &  Same as cols 35-39 but for Ti\\
(59) - (62) &         Cr              &                       &  Same as cols 35-39 but for Cr\\
(63) - (66) &         Ni              &                       &  Same as cols 35-39 but for Ni\\
(67) - (70) &         Zn              &                       &  Same as cols 35-39 but for Zn\\
(71) - (74) &         Y               &                       &  Same as cols 35-39 but for Y \\
(75) - (78) &         Ba              &                       &  Same as cols 35-39 but for Ba\\
\noalign{\smallskip}
\hline
\end{tabular}
\end{table}

\begin{table}[h]
\centering
\caption{
Measured equivalent widths and calculated elemental abundances for all 91 microlensing events (including OGLE-2013-BLG-0911S that is excluded from the final bulge sample, see Sect.~\ref{sec:ob130911}).\tablefootmark{$\dagger$}
\label{tab:ews}
}
\tiny
\begin{tabular}{cccccccc}
\hline\hline
\noalign{\smallskip}
Element                           &  
$\lambda$                         &
$\chi_{\rm l}$                    &
\multicolumn{2}{c}{star 1}      &
 $\cdots$                         &
\multicolumn{2}{c}{star 91}       \\
\noalign{\smallskip}
                                  &
[{\AA}]                           &
[eV]                              &
$W_{\rm \lambda}$                 &
$\epsilon (X)$                    &
  $\cdots$                        &
$W_{\rm \lambda}$                 &
$\epsilon (X)$                    \\
\noalign{\smallskip}
\hline
\noalign{\smallskip}
\vdots &
\vdots &
\vdots &
\vdots &
\vdots &
$\cdots$ &
\vdots &
\vdots \\
\noalign{\smallskip}
\hline
\end{tabular}
\tablefoot{
\tablefoottext{$\dagger$}{
For each line we give the $\log gf$ value, lower excitation energy ($\chi_{\rm l}$), equivalent width ($W_{\rm \lambda}$),
absolute abundance ($\log \epsilon (X)$).
The table is only available in electronic form at the CDS via anonymous ftp to 
\url{cdsarc.u-strasbg.fr (130.79.128.5)} or via 
\url{http://cdsweb.u-strasbg.fr/cgi-bin/qcat?J/A+A/XXX/AXX}.
}}
\end{table}

\clearpage
\onecolumn
\section{Microlensing light curves (Figure~\ref{fig:lightcurves})}

\begin{figure*}[ht]
\resizebox{\hsize}{!}{
\includegraphics[viewport=-10 40 515 535,clip]{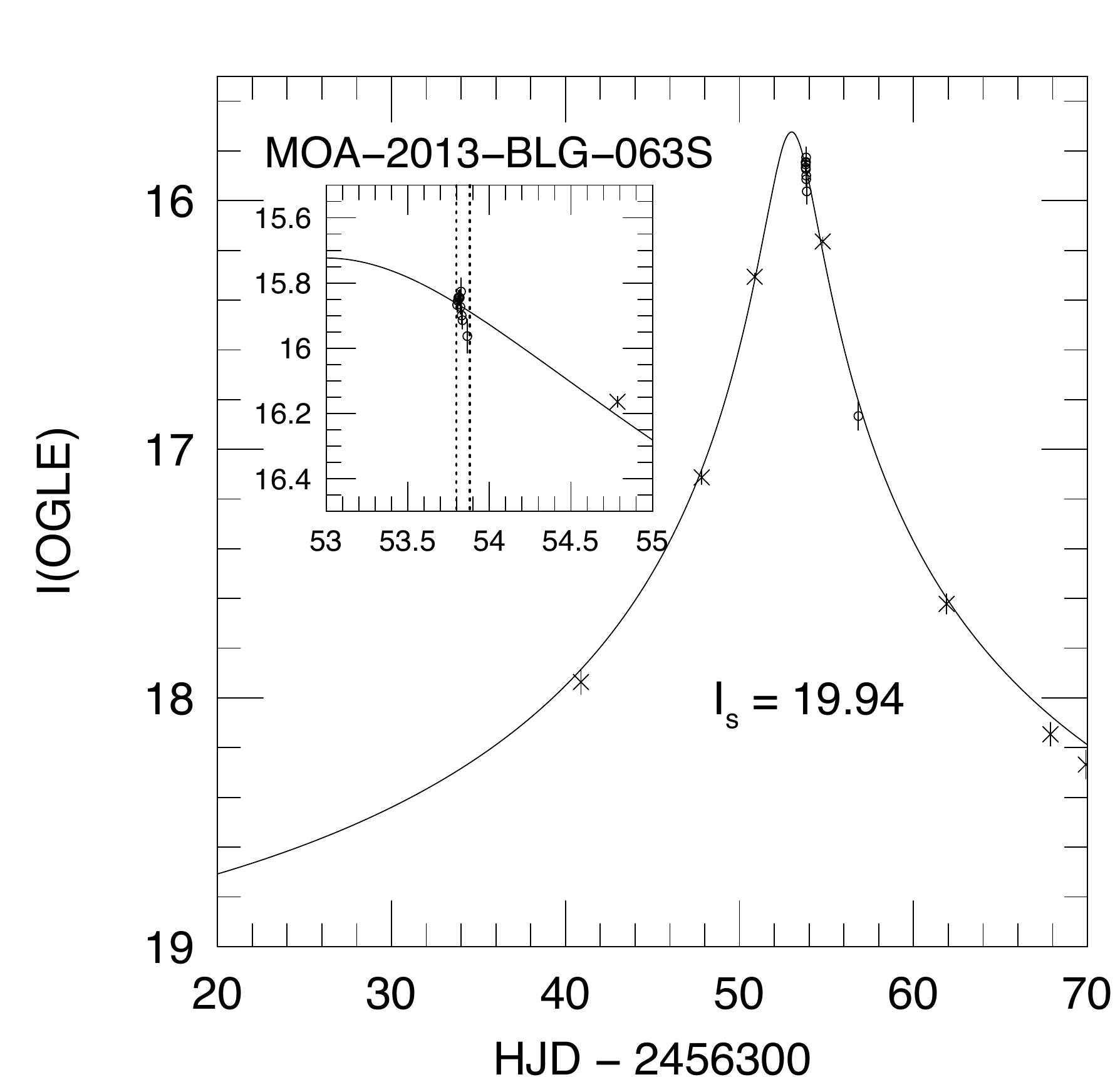}
\includegraphics[viewport= 67 40 515 535,clip]{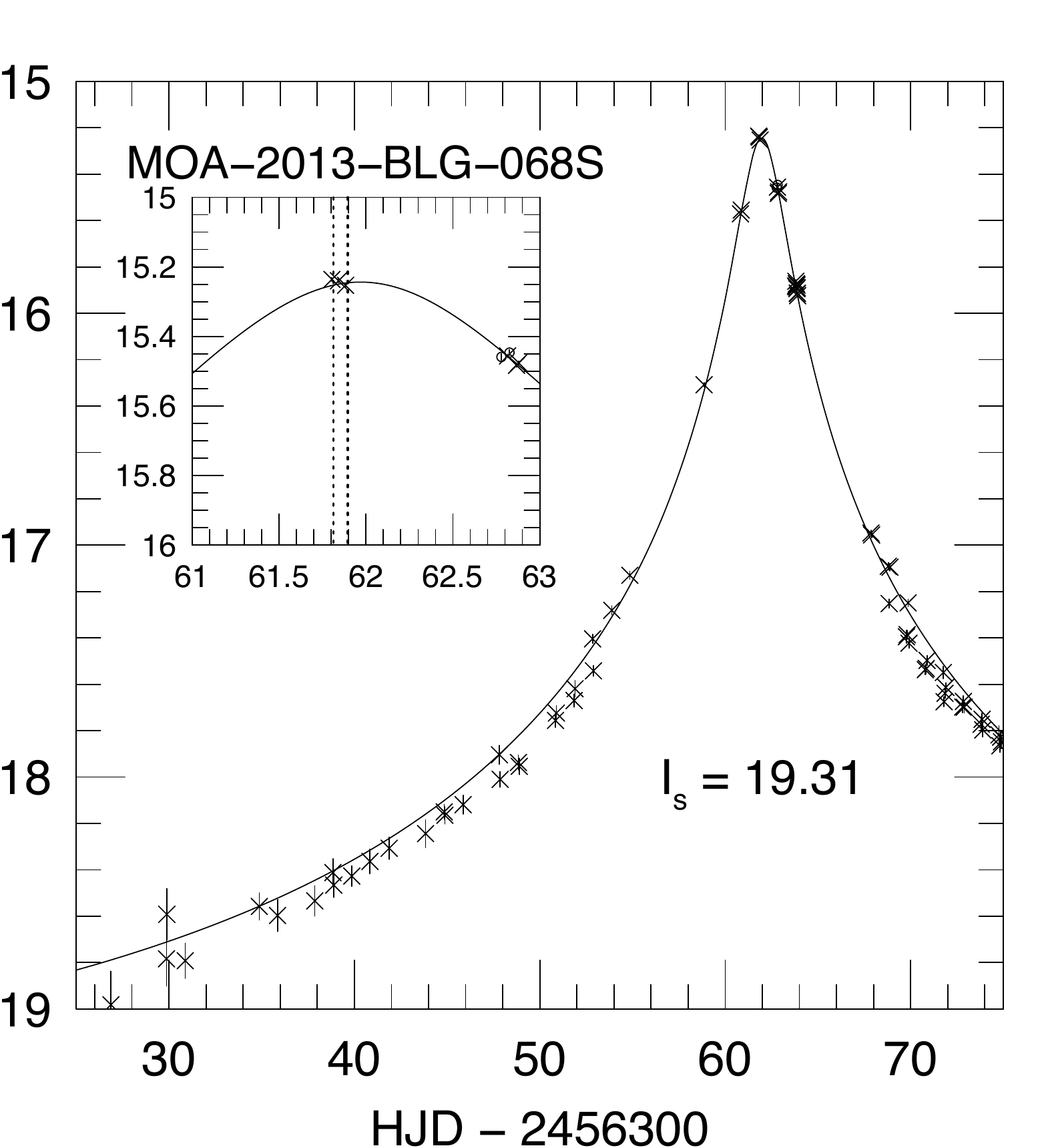}
\includegraphics[viewport= 67 40 515 535,clip]{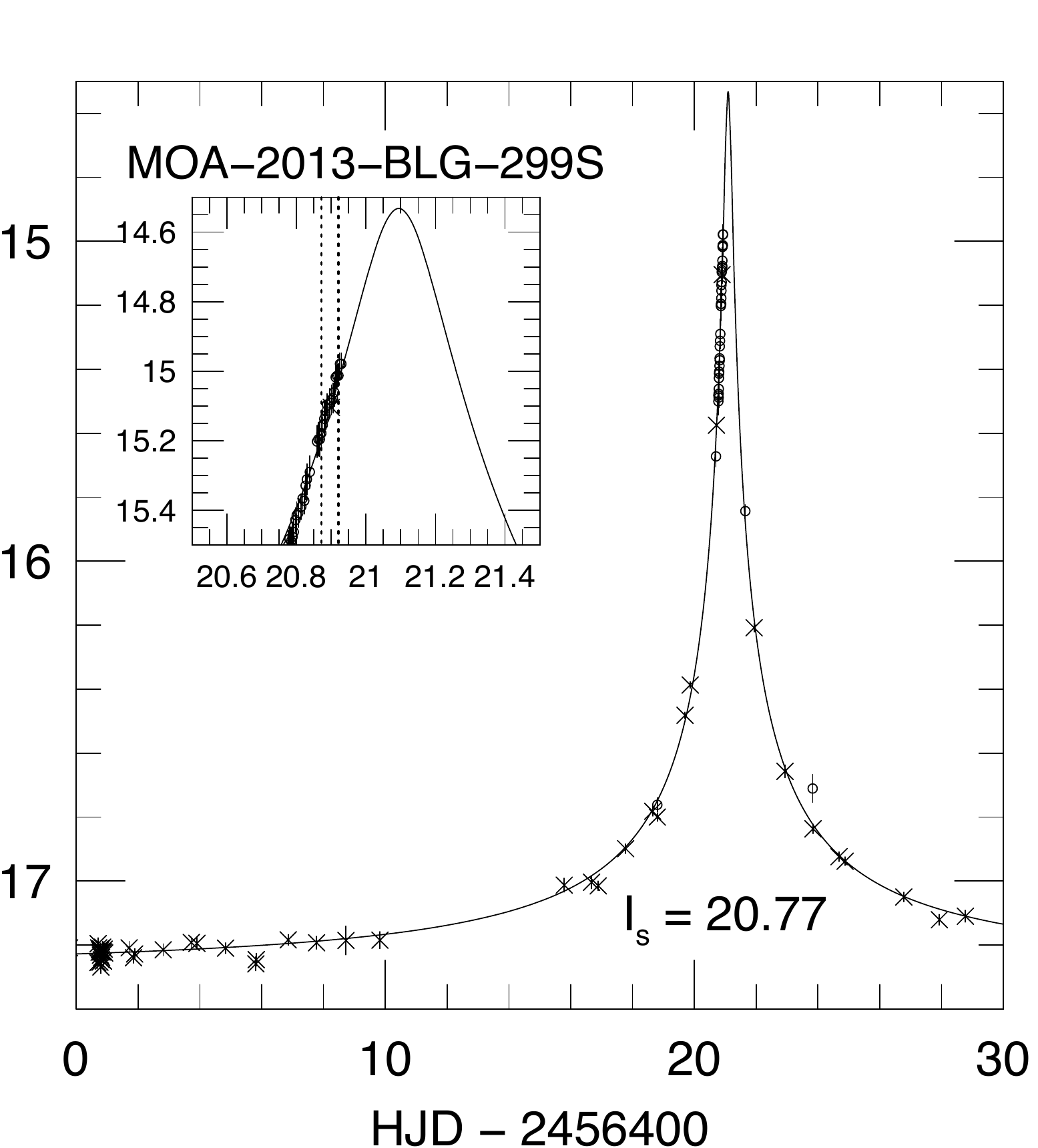}
\includegraphics[viewport= 67 40 525 535,clip]{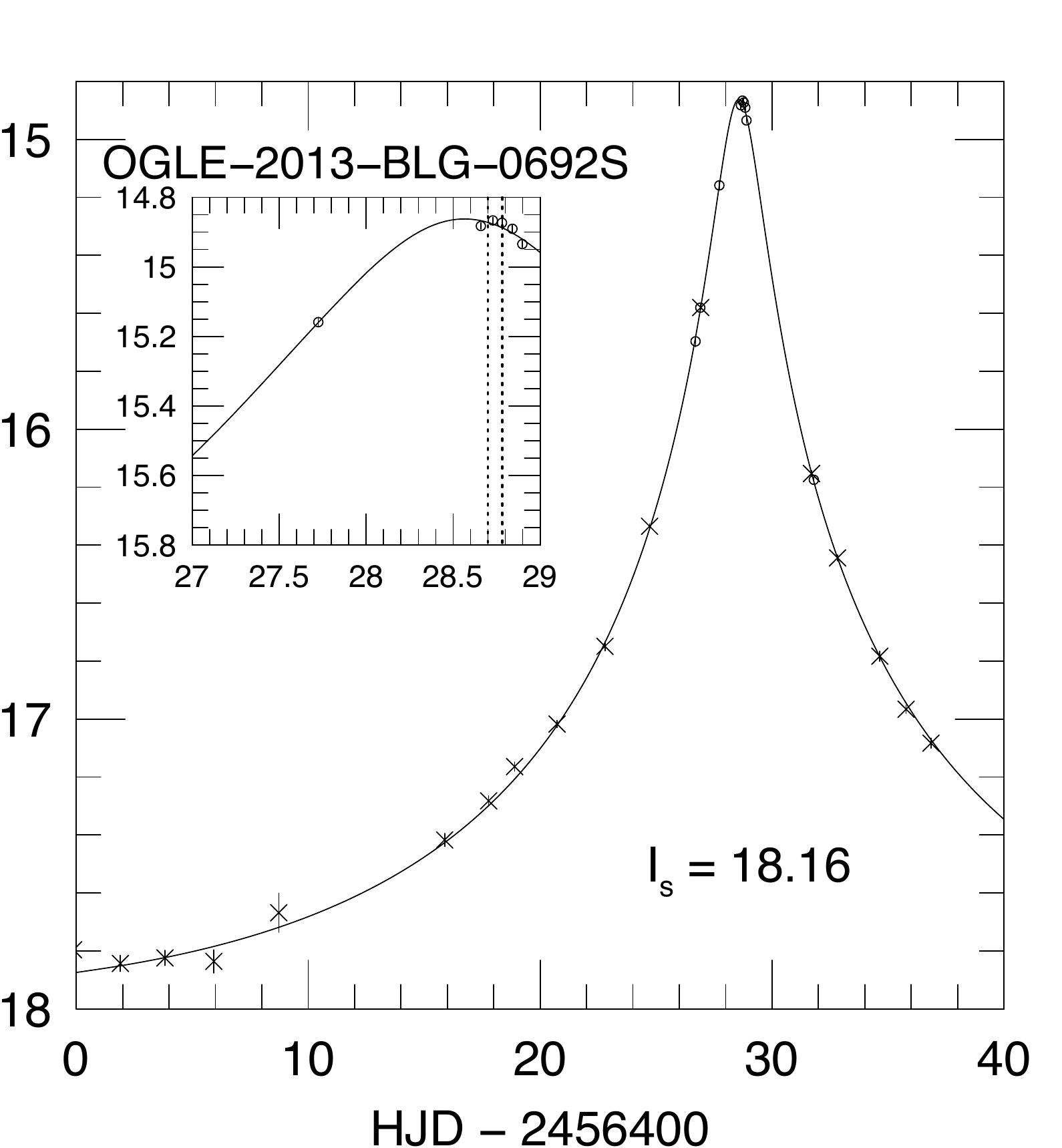}}
\resizebox{\hsize}{!}{
\includegraphics[viewport=-10 40 515 525,clip]{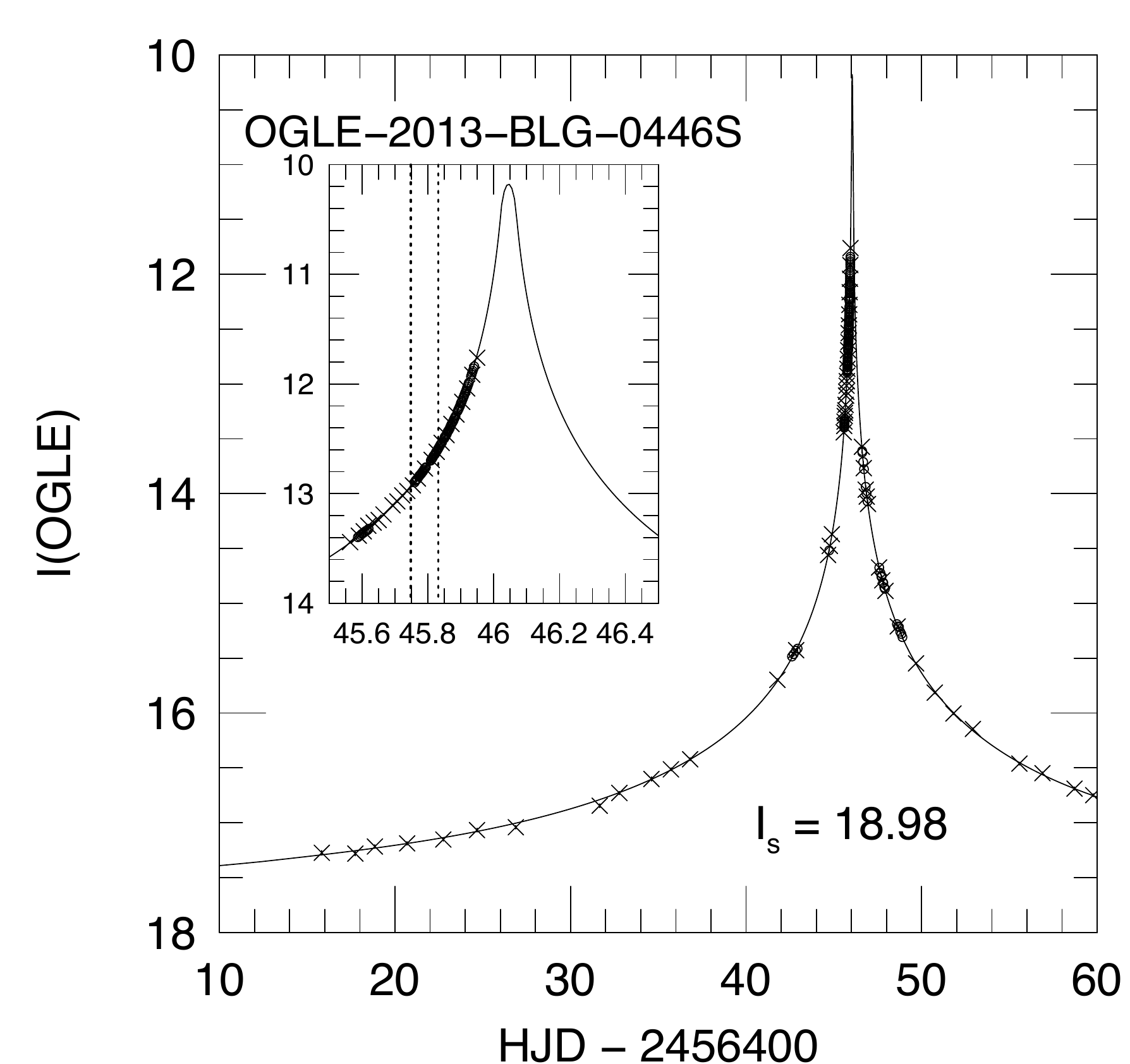}
\includegraphics[viewport= 67 40 515 525,clip]{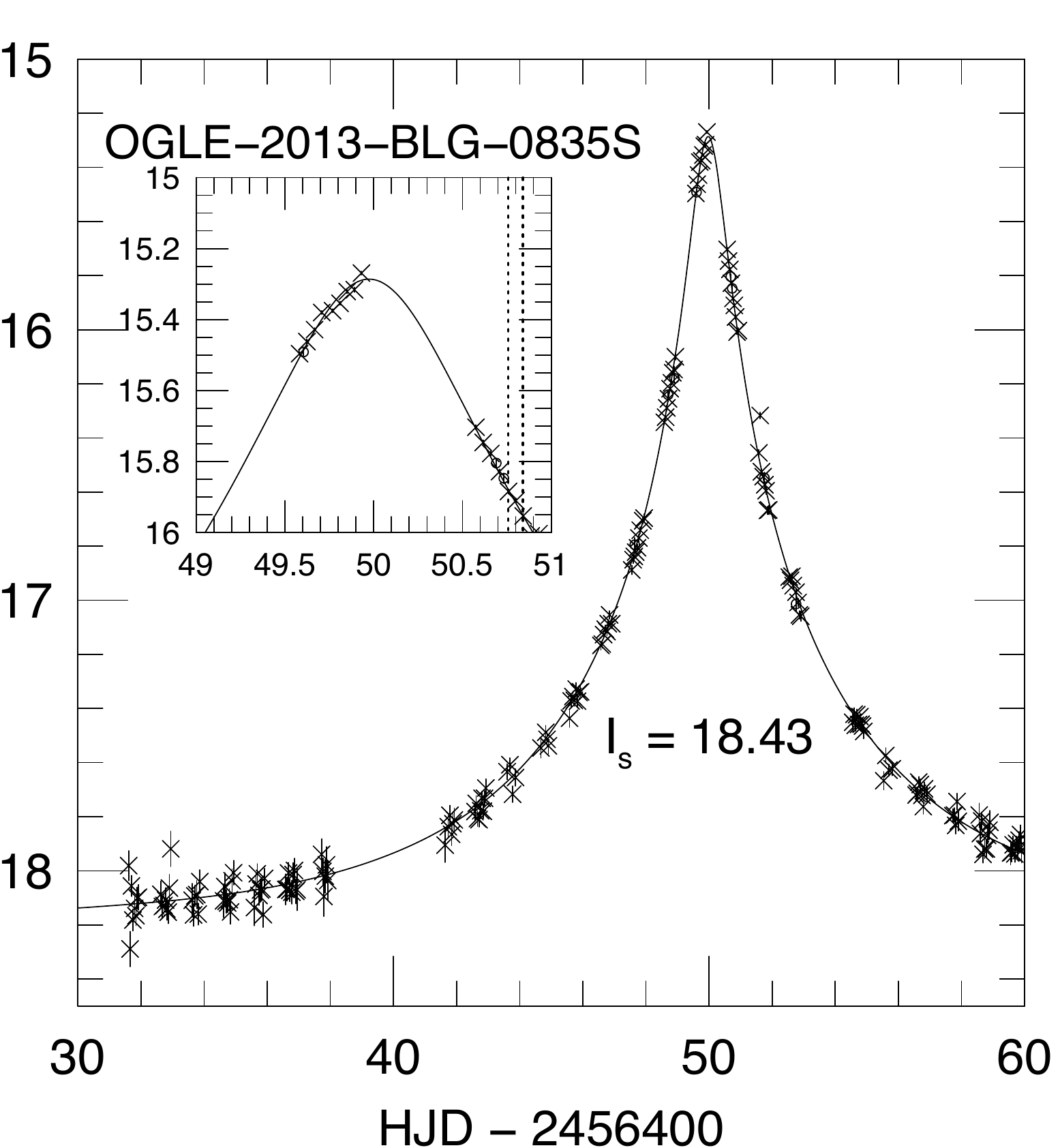}
\includegraphics[viewport= 67 40 515 525,clip]{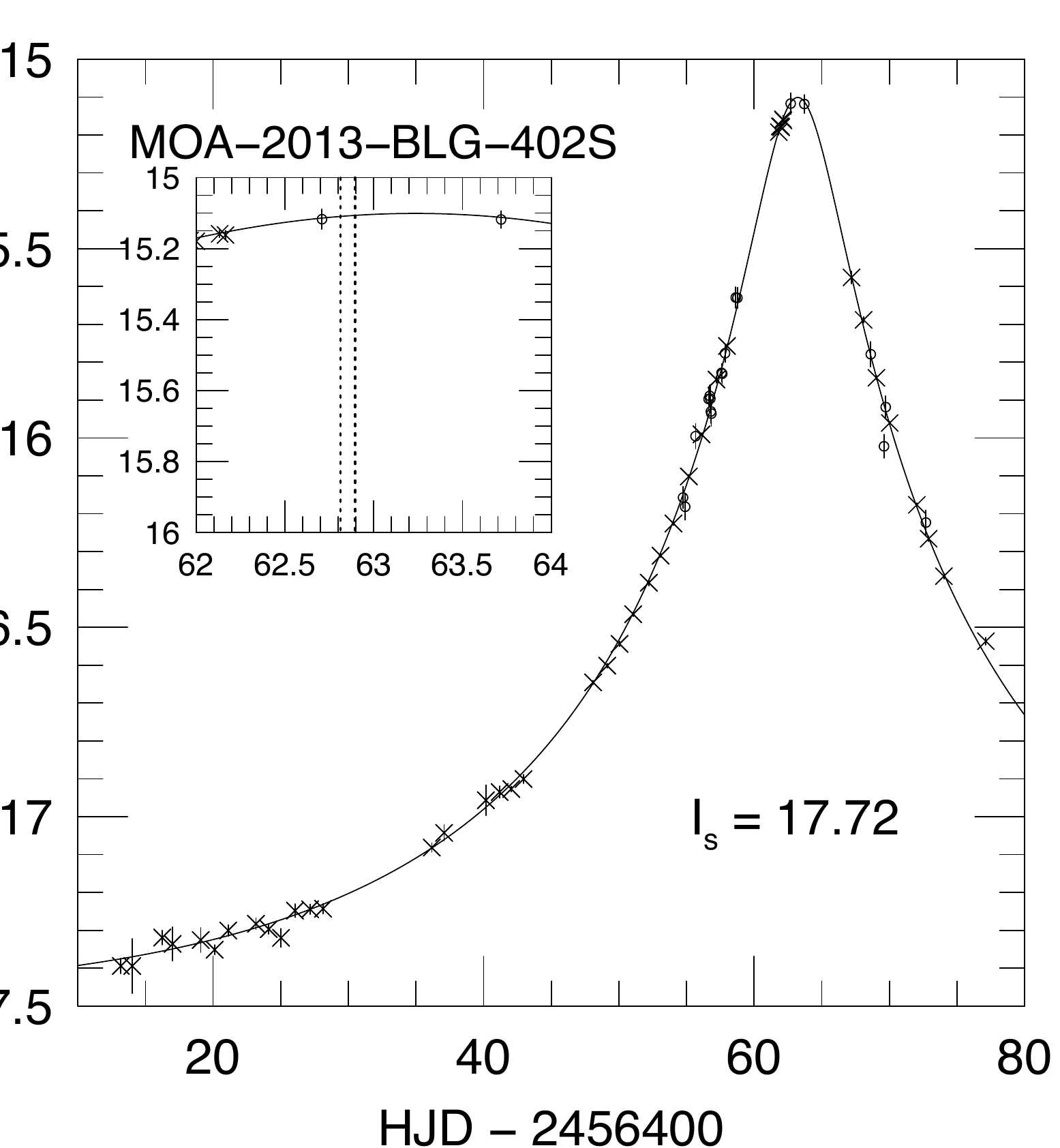}
\includegraphics[viewport= 67 40 525 525,clip]{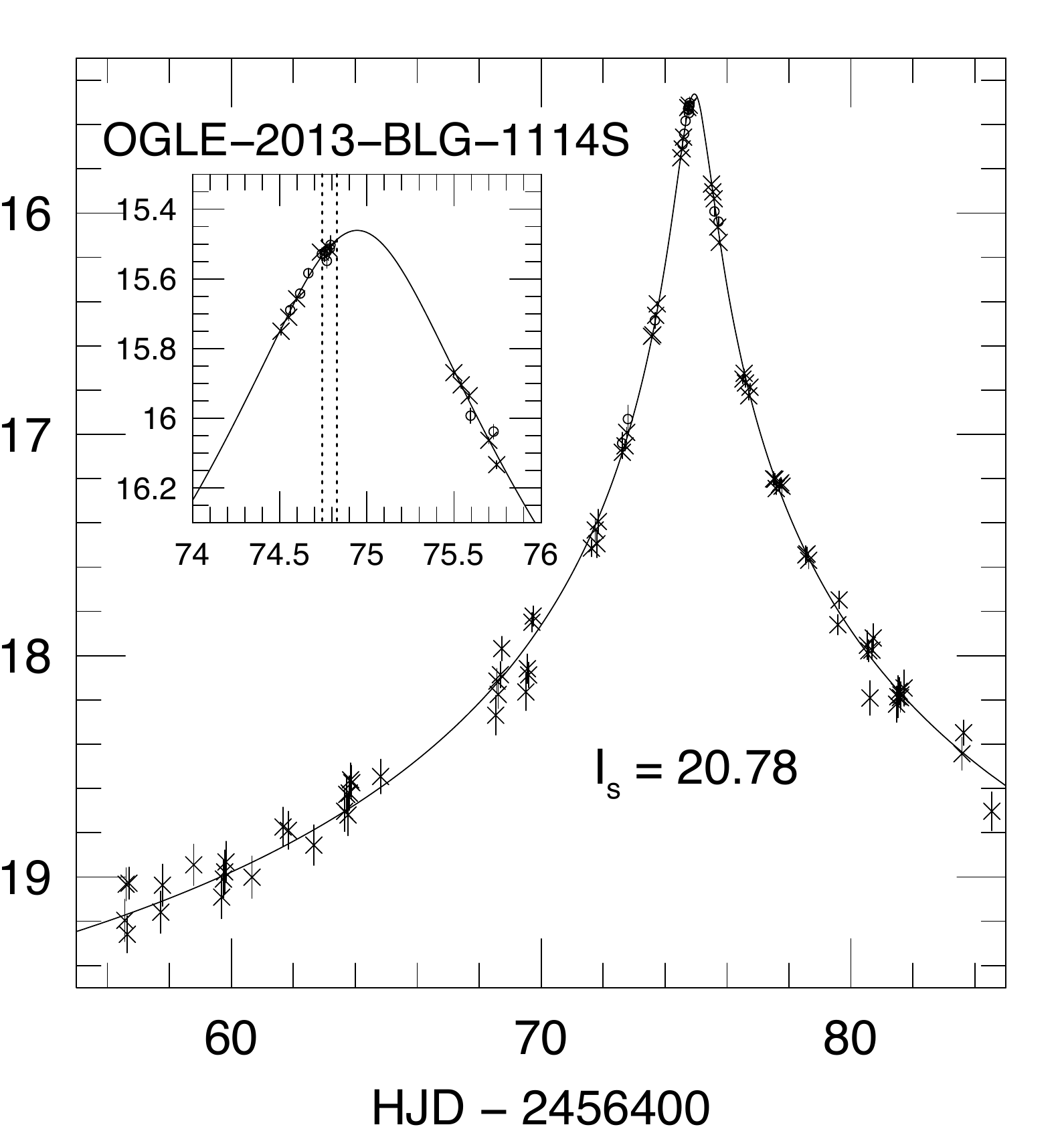}}
\resizebox{\hsize}{!}{
\includegraphics[viewport=-10 40 515 525,clip]{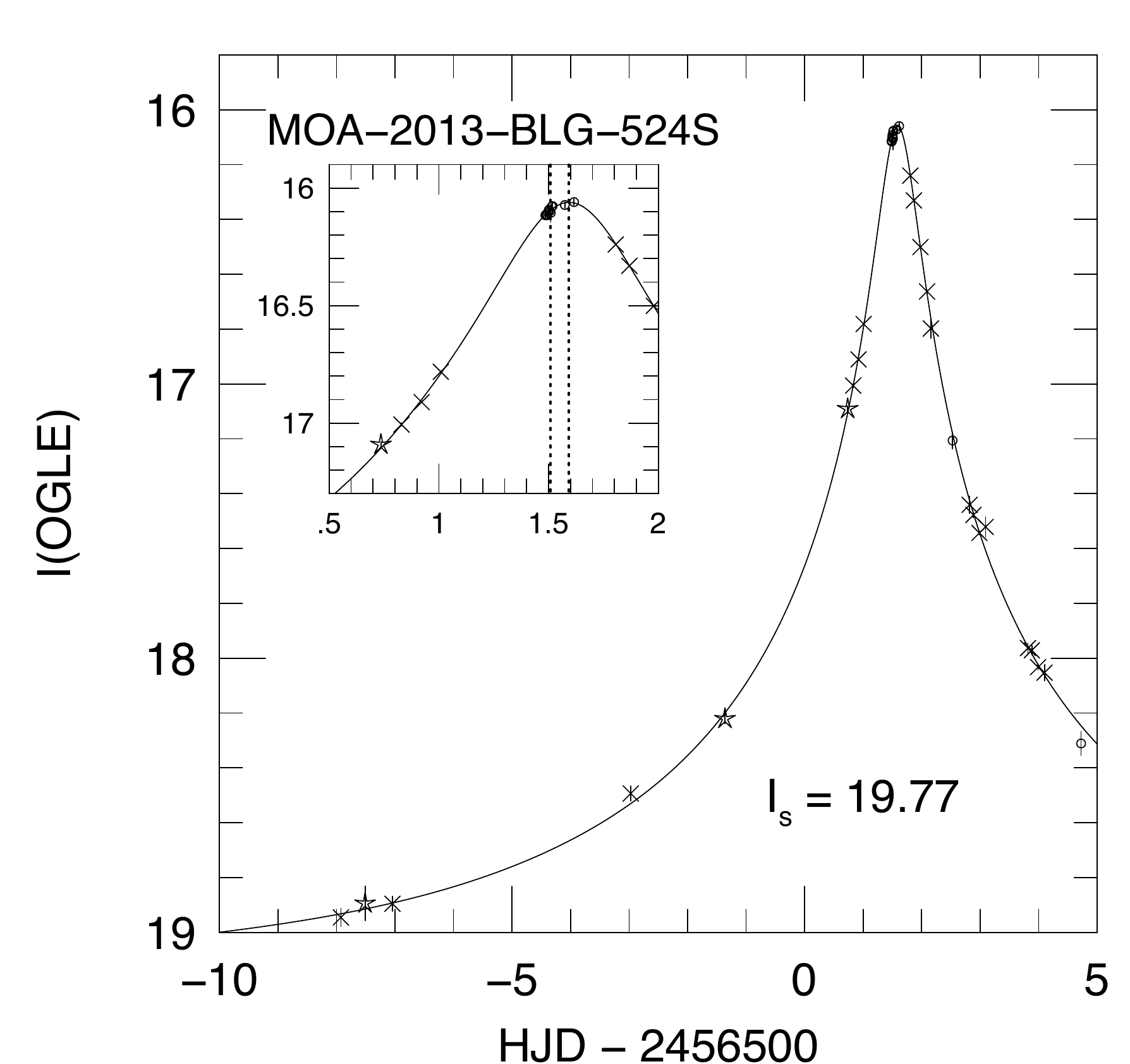}
\includegraphics[viewport= 67 40 515 525,clip]{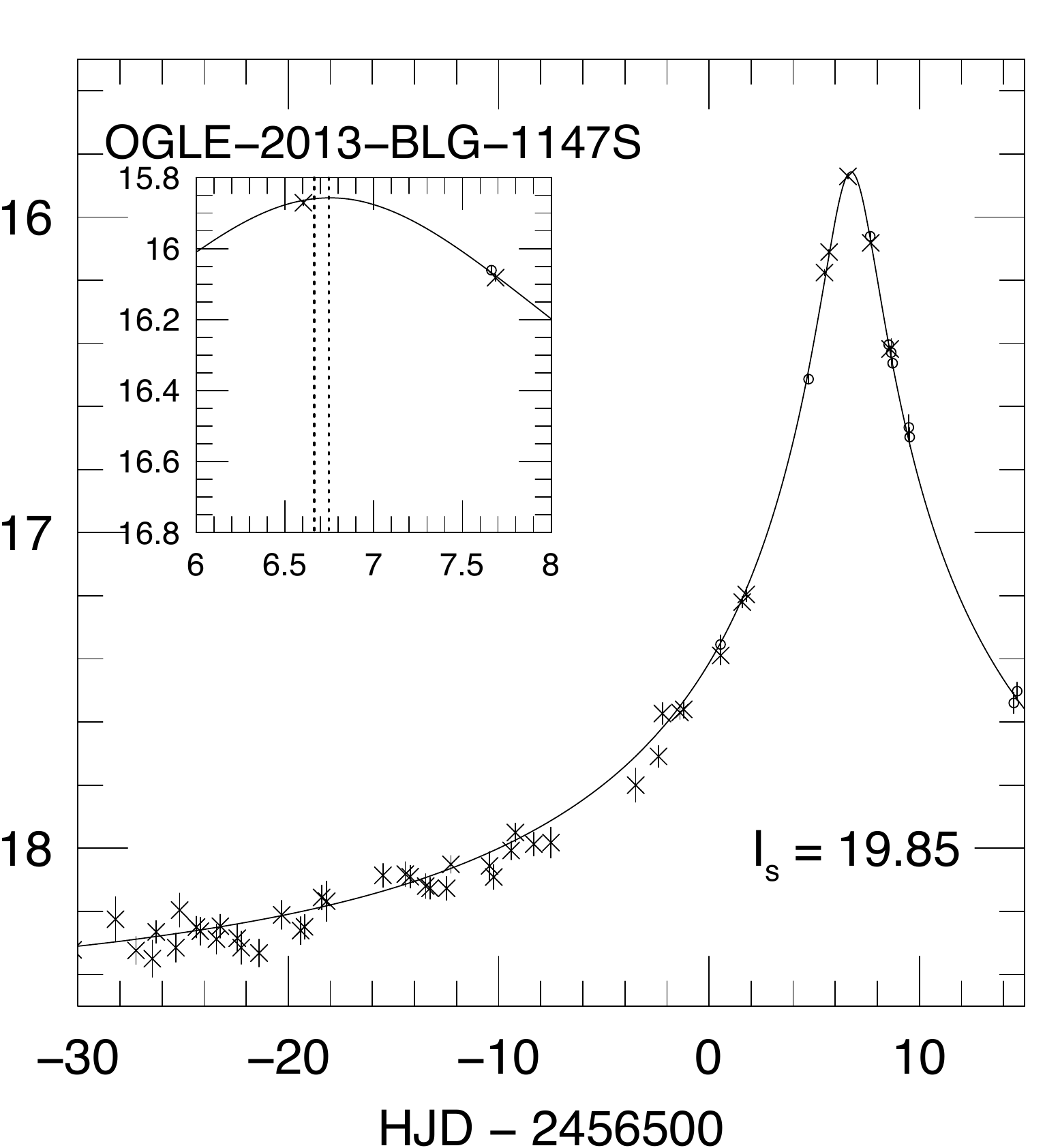}
\includegraphics[viewport= 67 40 515 525,clip]{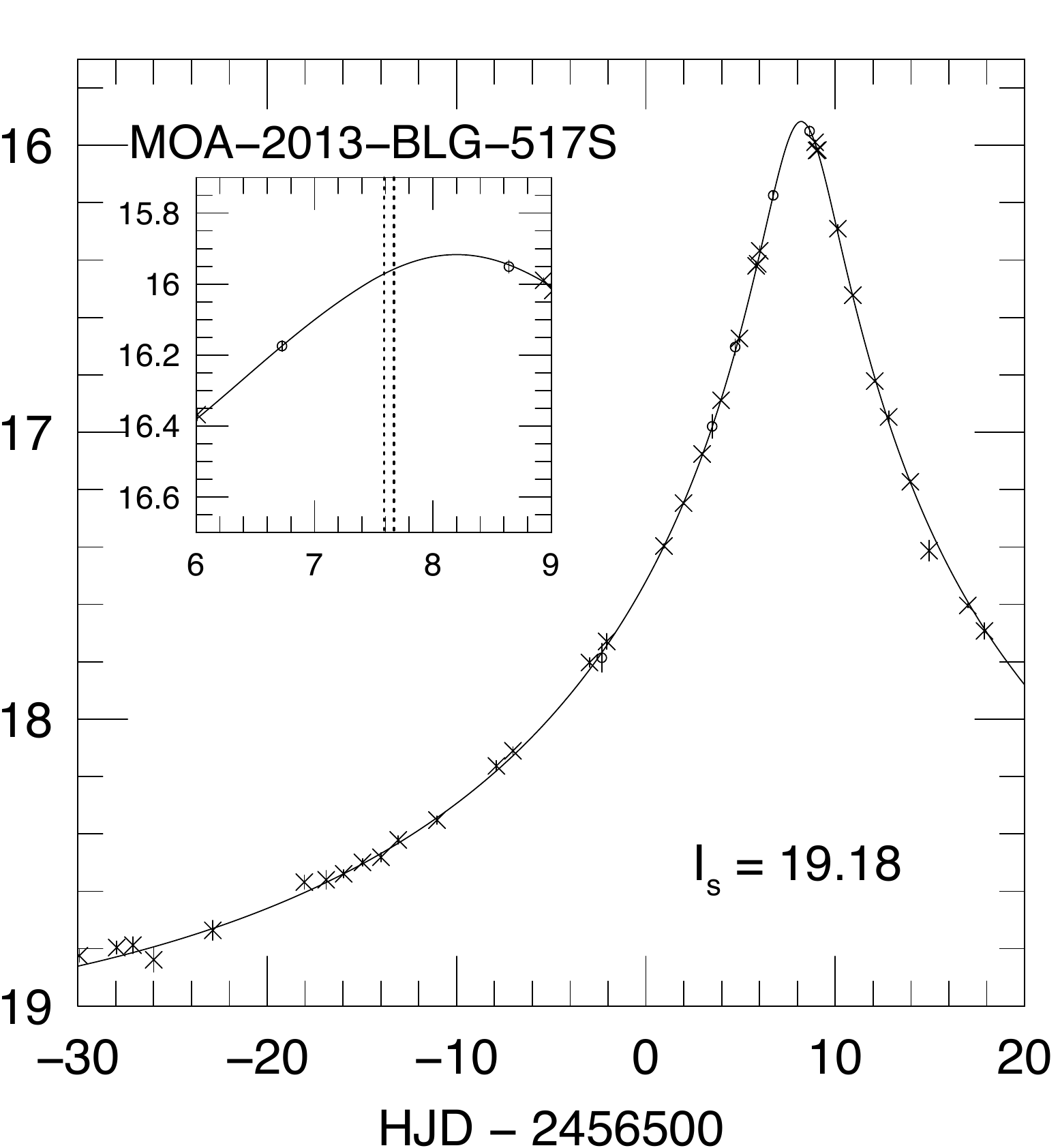}
\includegraphics[viewport= 67 40 525 525,clip]{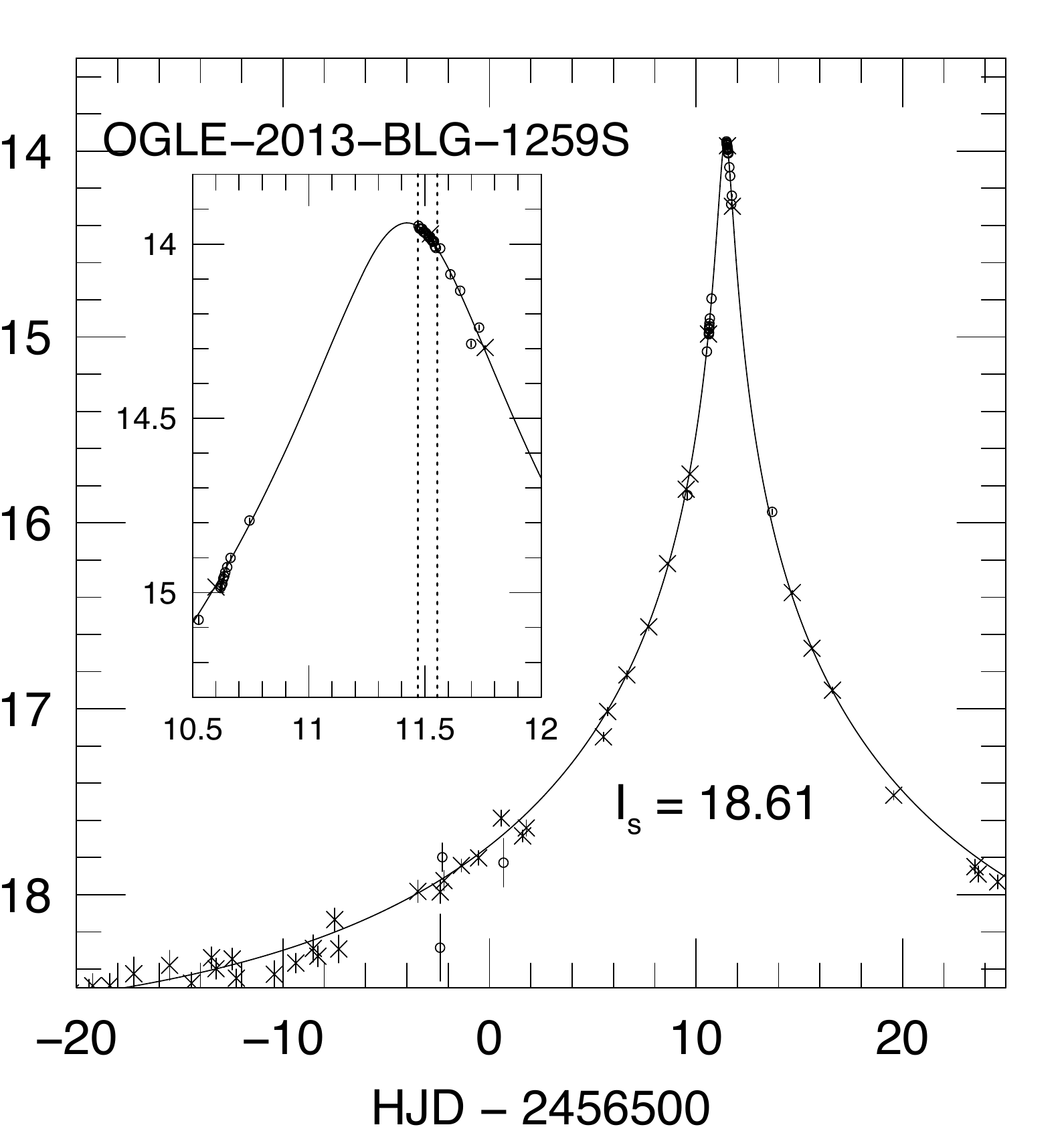}}
\resizebox{\hsize}{!}{
\includegraphics[viewport=-10 40 515 525,clip]{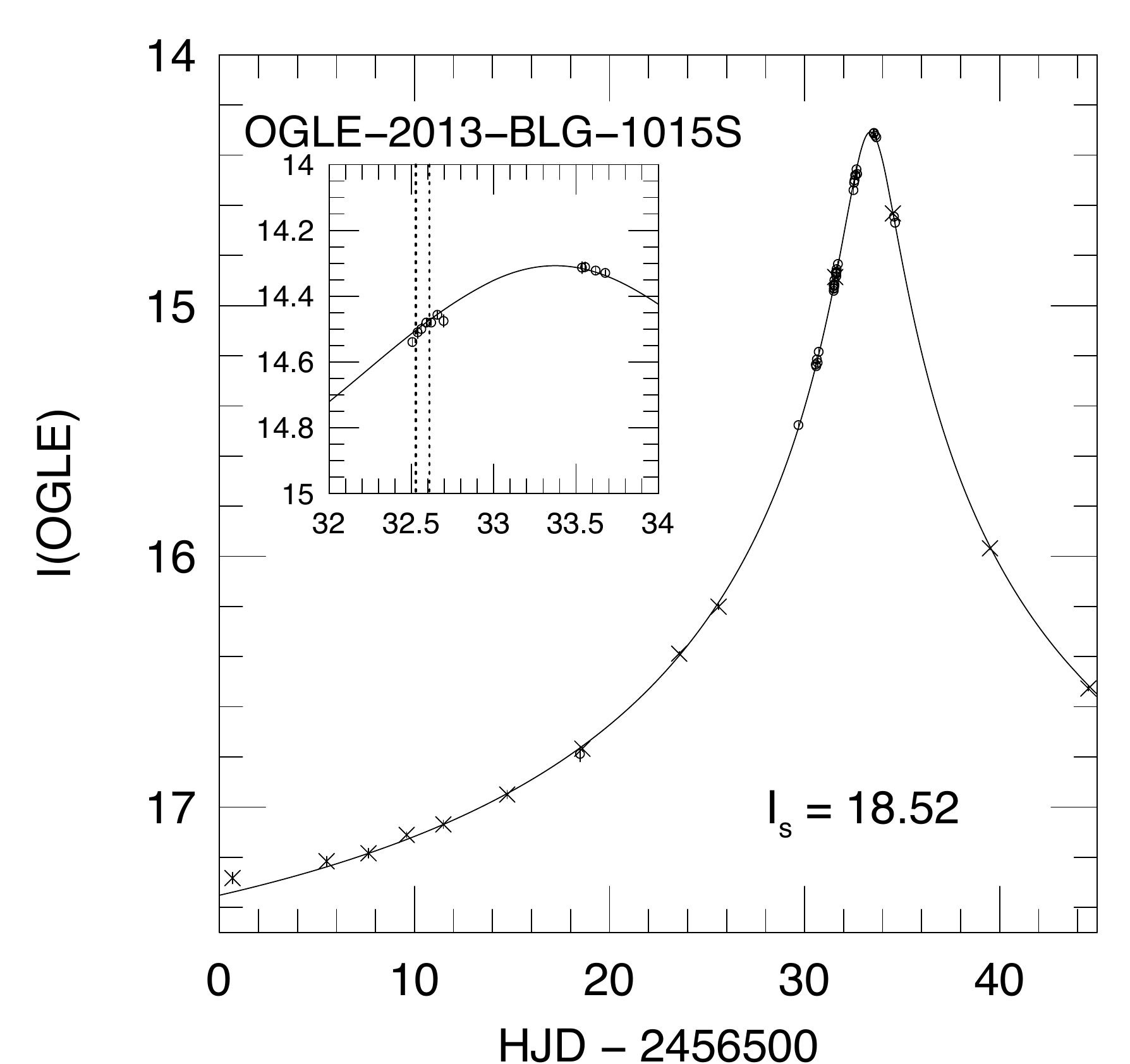}
\includegraphics[viewport= 67 40 515 525,clip]{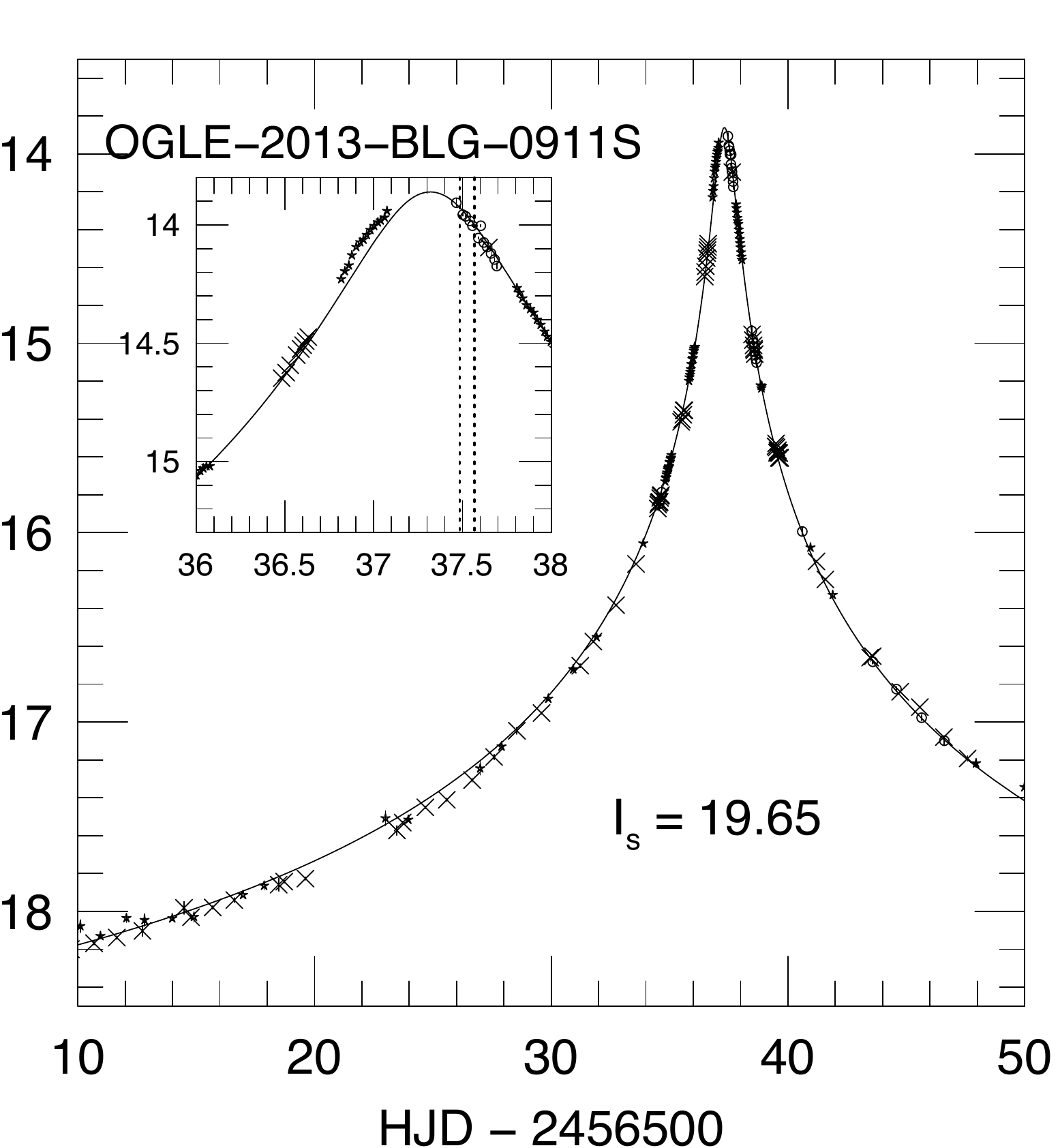}
\includegraphics[viewport= 67 40 515 525,clip]{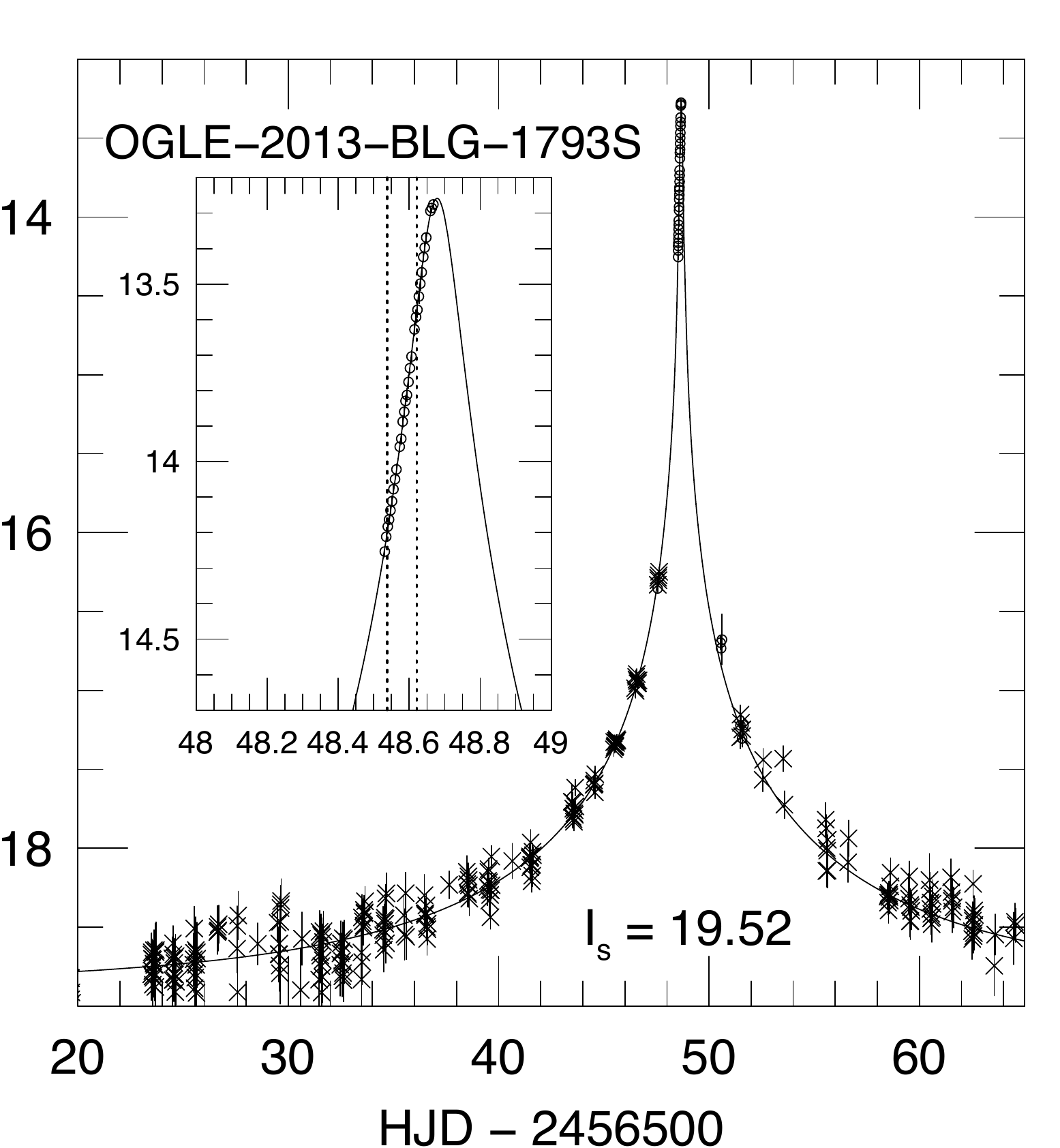}
\includegraphics[viewport= 67 40 525 525,clip]{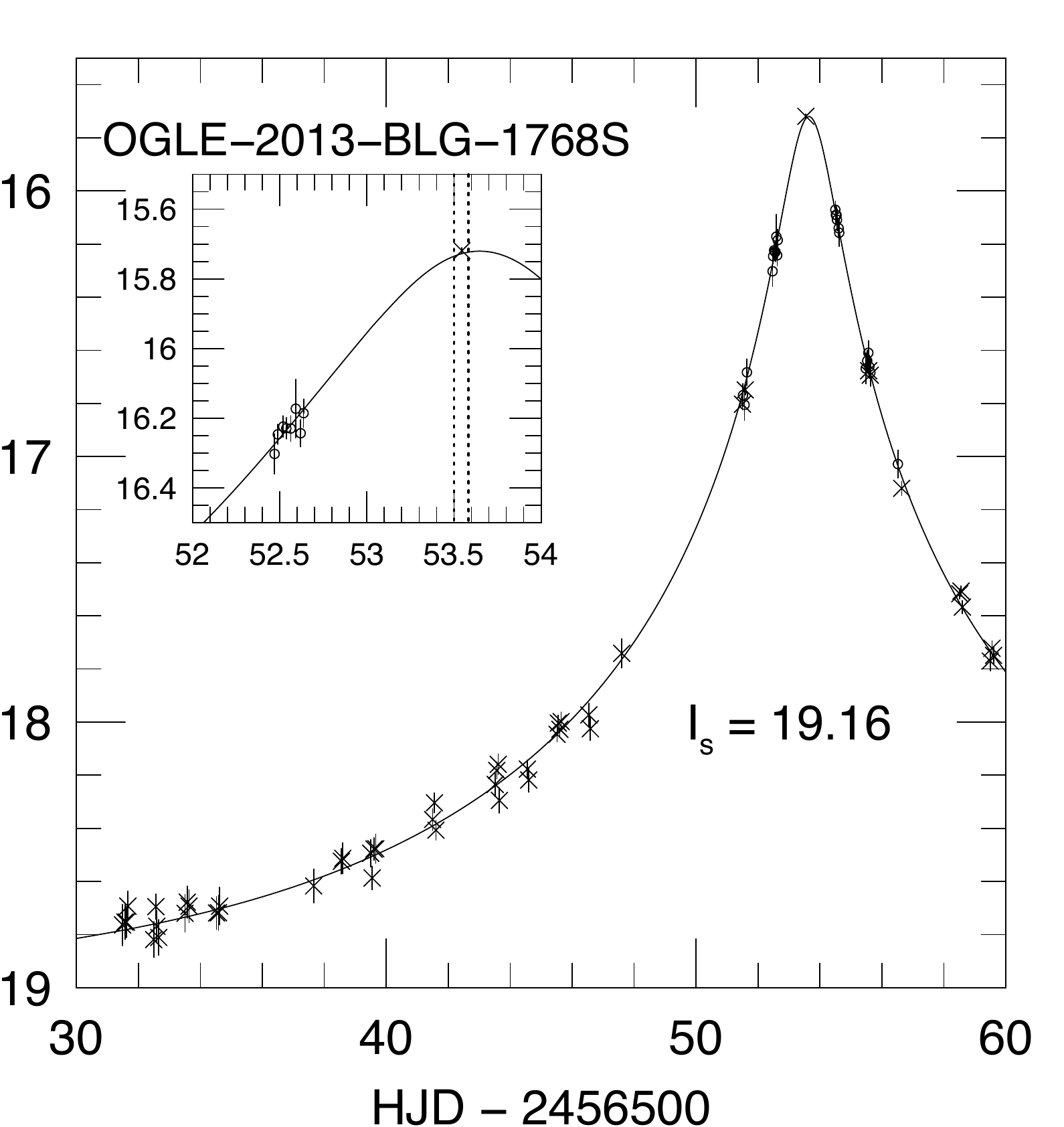}
}
\caption{Light curves for the 33 new microlensing events (including OGLE-2013-BLG-0911S that is excluded from the final bulge sample, see Sect.~\ref{sec:ob130911}). Each plot has a zoom window, showing the time intervals when the source stars were observed with high-resolution spectrographs. In each plot the un-lensed magnitude of the source star is also given ($I_{\rm S}$).
\label{fig:lightcurves}
}
\end{figure*}
\setcounter{figure}{0}    

\begin{figure*}[ht]
\resizebox{\hsize}{!}{
\includegraphics[viewport=-10 40 515 525,clip]{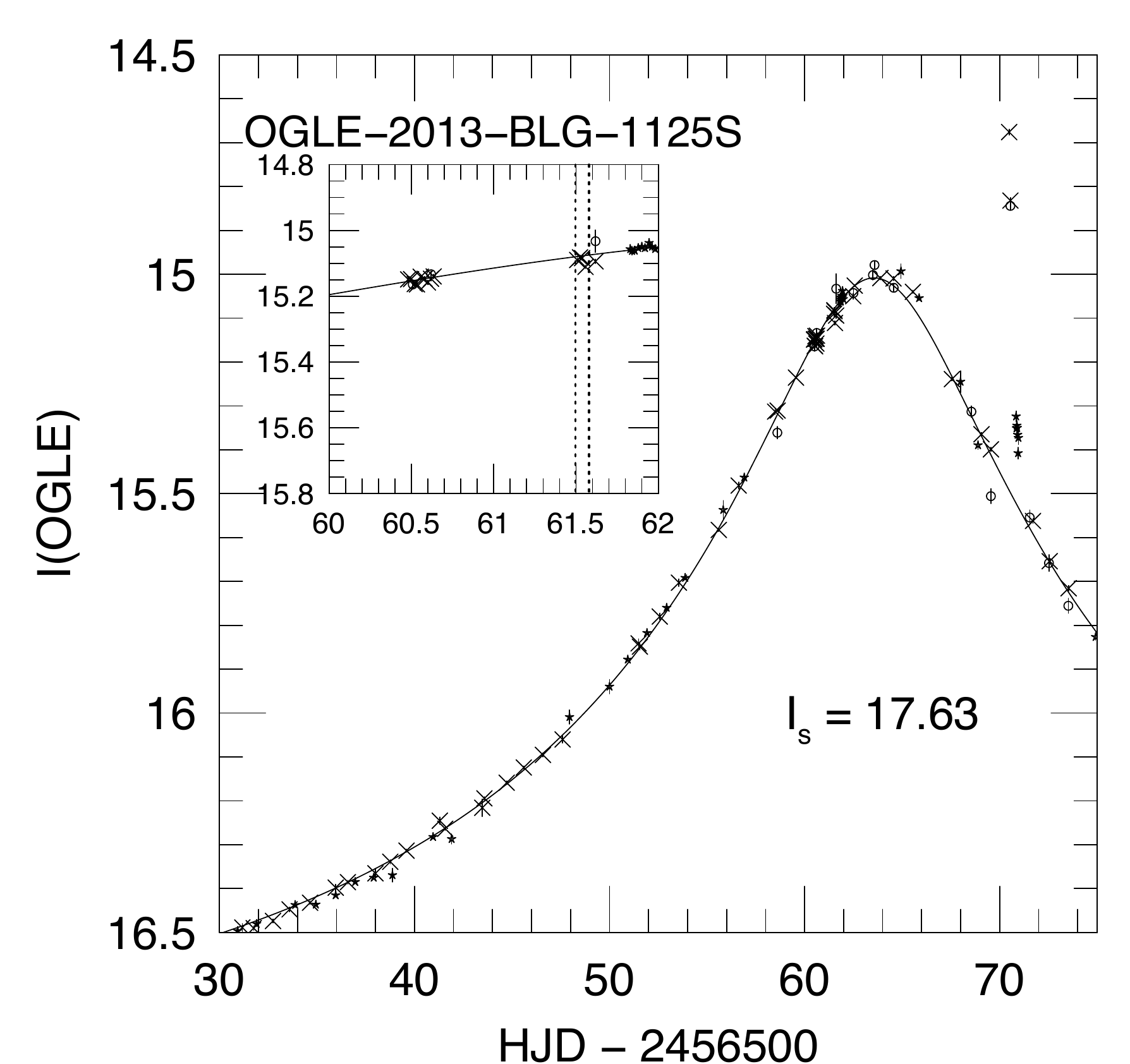}
\includegraphics[viewport= 67 40 515 525,clip]{lcb_mb13605.pdf}
\includegraphics[viewport= 67 40 515 525,clip]{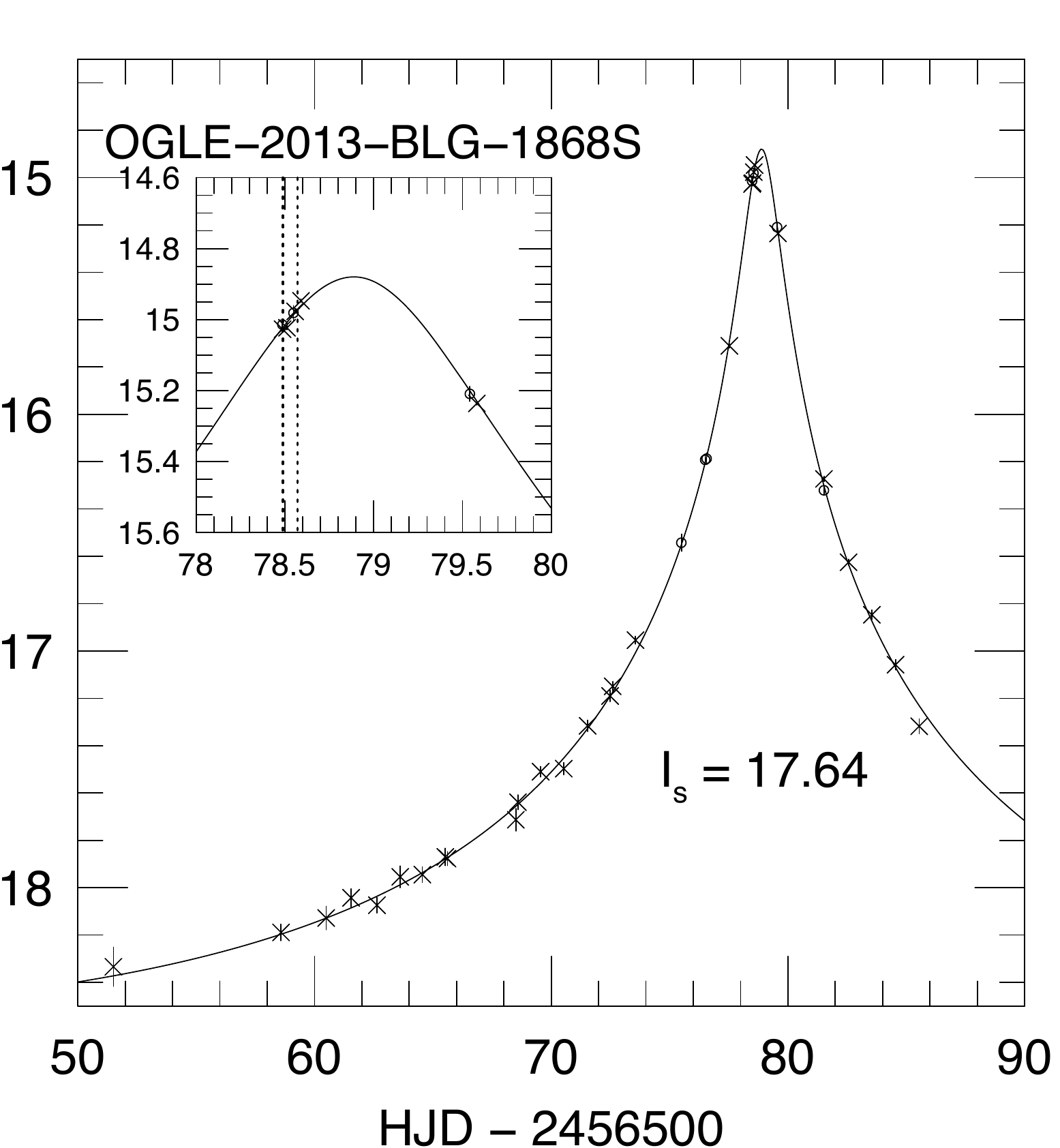}
\includegraphics[viewport= 67 40 525 525,clip]{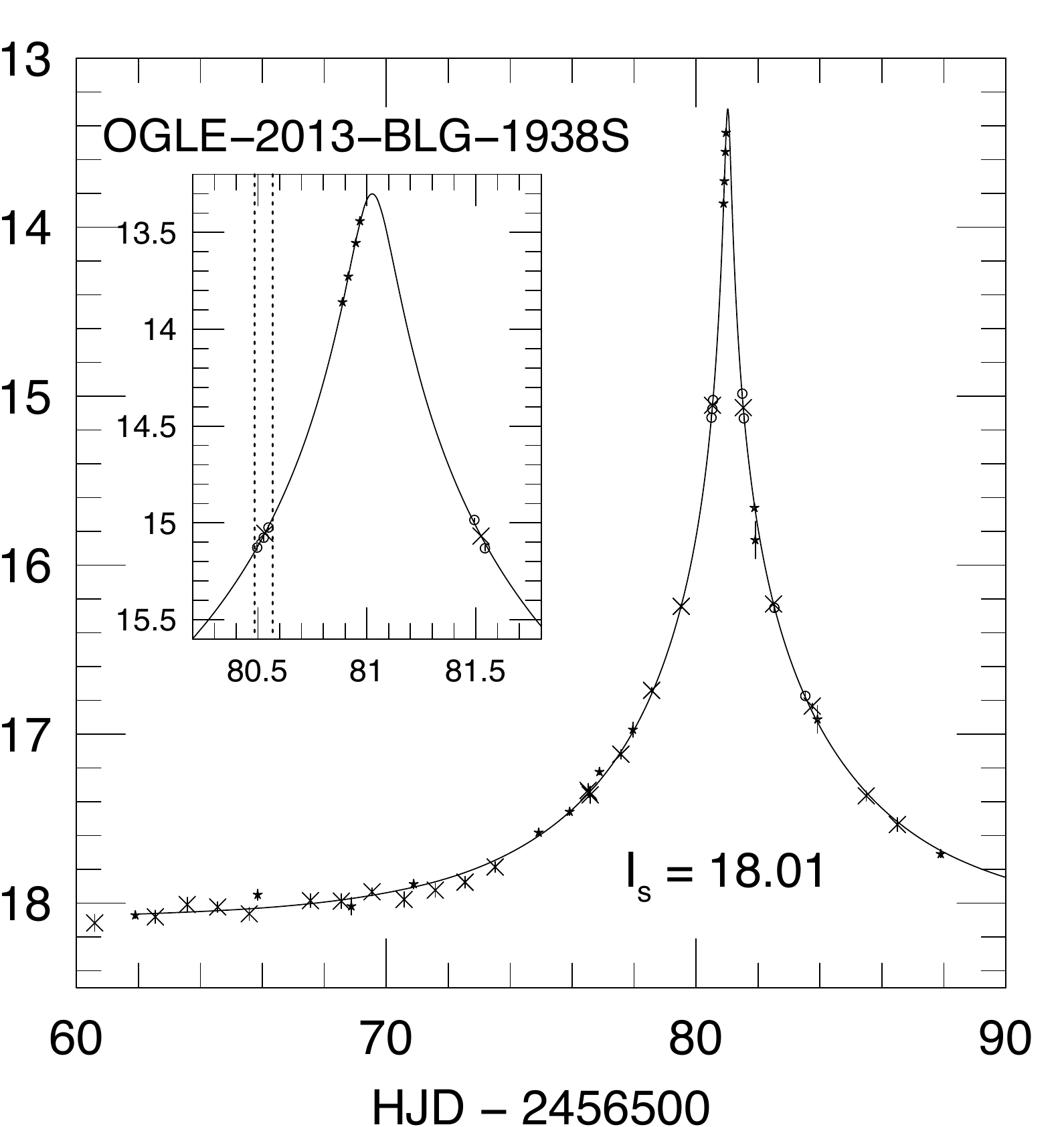}}
\resizebox{\hsize}{!}{
\includegraphics[viewport=-10 40 515 535,clip]{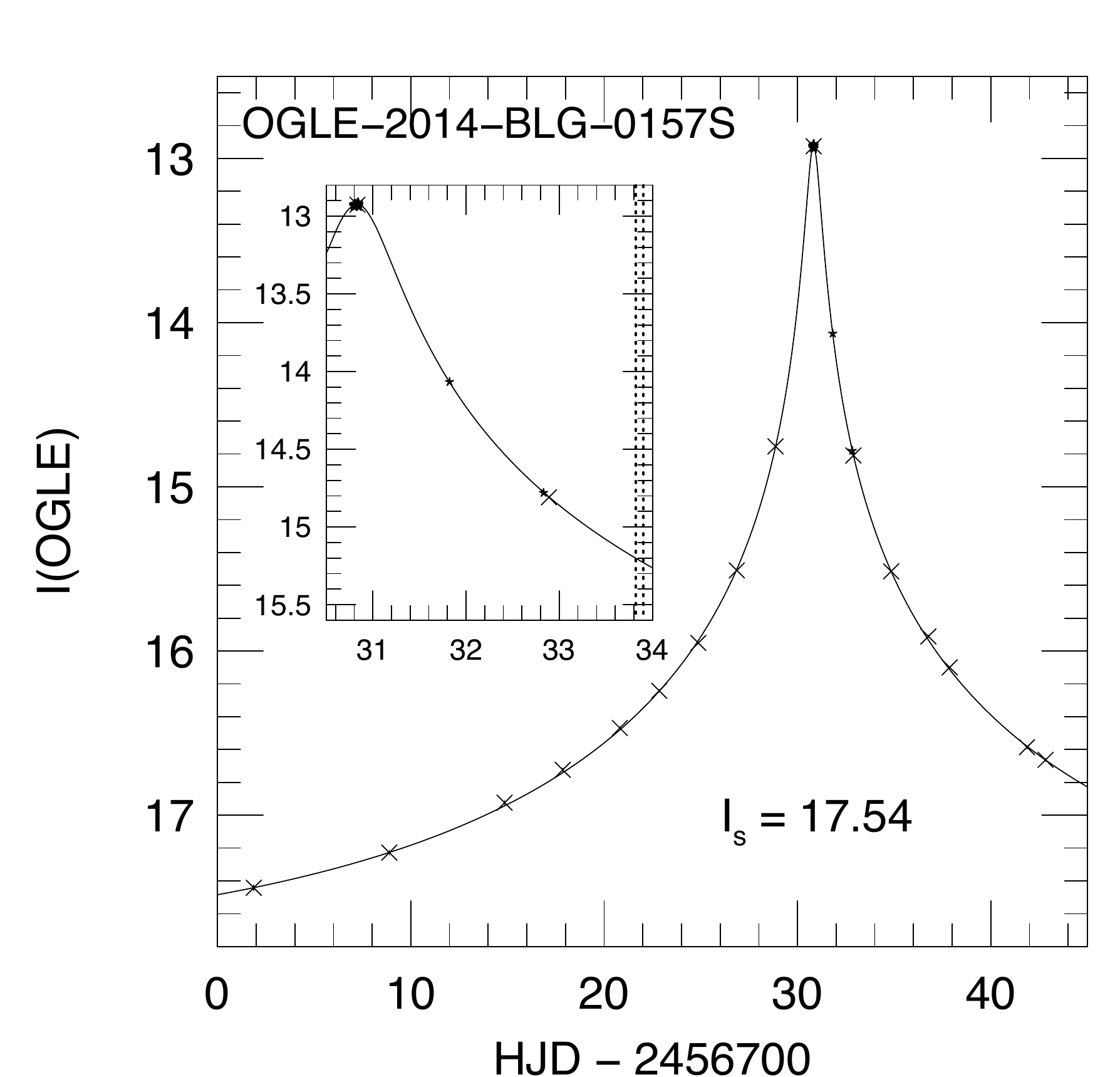}
\includegraphics[viewport= 67 40 515 535,clip]{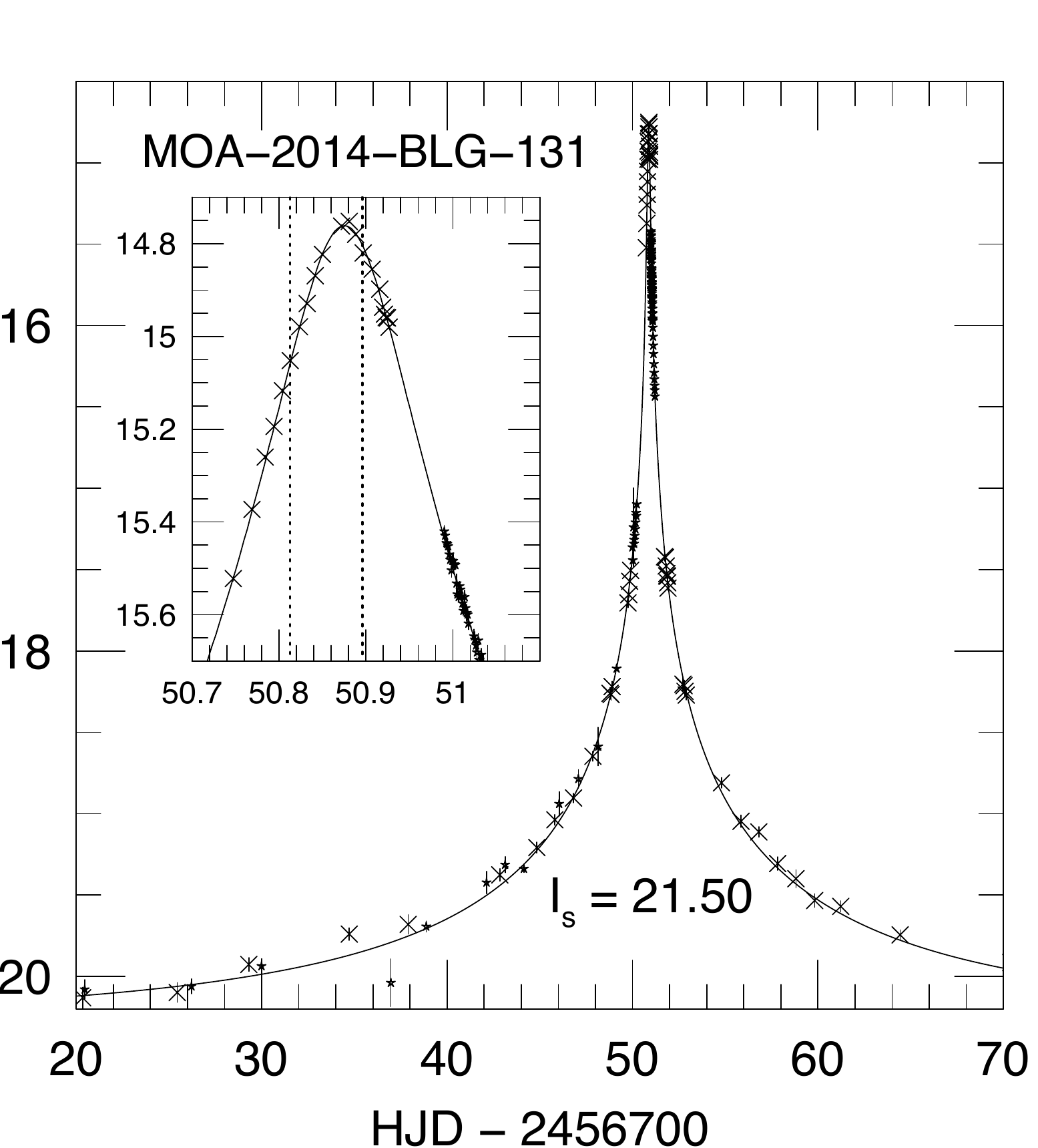}
\includegraphics[viewport= 67 40 515 535,clip]{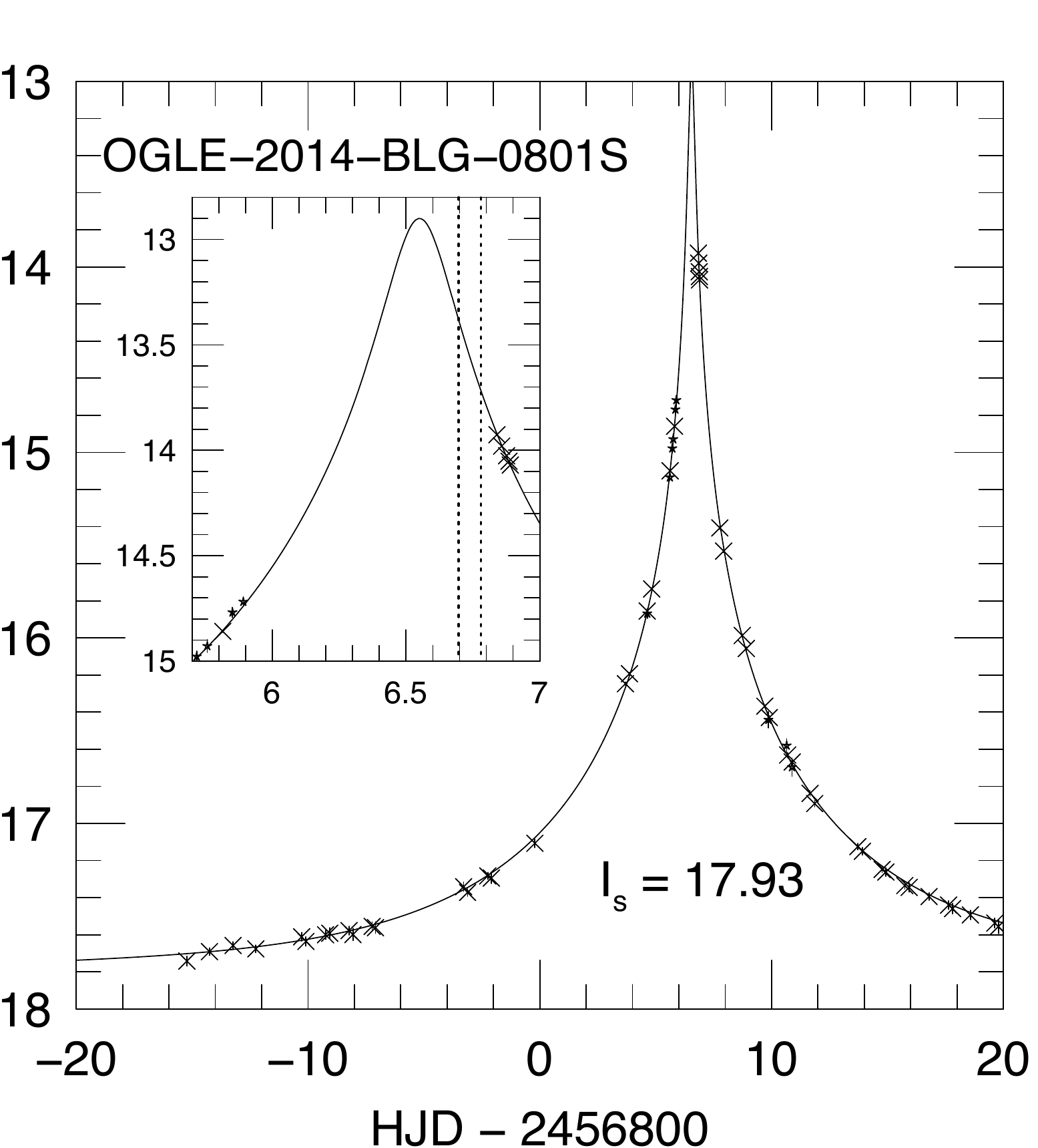}
\includegraphics[viewport= 67 40 525 535,clip]{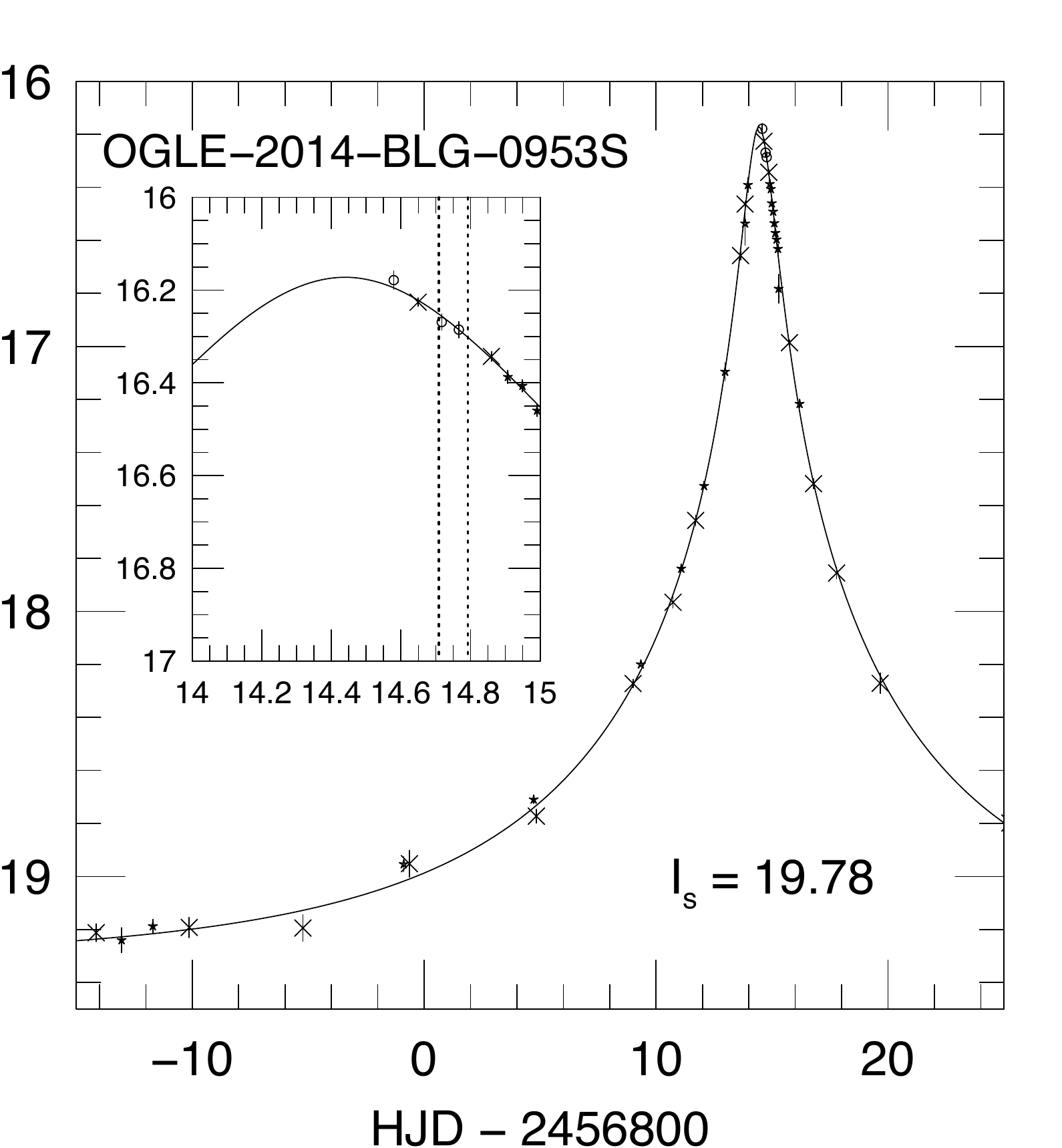}}
\resizebox{\hsize}{!}{
\includegraphics[viewport=-10 40 515 525,clip]{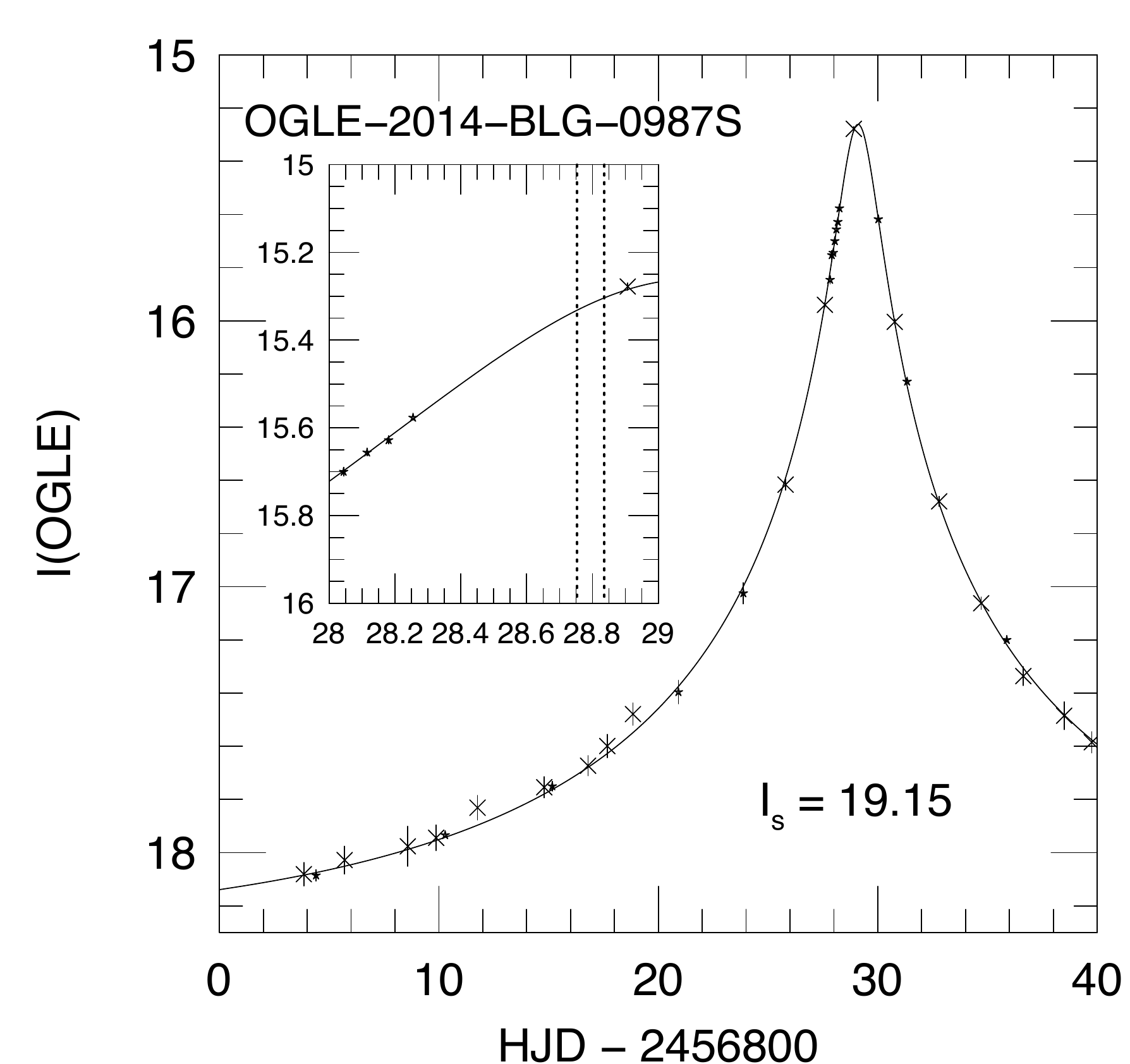}
\includegraphics[viewport= 67 40 515 525,clip]{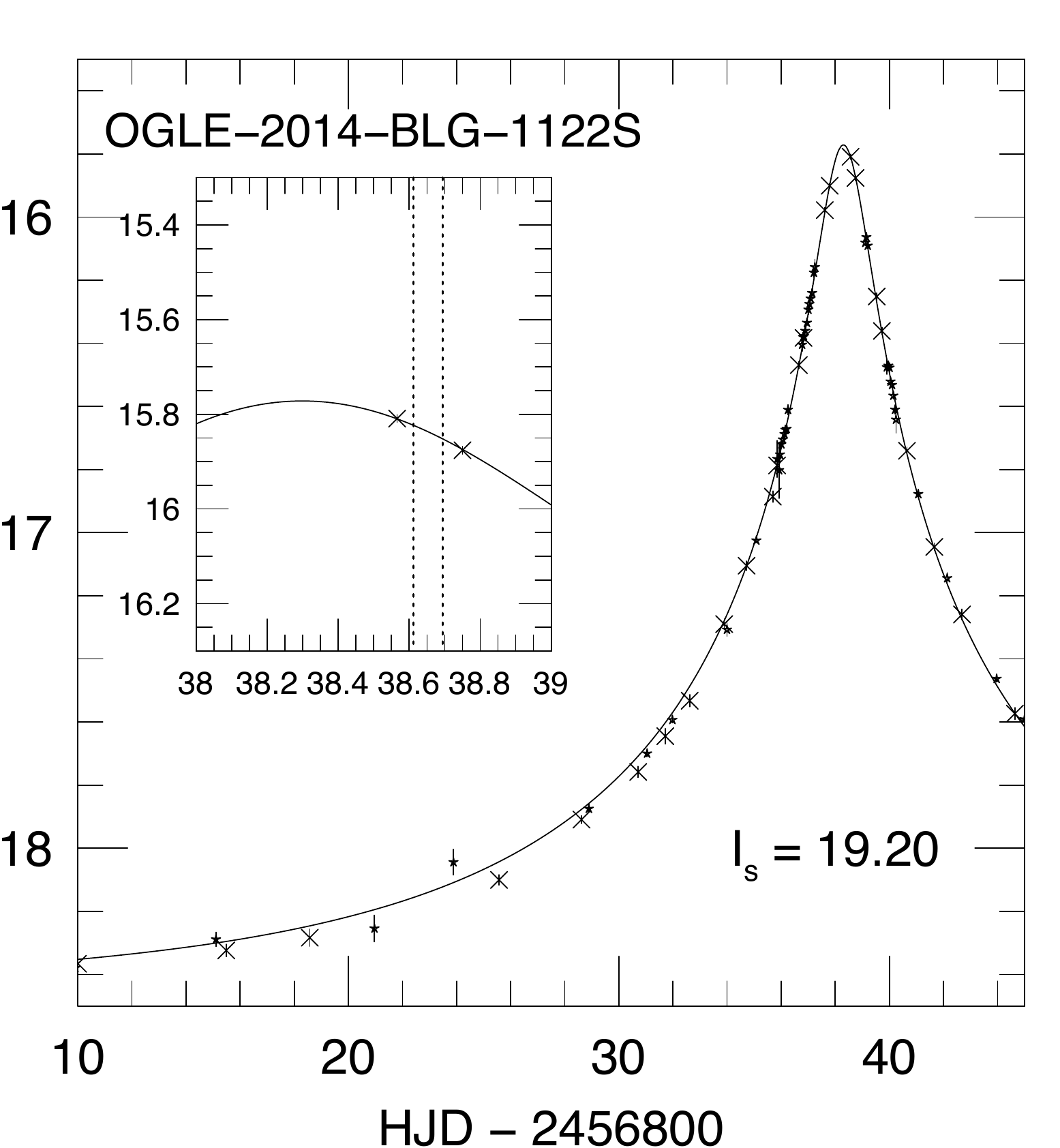}
\includegraphics[viewport= 67 40 515 525,clip]{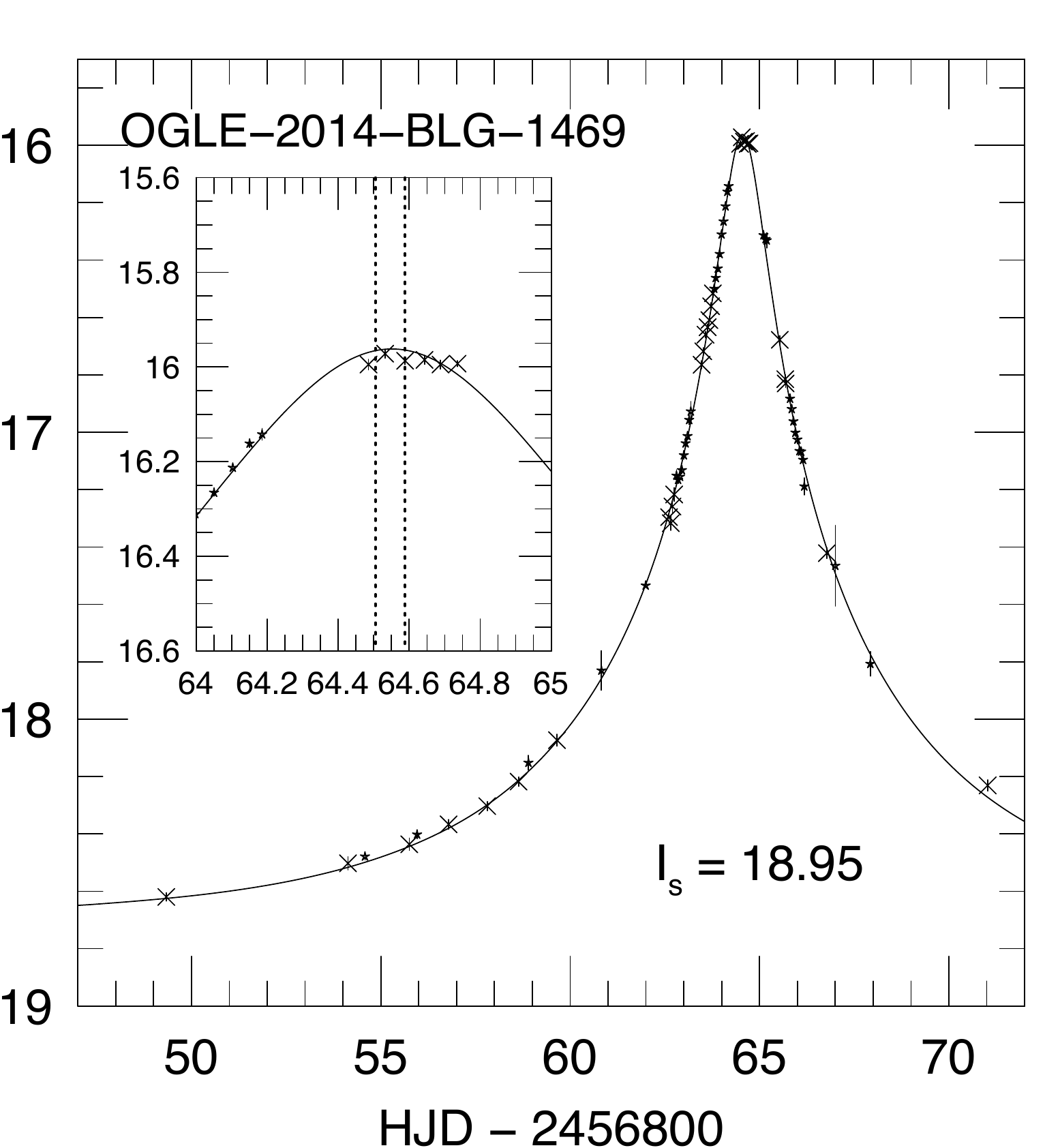}
\includegraphics[viewport= 67 40 525 525,clip]{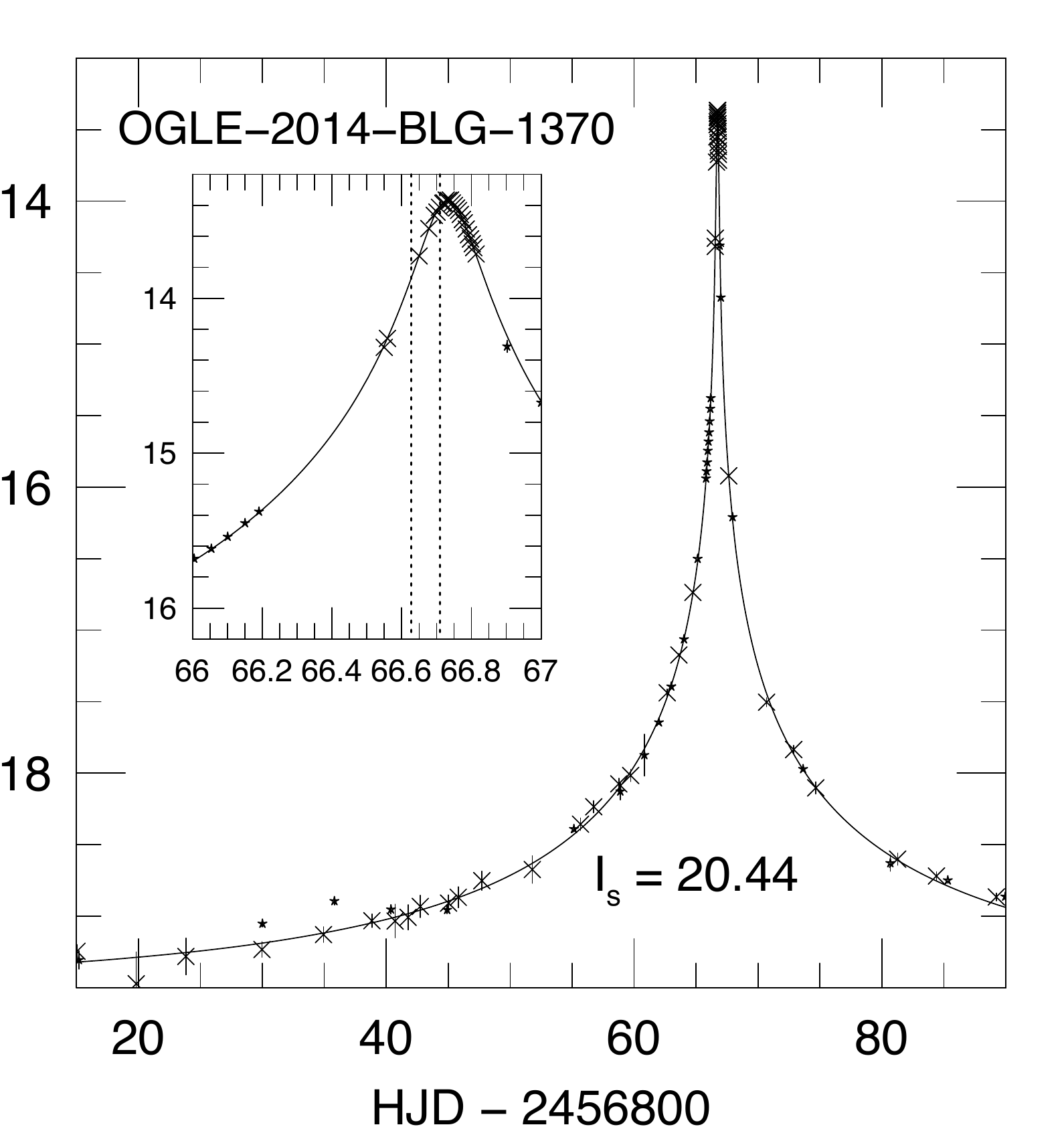}}
\resizebox{\hsize}{!}{
\includegraphics[viewport=-10 40 515 525,clip]{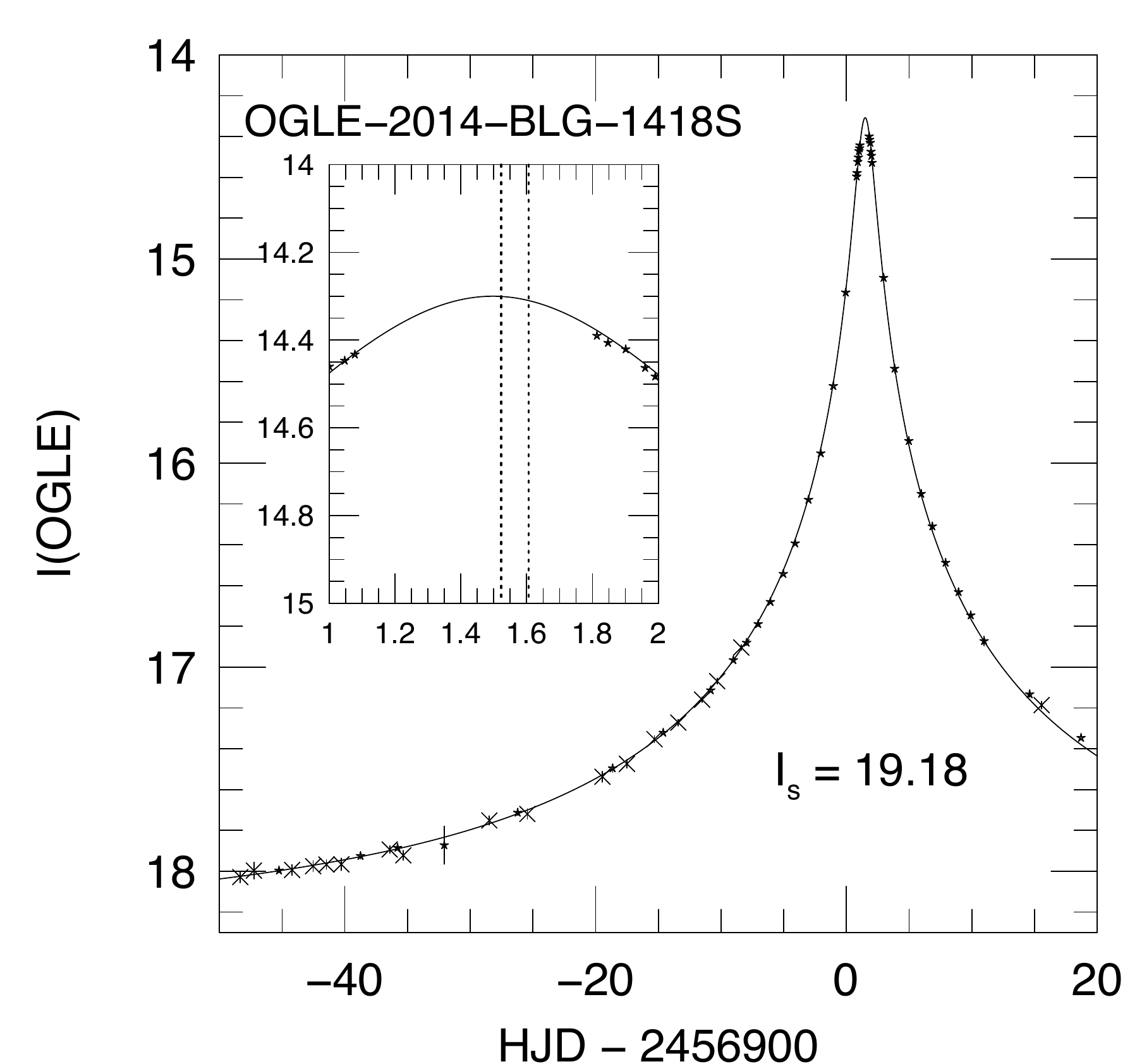}
\includegraphics[viewport= 67 40 515 525,clip]{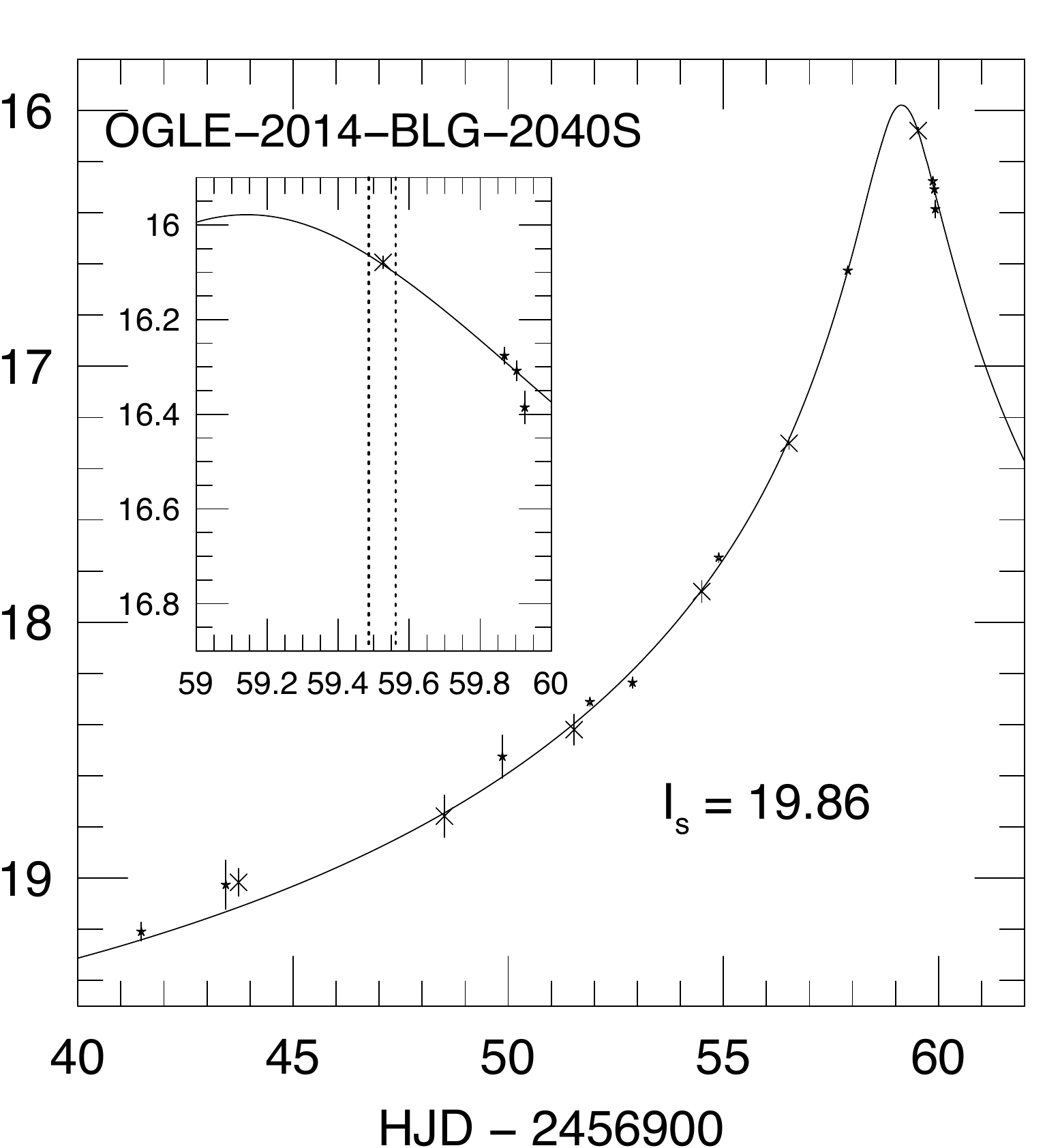}
\includegraphics[viewport= 67 40 515 525,clip]{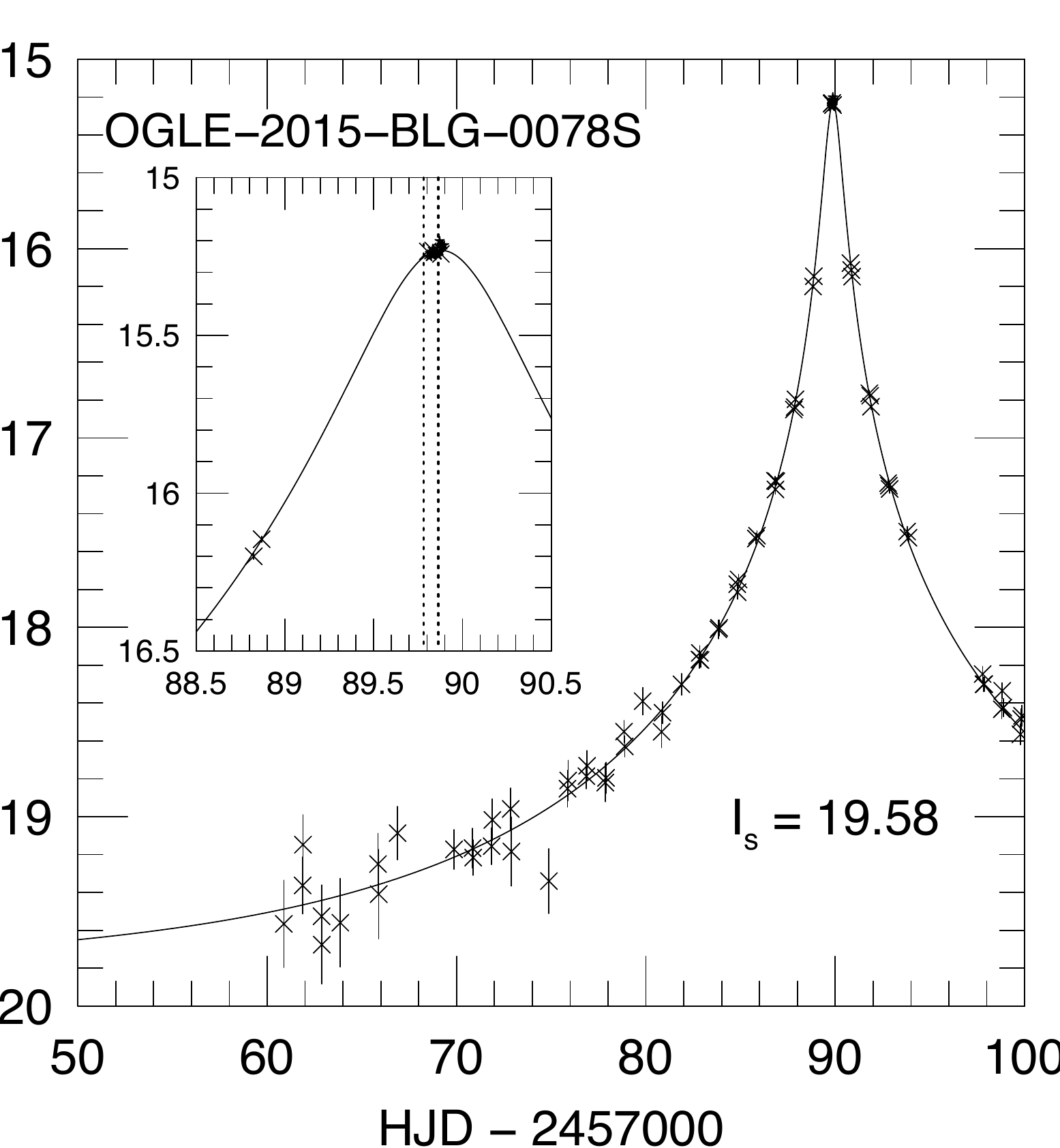}
\includegraphics[viewport= 67 40 525 40,clip]{lcb_ob131114.pdf}}
\resizebox{\hsize}{!}{
\includegraphics[viewport=-10 40 515 525,clip]{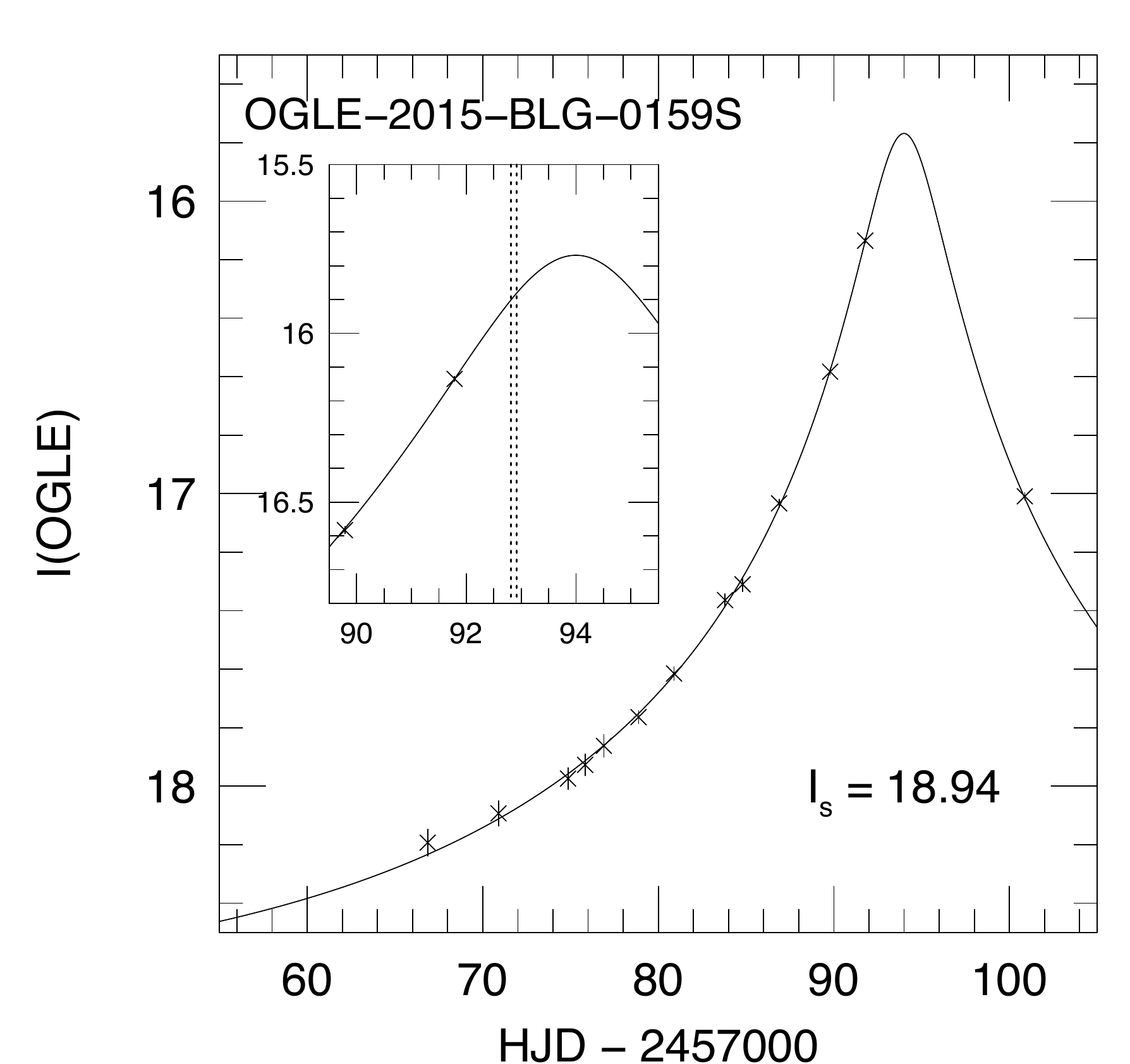}
\includegraphics[viewport= 67 40 515 525,clip]{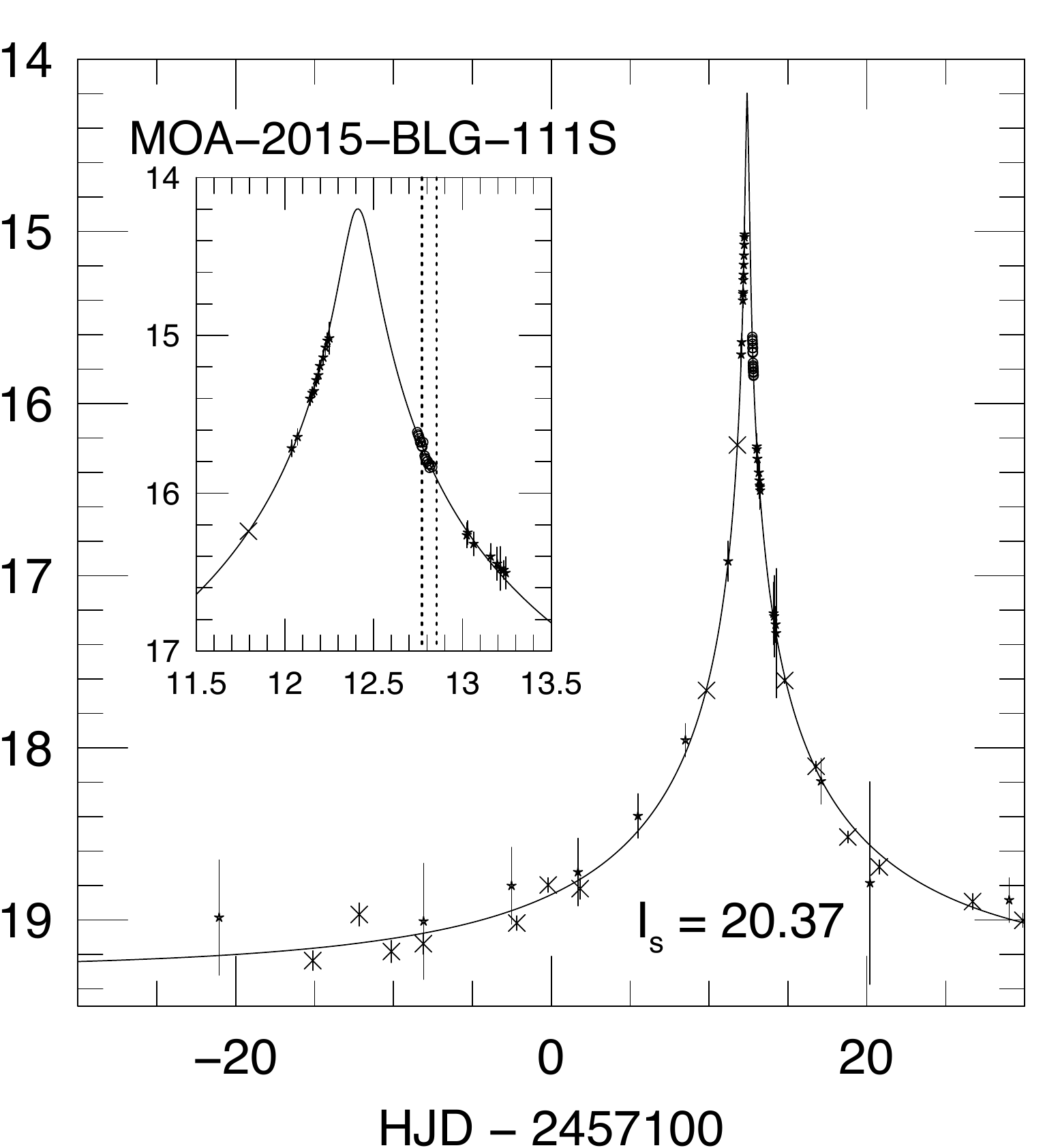}
\includegraphics[viewport= 67 40 515 40,clip]{lcb_ob131114.pdf}
\includegraphics[viewport= 67 40 525 40,clip]{lcb_ob131114.pdf}
}
\caption{\sl continued
}
\end{figure*}

\clearpage
\section{Age probability distributions and G functions (Figure~\ref{fig:agefunctions})}
\label{sec:agefunctions}

\begin{figure*}[ht]
\centering
\resizebox{\hsize}{!}{
\includegraphics[viewport= 0 0 648 648,clip]{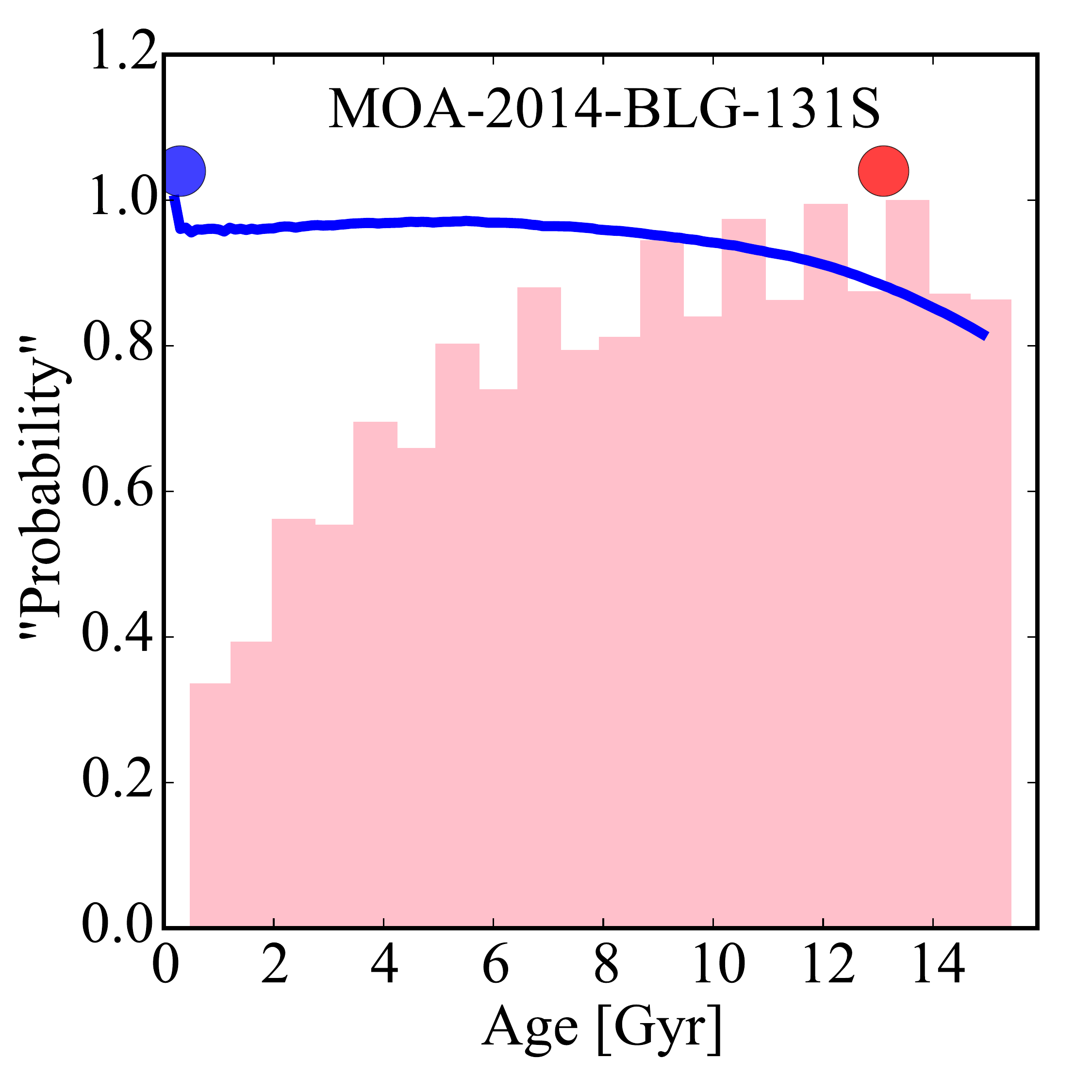}
\includegraphics[viewport= 93 0 648 648,clip]{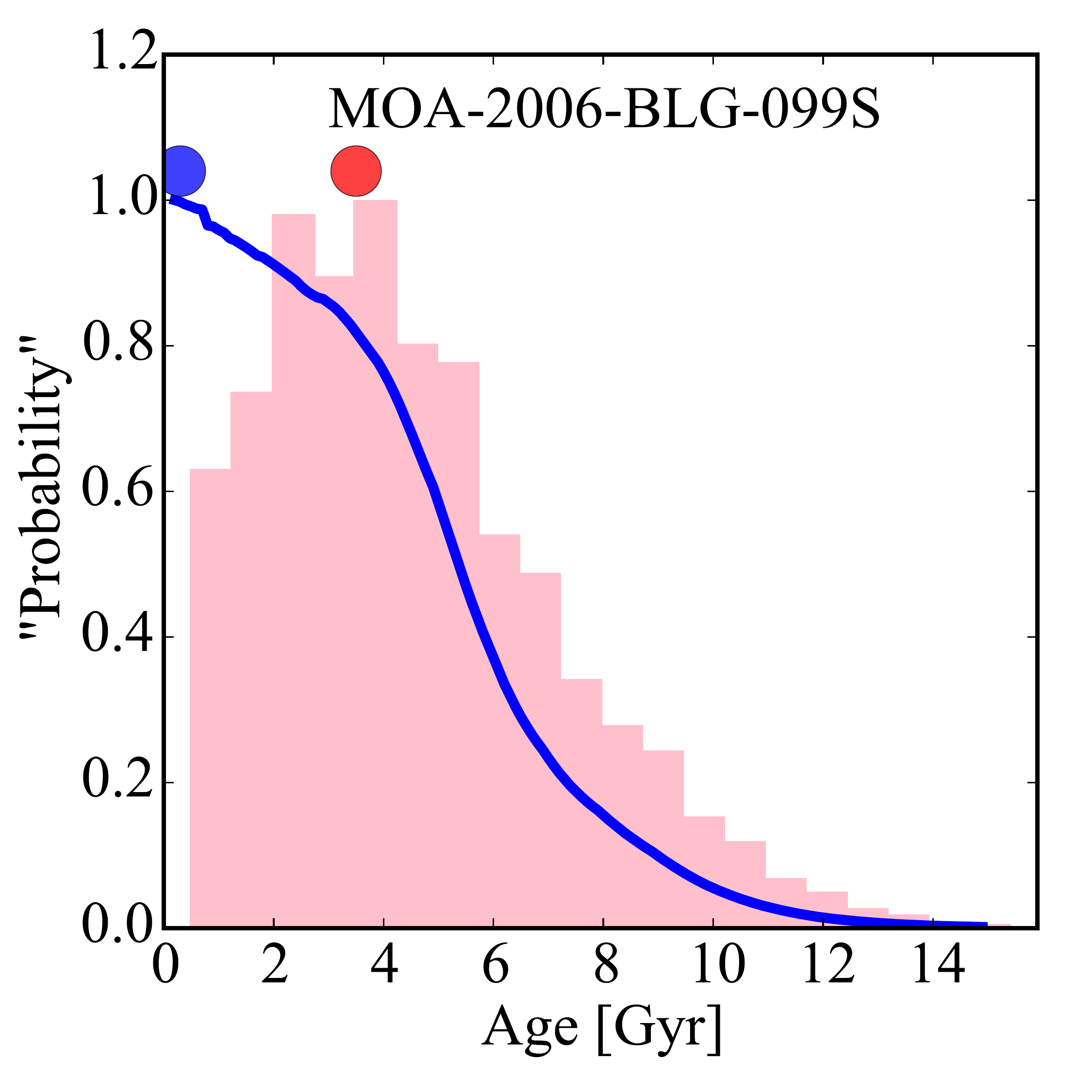}
\includegraphics[viewport= 93 0 648 648,clip]{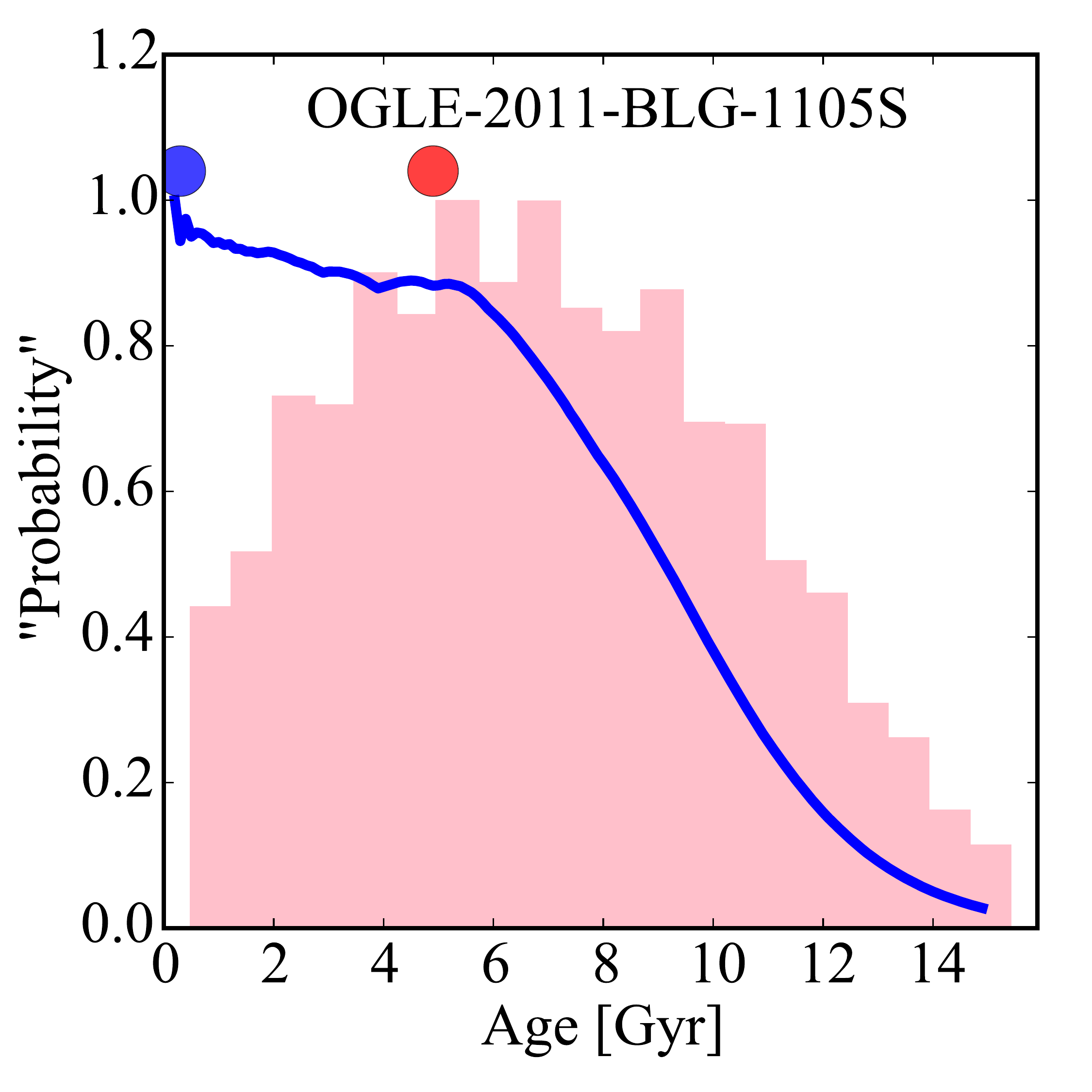}
\includegraphics[viewport= 93 0 648 648,clip]{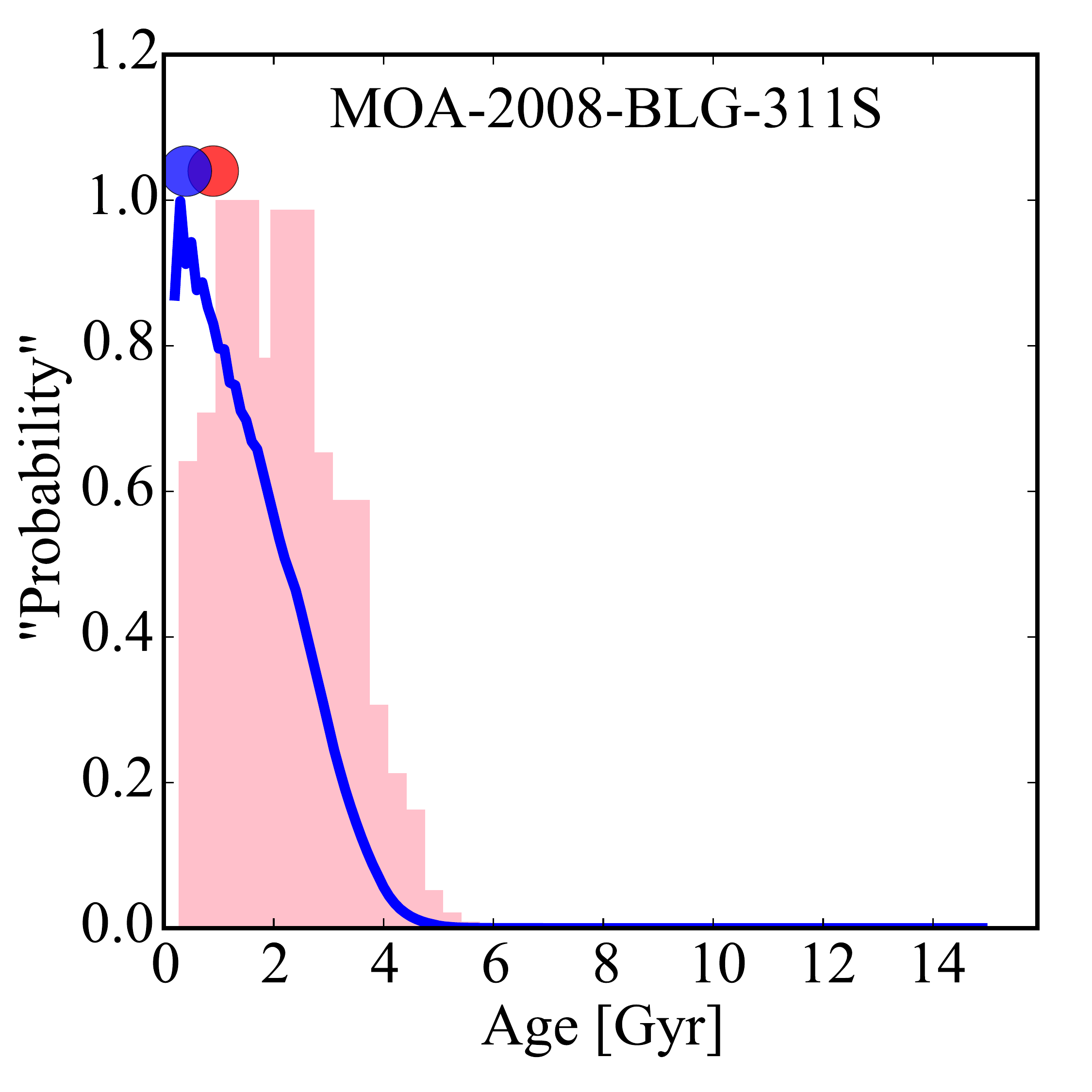}
\includegraphics[viewport= 93 0 648 648,clip]{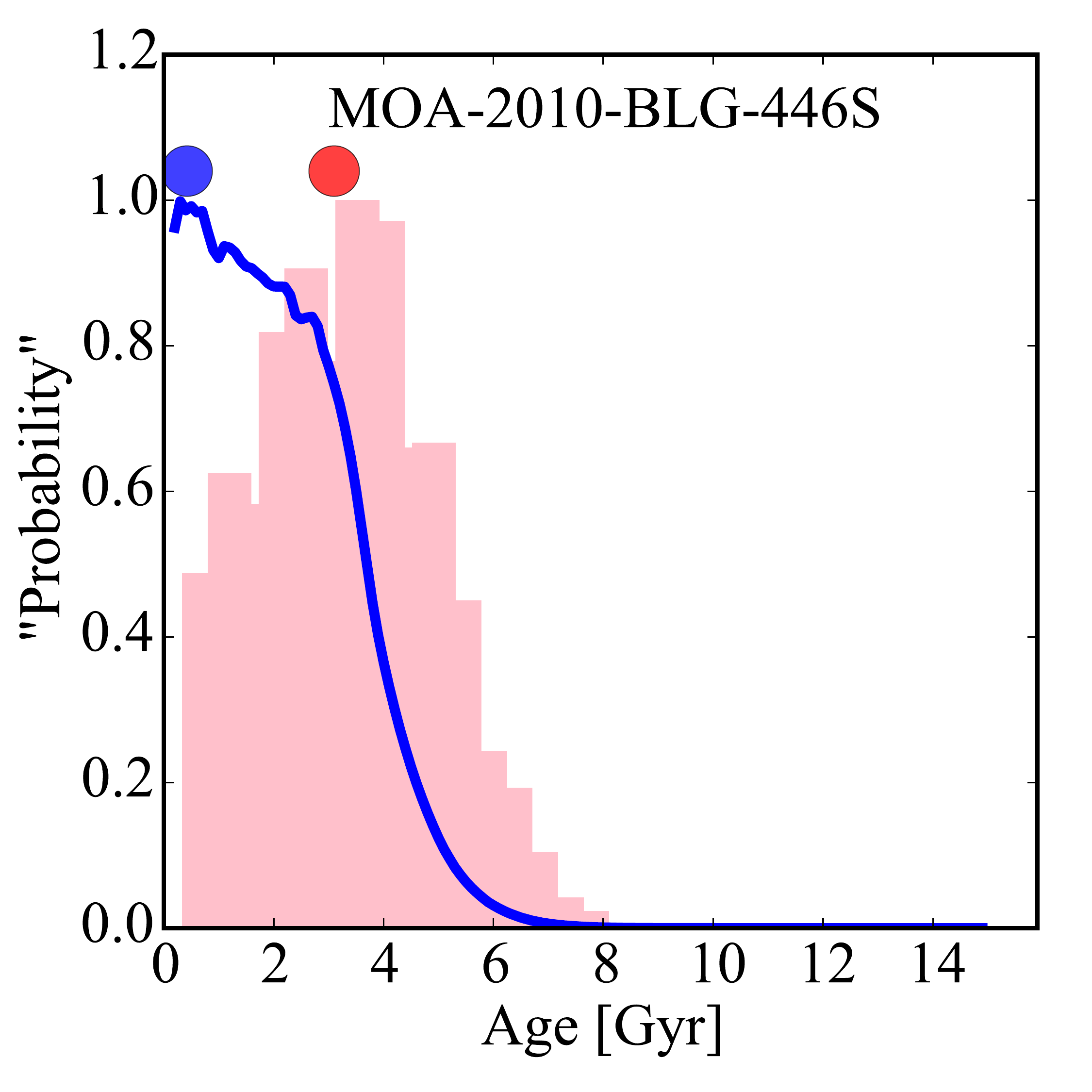}}
\resizebox{\hsize}{!}{
\includegraphics[viewport= 0 0 648 648,clip]{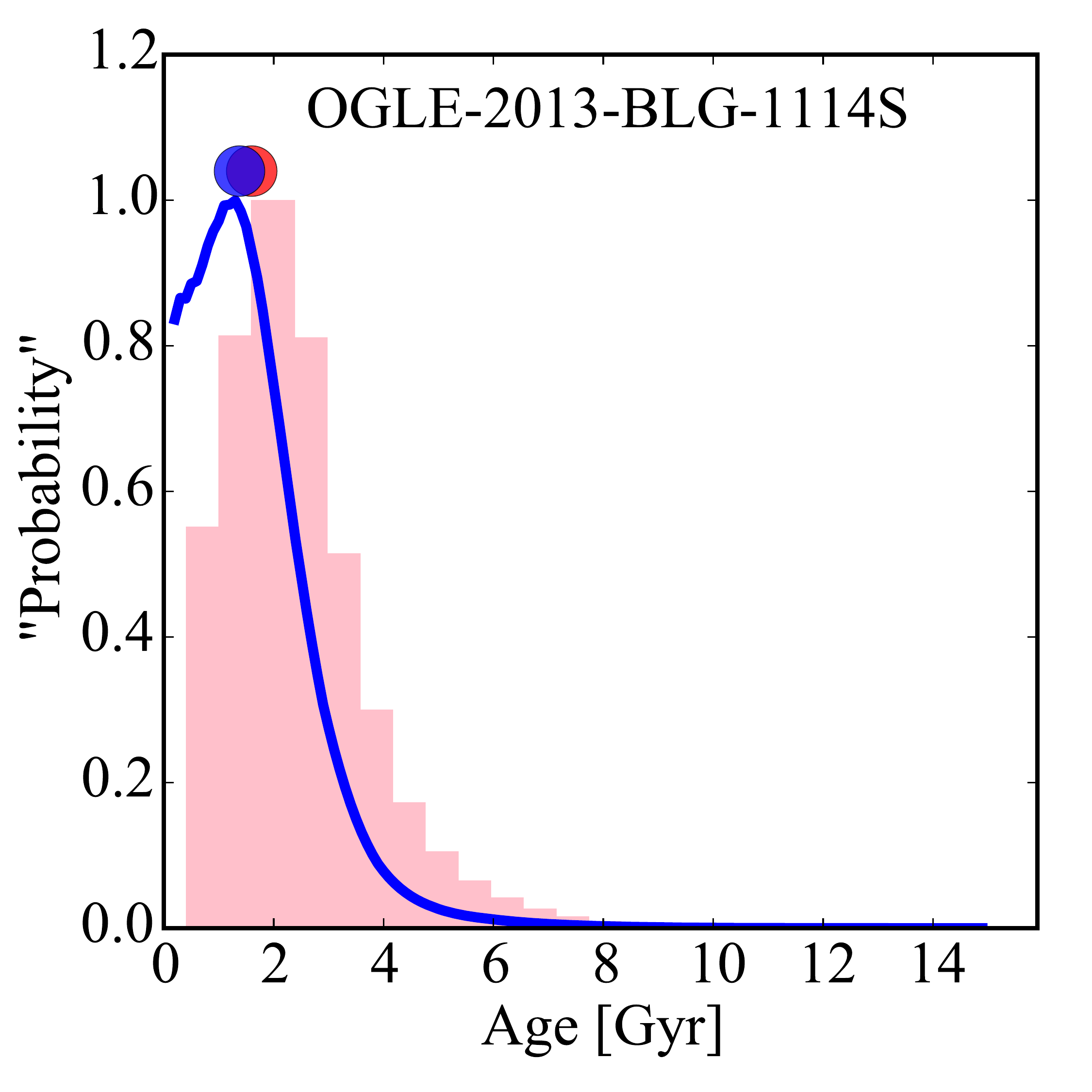}
\includegraphics[viewport= 93 0 648 648,clip]{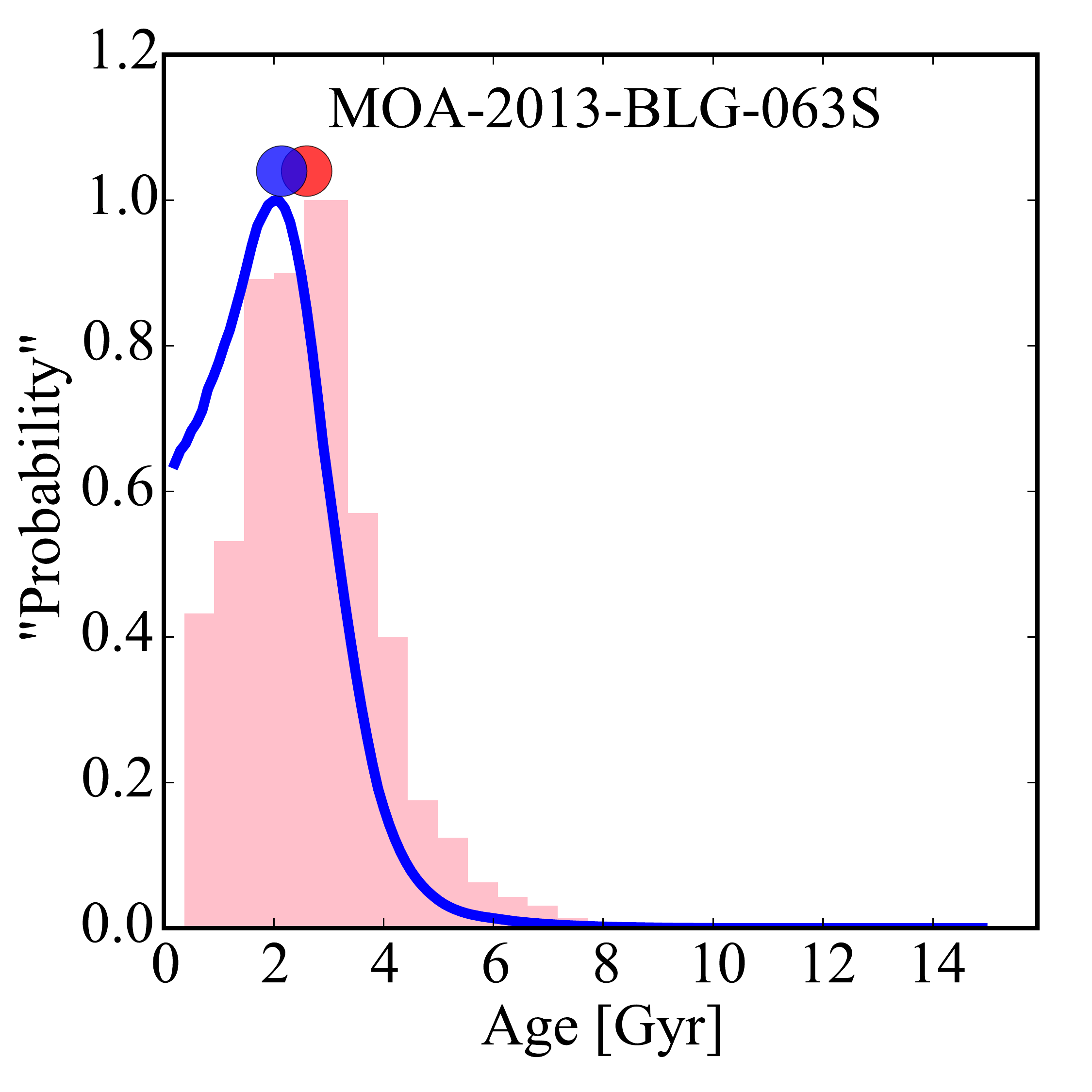}
\includegraphics[viewport= 93 0 648 648,clip]{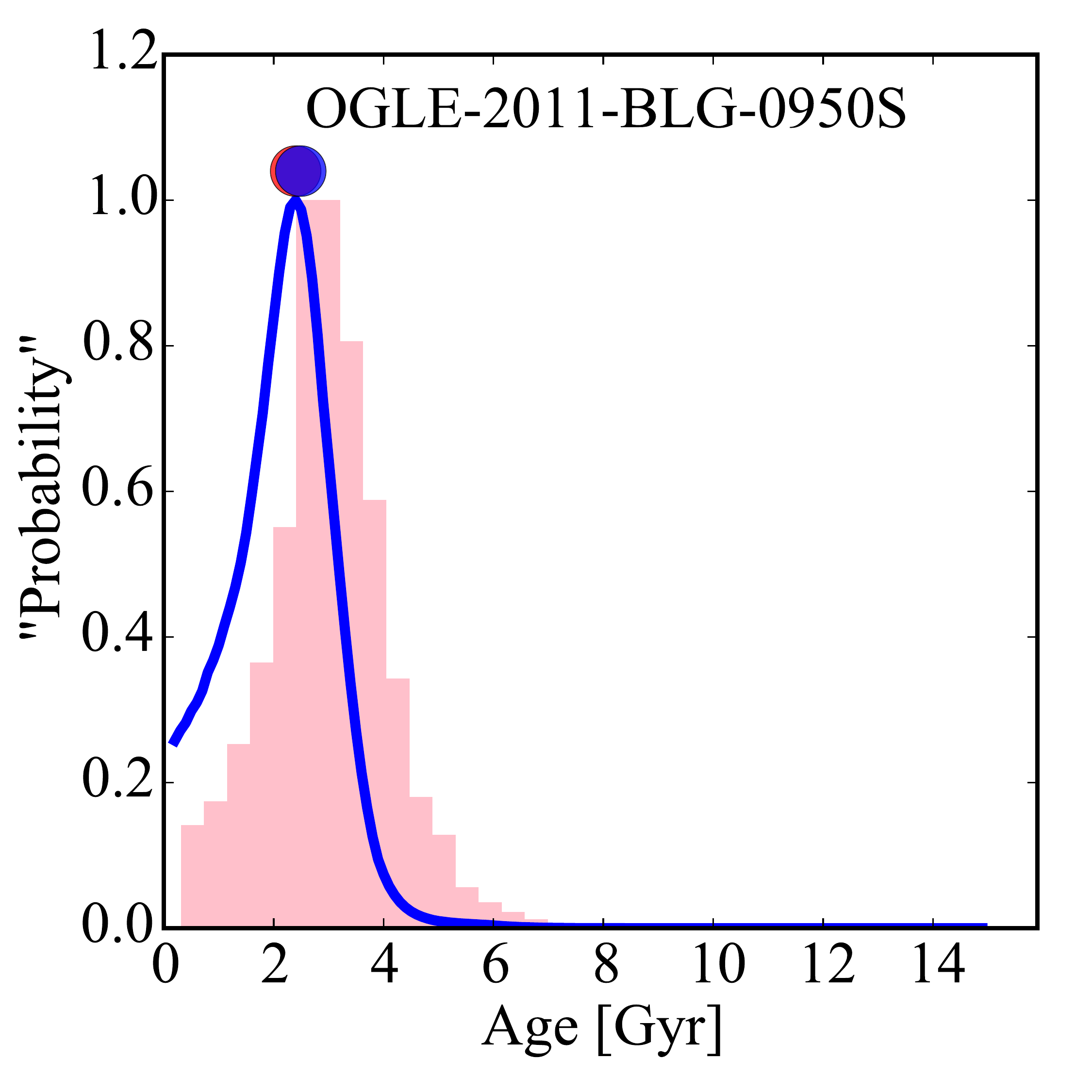}
\includegraphics[viewport= 93 0 648 648,clip]{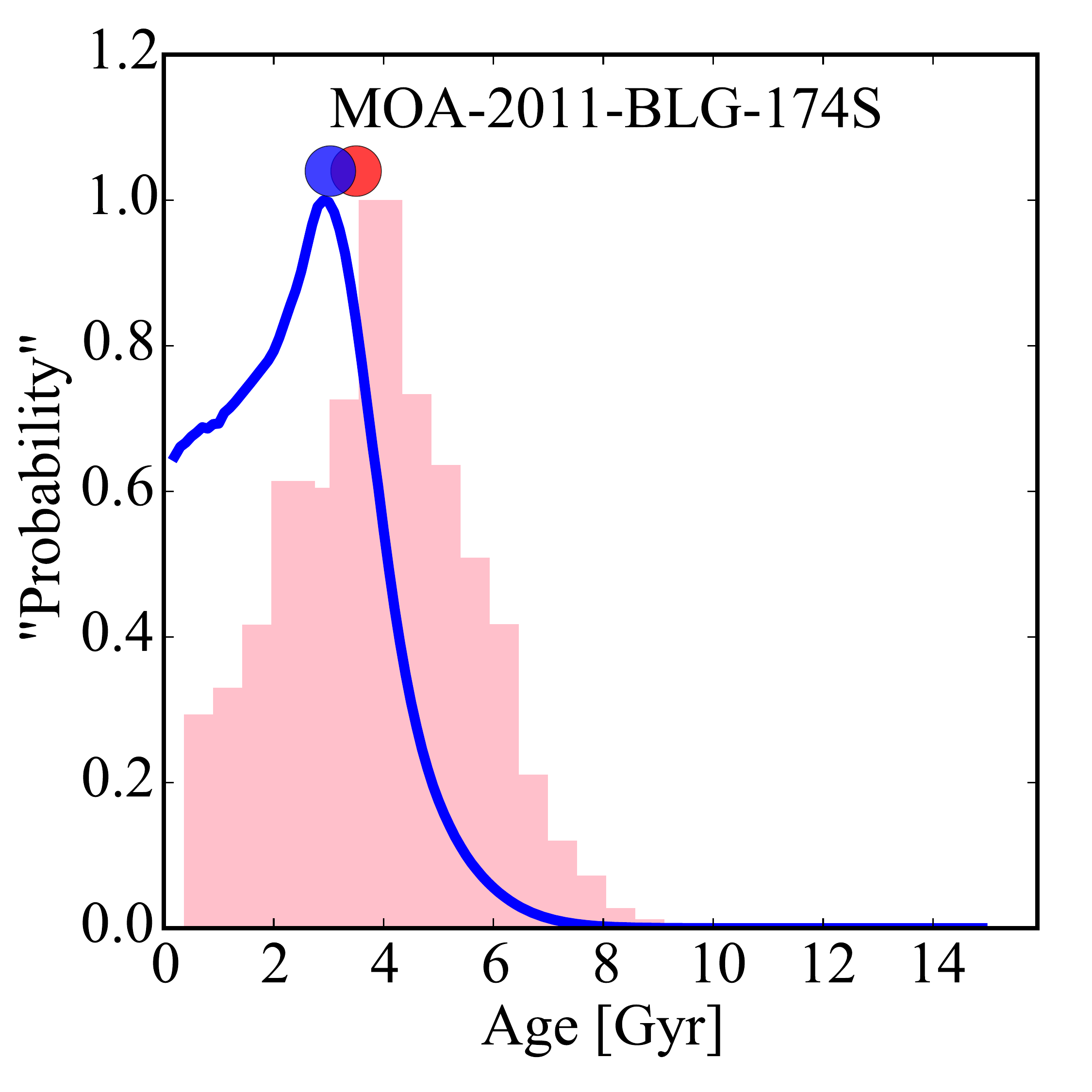}
\includegraphics[viewport= 93 0 648 648,clip]{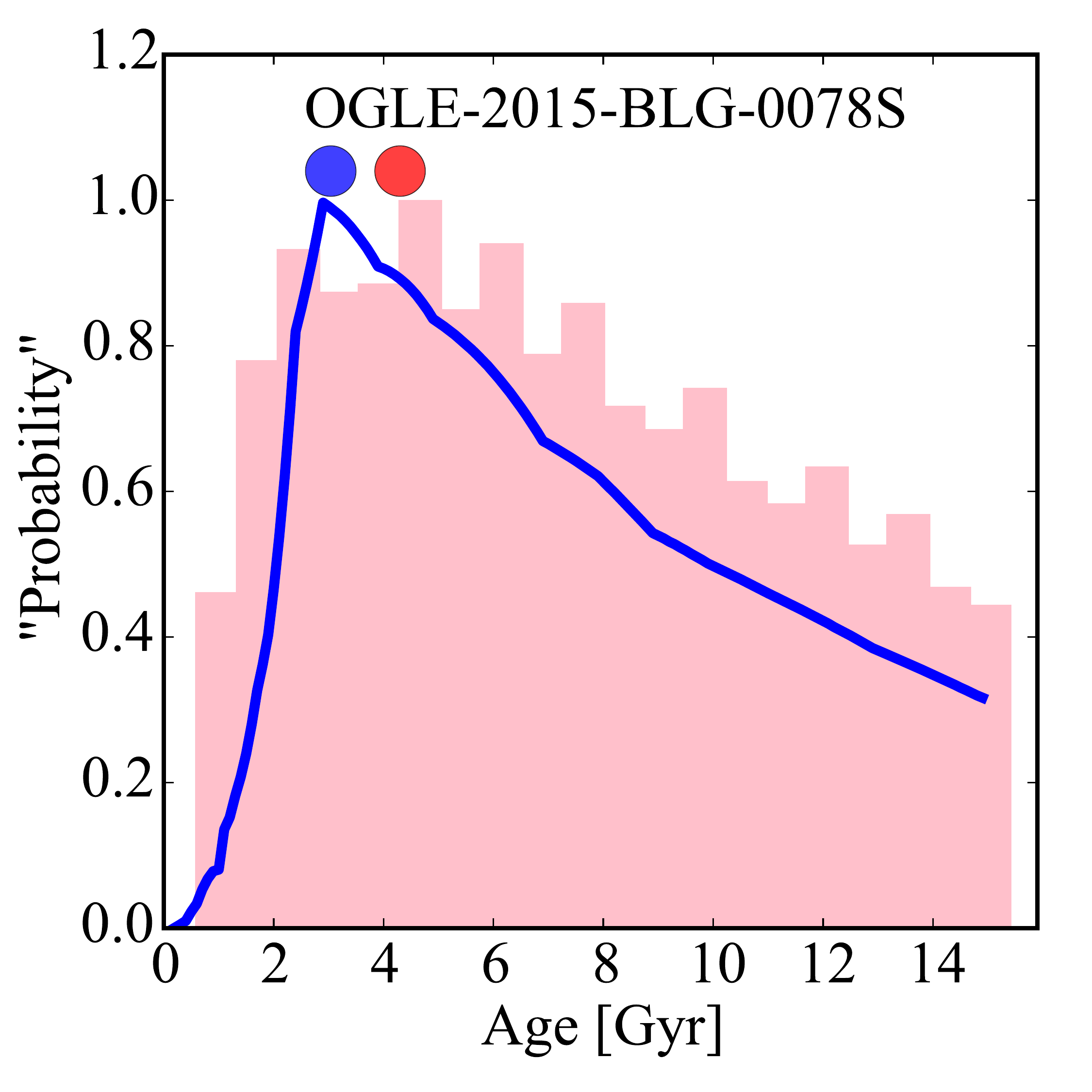}}
\resizebox{\hsize}{!}{
\includegraphics[viewport= 0 0 648 648,clip]{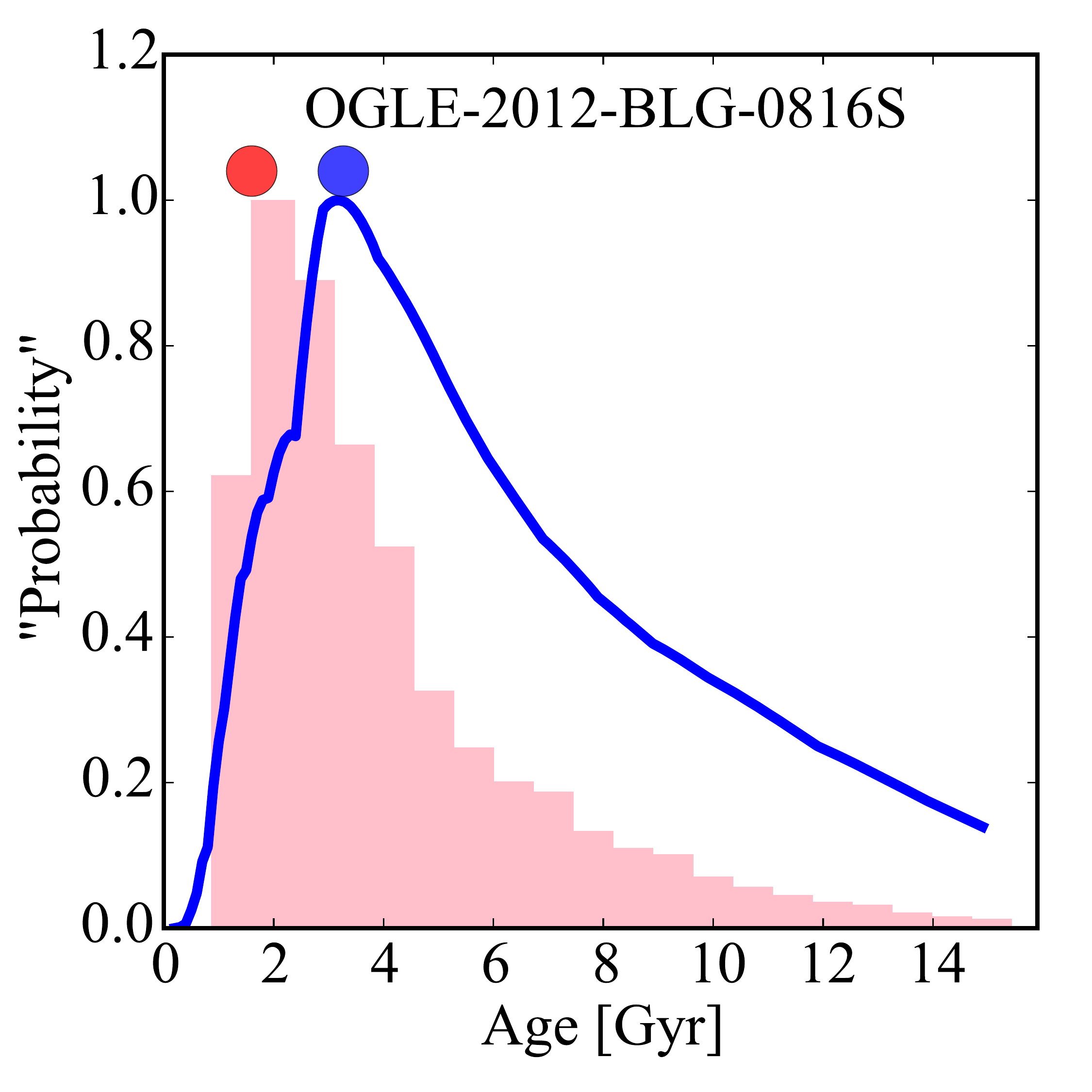}
\includegraphics[viewport= 93 0 648 648,clip]{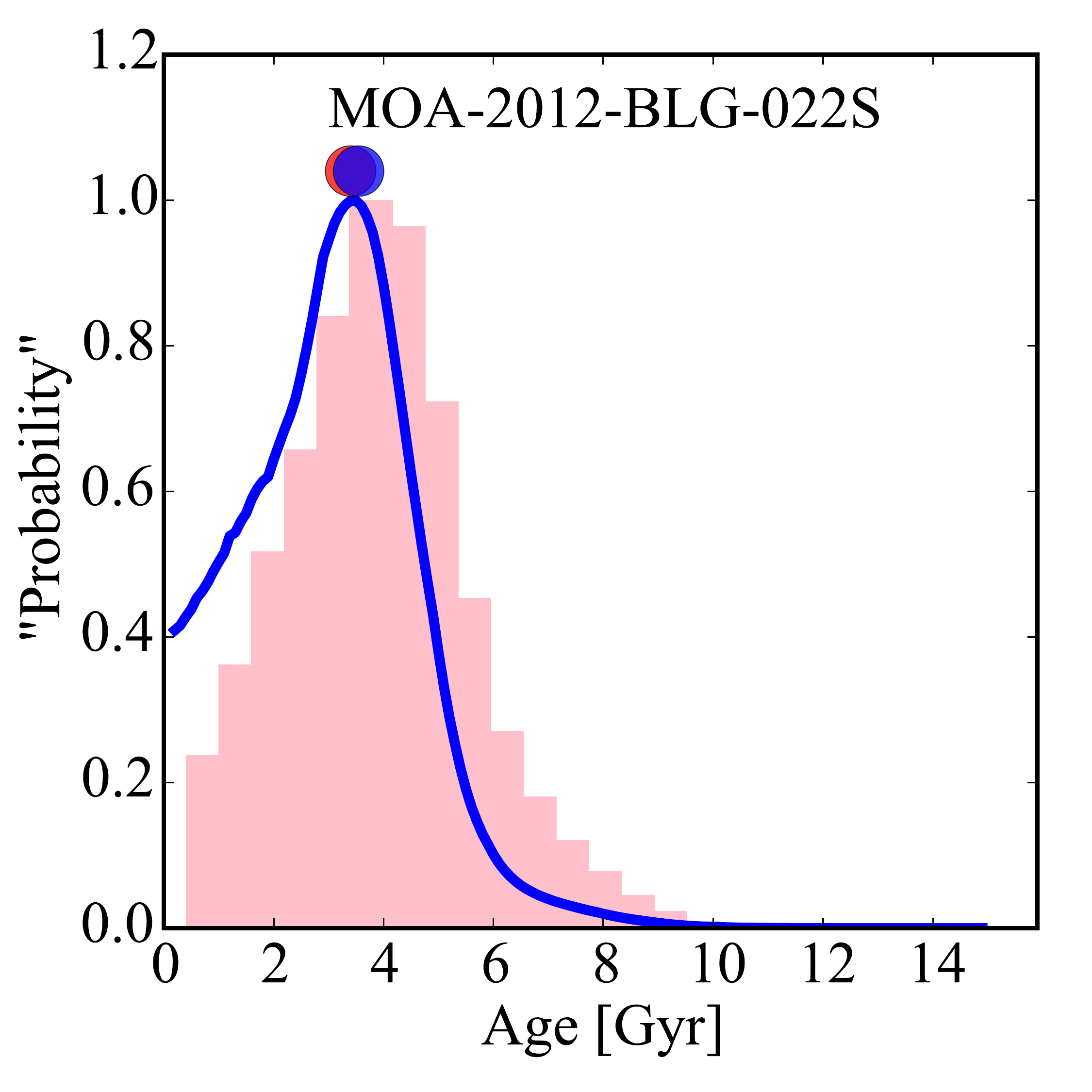}
\includegraphics[viewport= 93 0 648 648,clip]{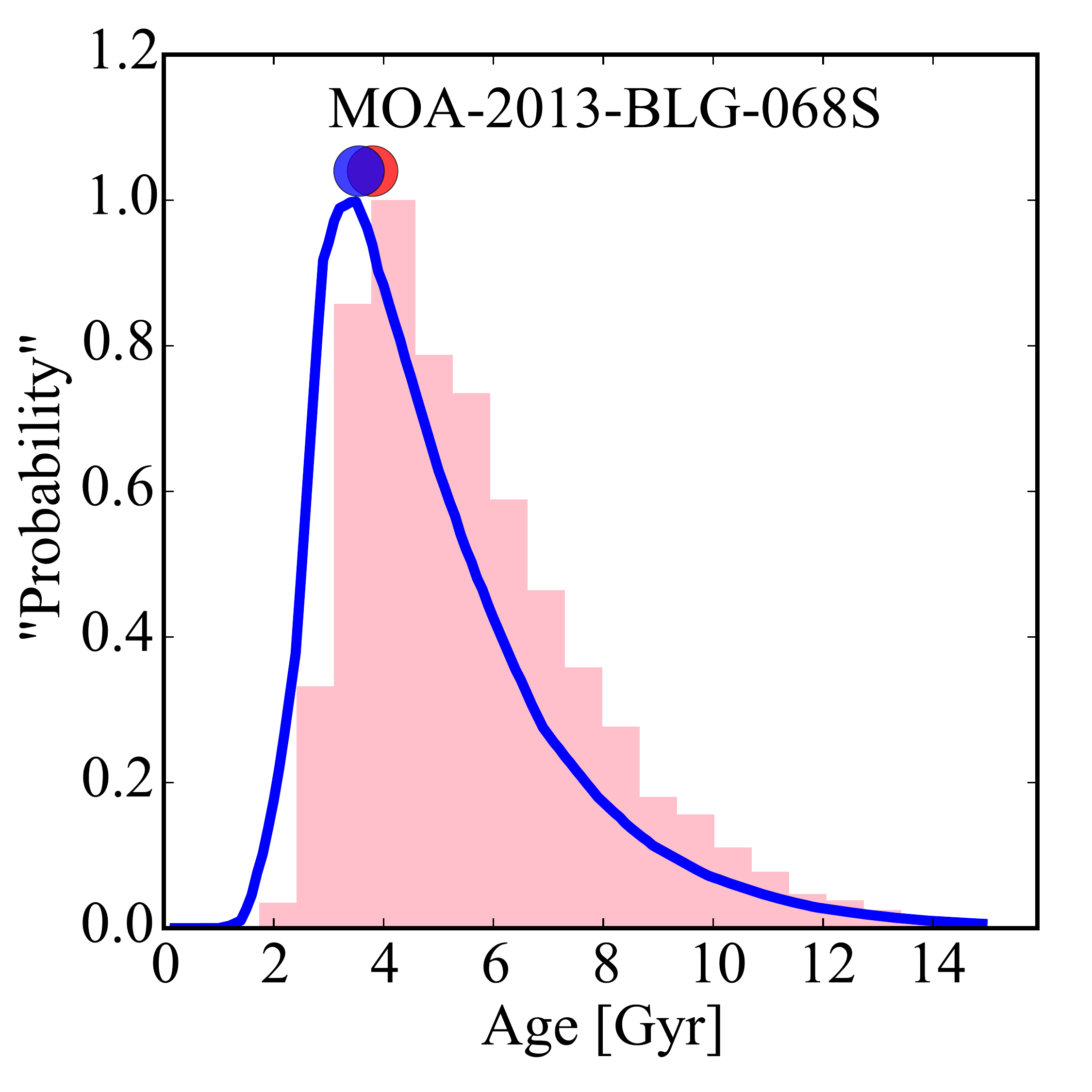}
\includegraphics[viewport= 93 0 648 648,clip]{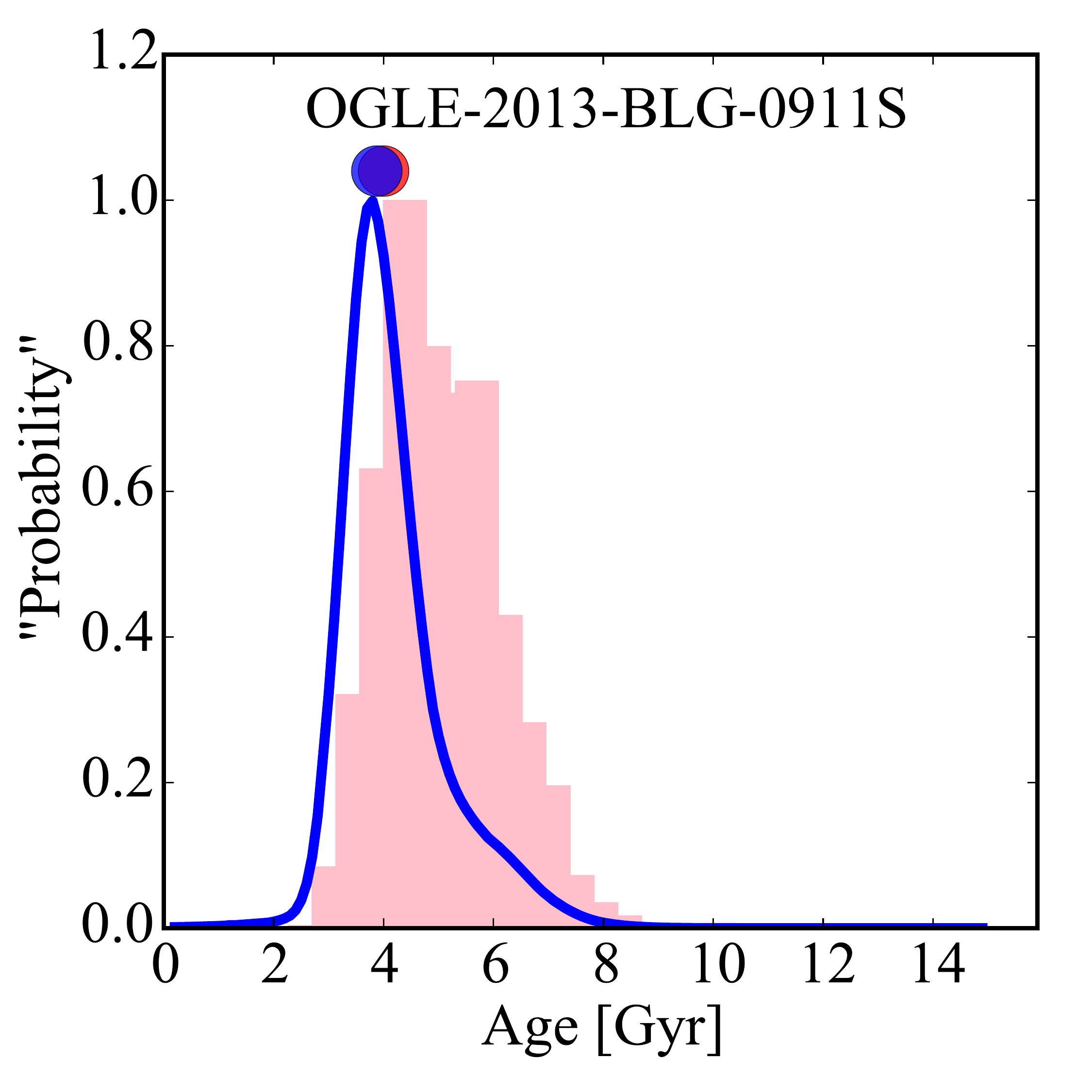}
\includegraphics[viewport= 93 0 648 648,clip]{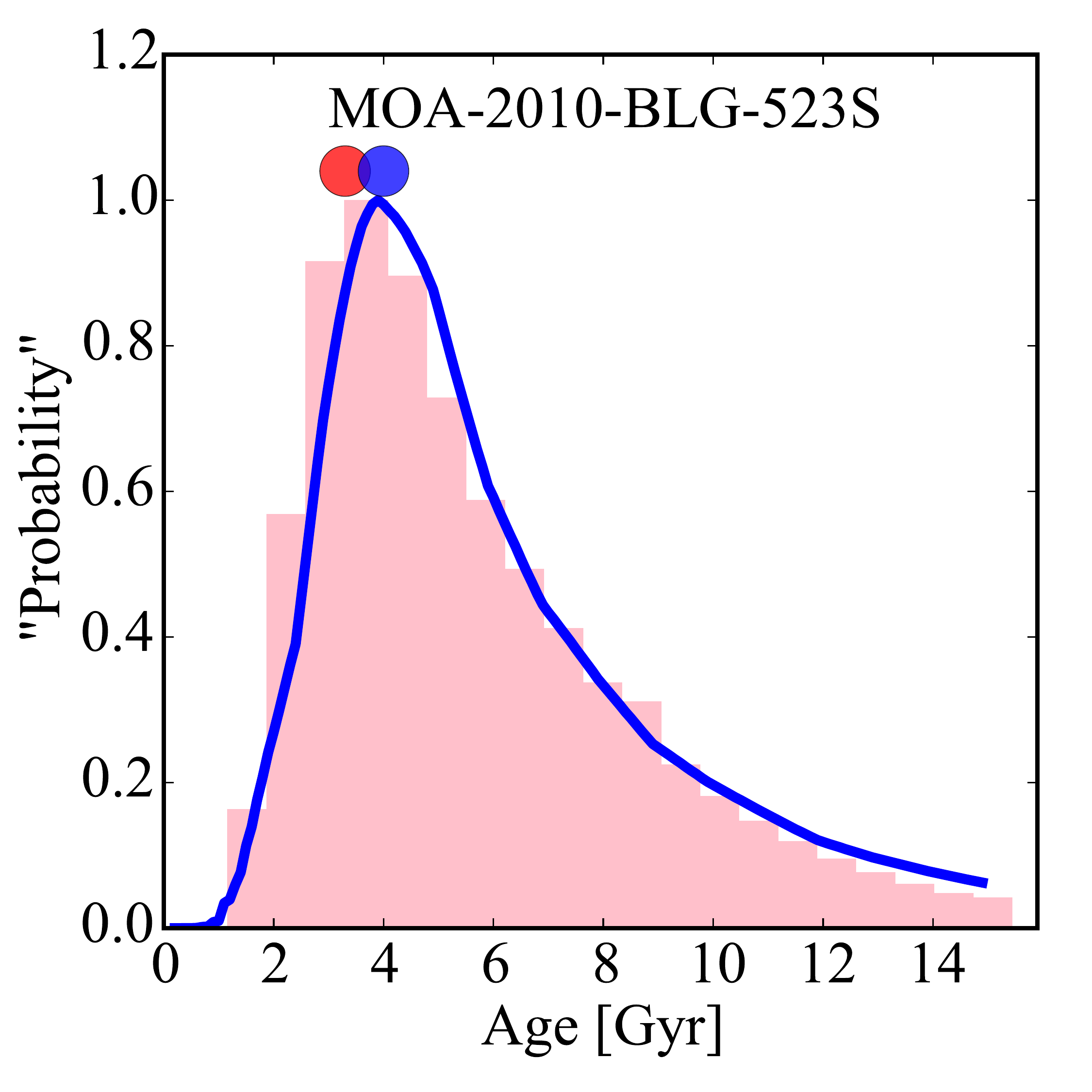}}
\resizebox{\hsize}{!}{
\includegraphics[viewport= 0 0 648 648,clip]{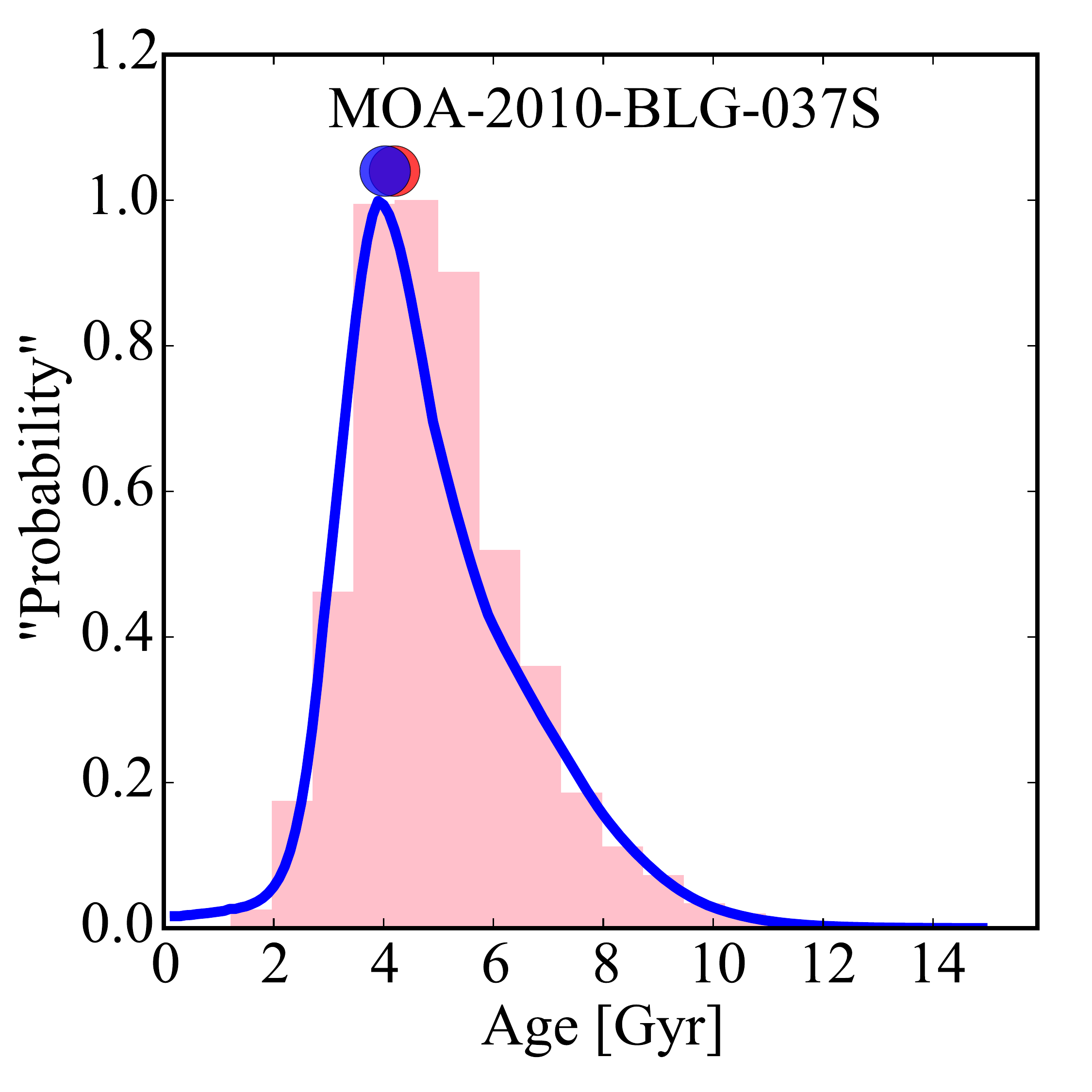}
\includegraphics[viewport= 93 0 648 648,clip]{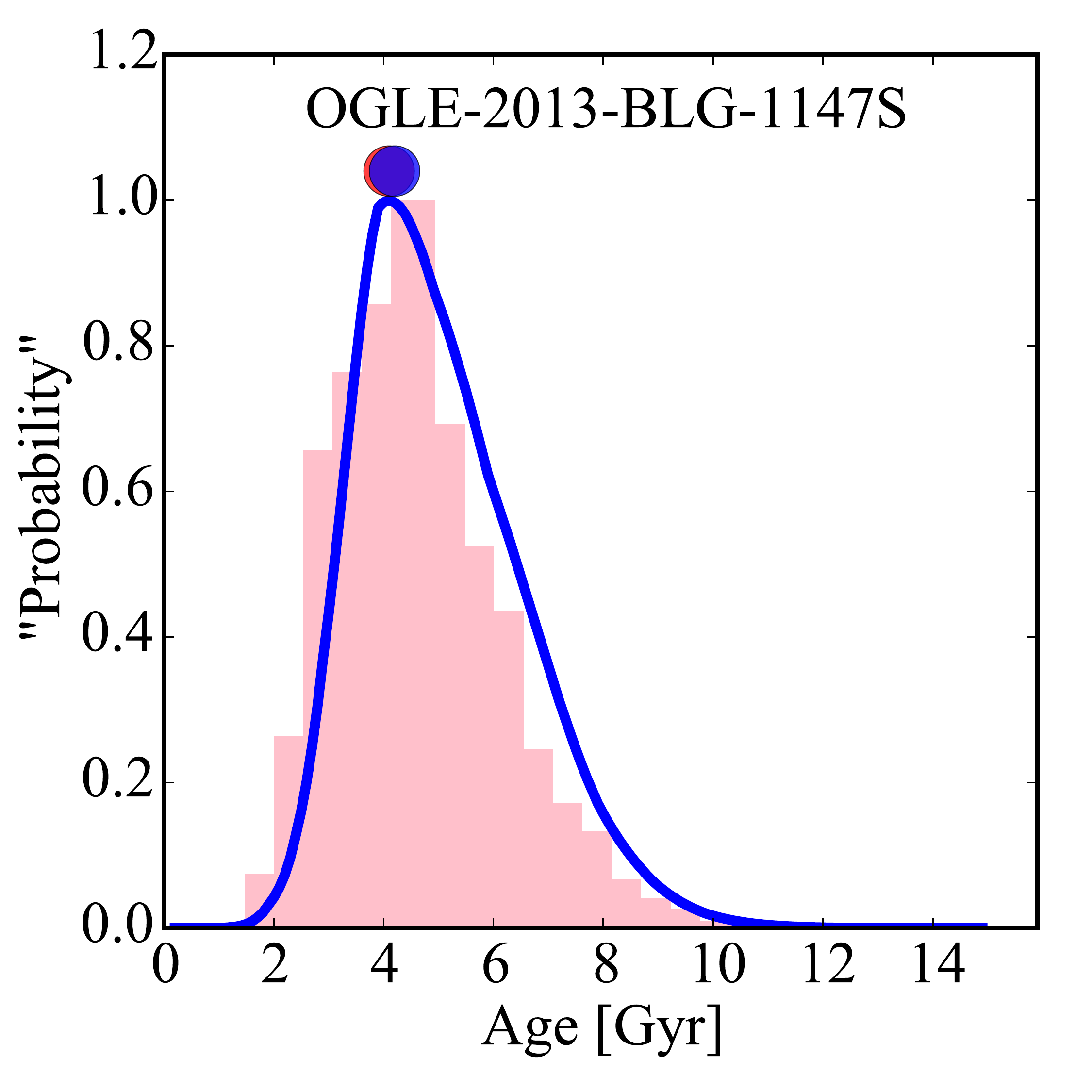}
\includegraphics[viewport= 93 0 648 648,clip]{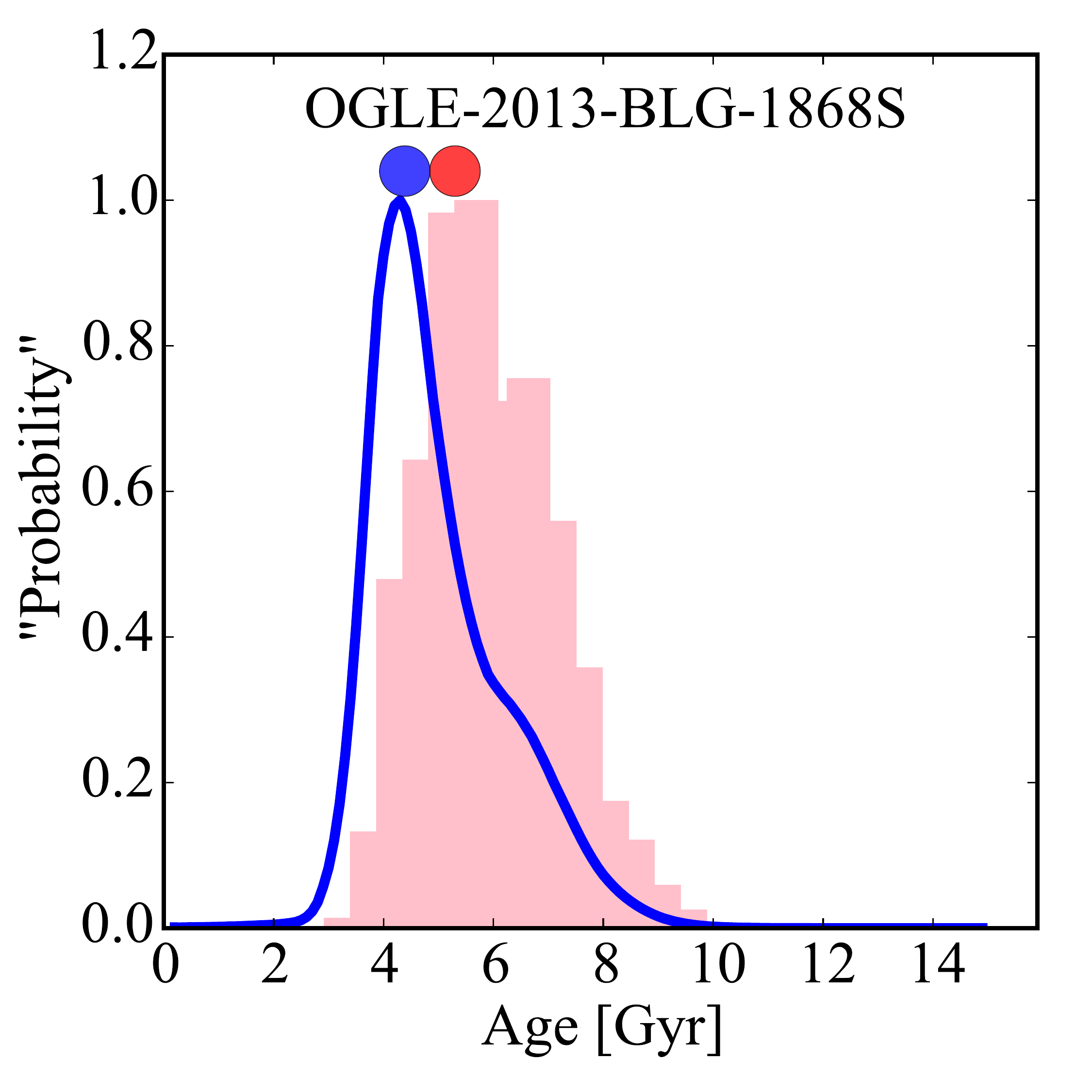}
\includegraphics[viewport= 93 0 648 648,clip]{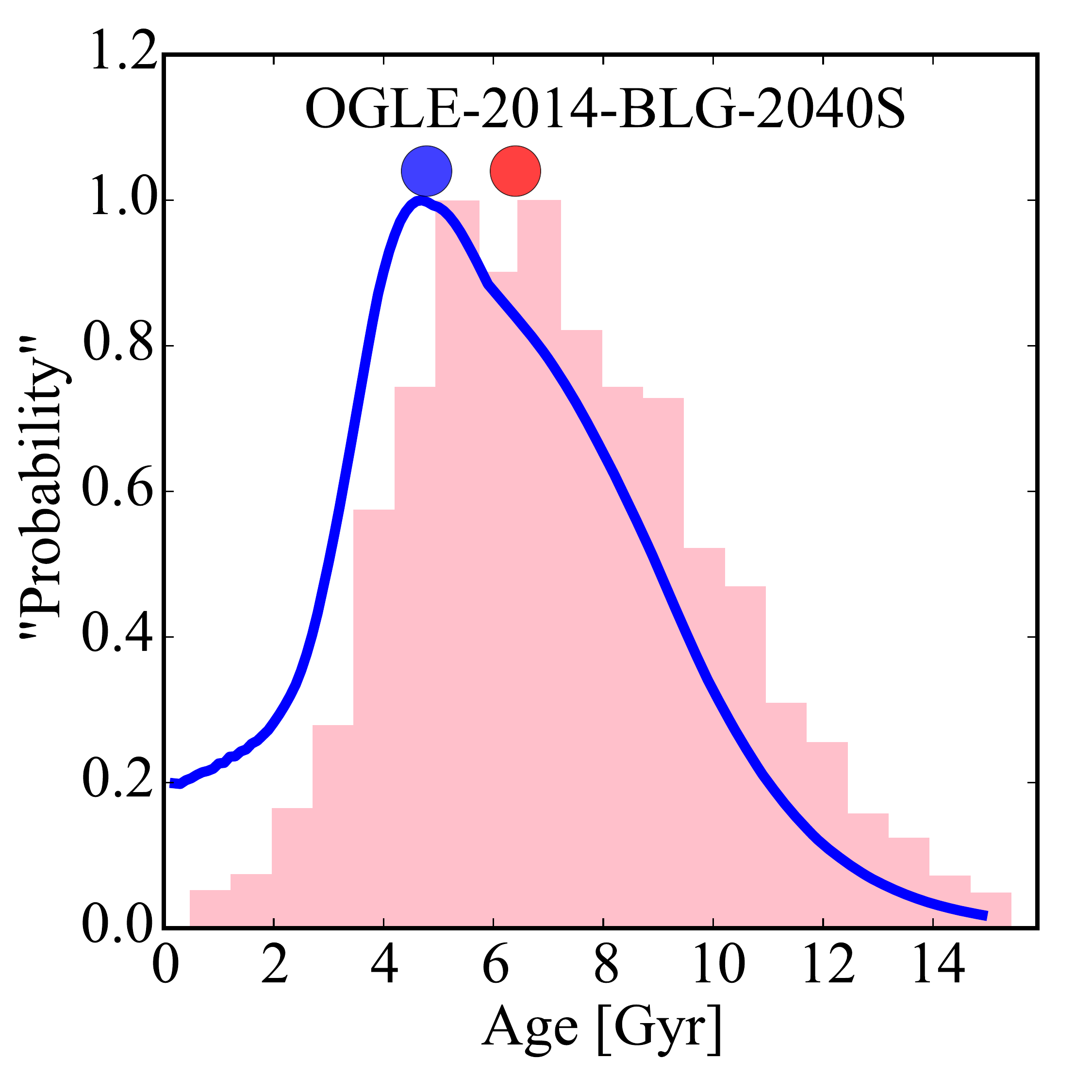}
\includegraphics[viewport= 93 0 648 648,clip]{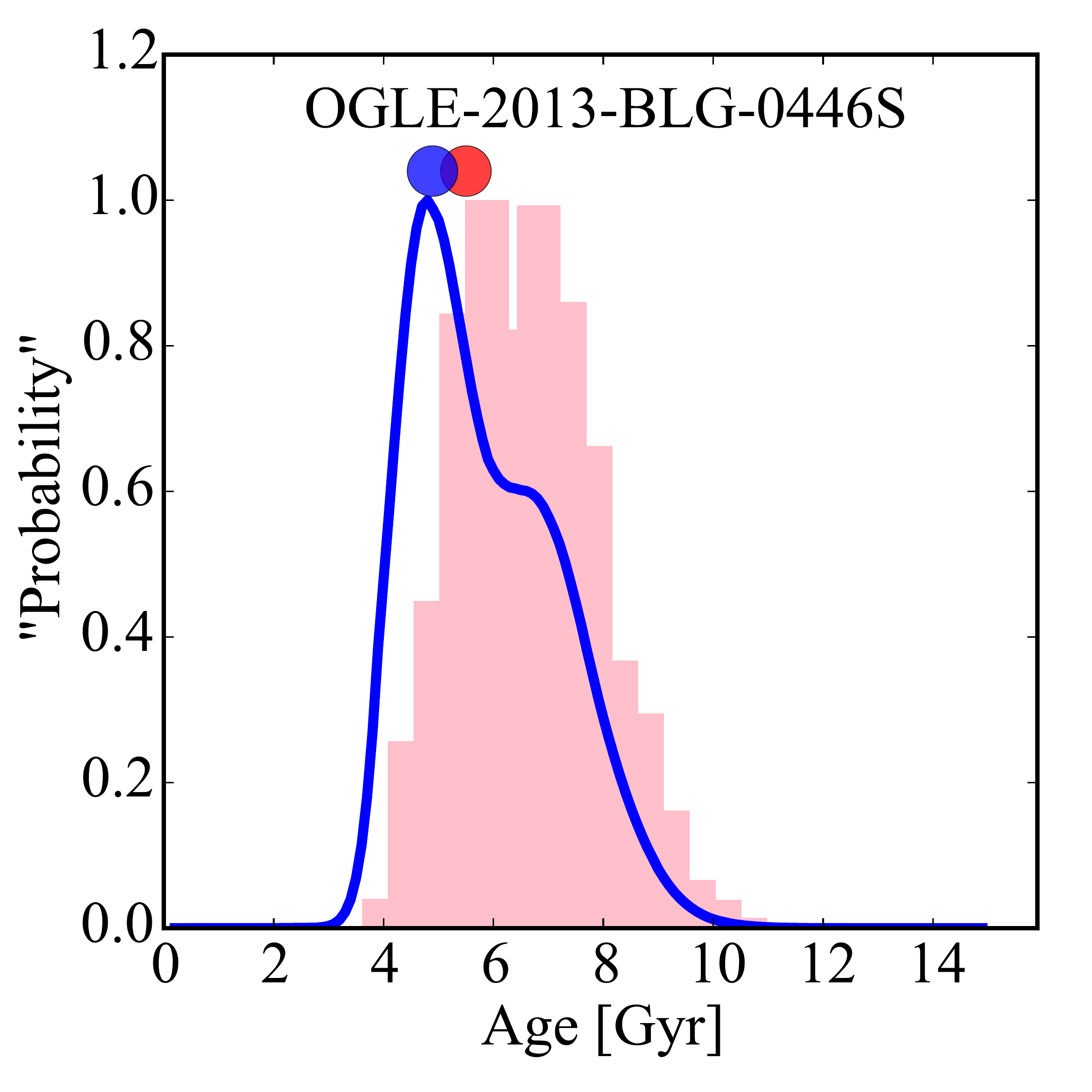}}
\resizebox{\hsize}{!}{
\includegraphics[viewport= 0 0 648 648,clip]{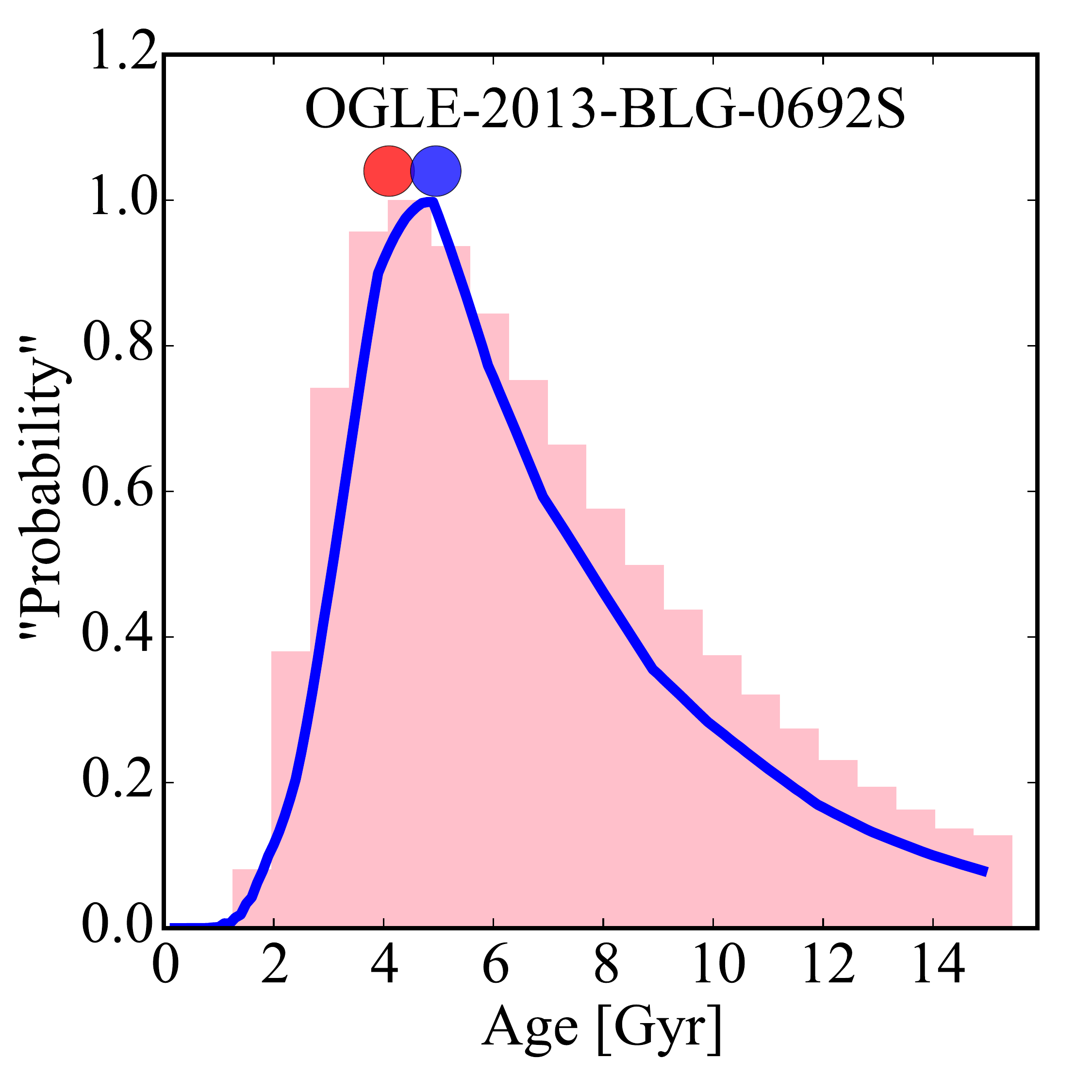}
\includegraphics[viewport= 93 0 648 648,clip]{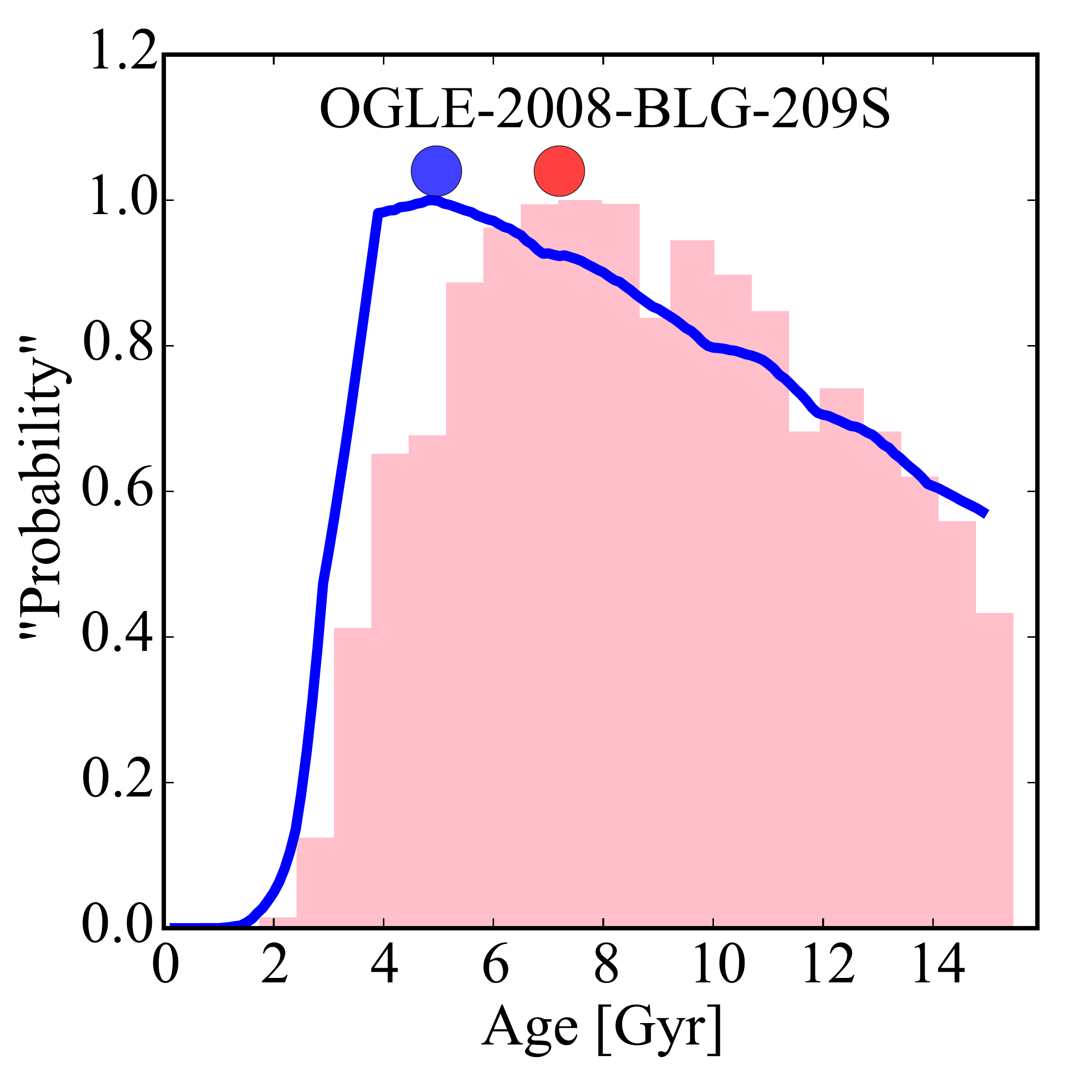}
\includegraphics[viewport= 93 0 648 648,clip]{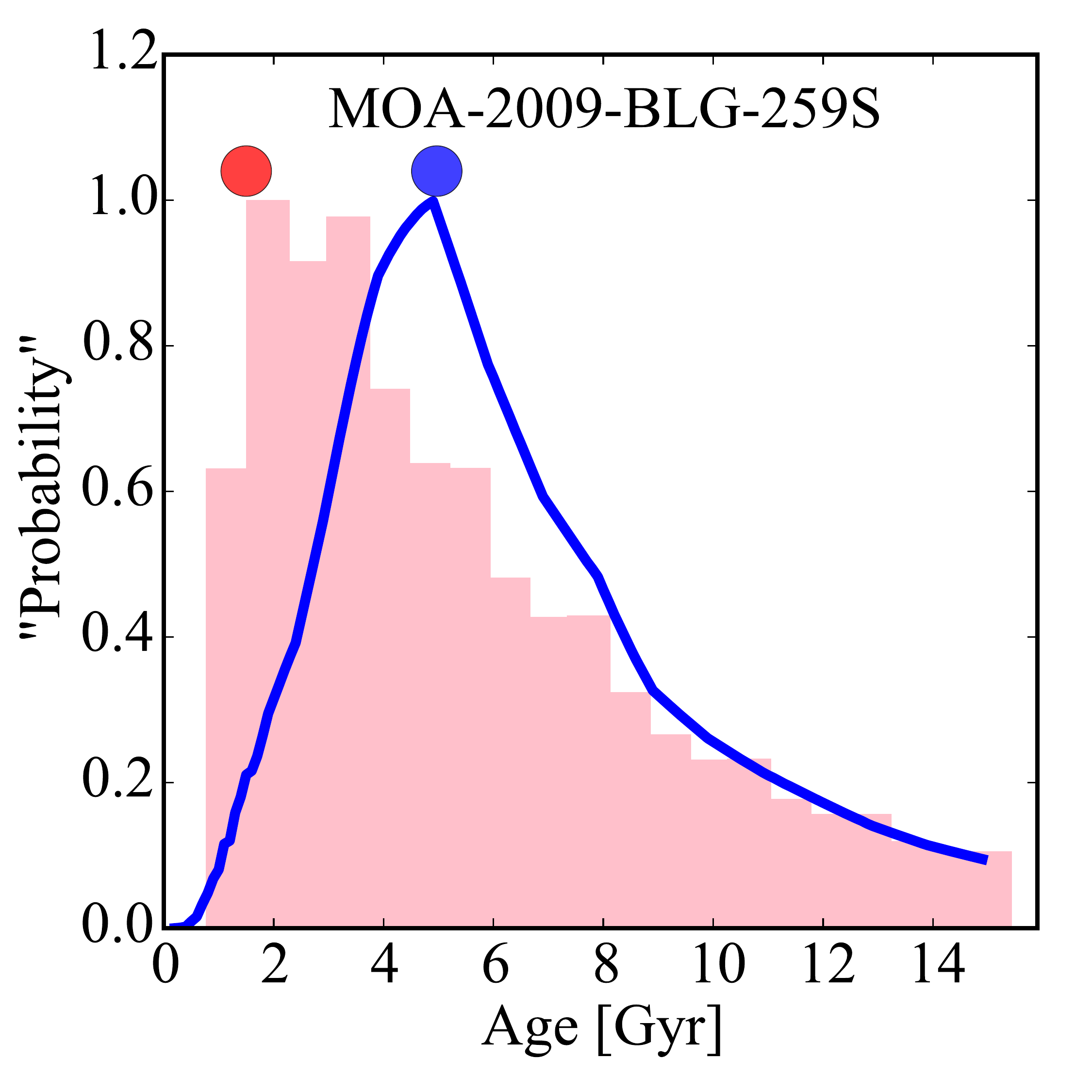}
\includegraphics[viewport= 93 0 648 648,clip]{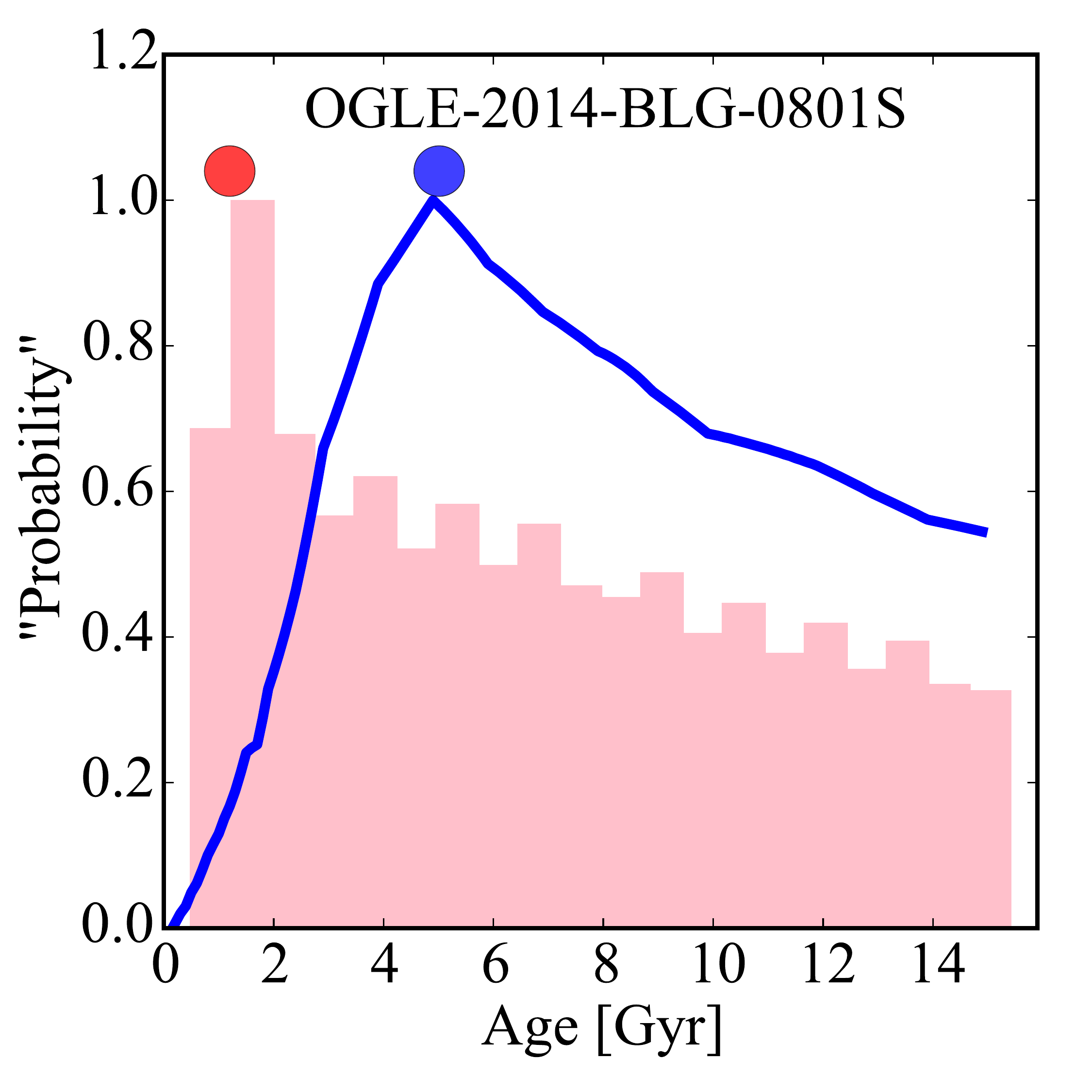}
\includegraphics[viewport= 93 0 648 648,clip]{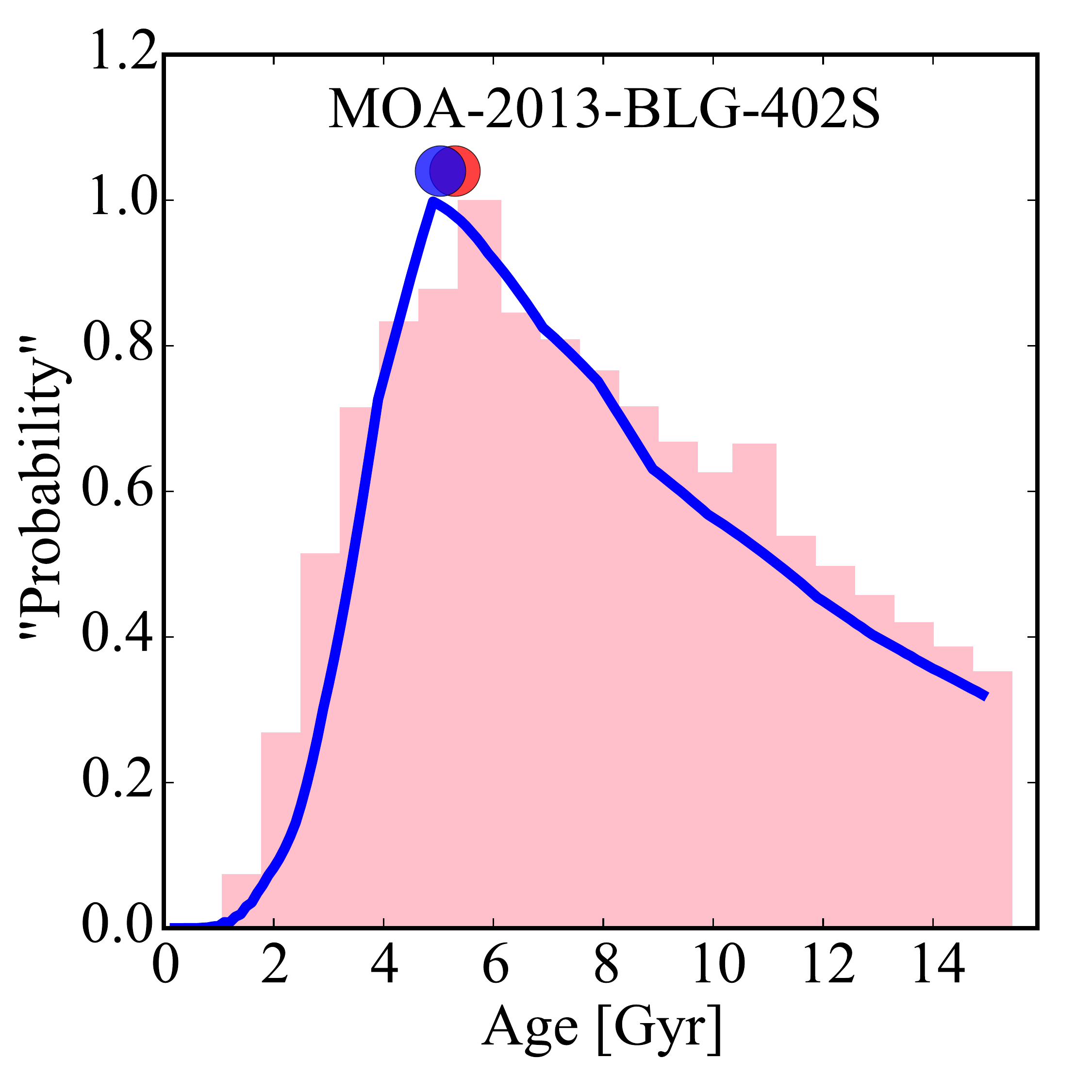}}
\caption{The red histograms in the figures below show the individual age probability distribution functions that were used to estimate the ages for all 91 stars. The estimated age is indicated by the red circle. The figures also show the individual G functions (blue lines) for the alternative age determination method, and the estimated ages from this method are marked by blue circles. Each plot contains the name of the target, and the plots have been sorted by estimated age (low to high).
\label{fig:agefunctions}
}
\end{figure*}

\setcounter{figure}{0}    
\begin{figure*}[ht]
\resizebox{\hsize}{!}{
\includegraphics[viewport= 0 0 648 648,clip]{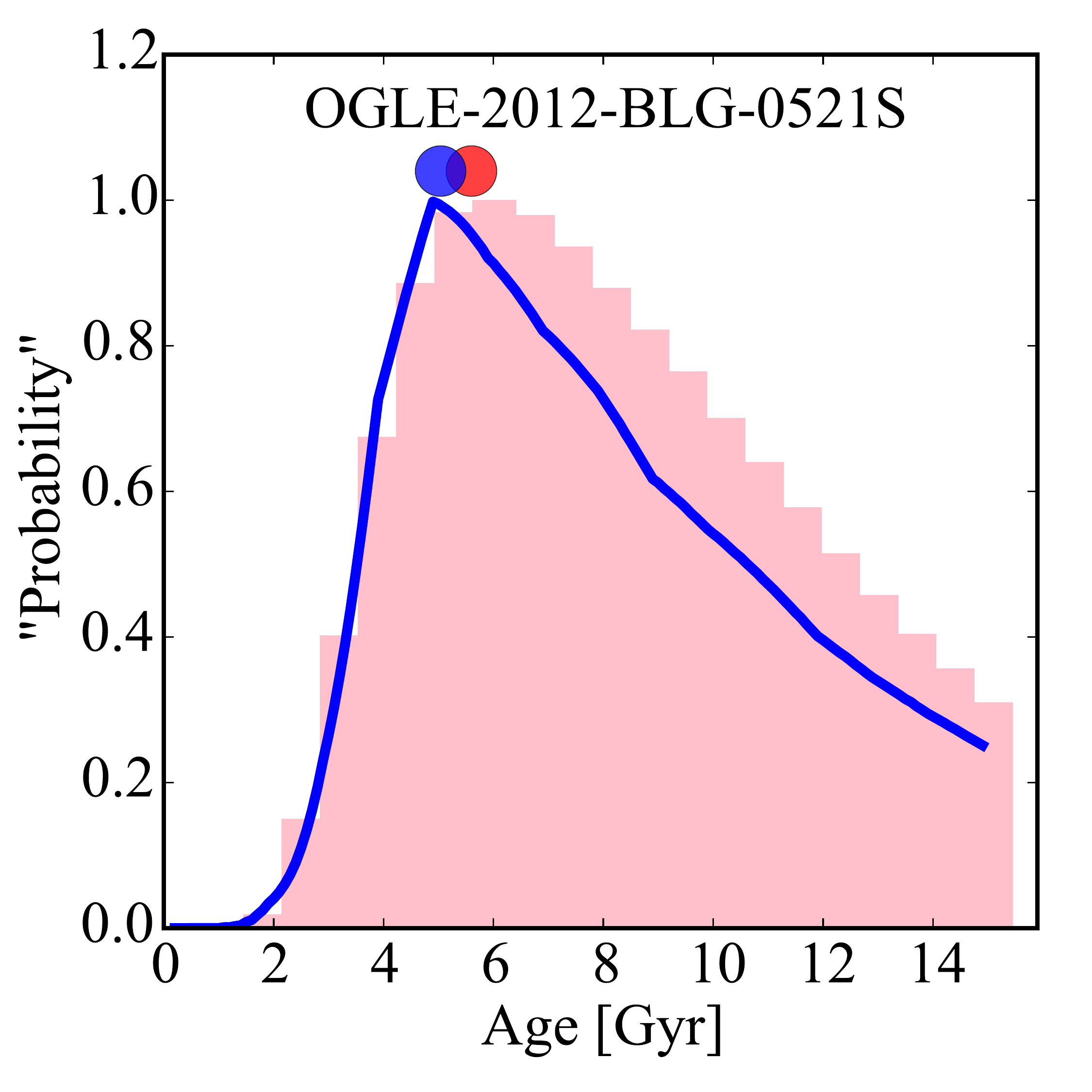}
\includegraphics[viewport= 93 0 648 648,clip]{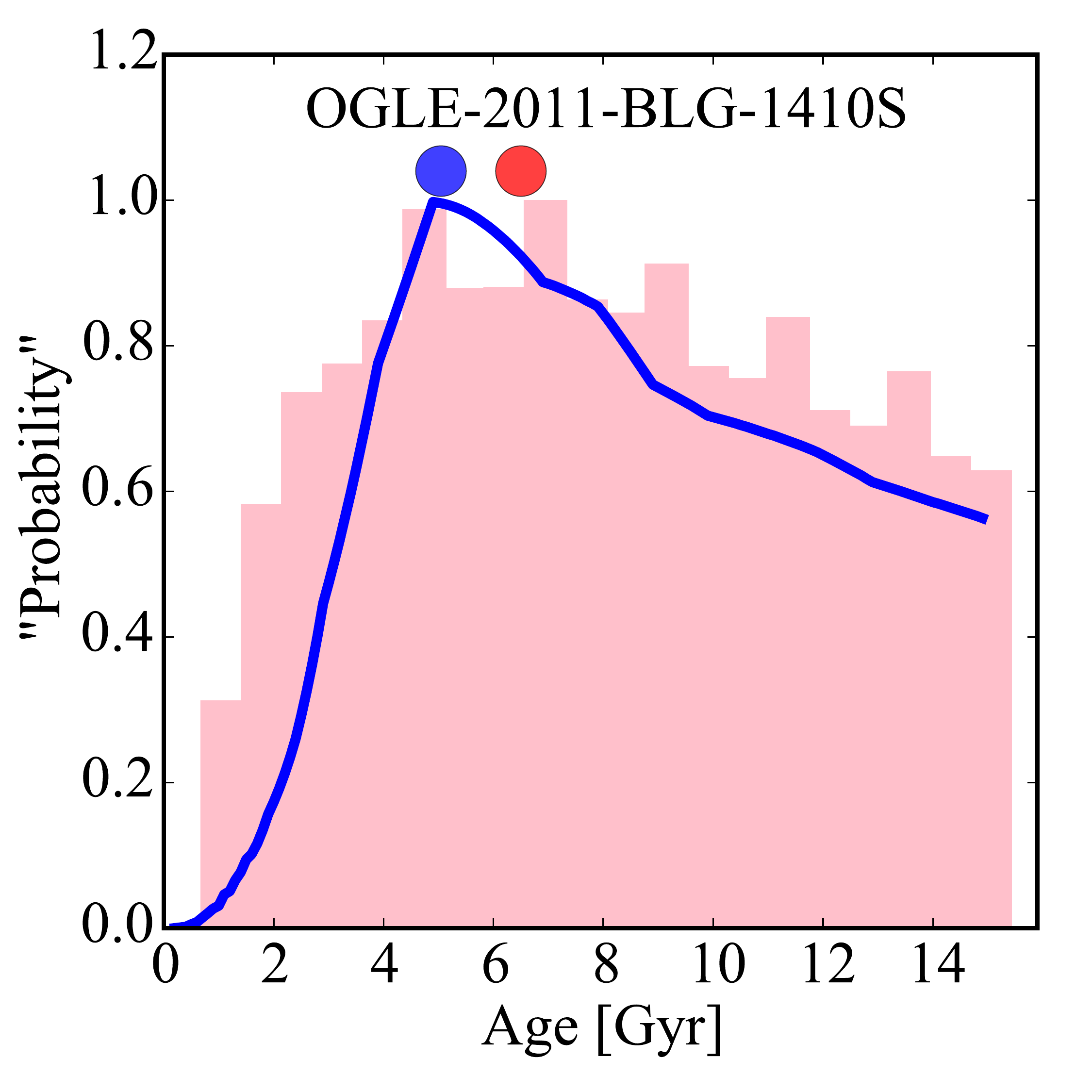}
\includegraphics[viewport= 93 0 648 648,clip]{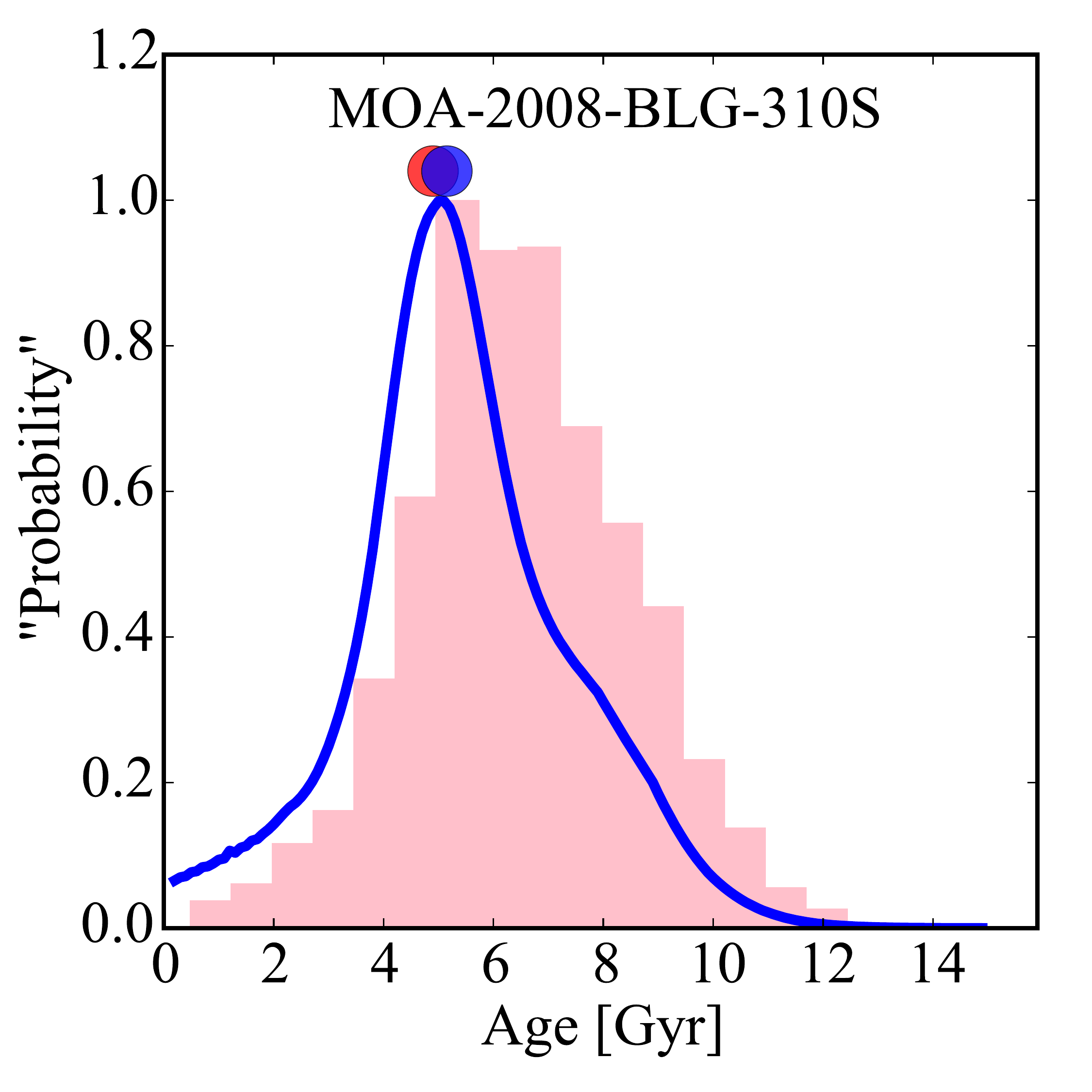}
\includegraphics[viewport= 93 0 648 648,clip]{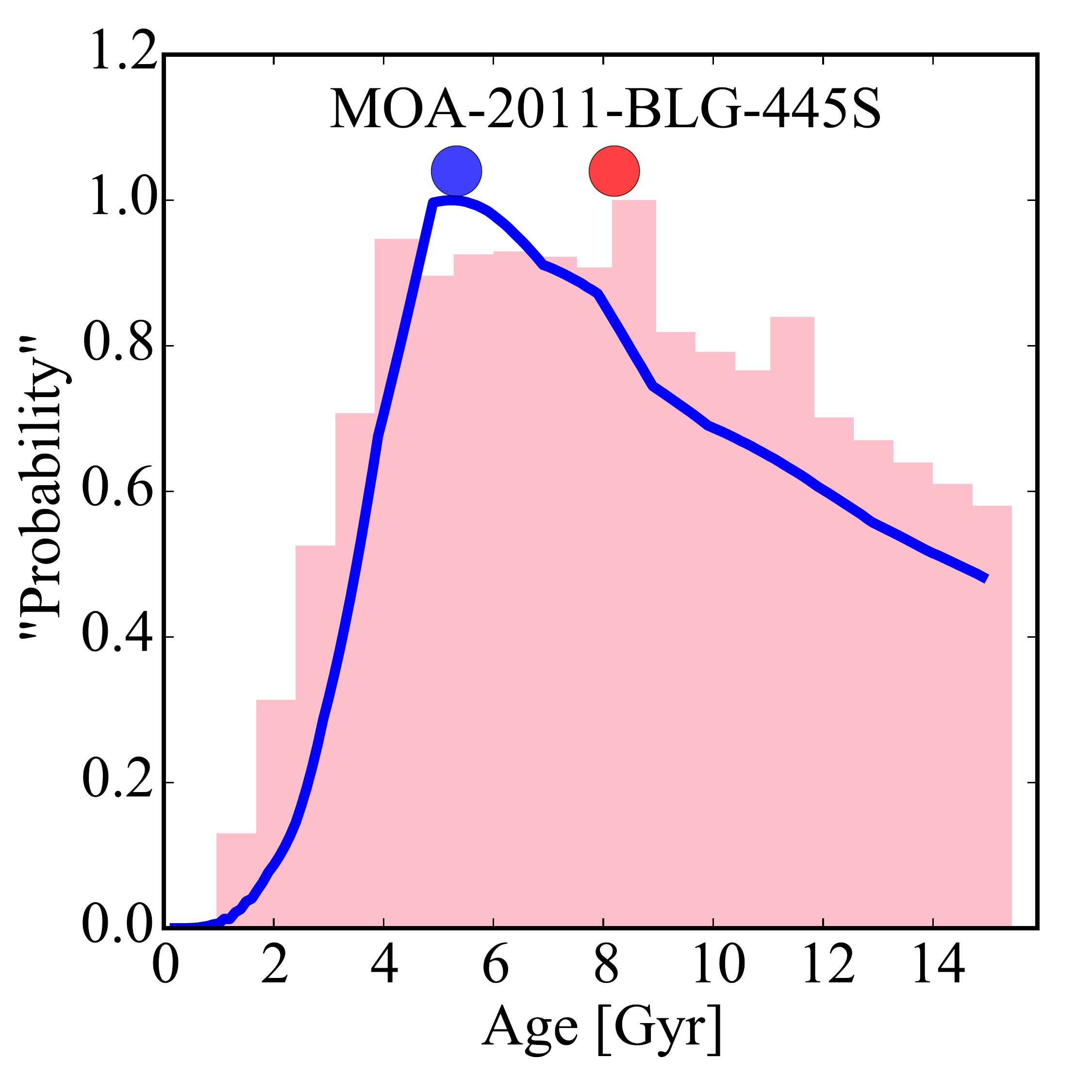}
\includegraphics[viewport= 93 0 648 648,clip]{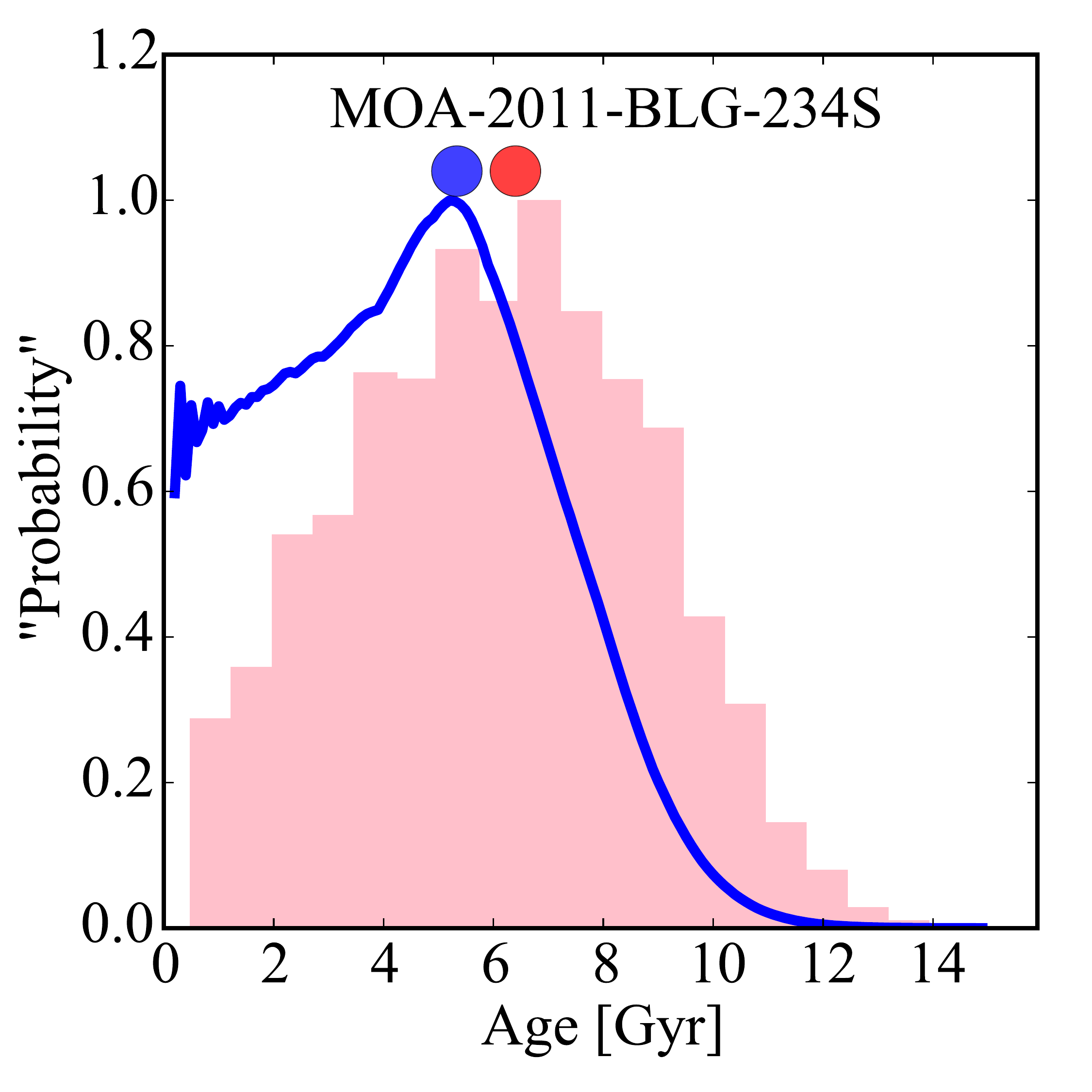}}
\resizebox{\hsize}{!}{
\includegraphics[viewport= 0 0 648 648,clip]{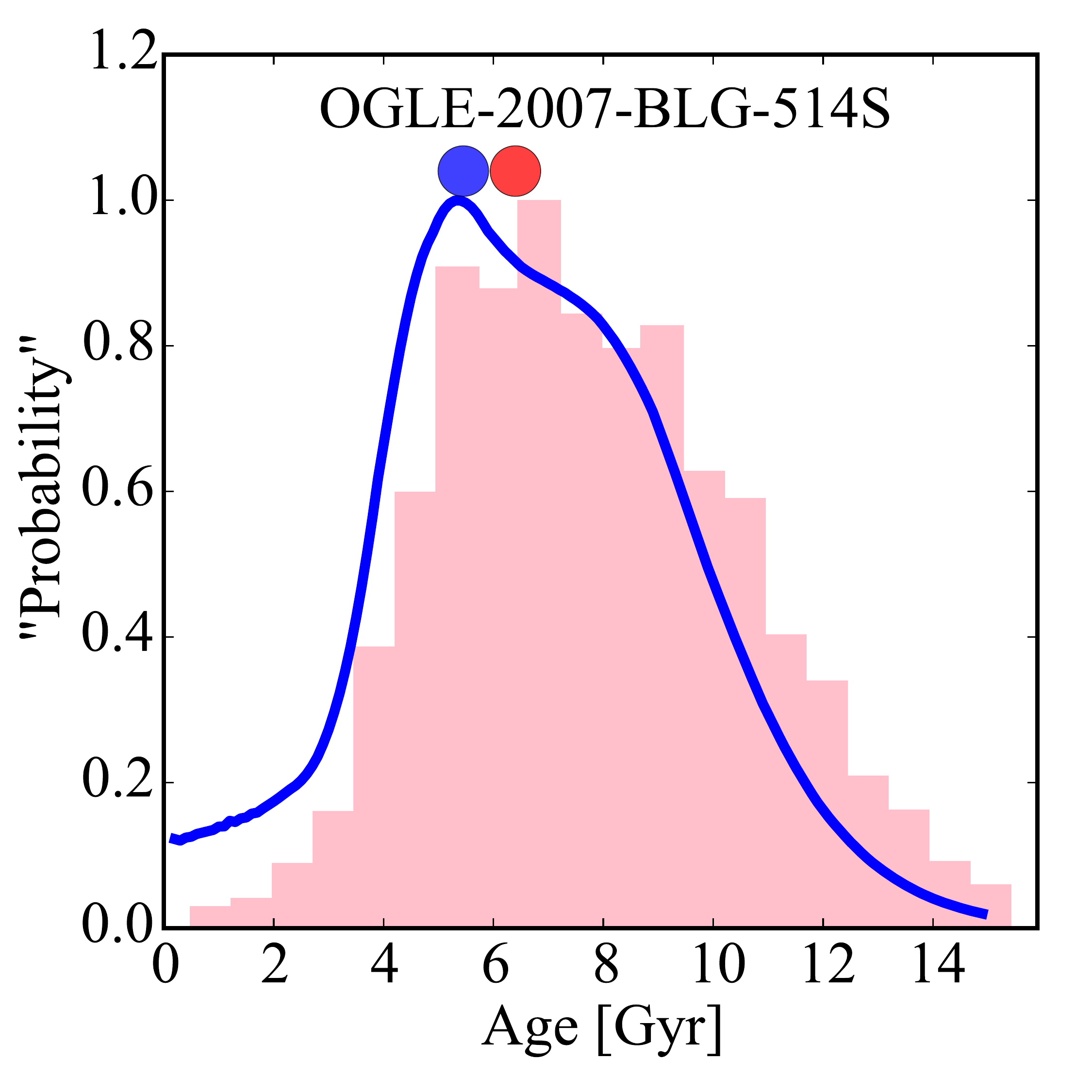}
\includegraphics[viewport= 93 0 648 648,clip]{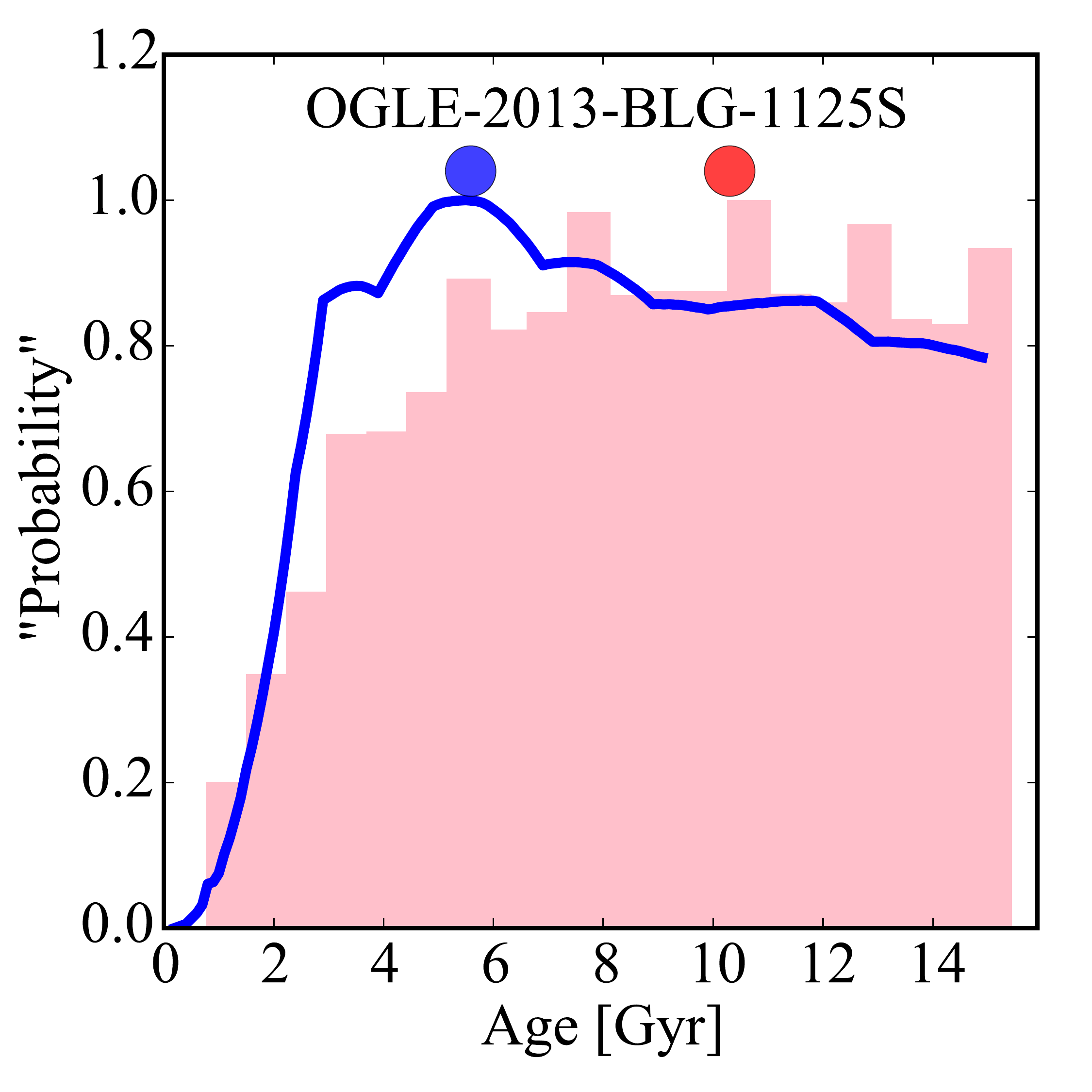}
\includegraphics[viewport= 93 0 648 648,clip]{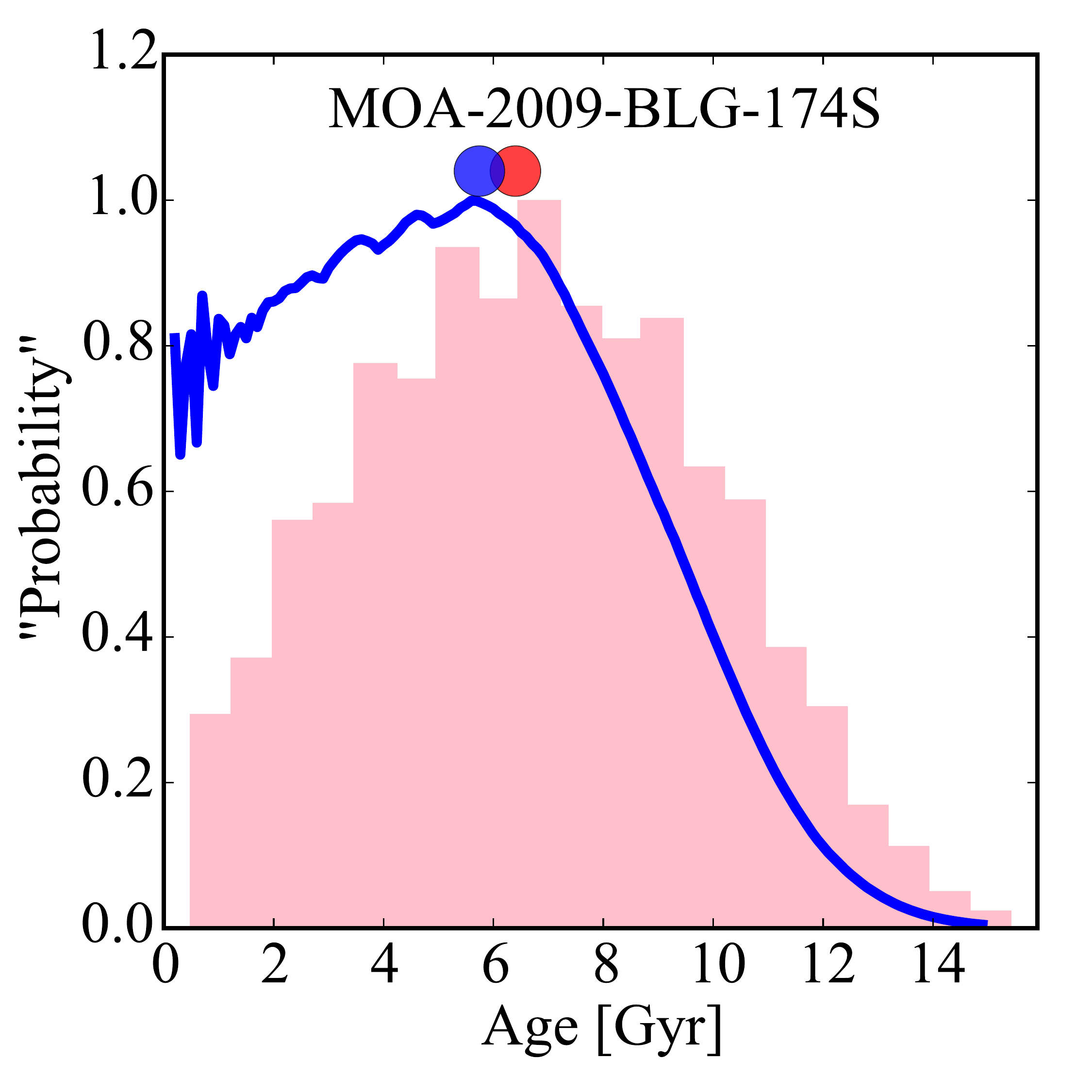}
\includegraphics[viewport= 93 0 648 648,clip]{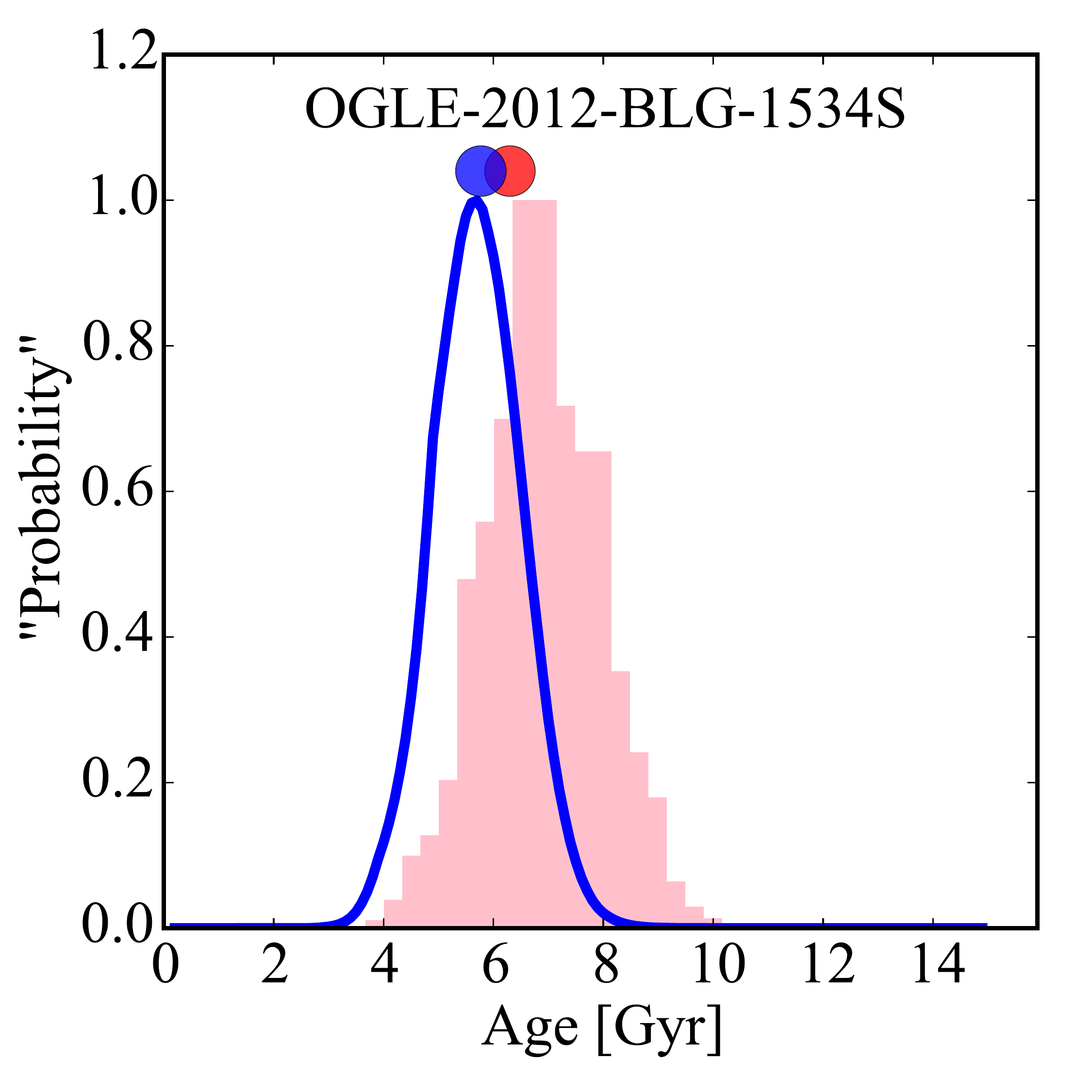}
\includegraphics[viewport= 93 0 648 648,clip]{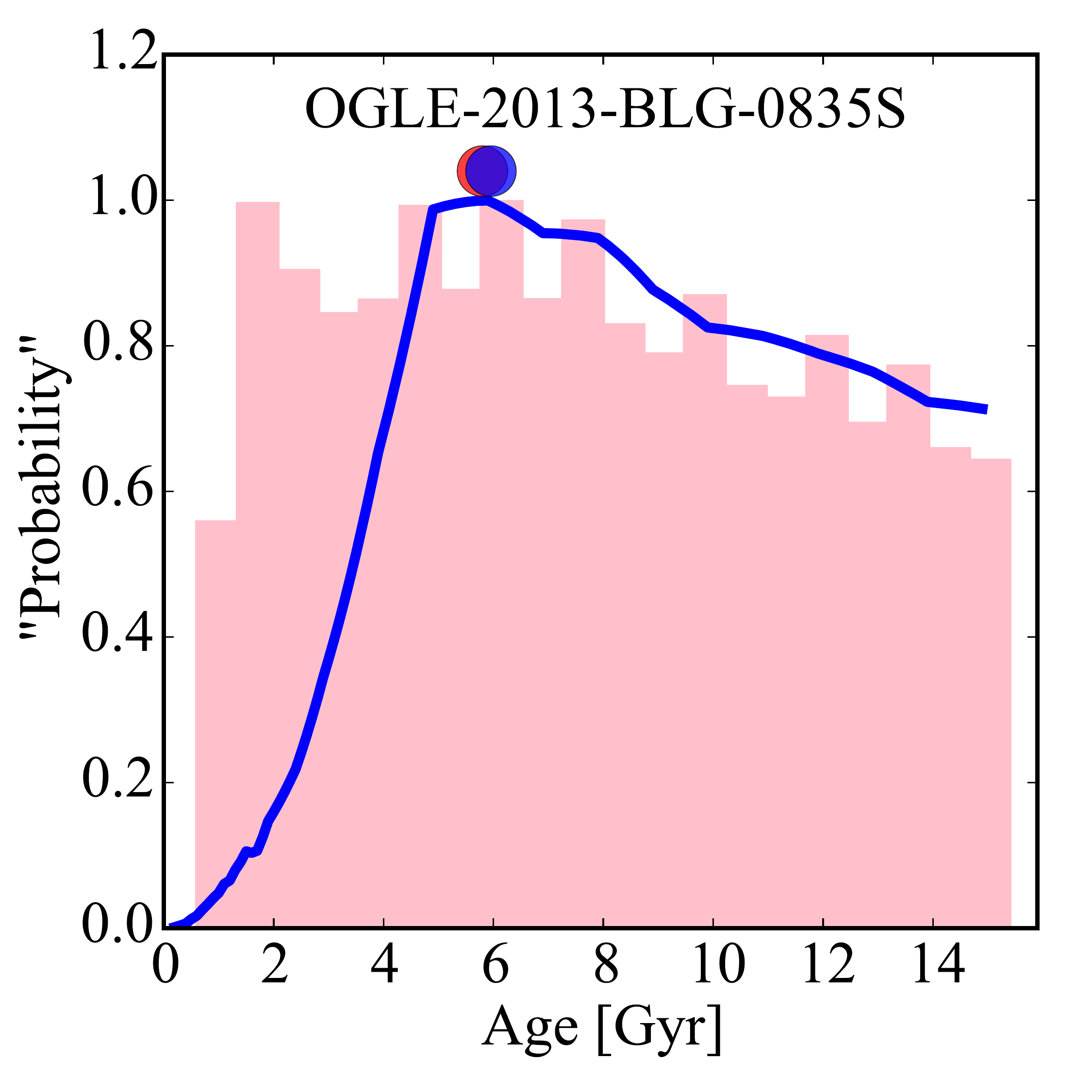}}
\resizebox{\hsize}{!}{
\includegraphics[viewport= 0 0 648 648,clip]{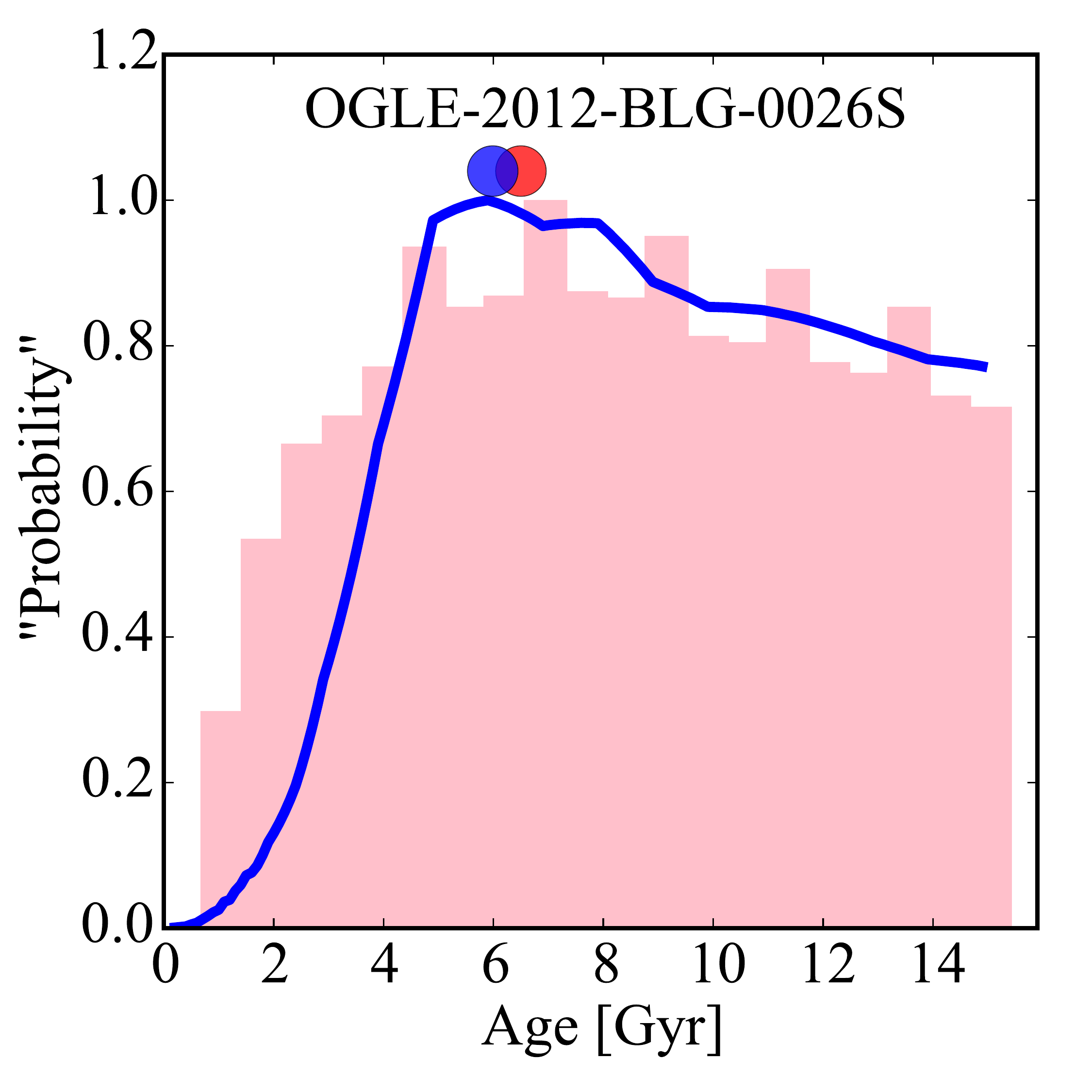}
\includegraphics[viewport= 93 0 648 648,clip]{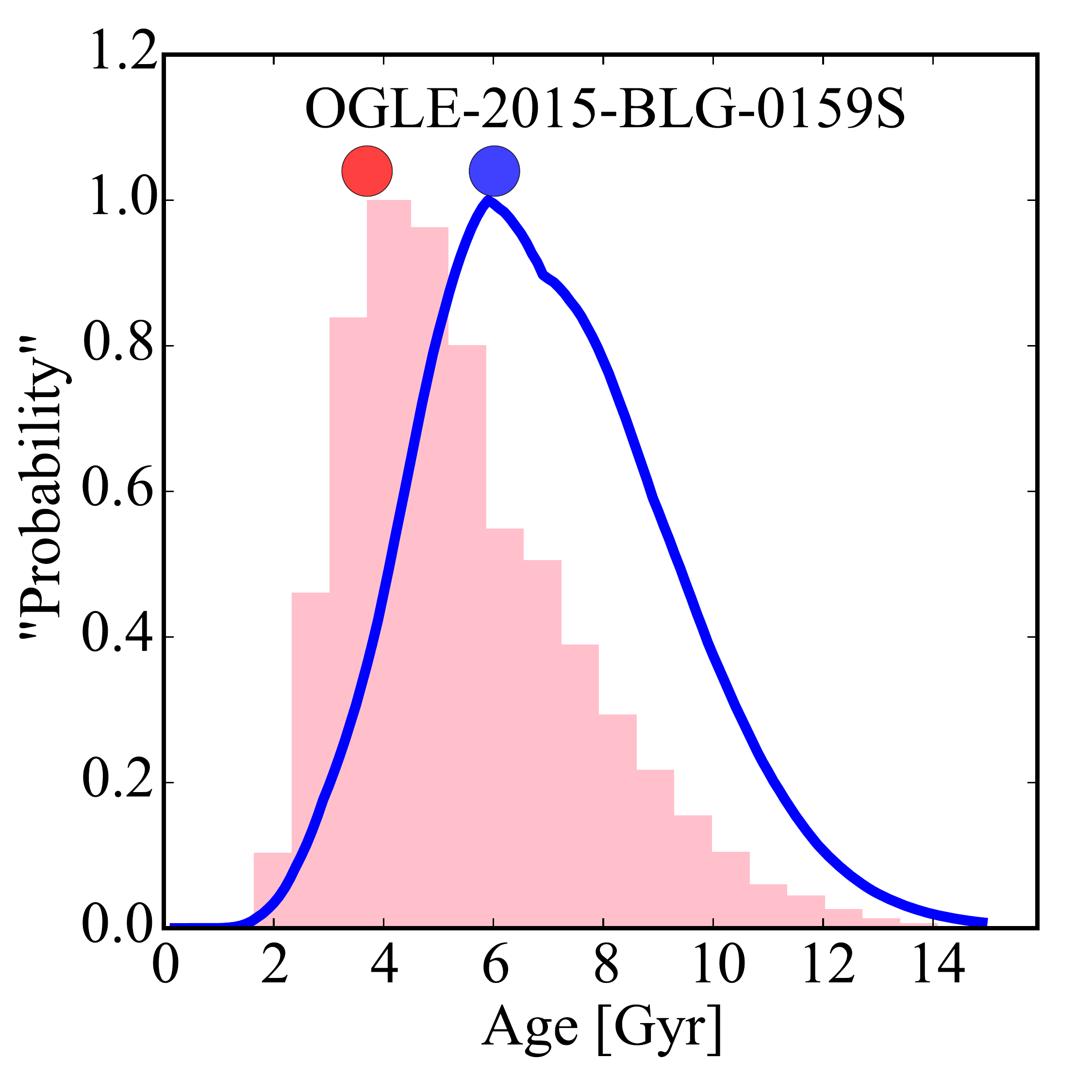}
\includegraphics[viewport= 93 0 648 648,clip]{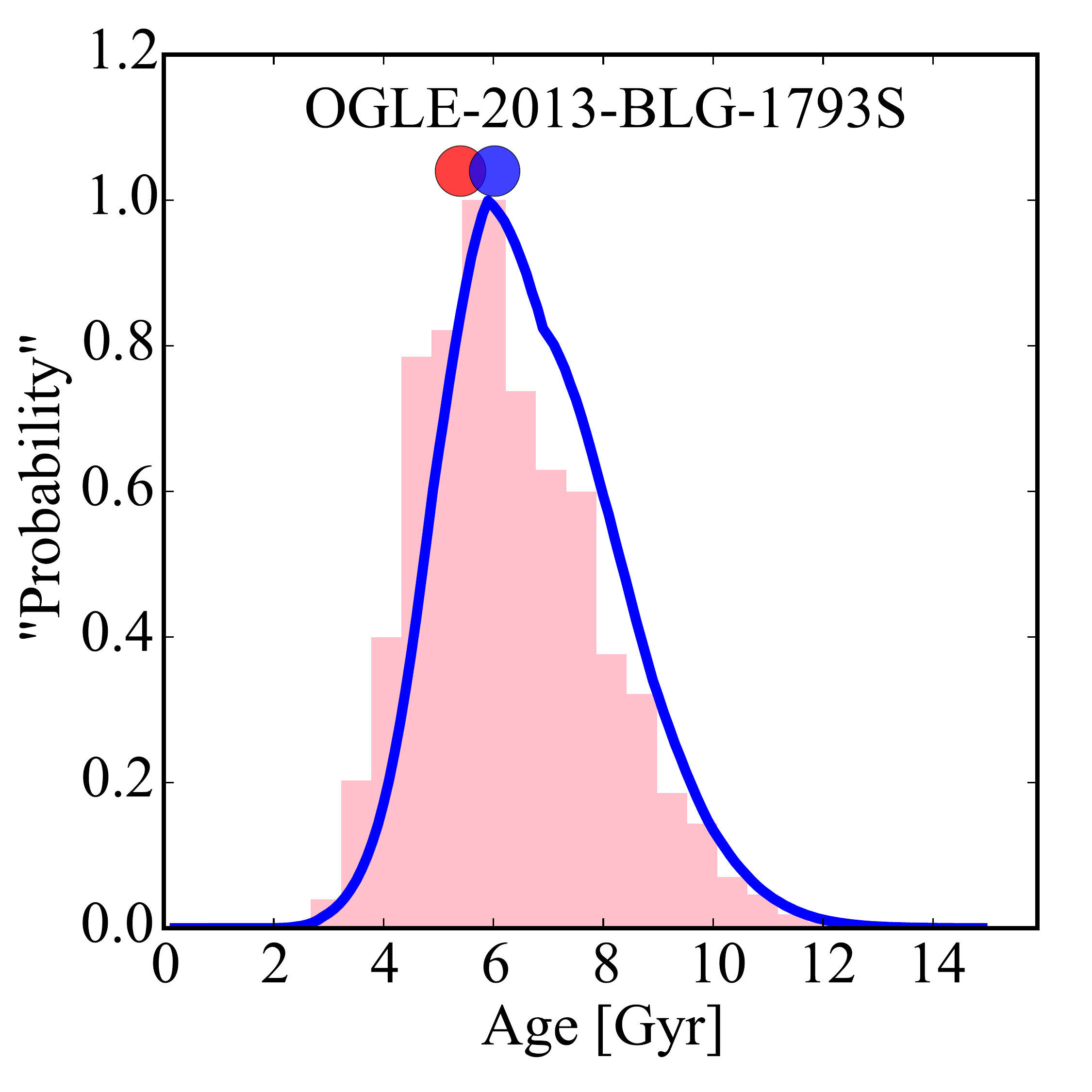}
\includegraphics[viewport= 93 0 648 648,clip]{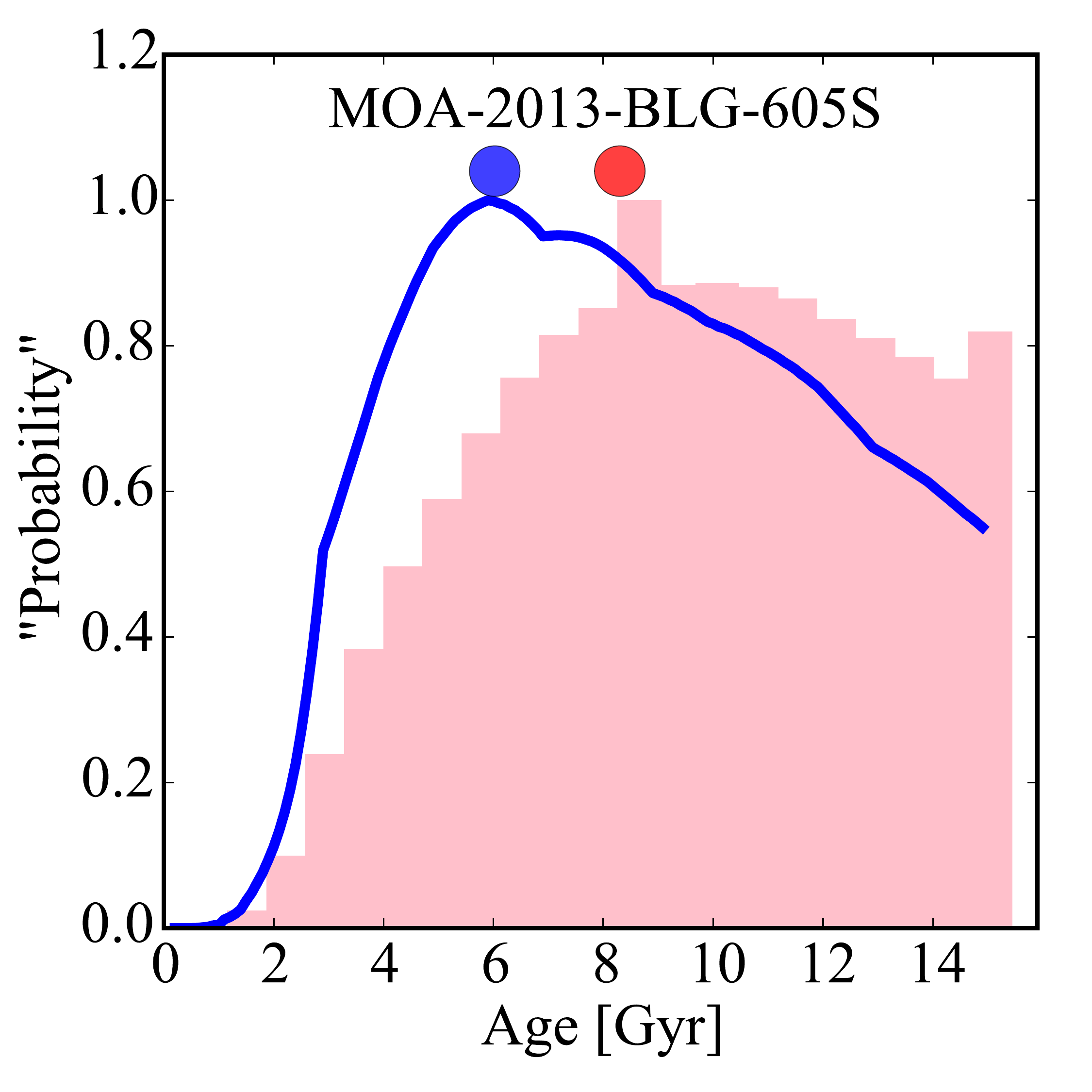}
\includegraphics[viewport= 93 0 648 648,clip]{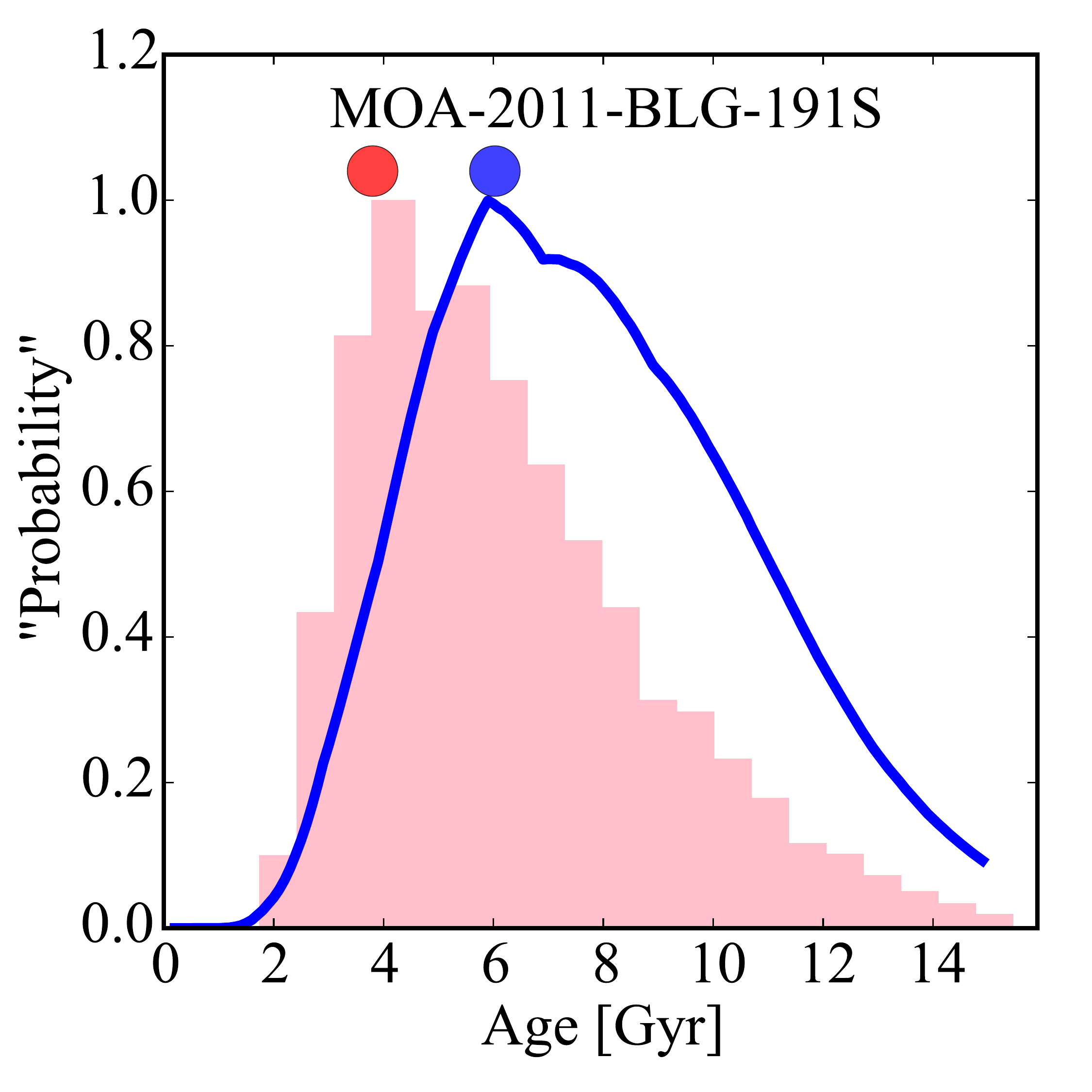}}
\resizebox{\hsize}{!}{
\includegraphics[viewport= 0 0 648 648,clip]{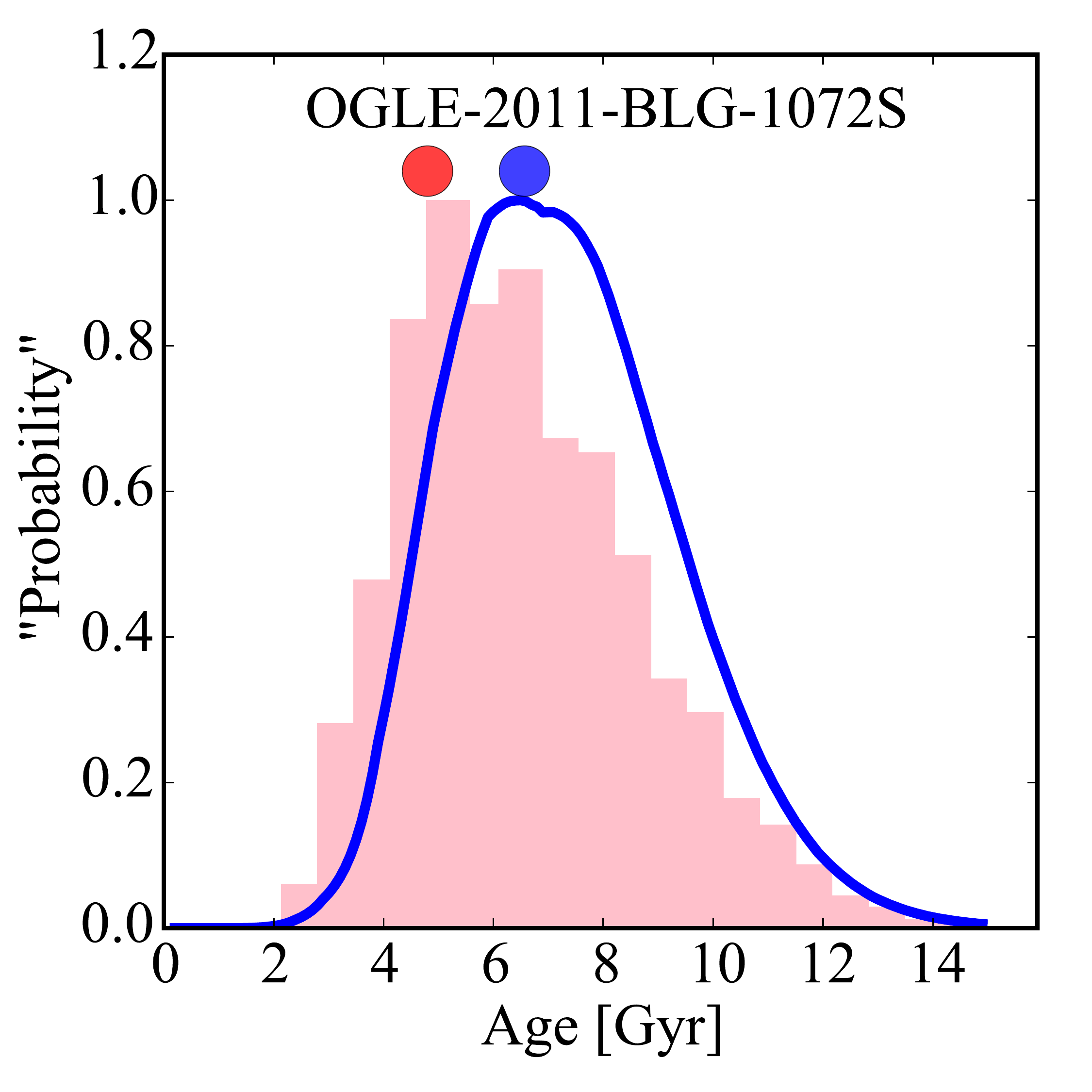}
\includegraphics[viewport= 93 0 648 648,clip]{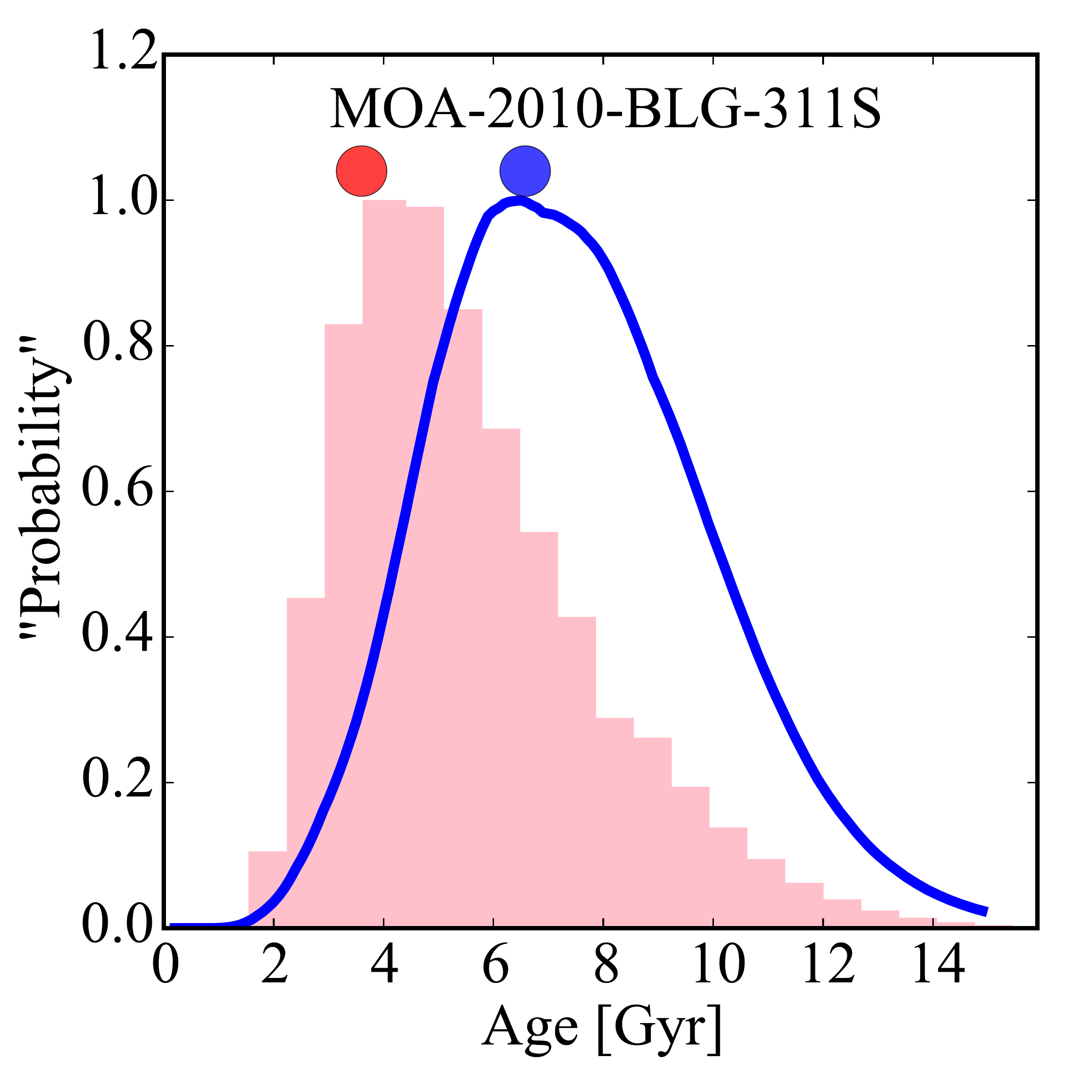}
\includegraphics[viewport= 93 0 648 648,clip]{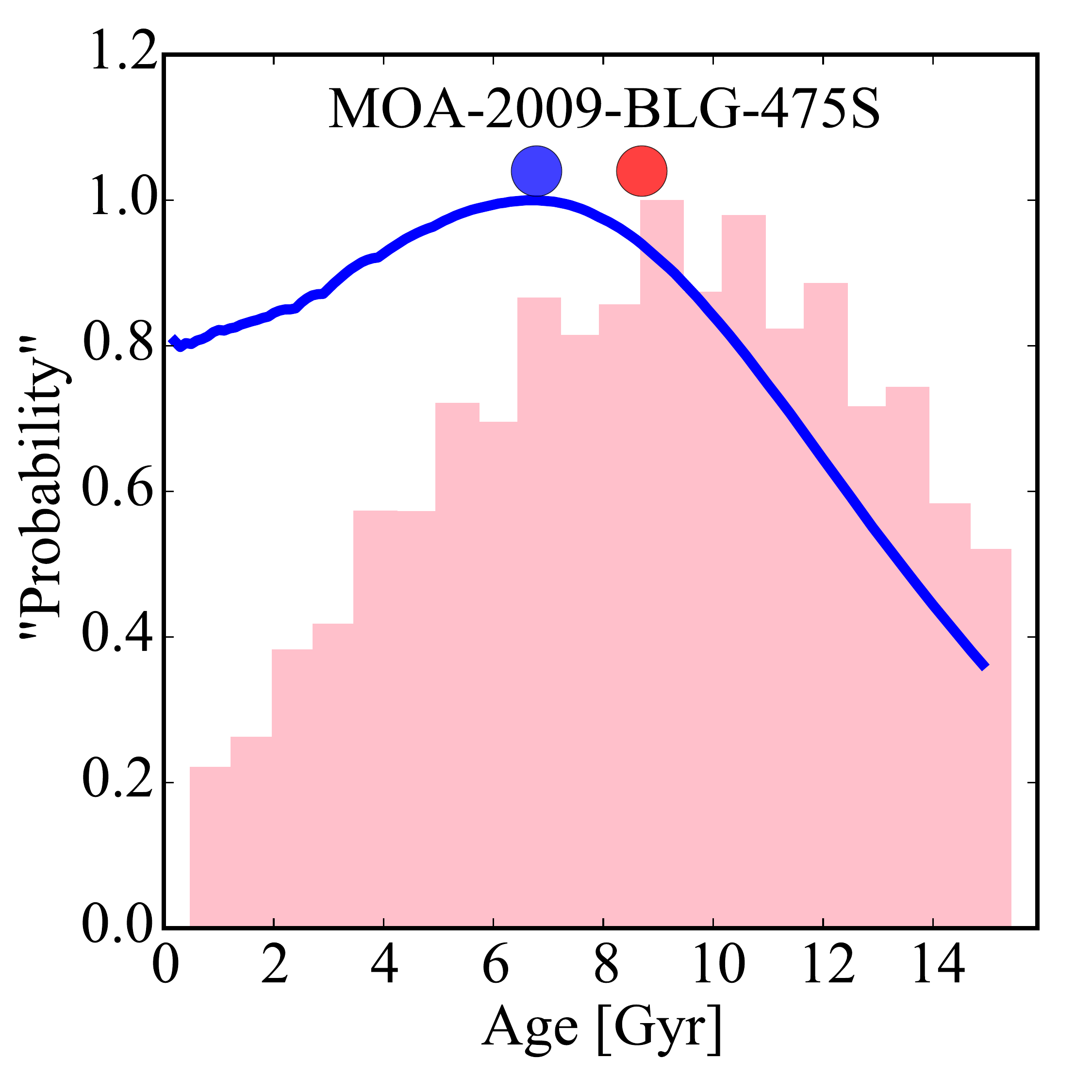}
\includegraphics[viewport= 93 0 648 648,clip]{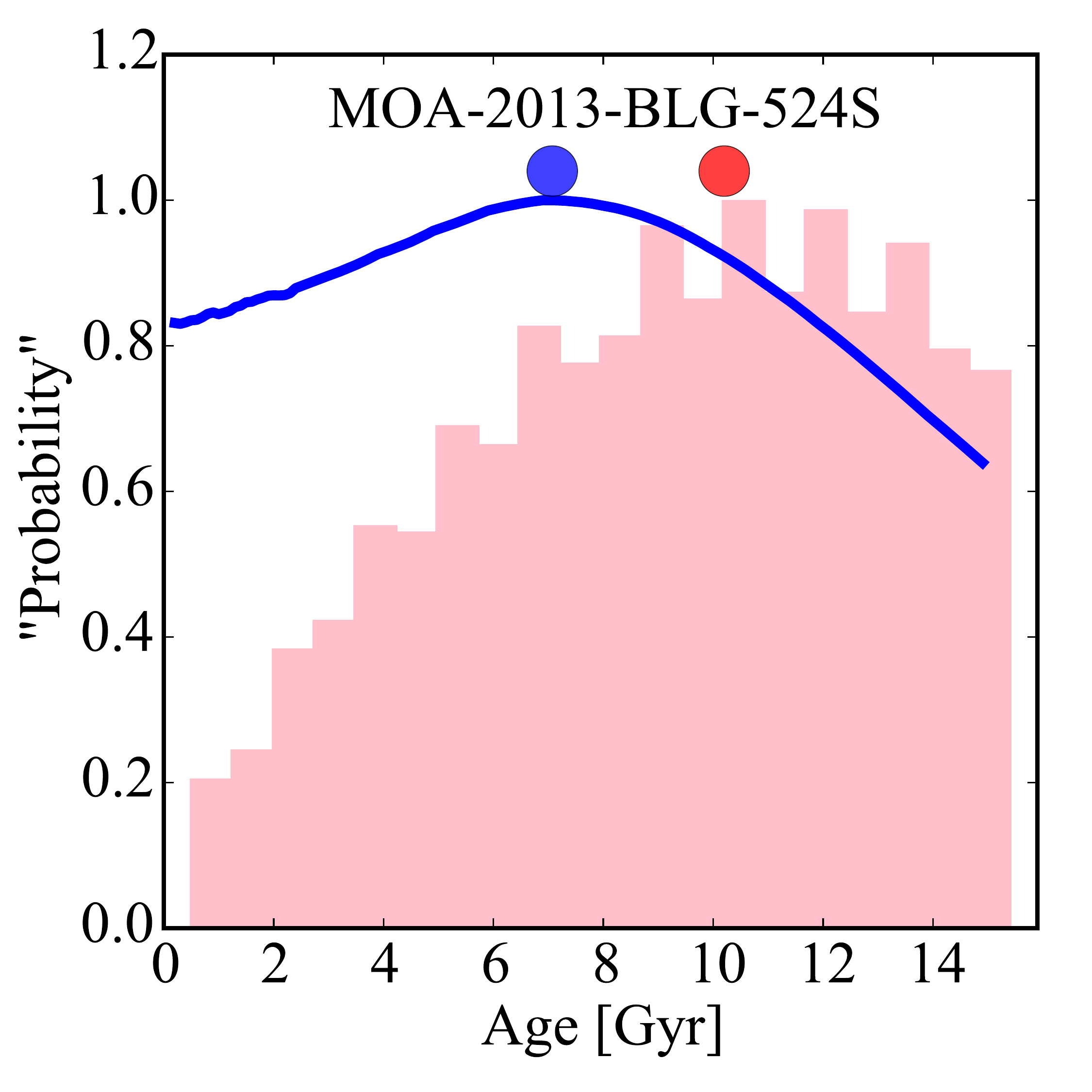}
\includegraphics[viewport= 93 0 648 648,clip]{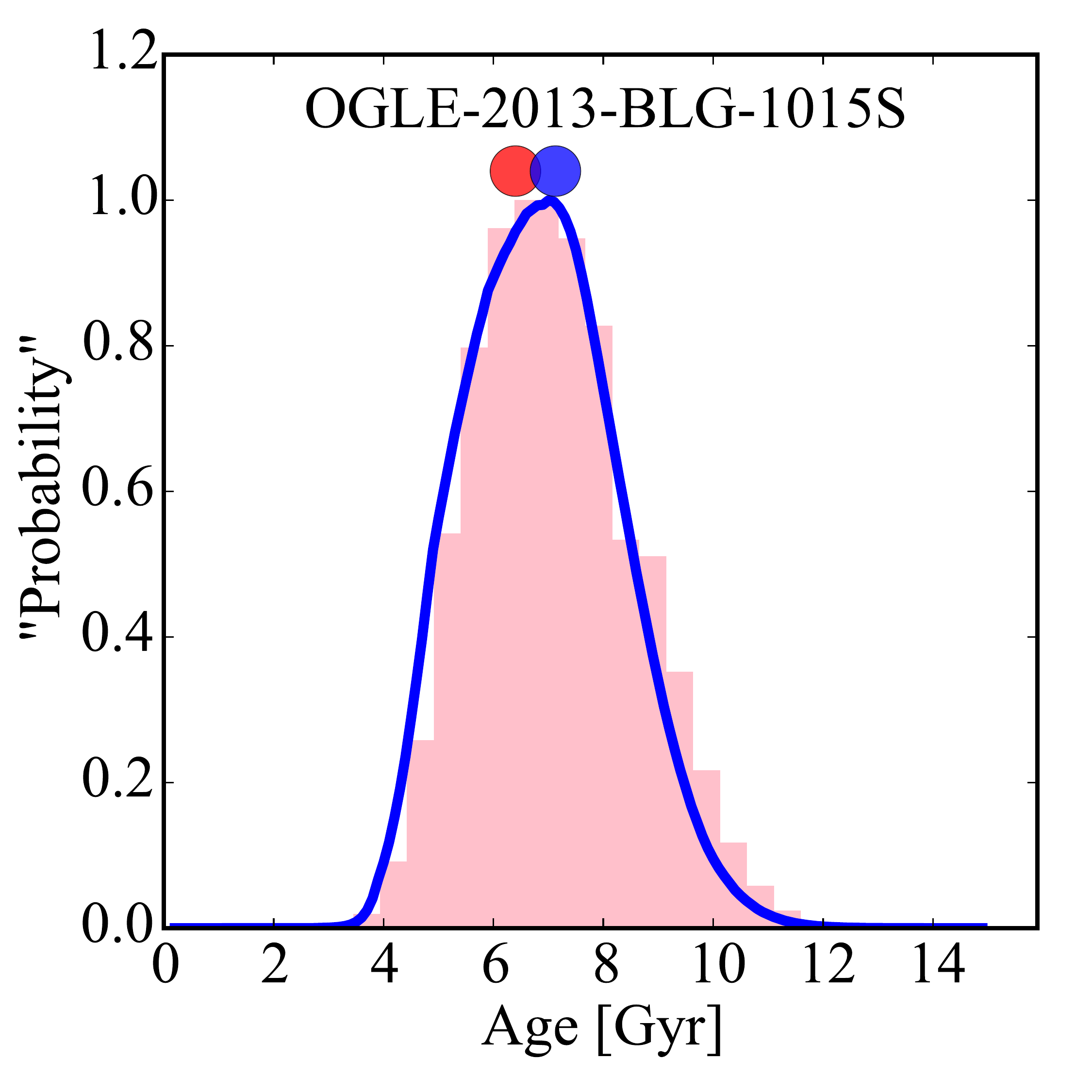}}
\resizebox{\hsize}{!}{
\includegraphics[viewport= 0 0 648 648,clip]{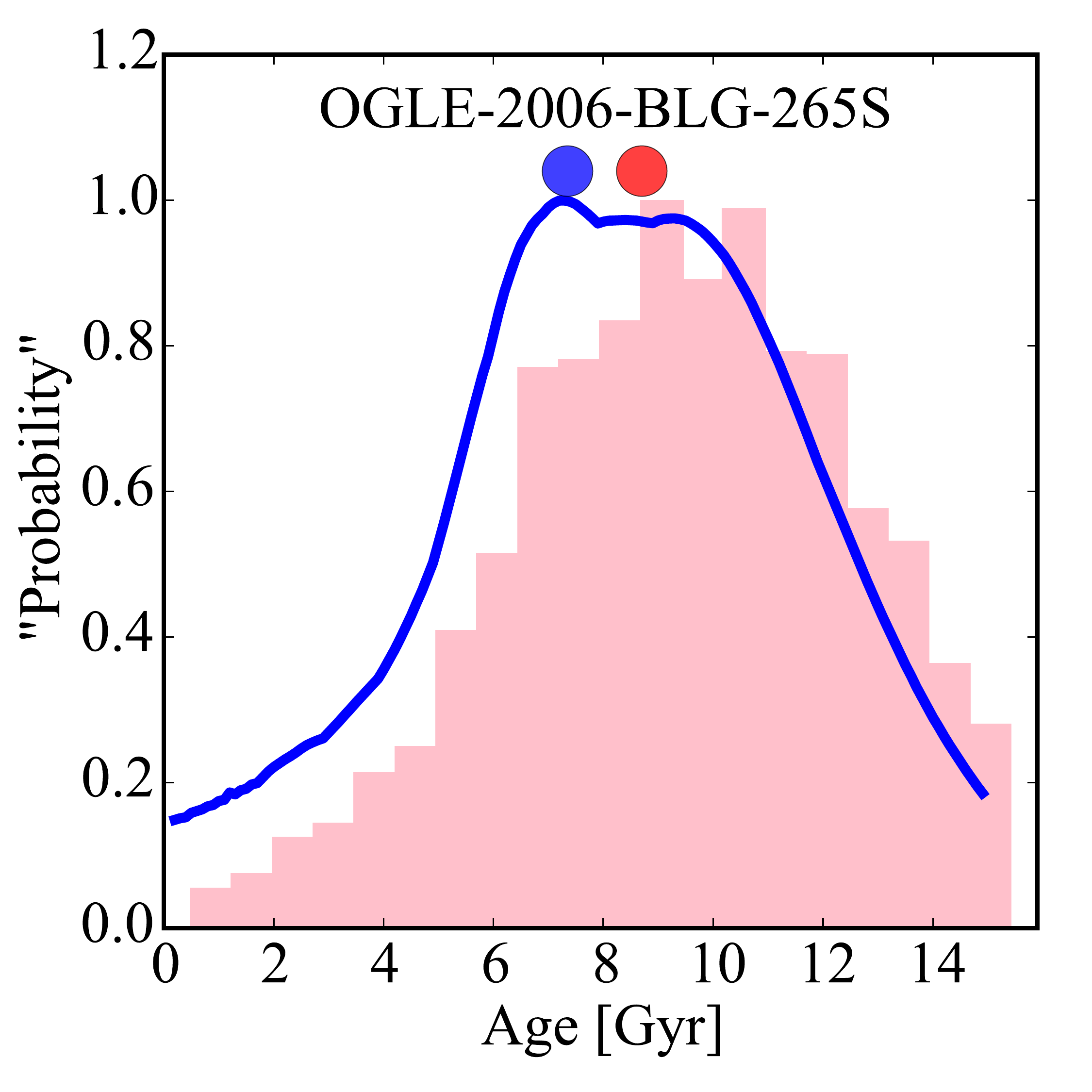}
\includegraphics[viewport= 93 0 648 648,clip]{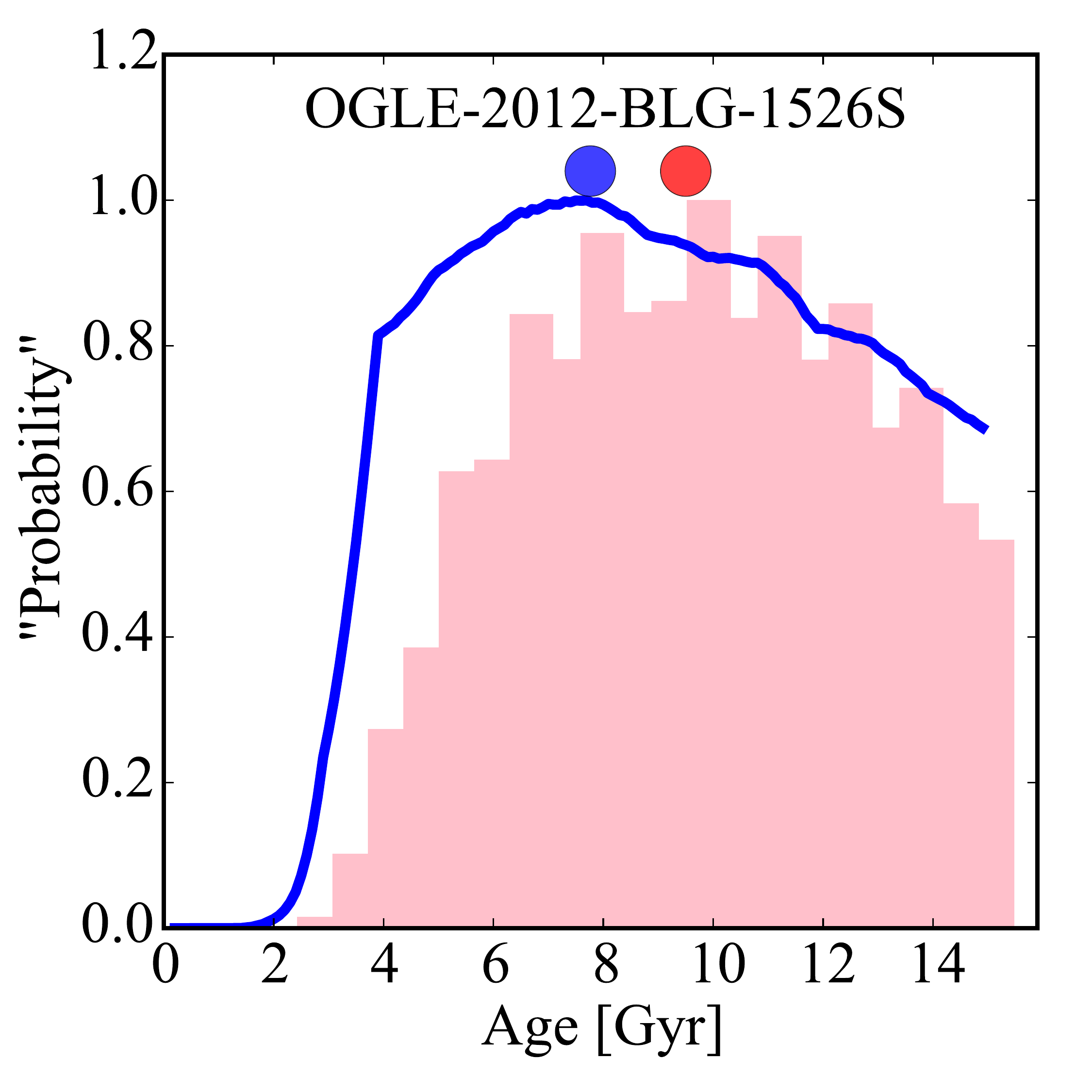}
\includegraphics[viewport= 93 0 648 648,clip]{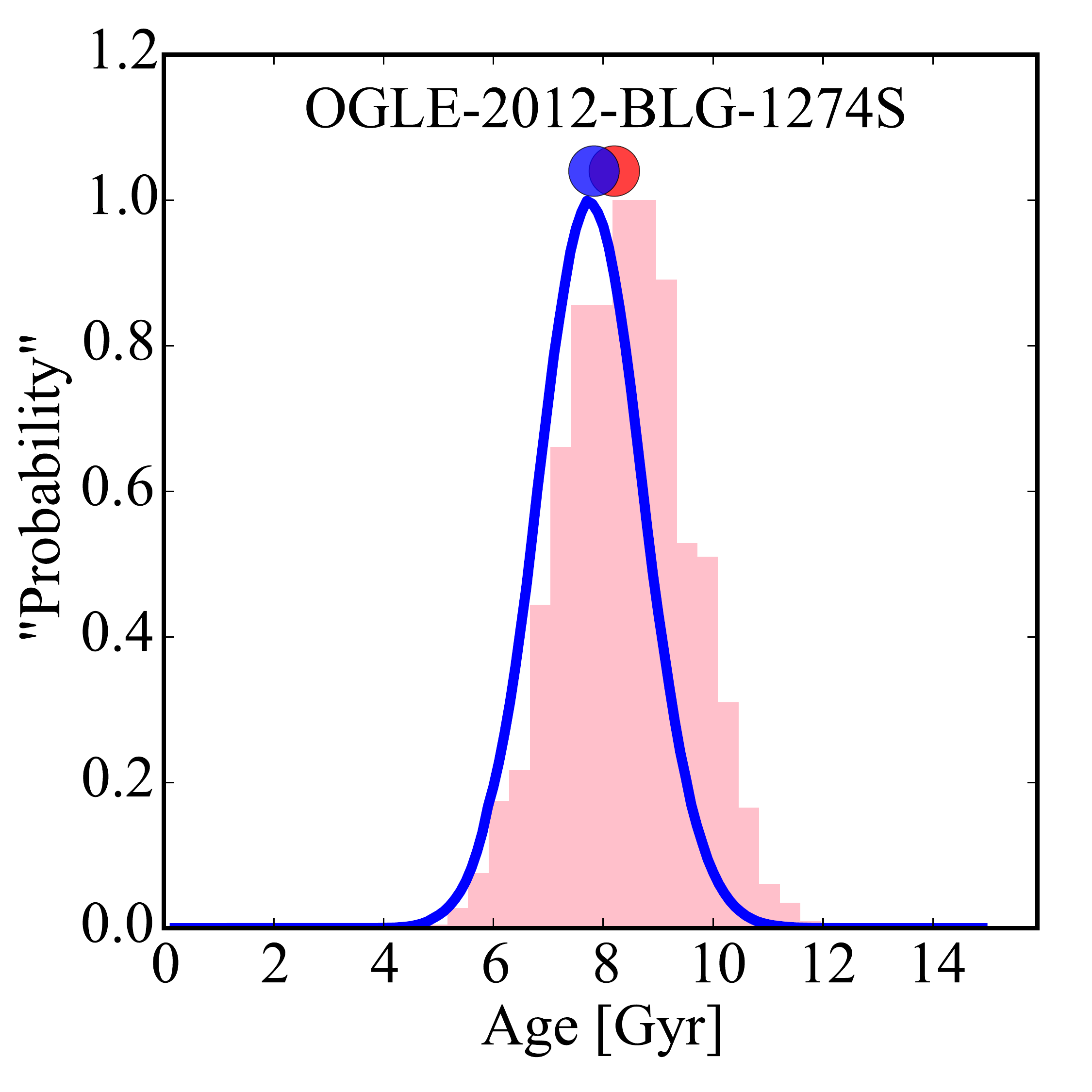}
\includegraphics[viewport= 93 0 648 648,clip]{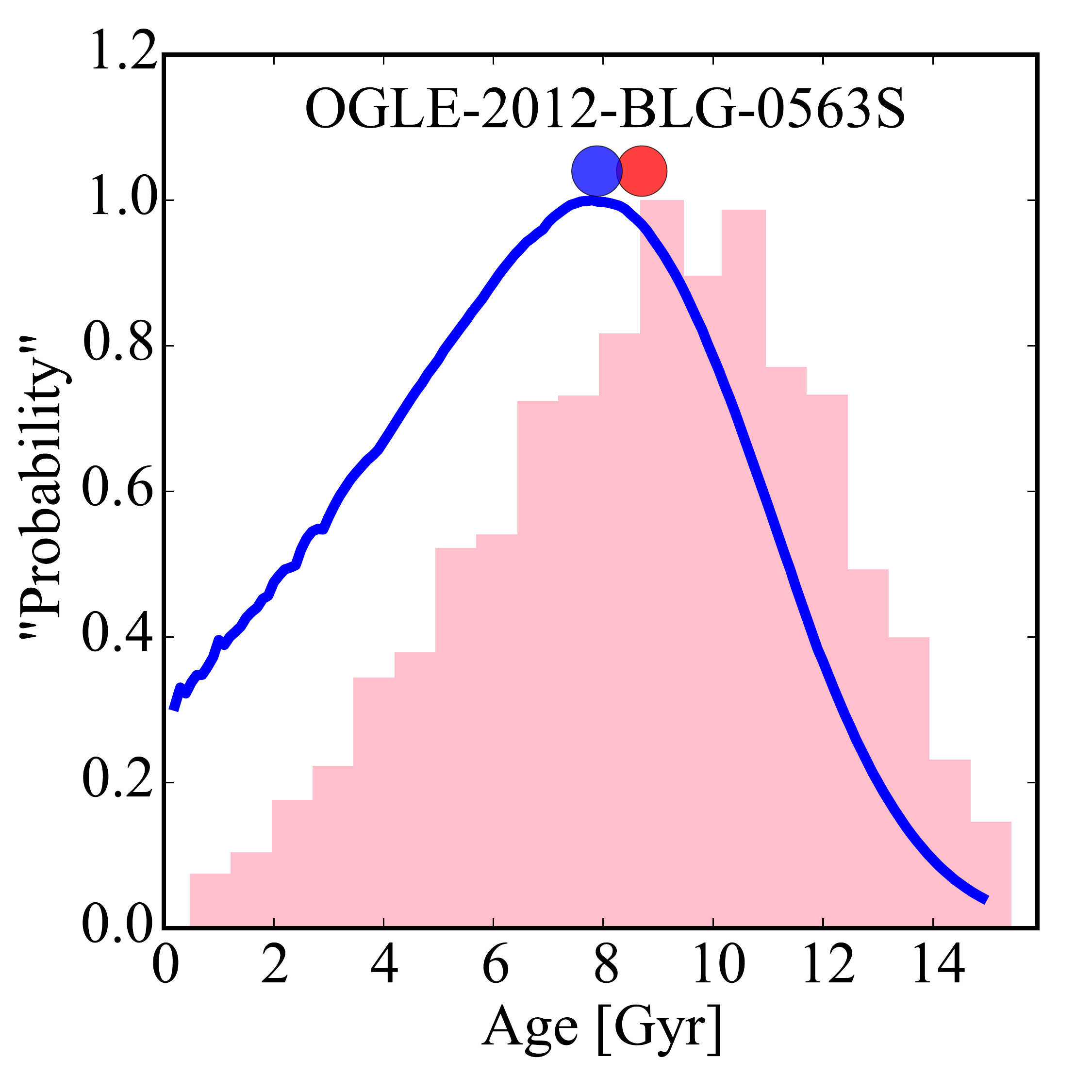}
\includegraphics[viewport= 93 0 648 648,clip]{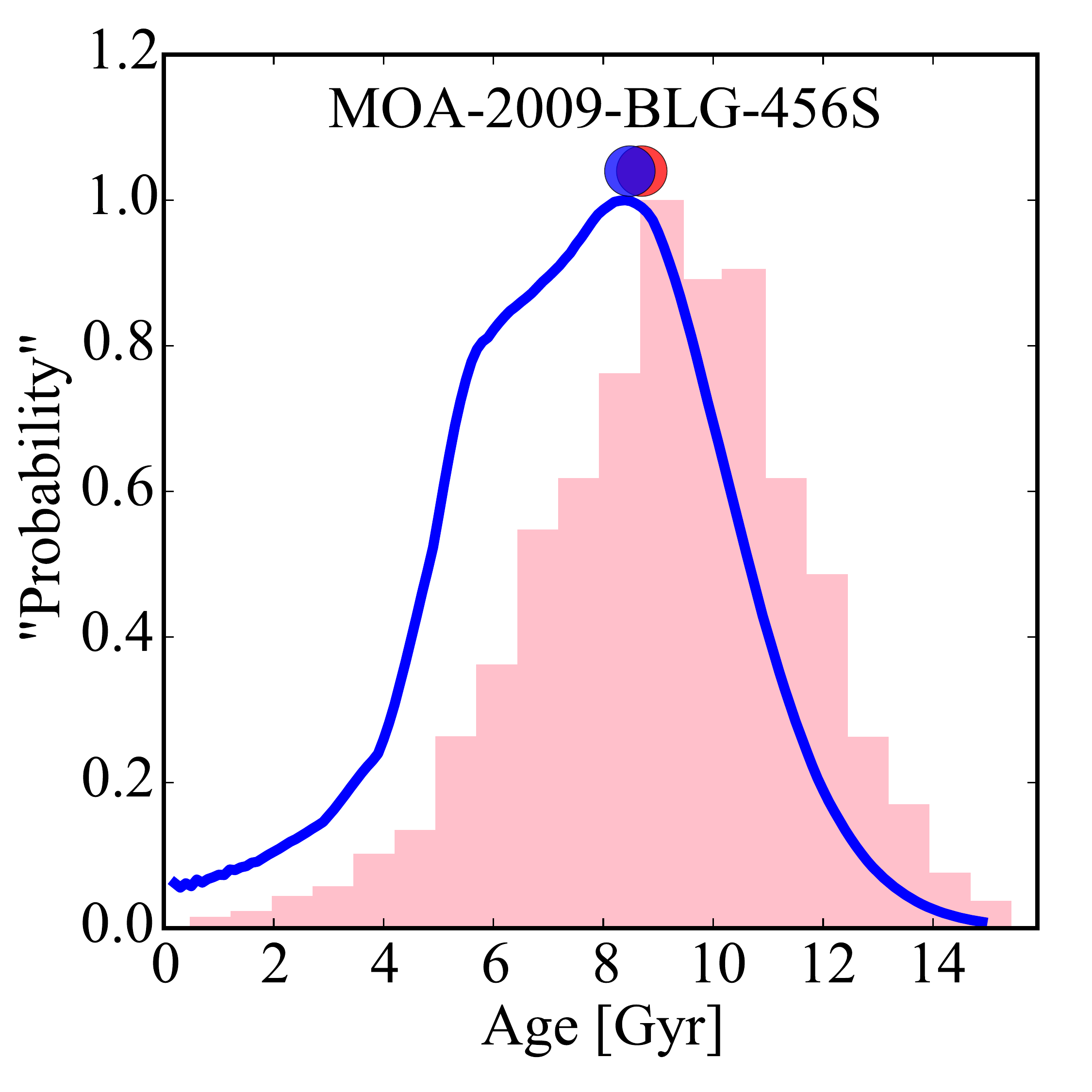}}
\caption{\sl continued
}
\end{figure*}

\setcounter{figure}{0}    
\begin{figure*}[ht]
\resizebox{\hsize}{!}{
\includegraphics[viewport= 0 0 648 648,clip]{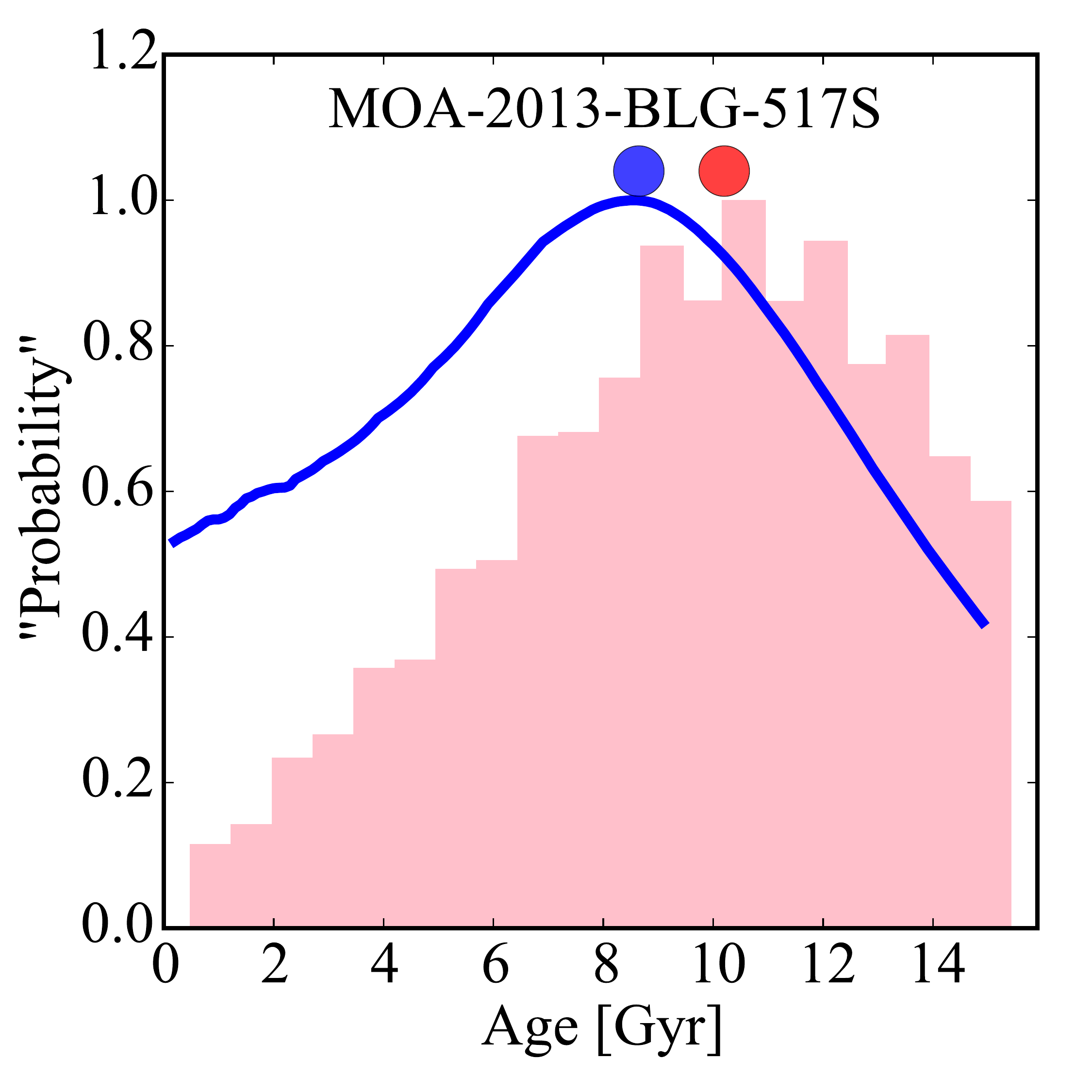}
\includegraphics[viewport= 93 0 648 648,clip]{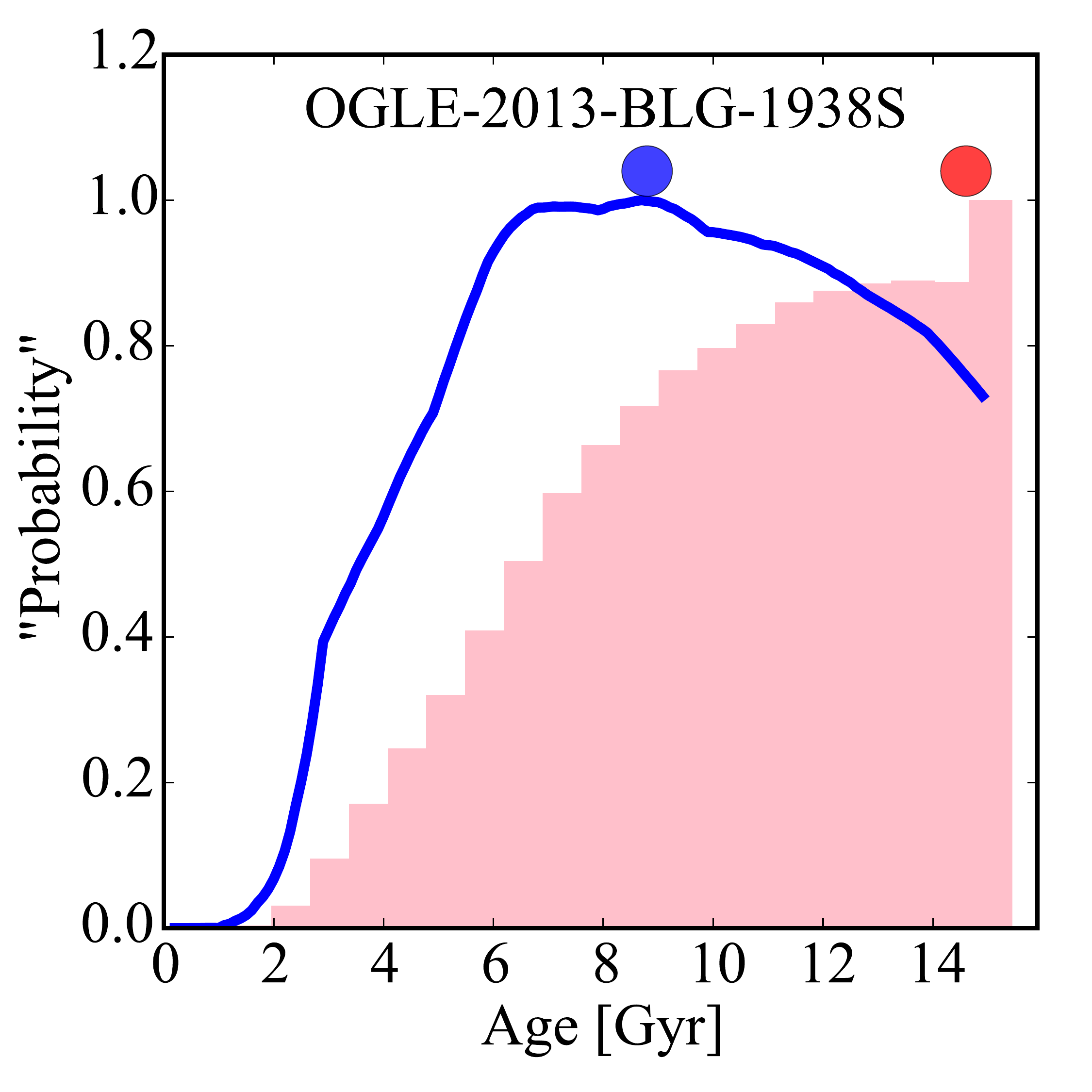}
\includegraphics[viewport= 93 0 648 648,clip]{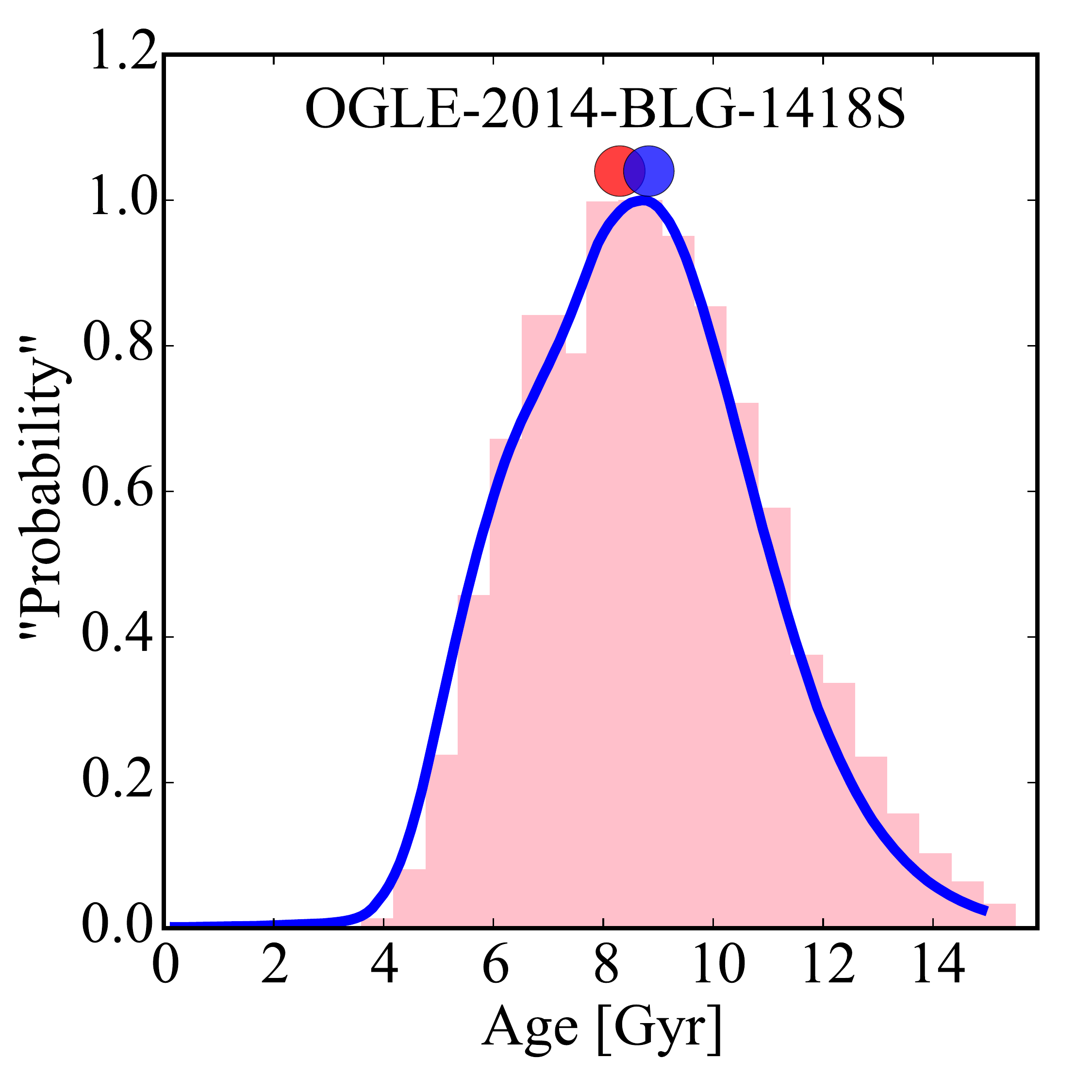}
\includegraphics[viewport= 93 0 648 648,clip]{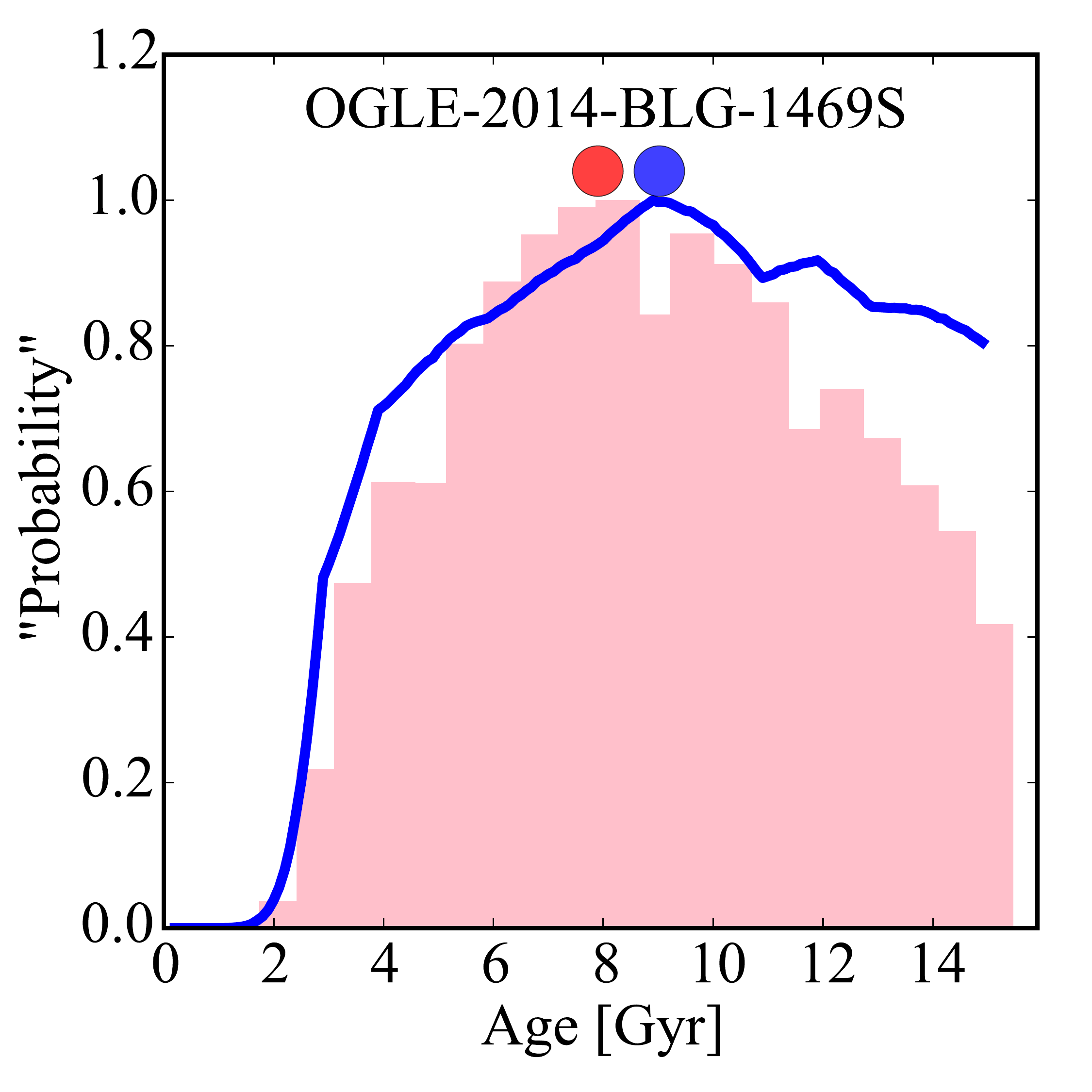}
\includegraphics[viewport= 93 0 648 648,clip]{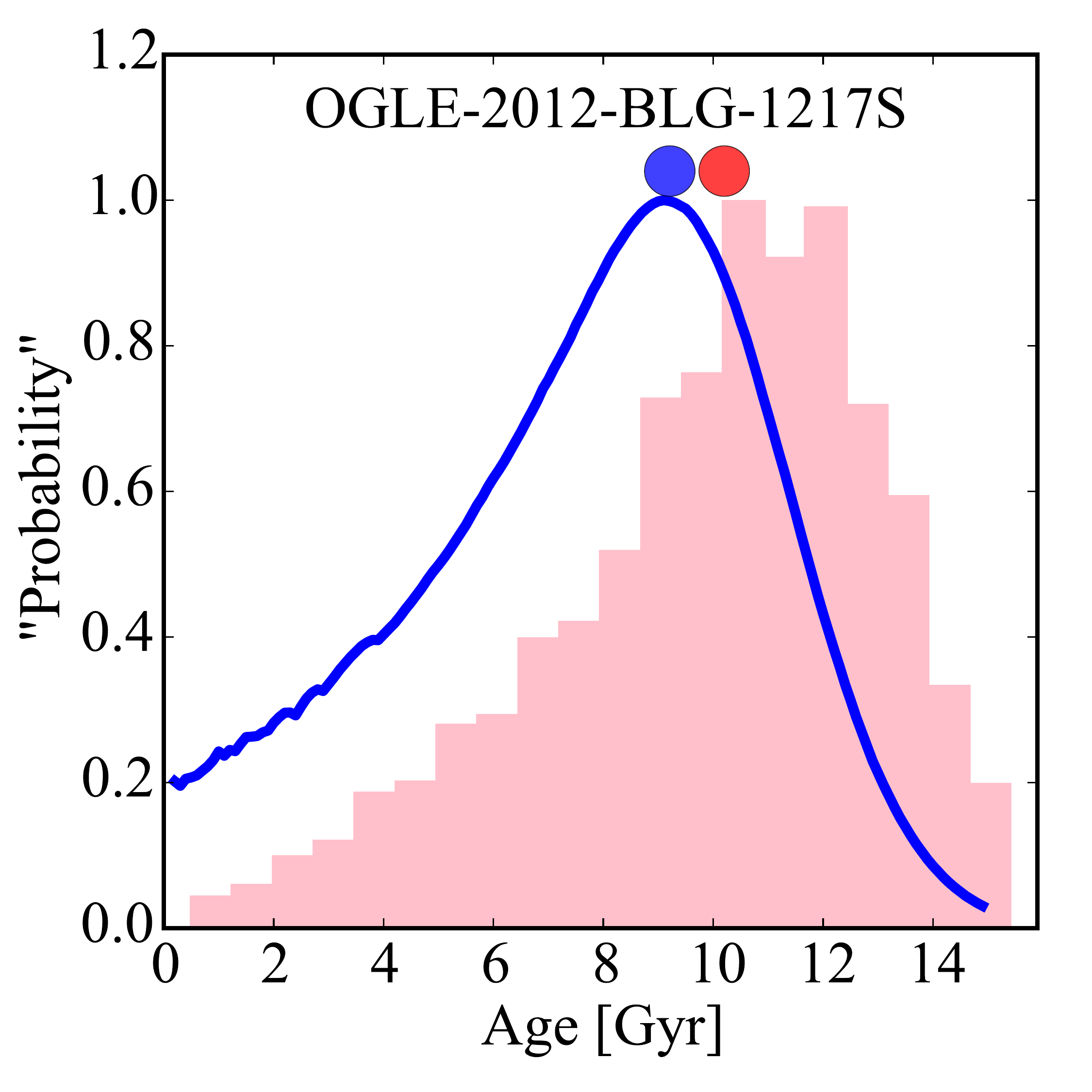}}
\resizebox{\hsize}{!}{
\includegraphics[viewport= 0 0 648 648,clip]{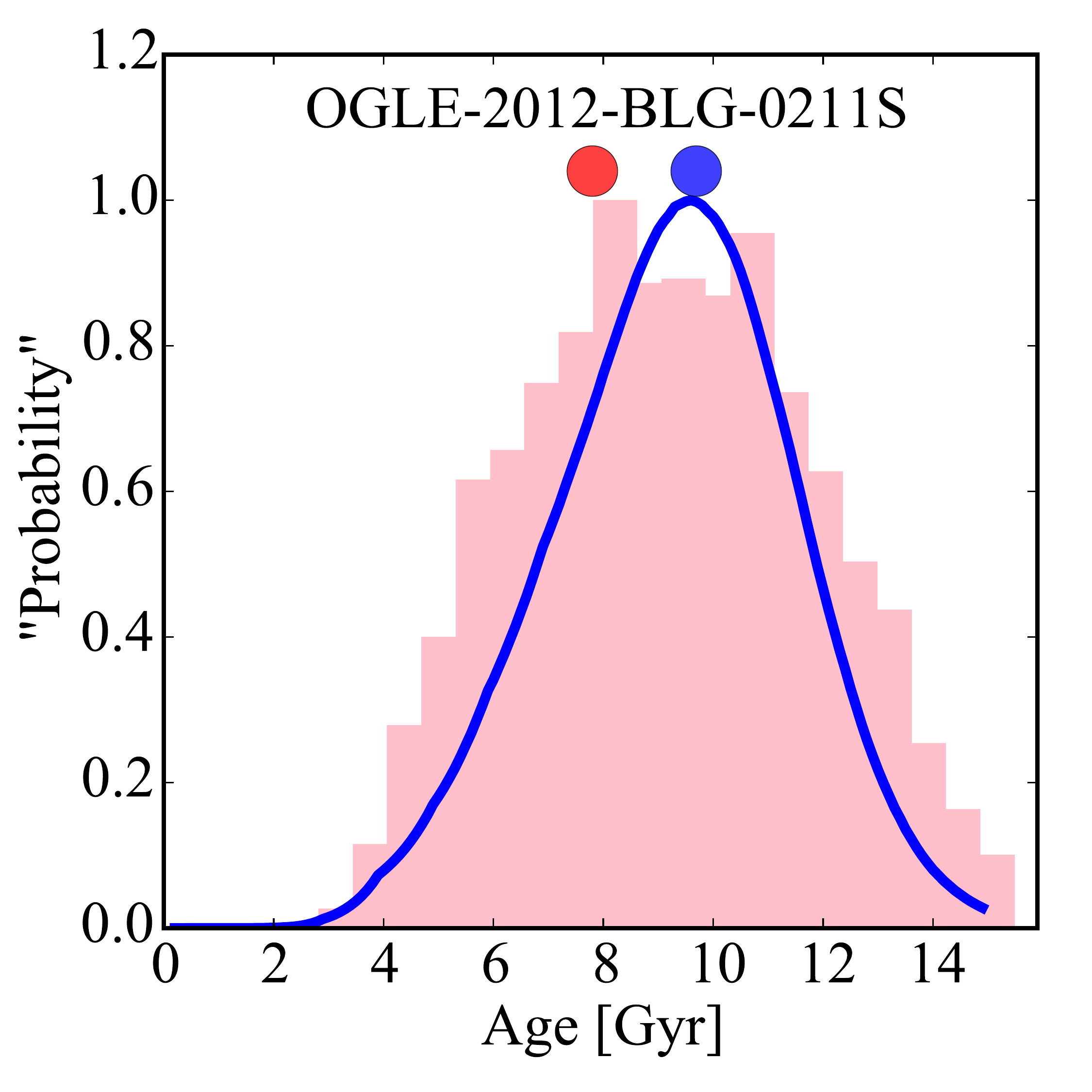}
\includegraphics[viewport= 93 0 648 648,clip]{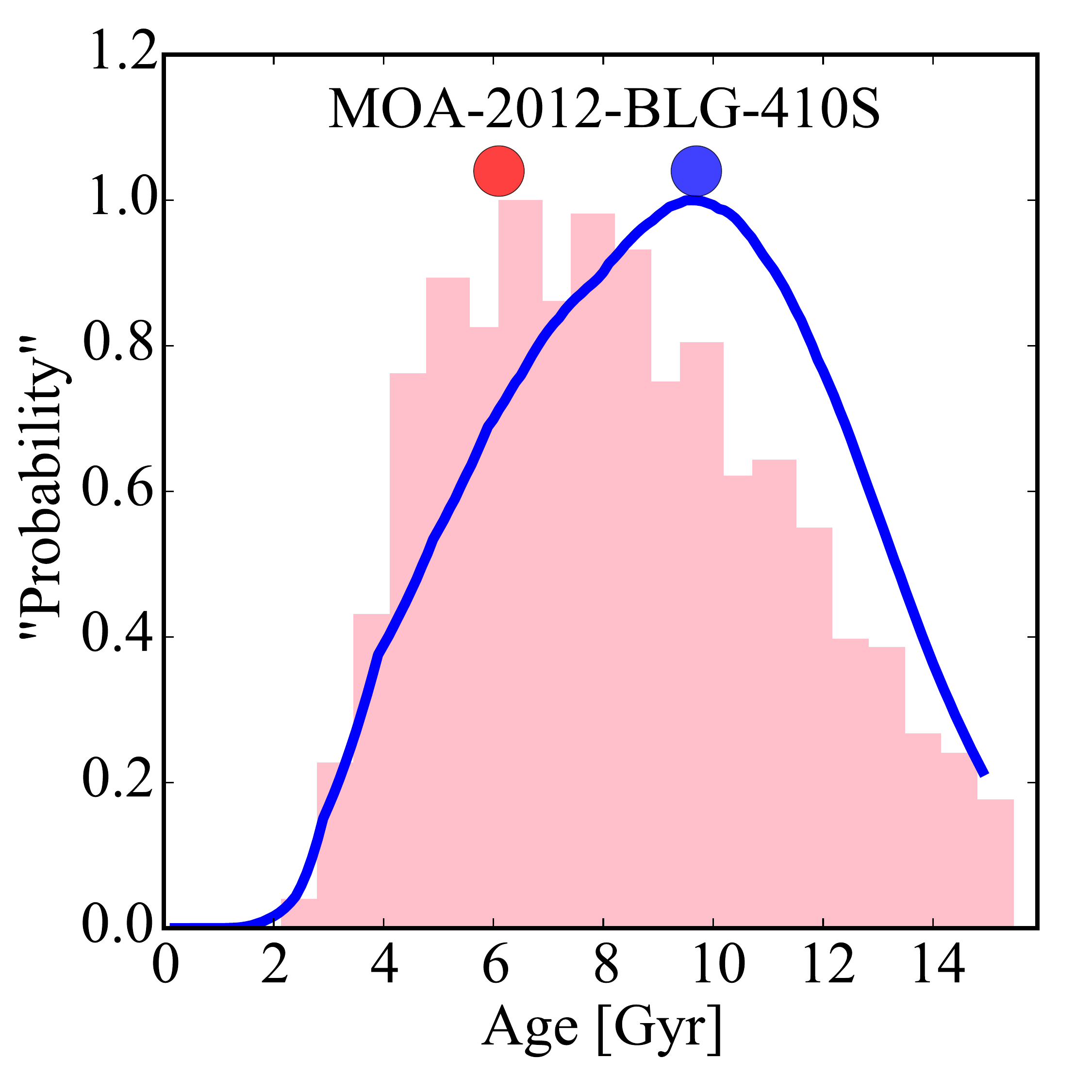}
\includegraphics[viewport= 93 0 648 648,clip]{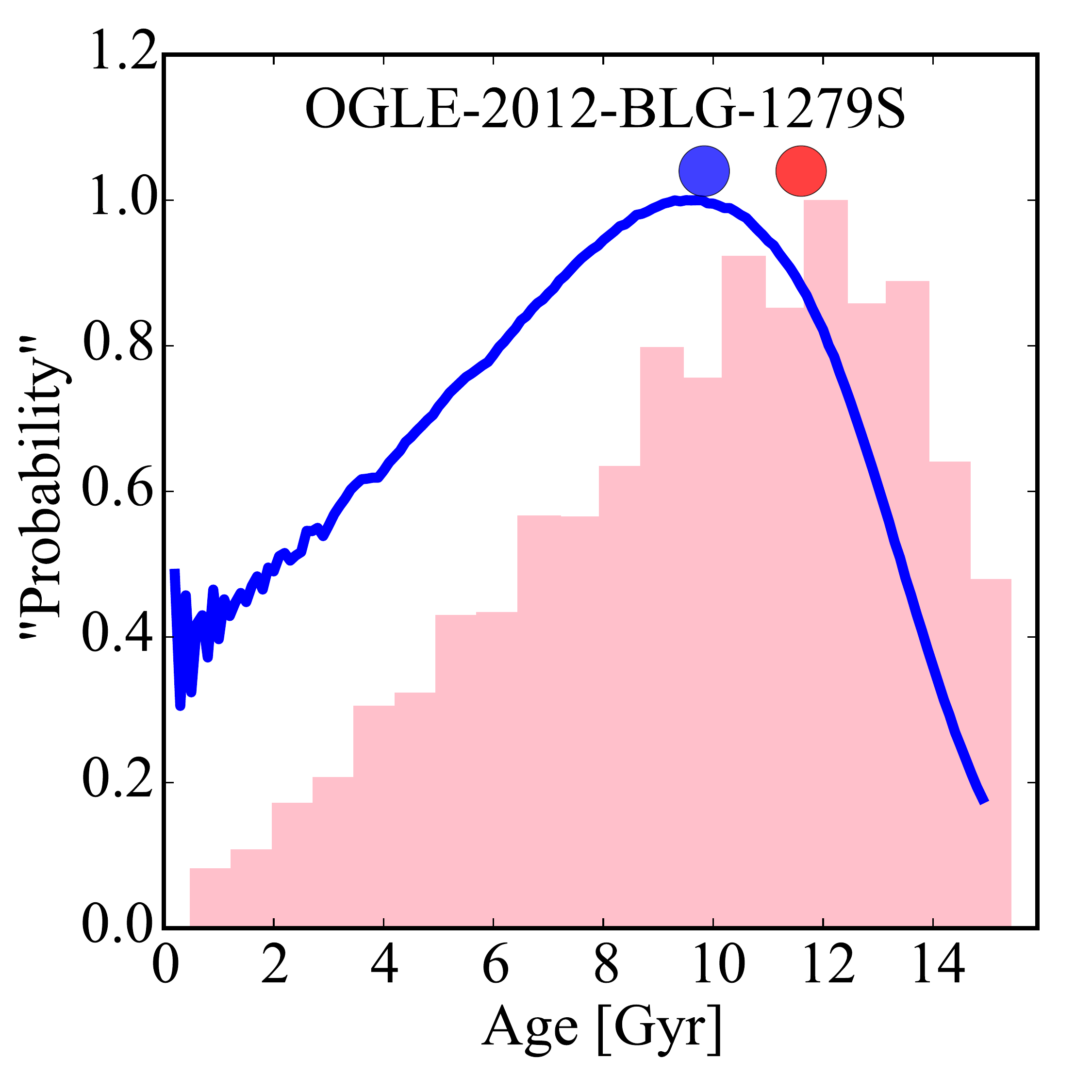}
\includegraphics[viewport= 93 0 648 648,clip]{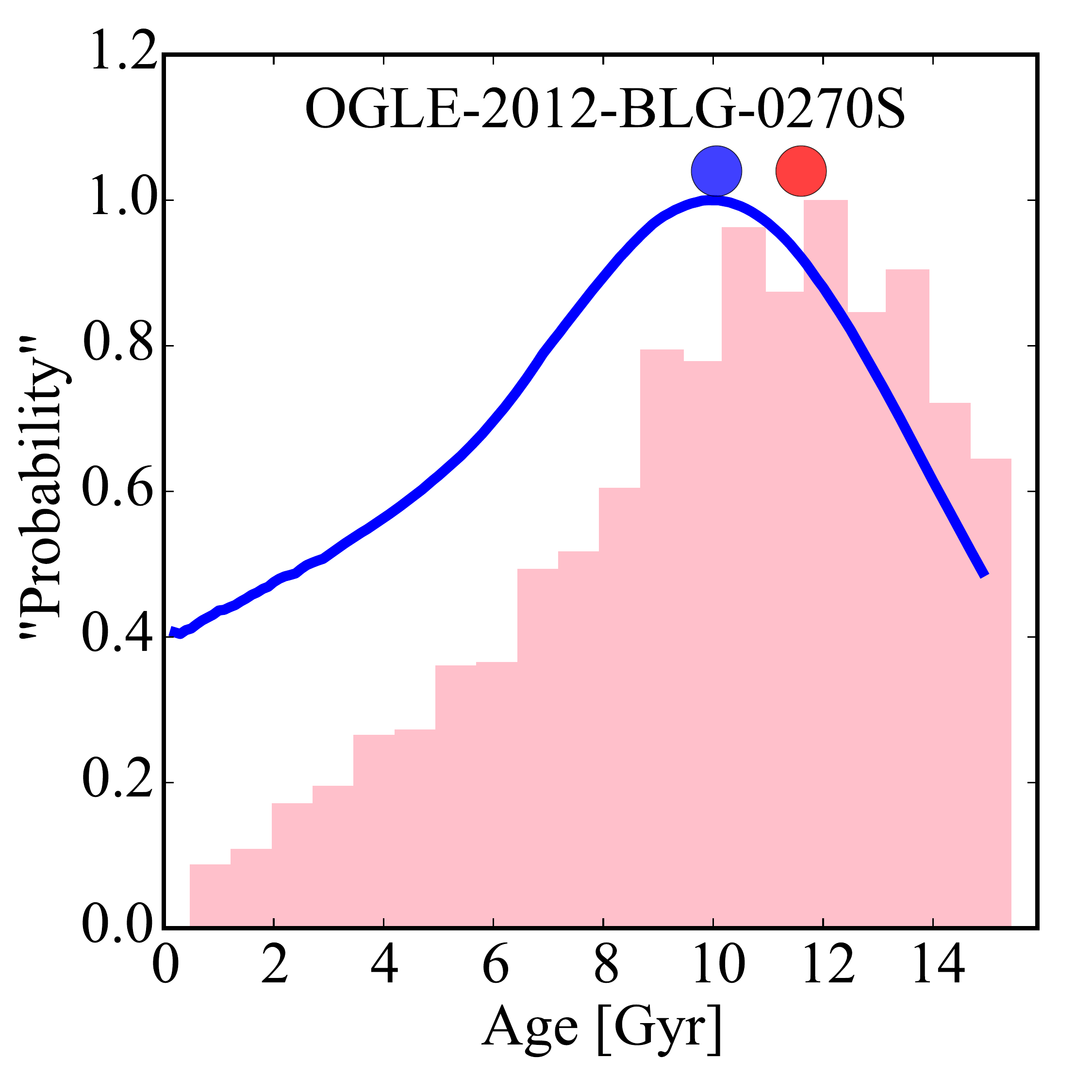}
\includegraphics[viewport= 93 0 648 648,clip]{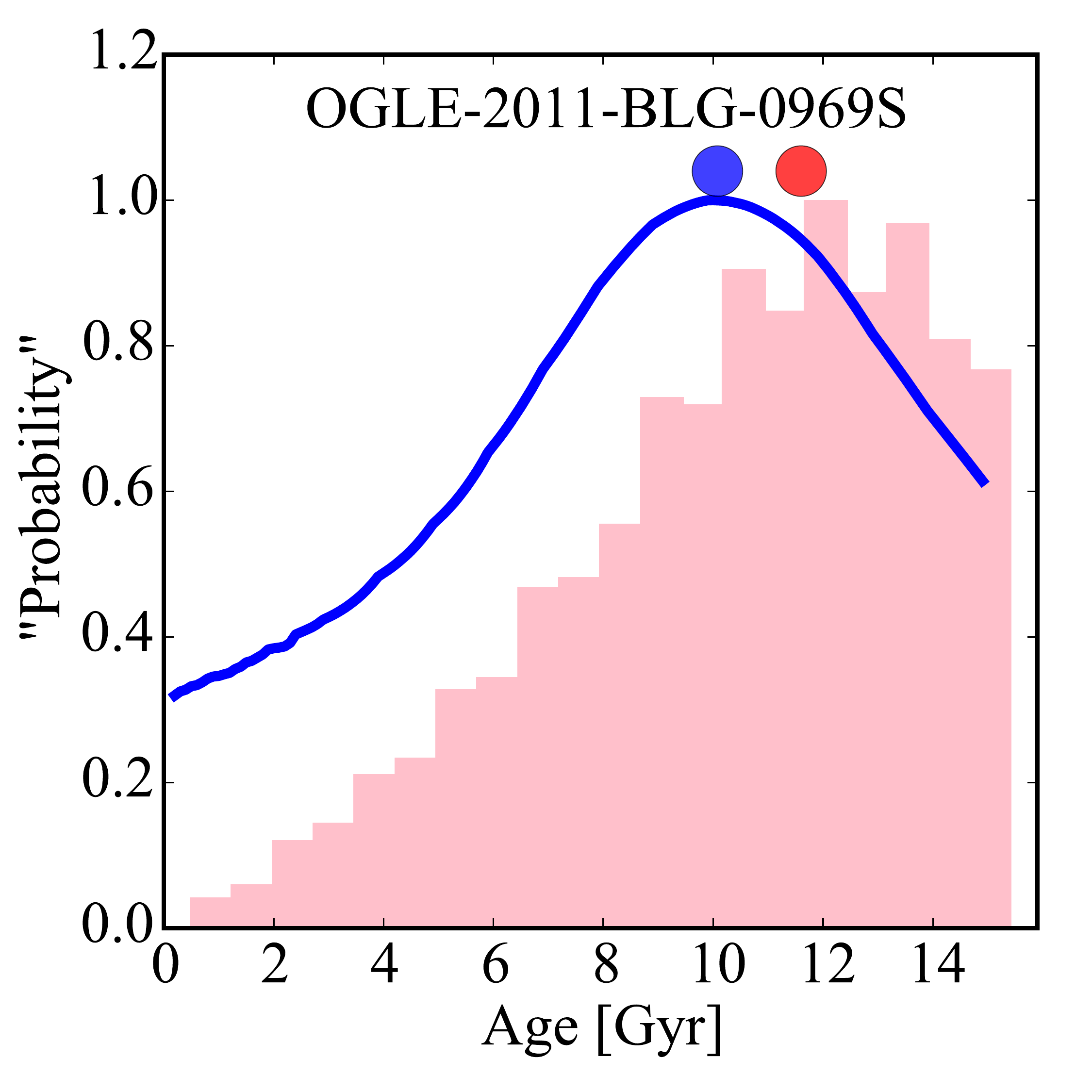}}
\resizebox{\hsize}{!}{
\includegraphics[viewport= 0 0 648 648,clip]{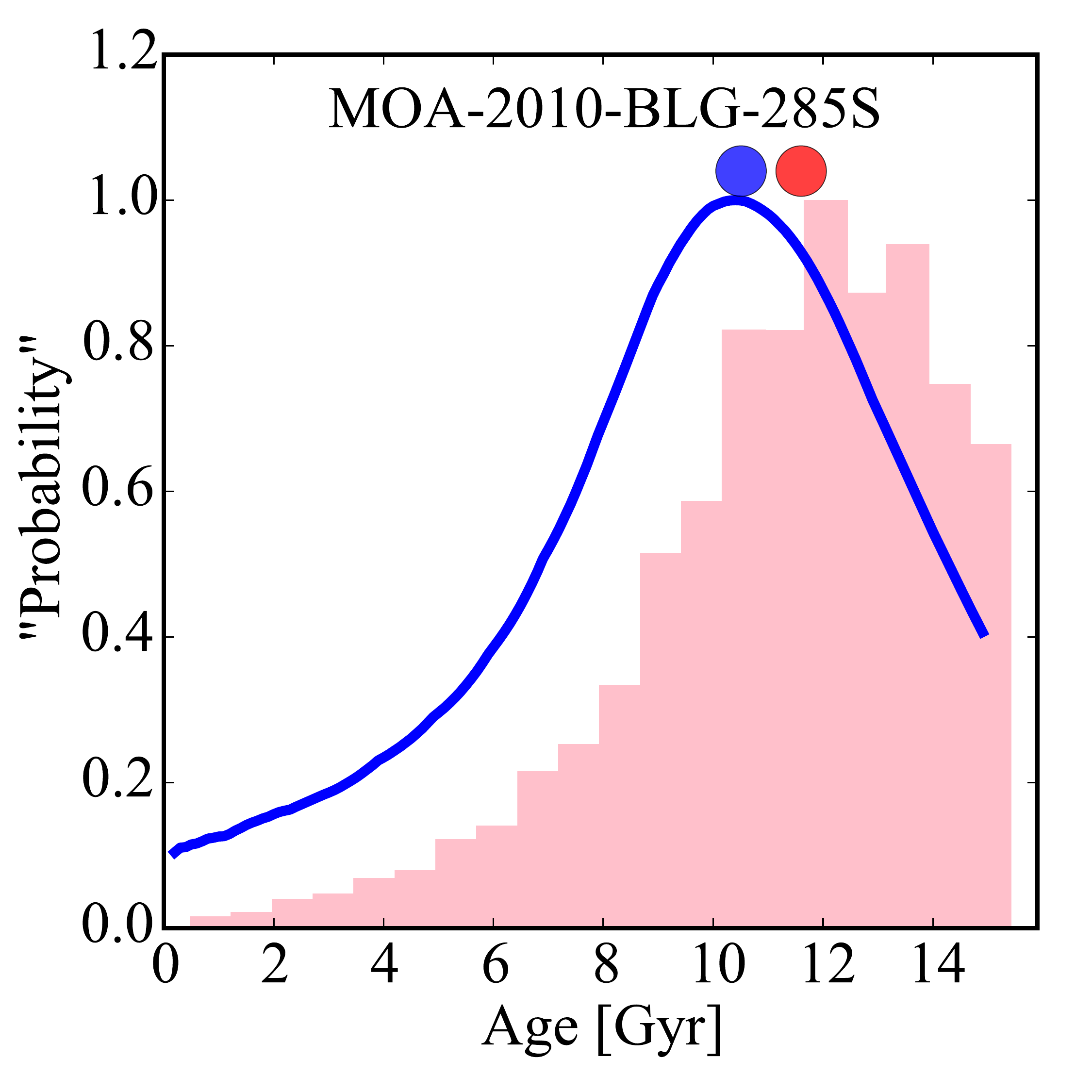}
\includegraphics[viewport= 93 0 648 648,clip]{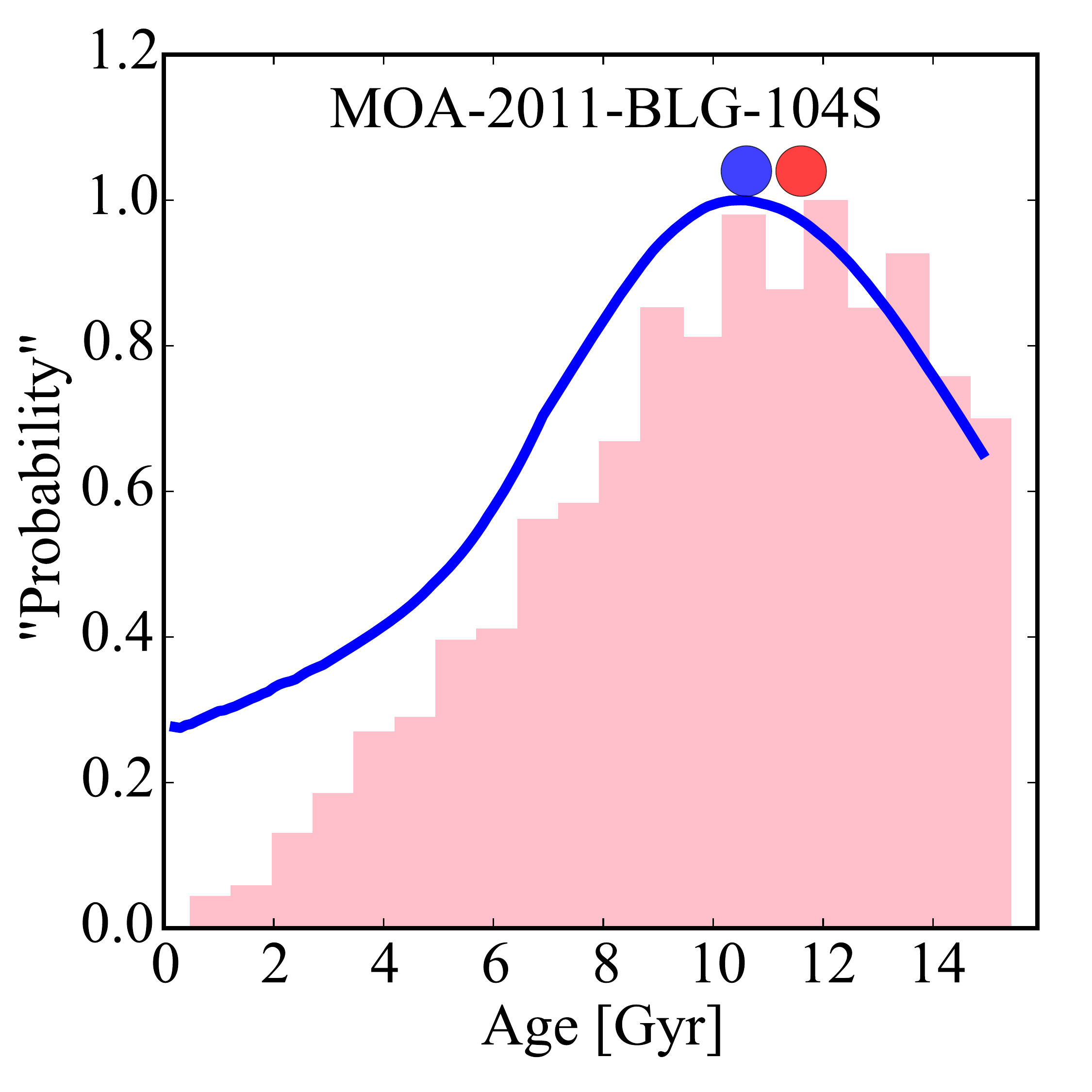}
\includegraphics[viewport= 93 0 648 648,clip]{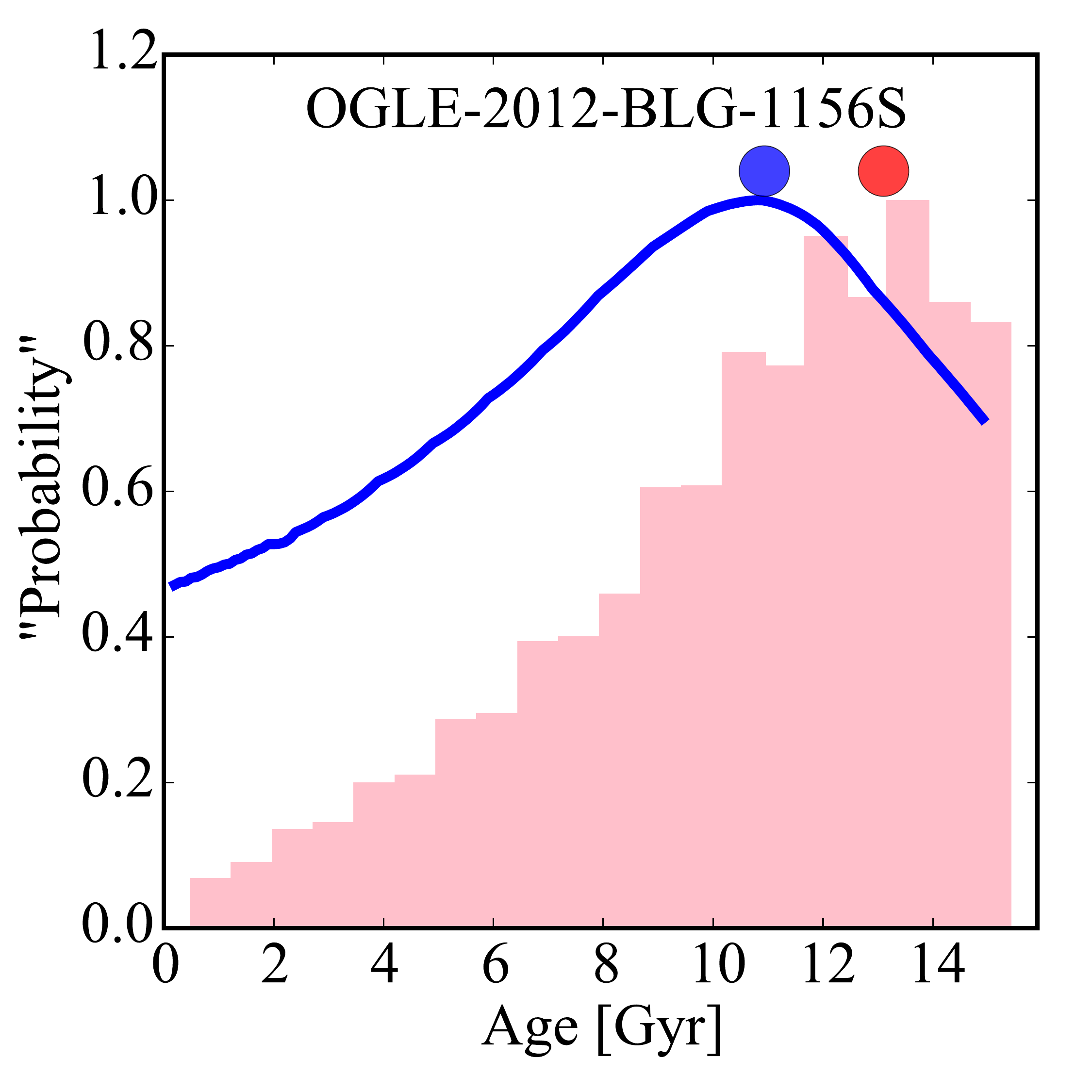}
\includegraphics[viewport= 93 0 648 648,clip]{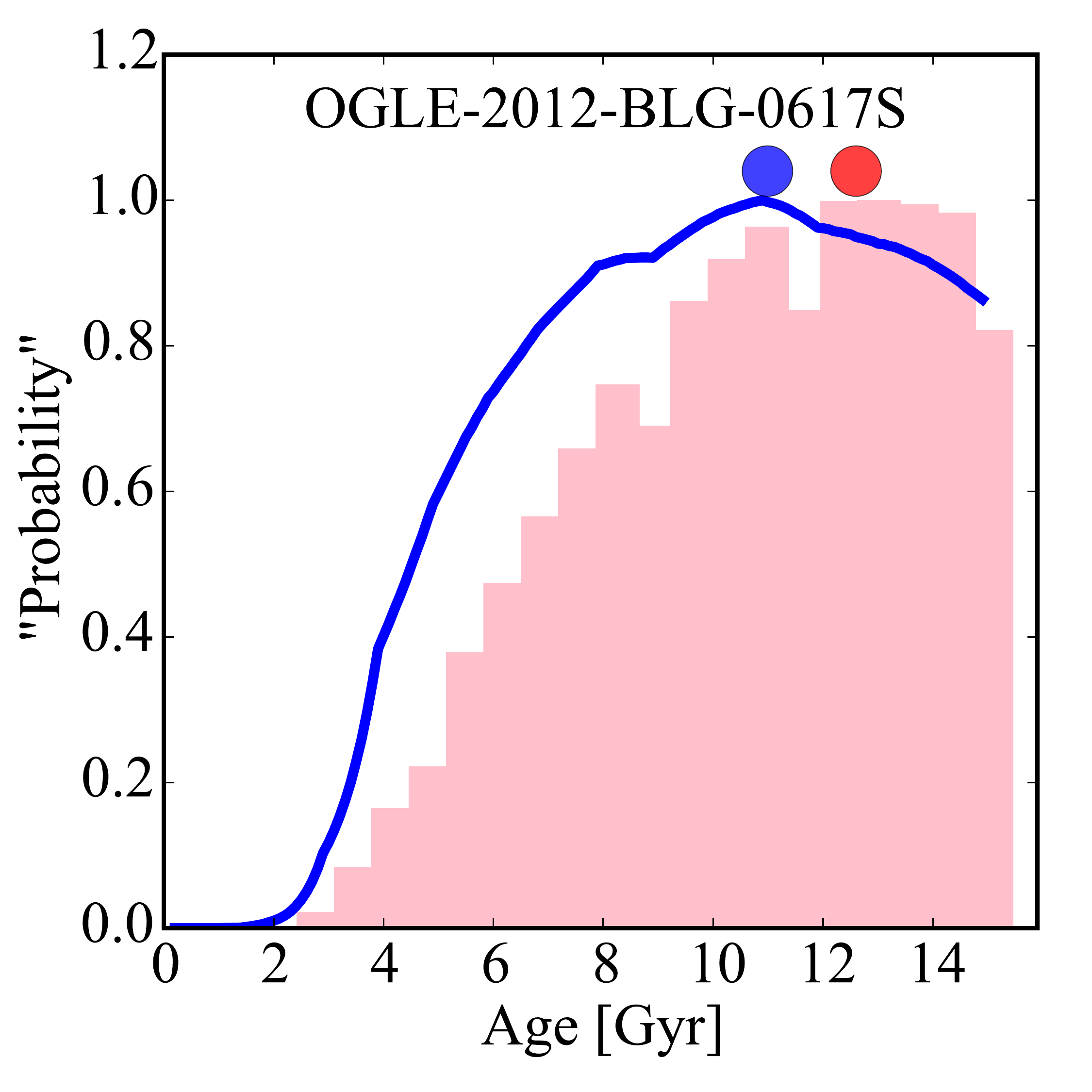}
\includegraphics[viewport= 93 0 648 648,clip]{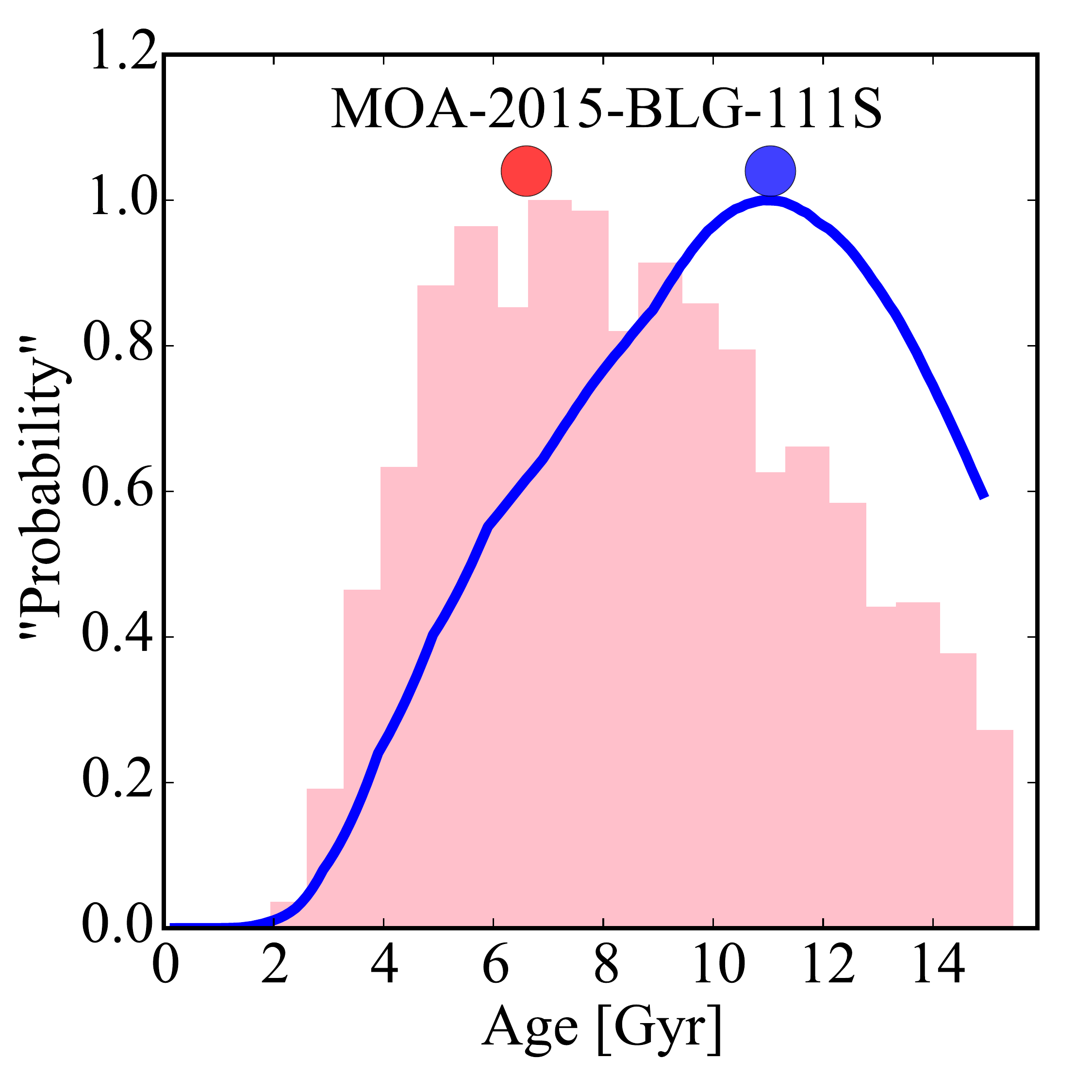}}
\resizebox{\hsize}{!}{
\includegraphics[viewport= 0 0 648 648,clip]{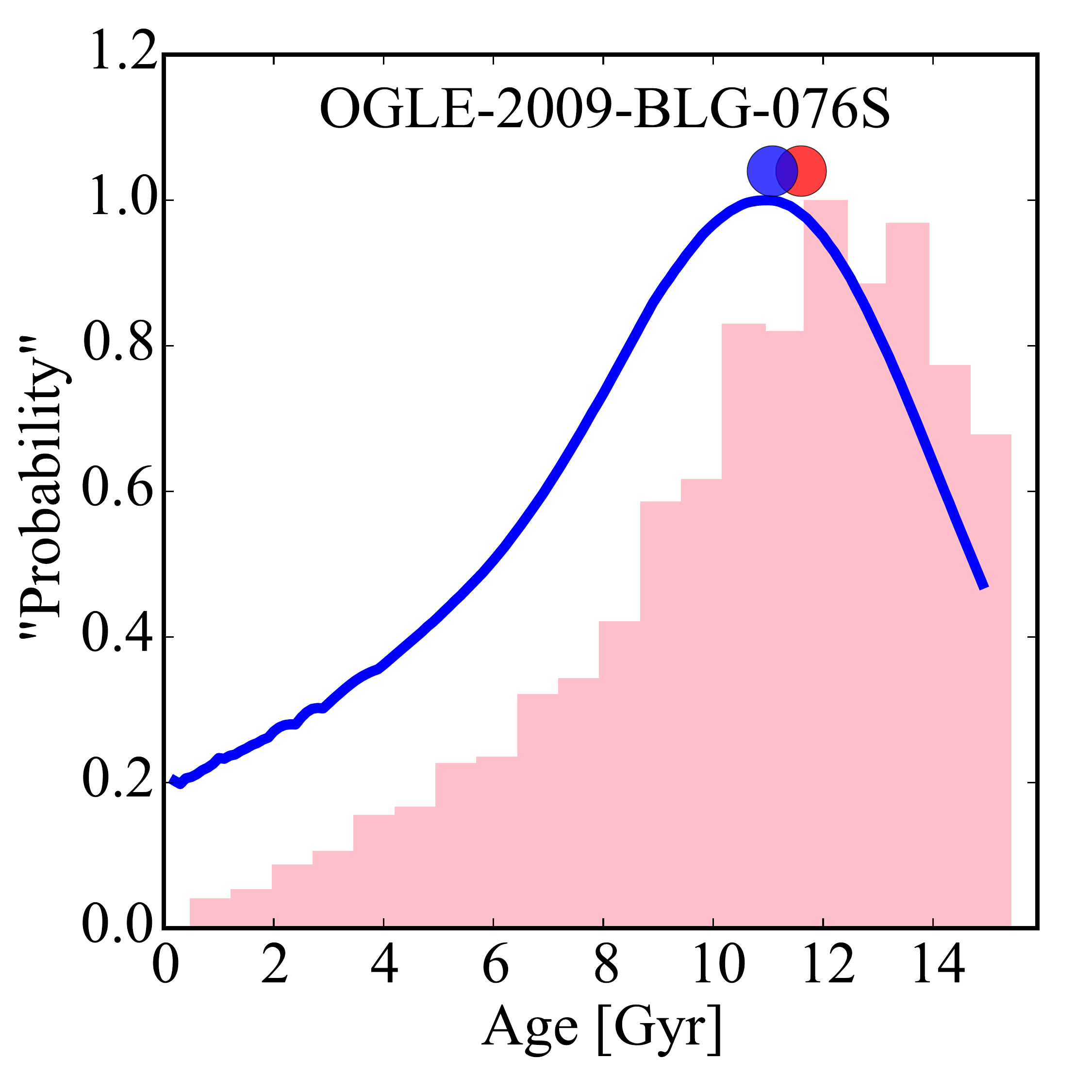}
\includegraphics[viewport= 93 0 648 648,clip]{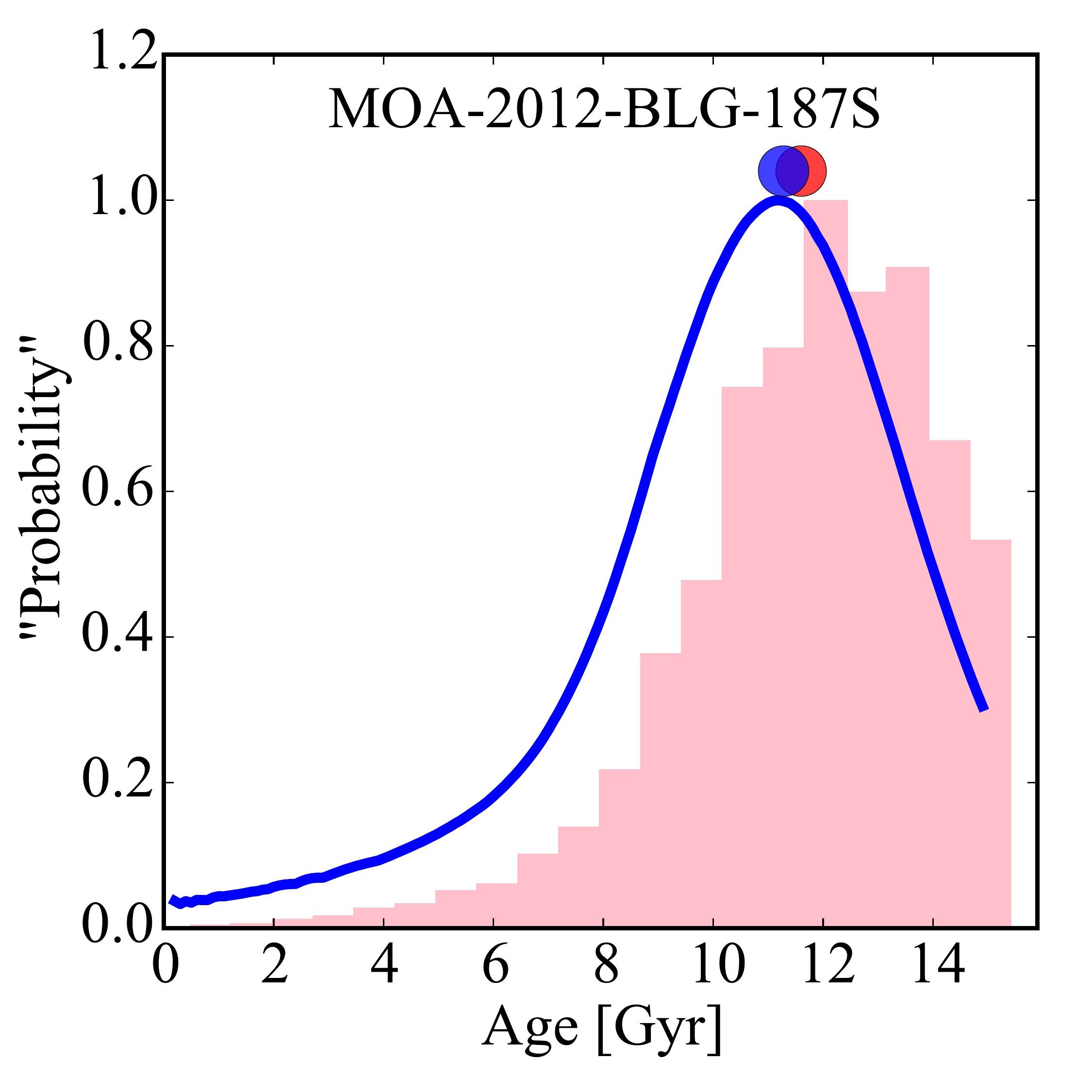}
\includegraphics[viewport= 93 0 648 648,clip]{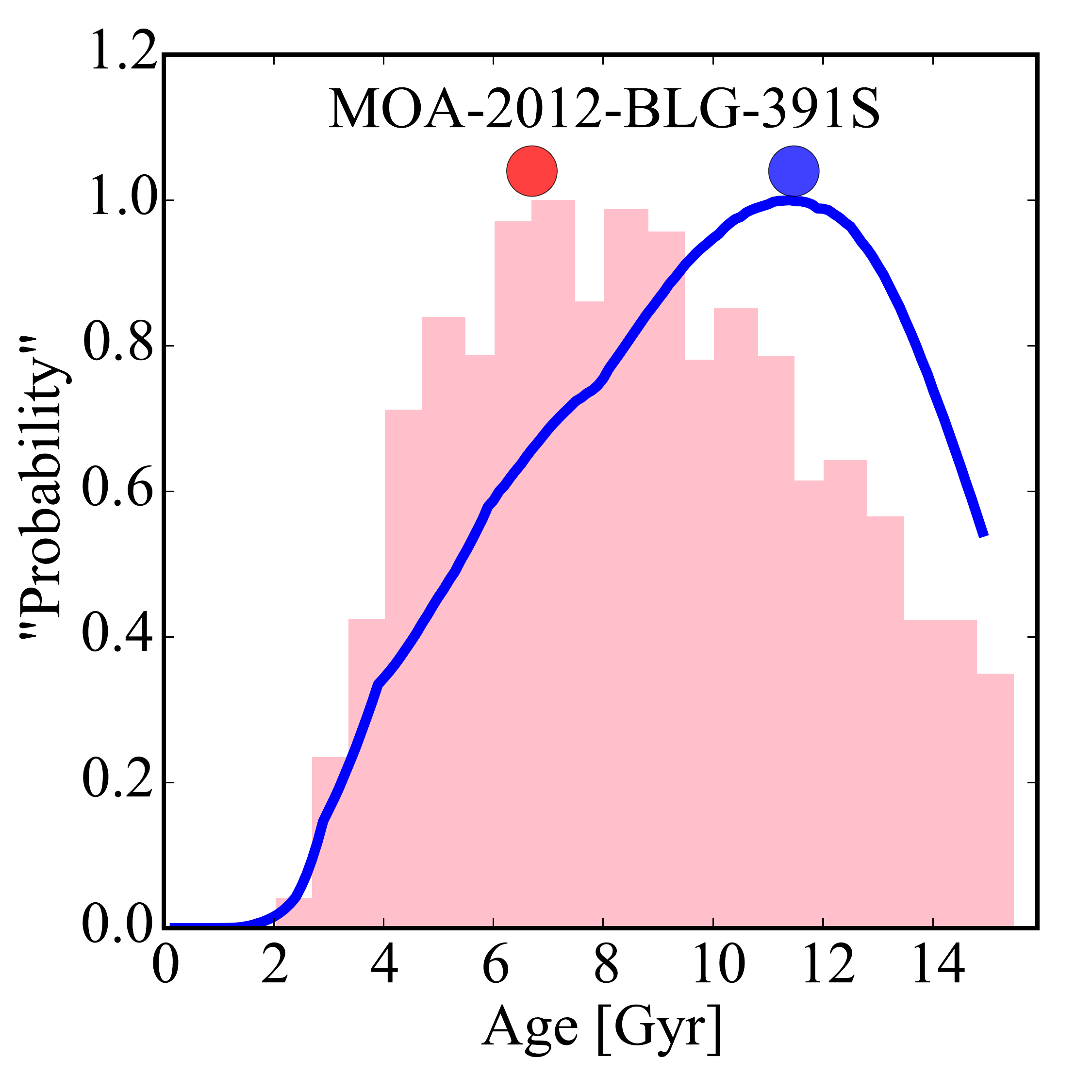}
\includegraphics[viewport= 93 0 648 648,clip]{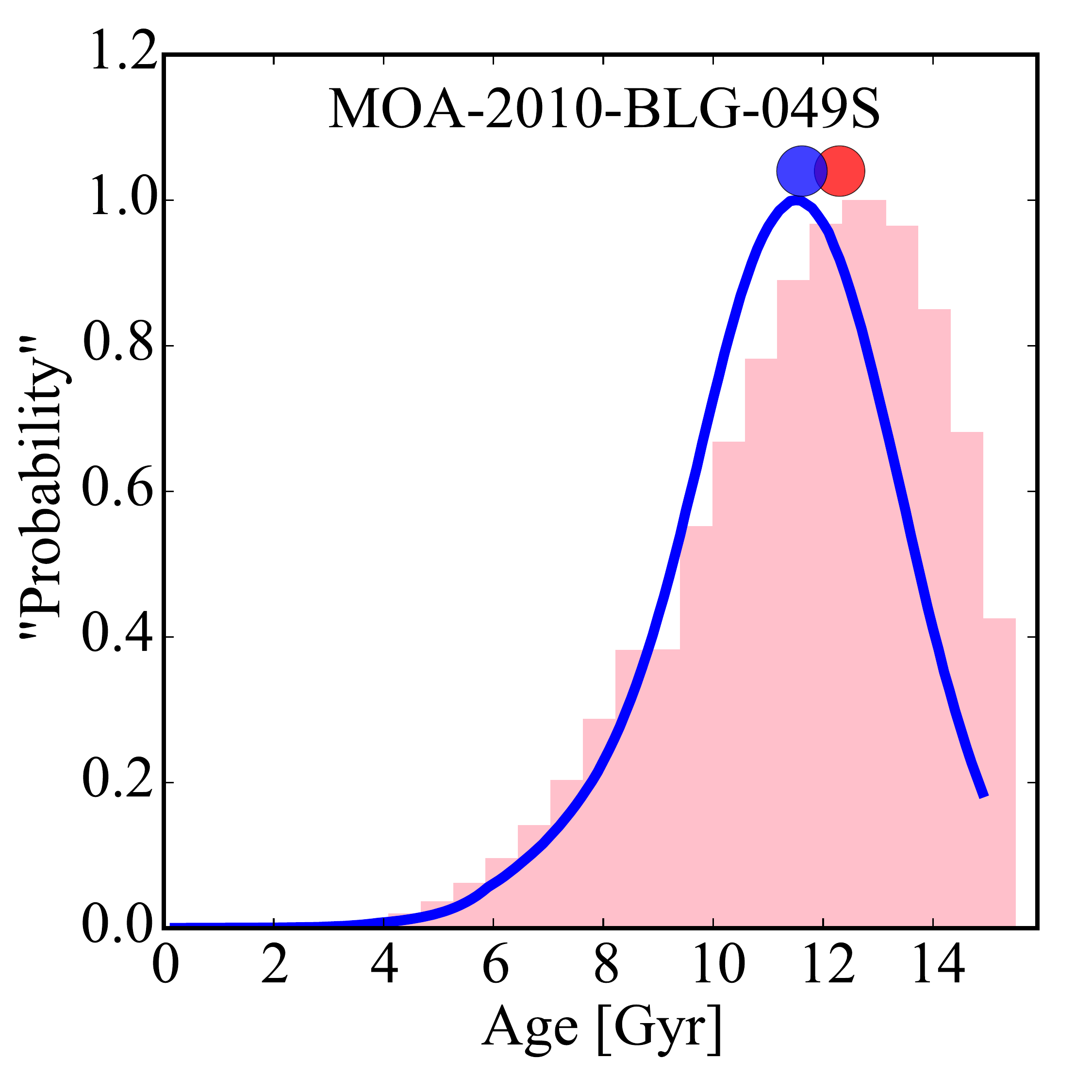}
\includegraphics[viewport= 93 0 648 648,clip]{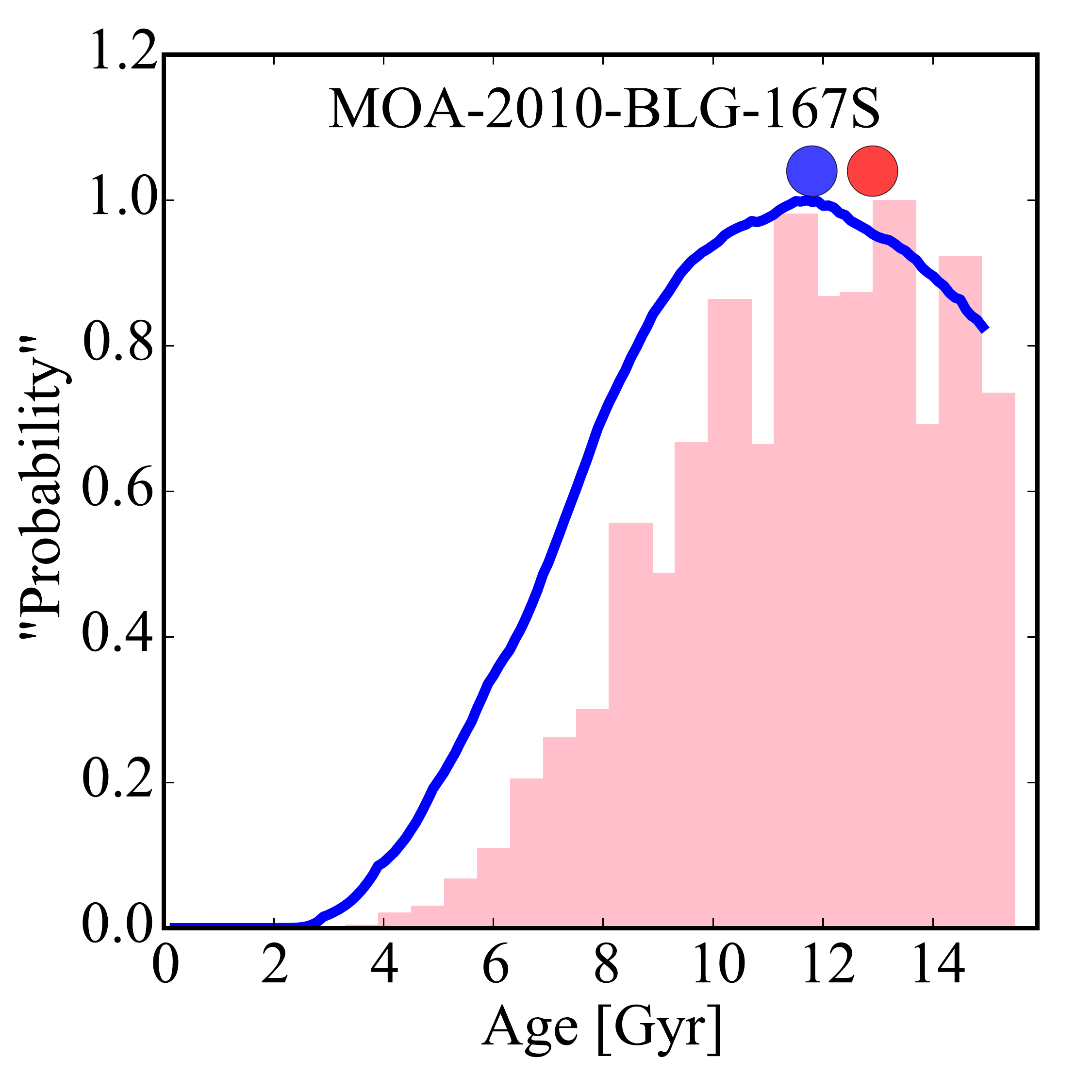}}
\resizebox{\hsize}{!}{
\includegraphics[viewport= 0 0 648 648,clip]{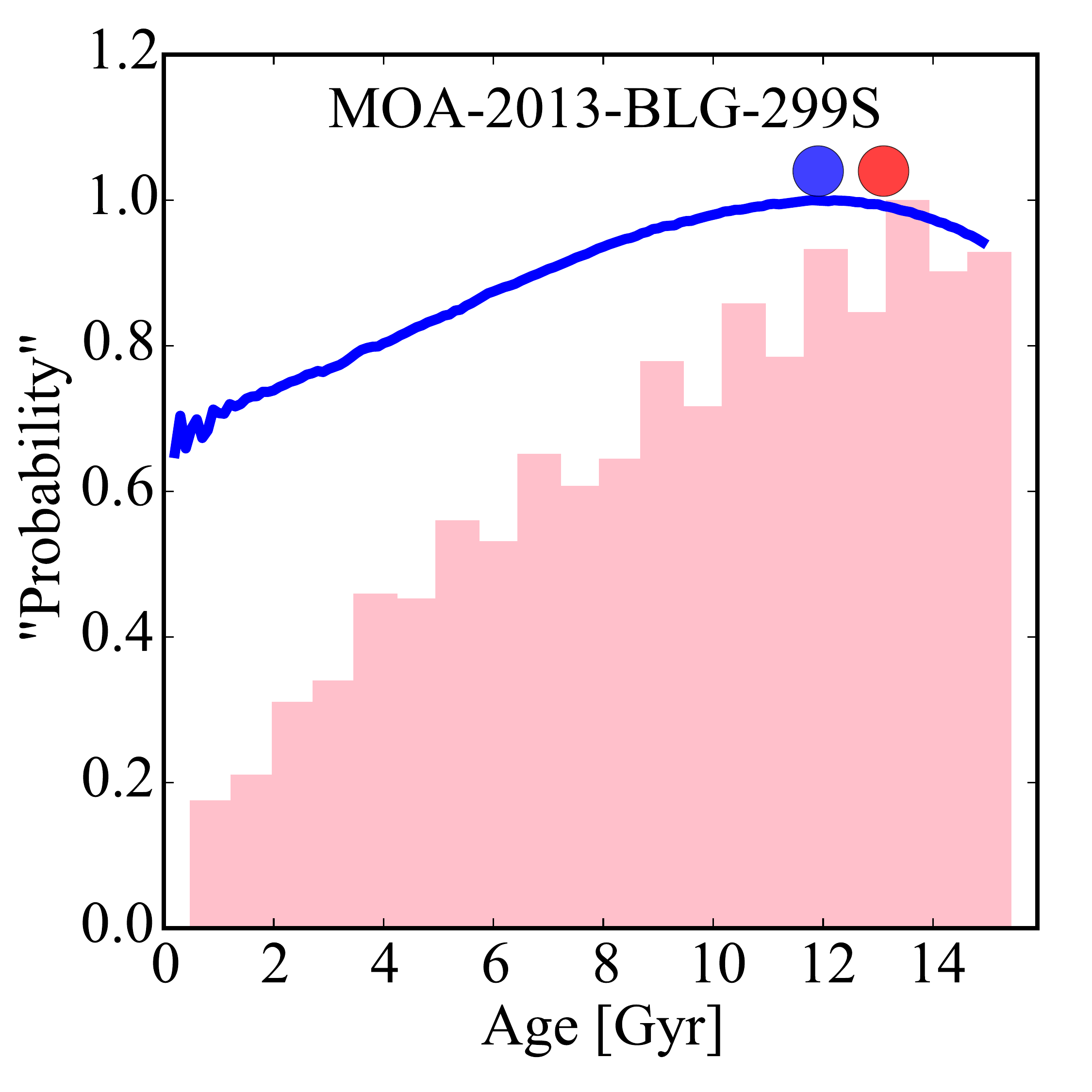}
\includegraphics[viewport= 93 0 648 648,clip]{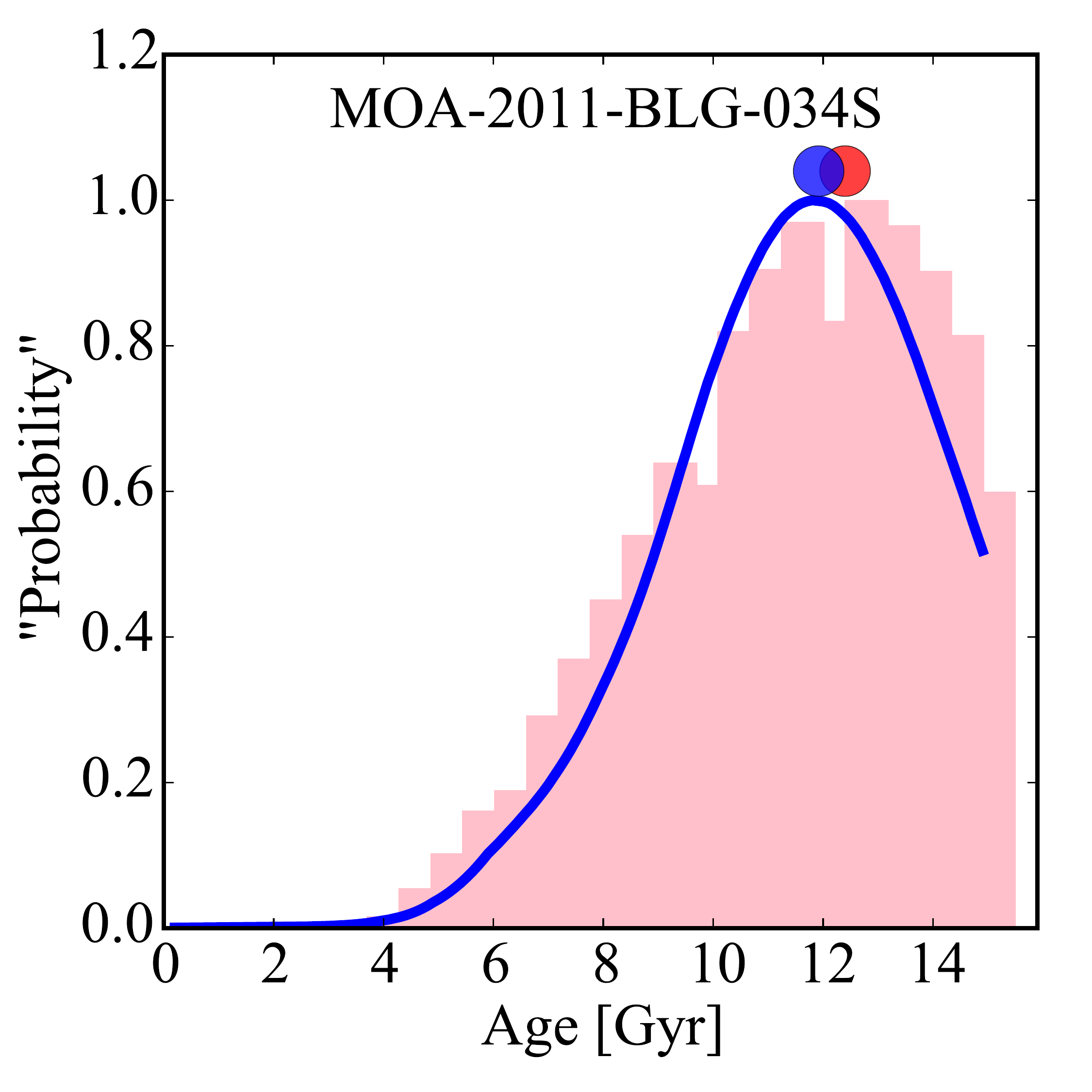}
\includegraphics[viewport= 93 0 648 648,clip]{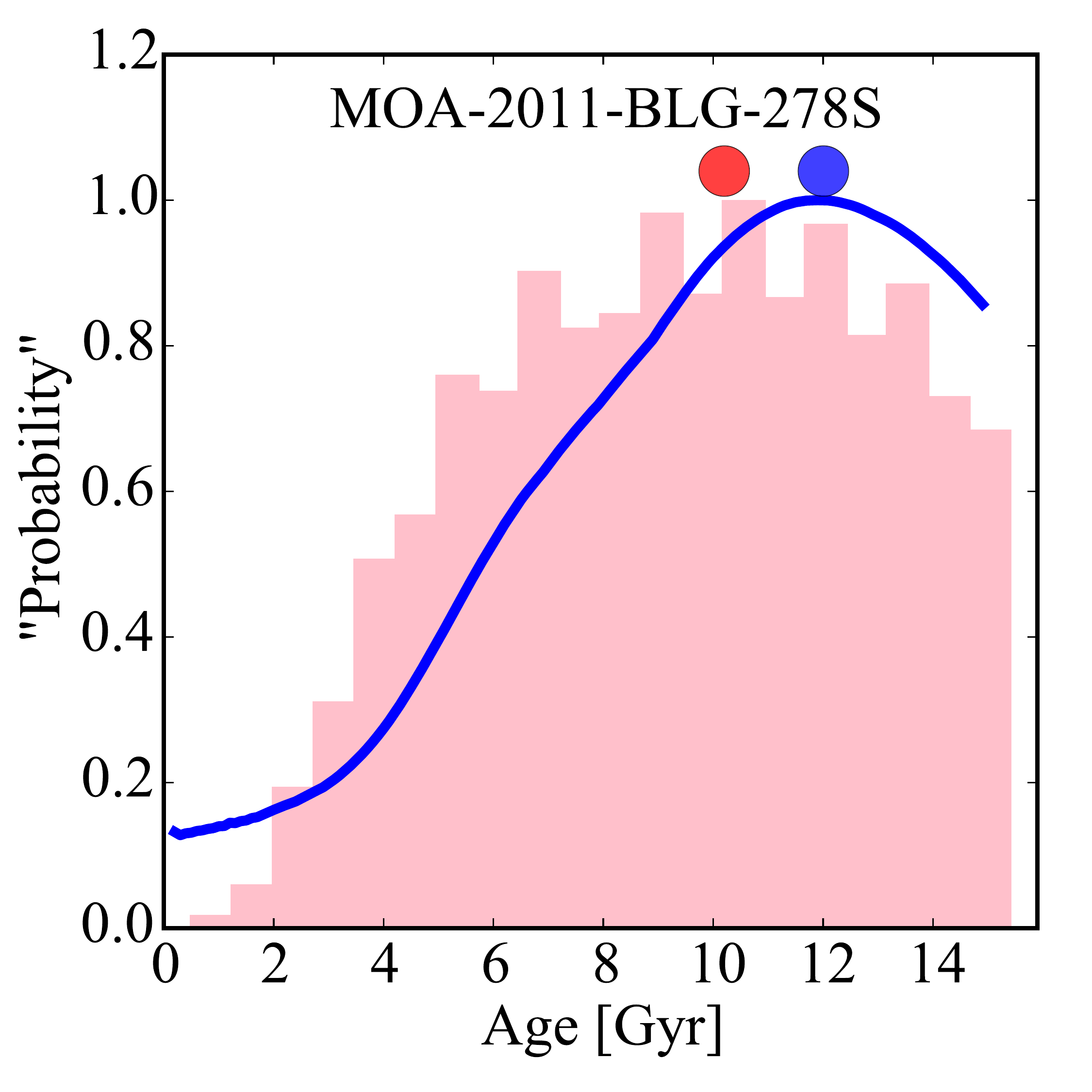}
\includegraphics[viewport= 93 0 648 648,clip]{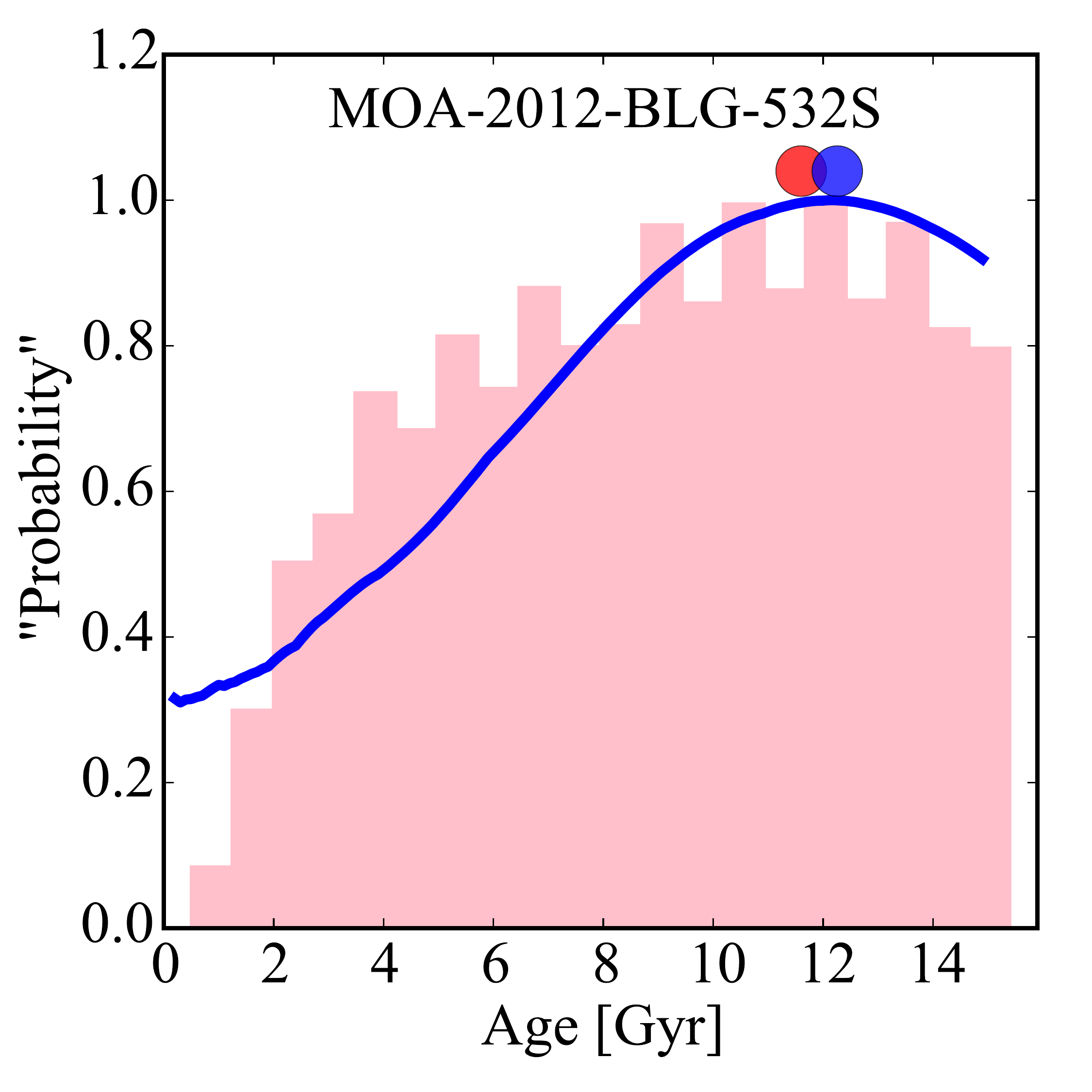}
\includegraphics[viewport= 93 0 648 648,clip]{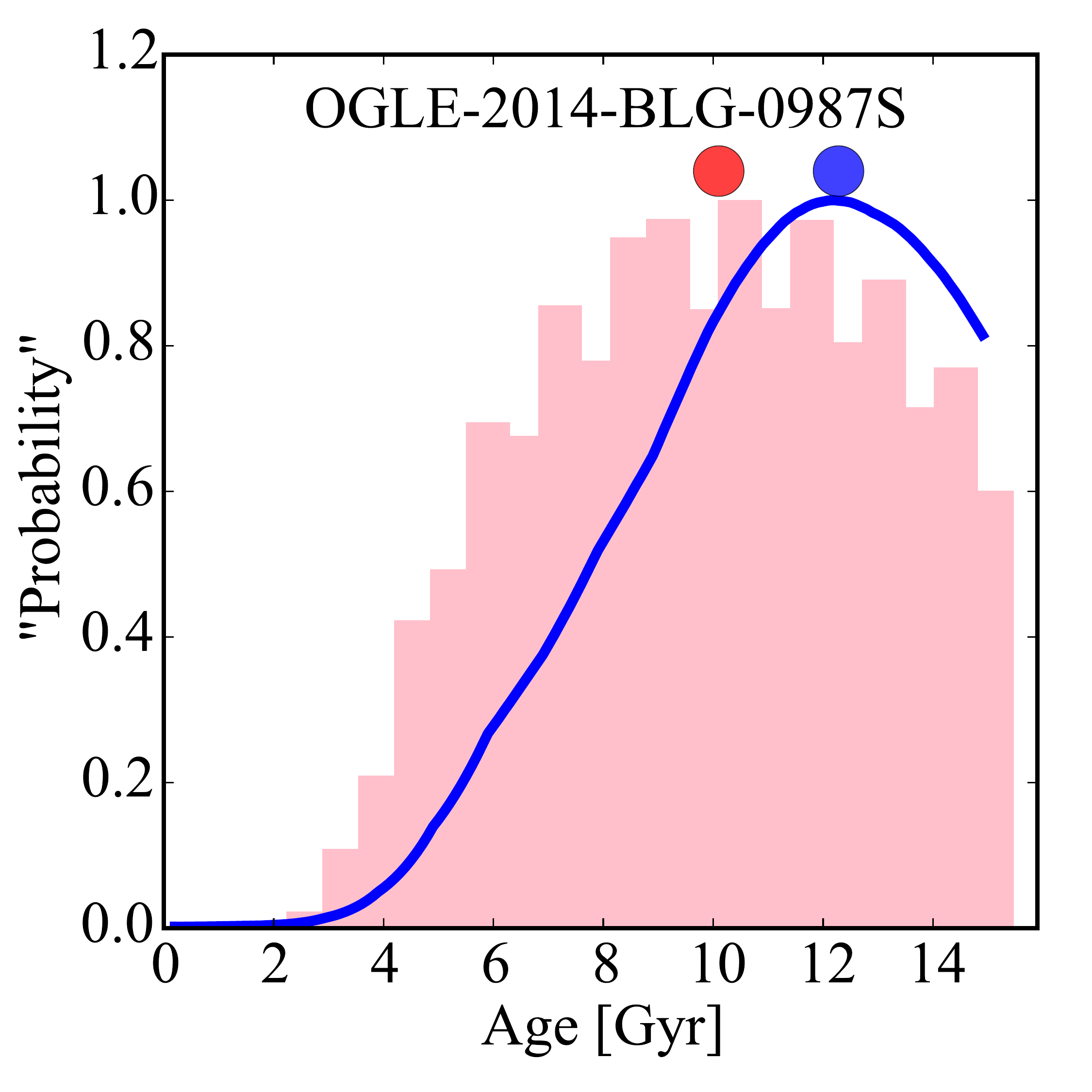}}
\caption{\sl continued
}
\end{figure*}

\setcounter{figure}{0}    
\begin{figure*}[ht]
\resizebox{\hsize}{!}{
\includegraphics[viewport= 0 0 648 648,clip]{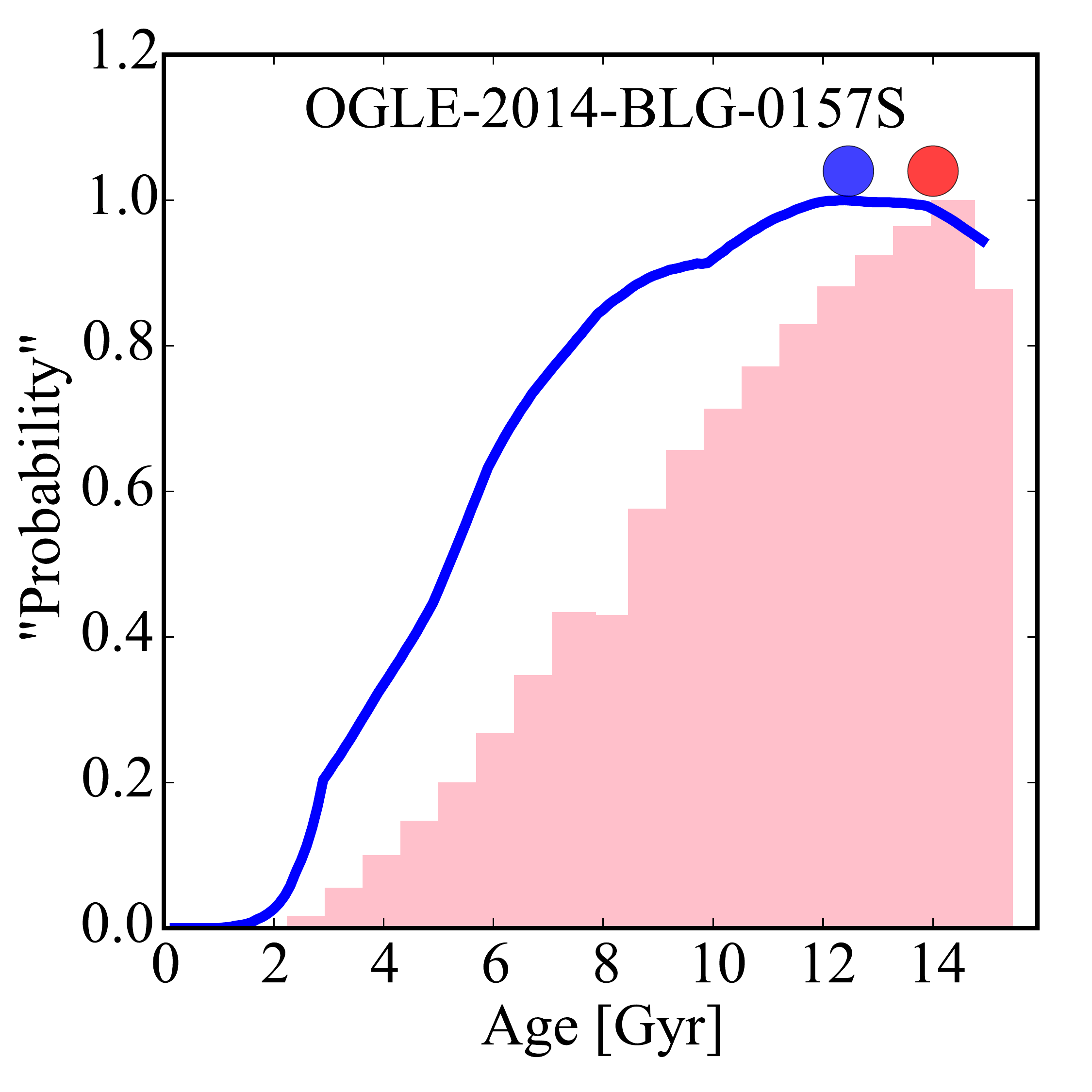}
\includegraphics[viewport= 93 0 648 648,clip]{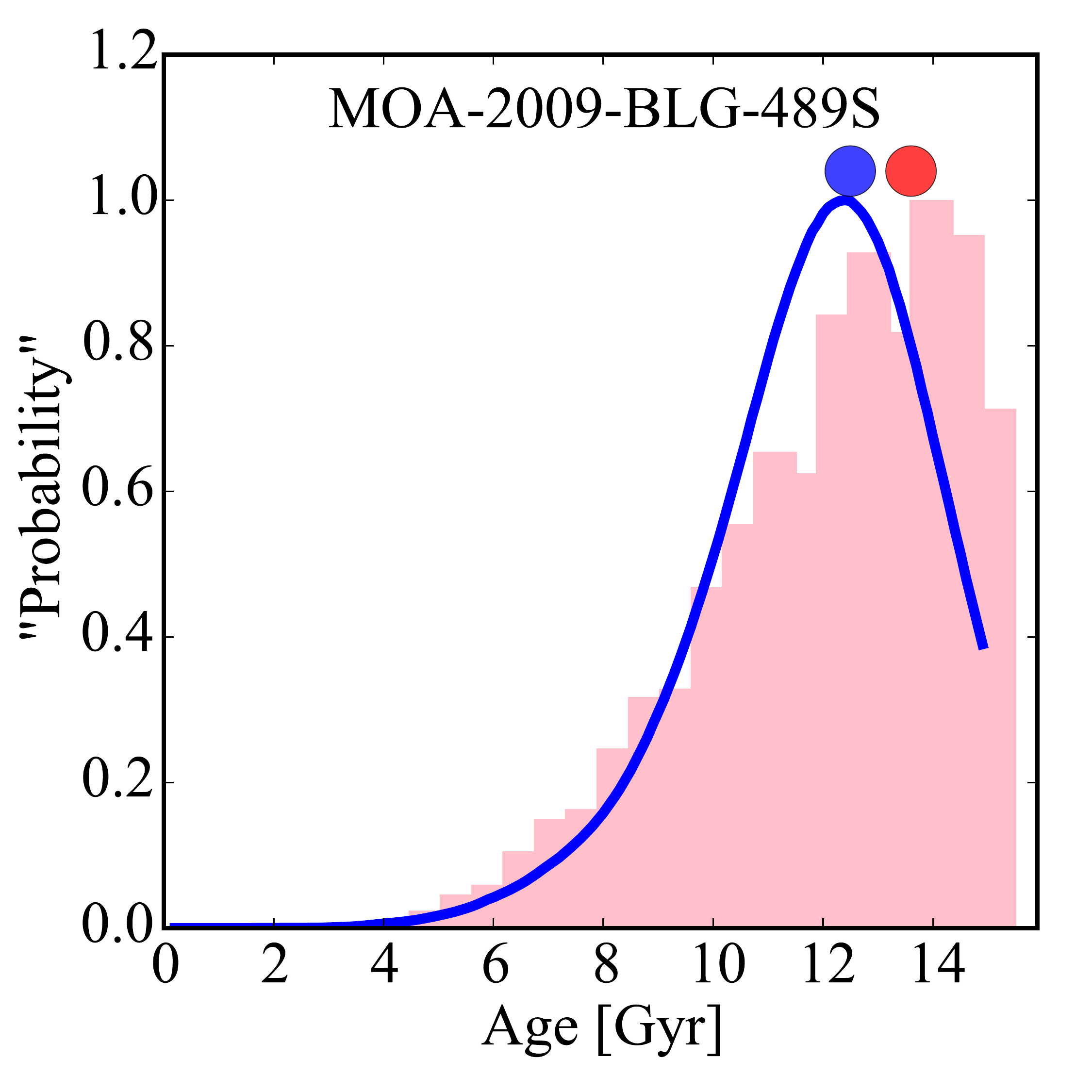}
\includegraphics[viewport= 93 0 648 648,clip]{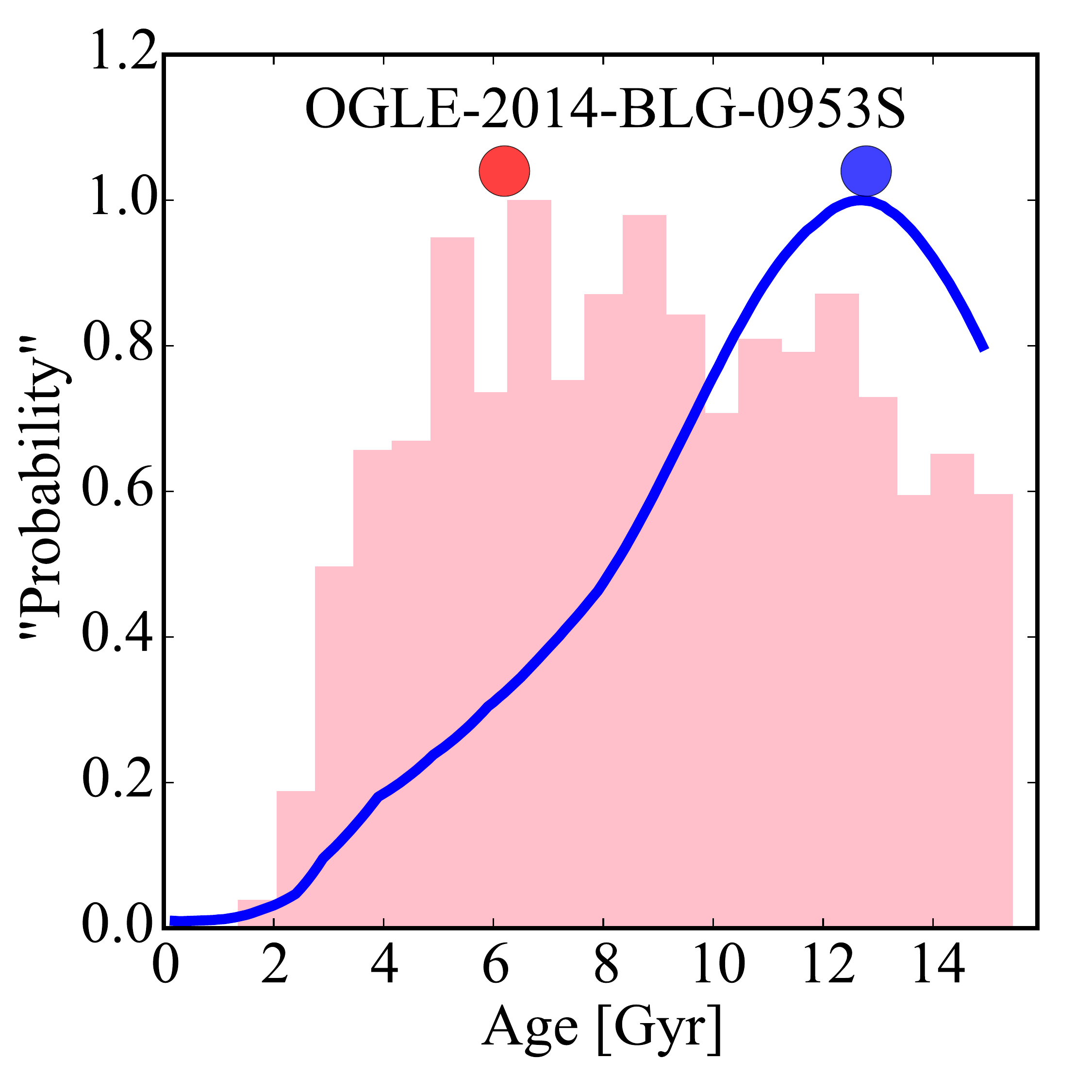}
\includegraphics[viewport= 93 0 648 648,clip]{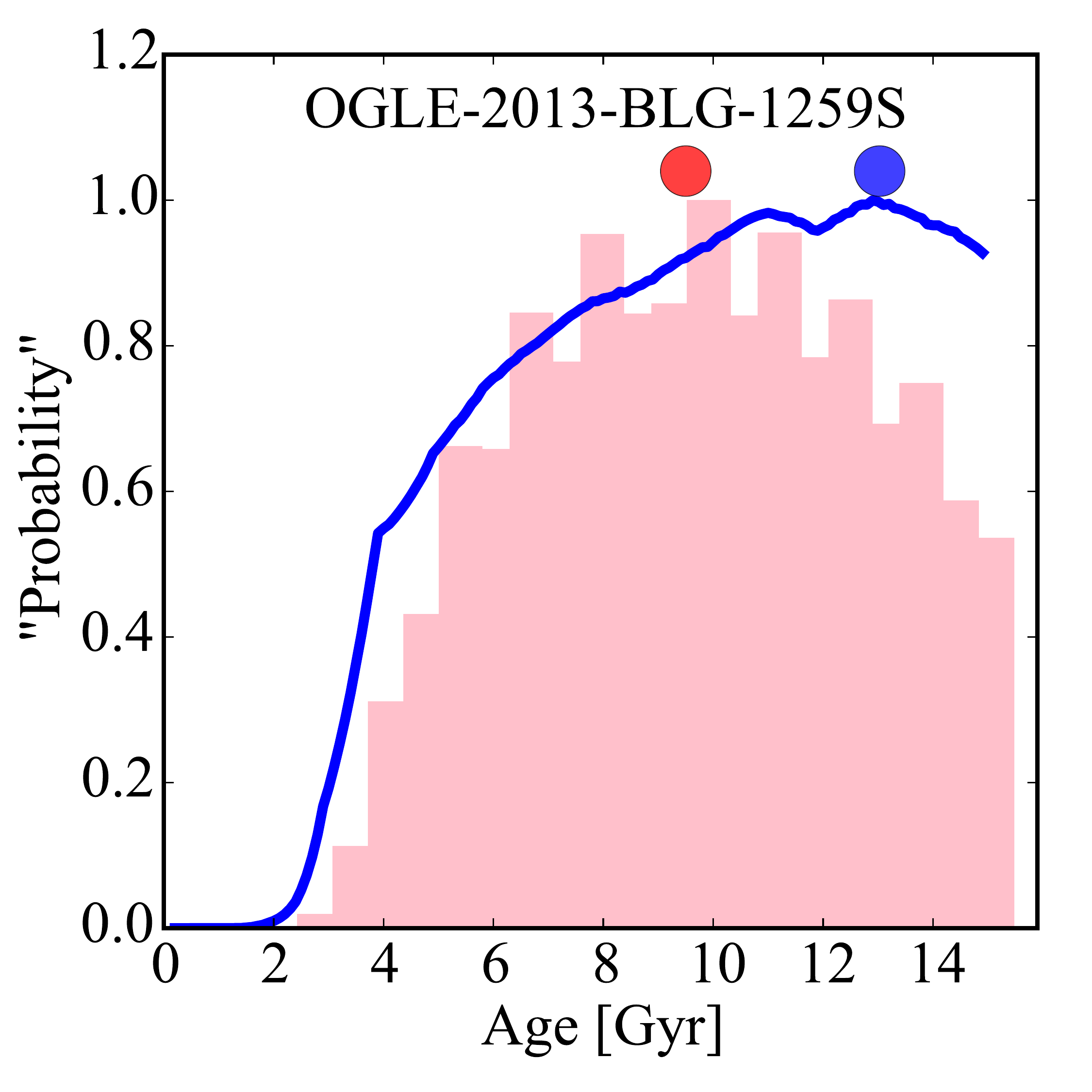}
\includegraphics[viewport= 93 0 648 648,clip]{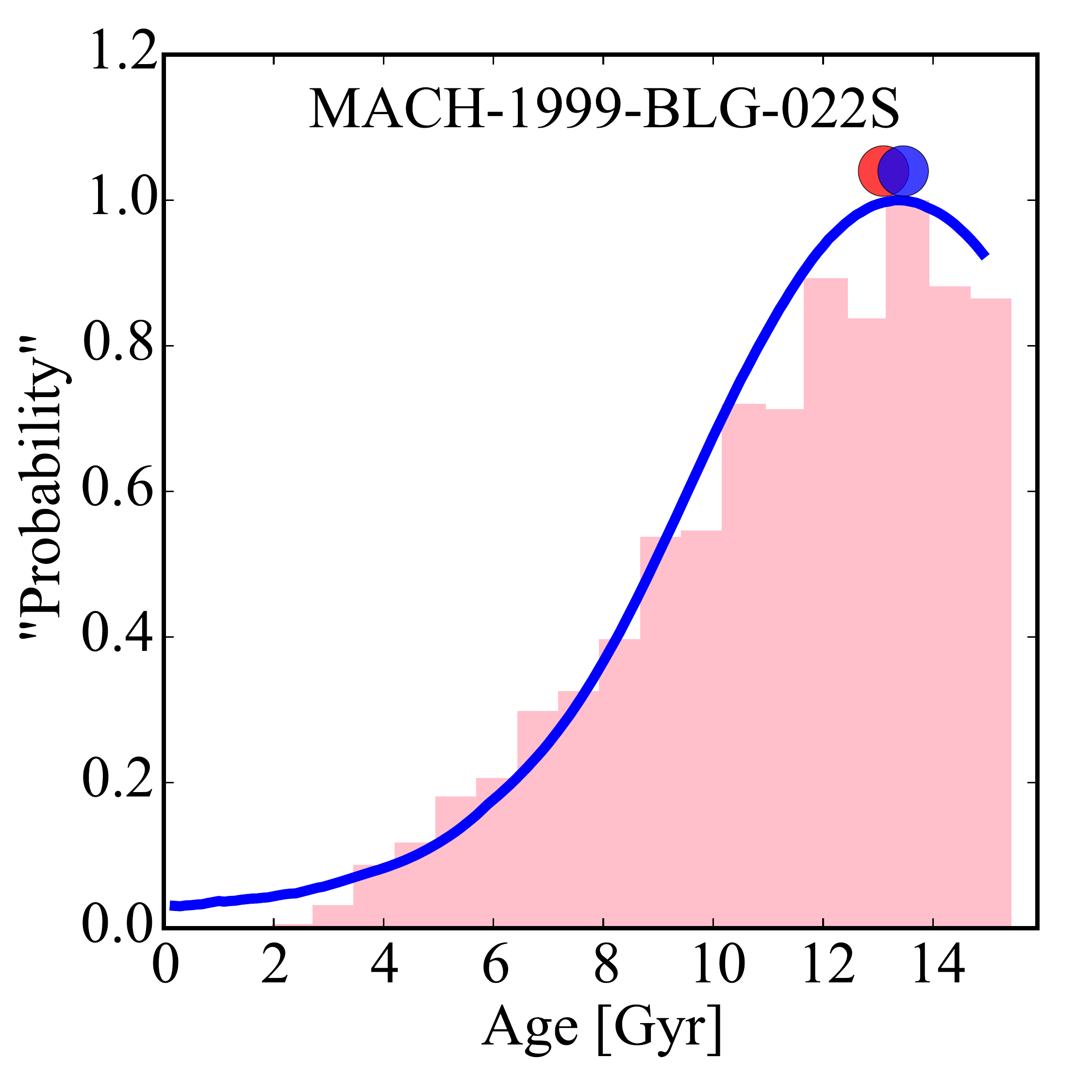}}
\resizebox{\hsize}{!}{
\includegraphics[viewport= 0 0 648 648,clip]{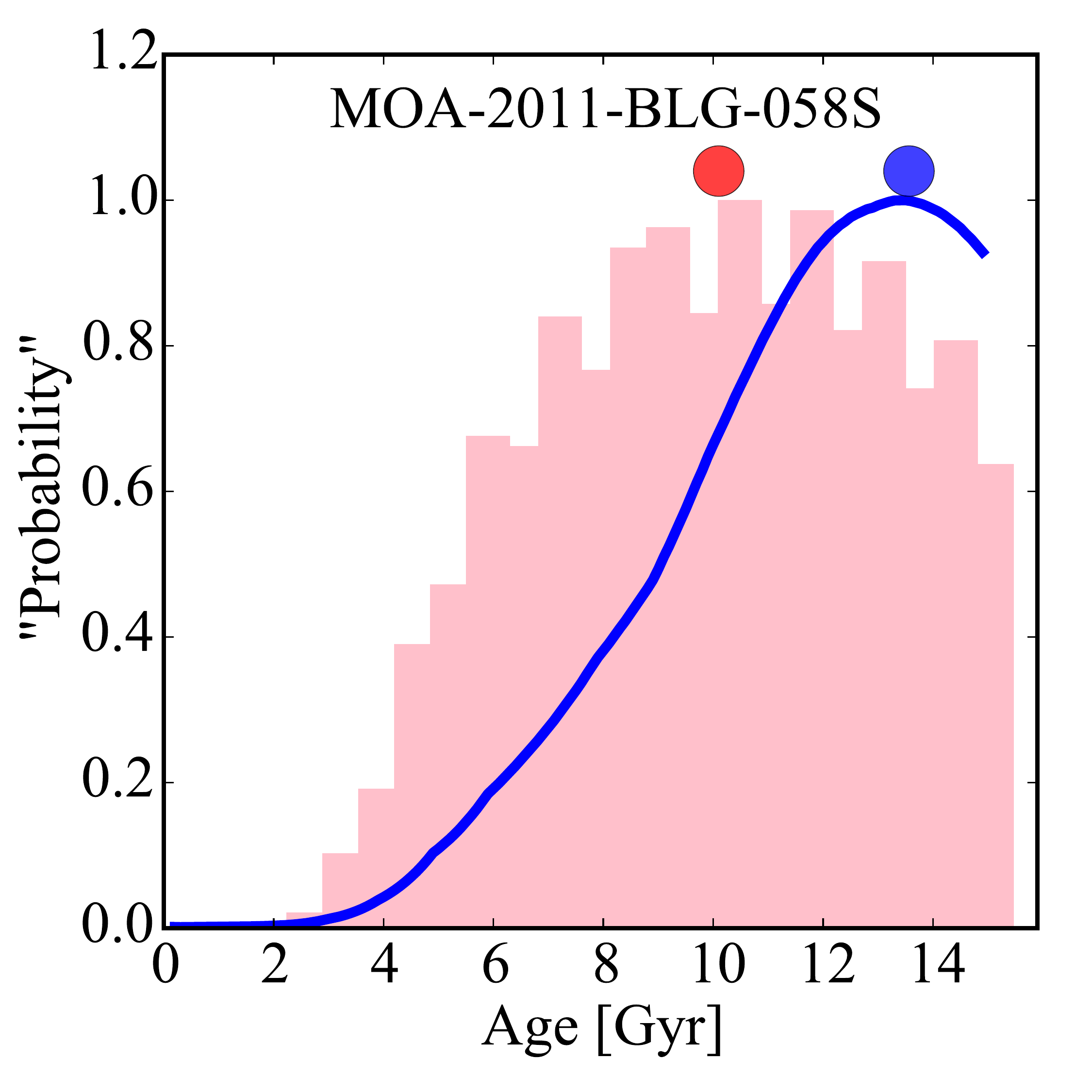}
\includegraphics[viewport= 93 0 648 648,clip]{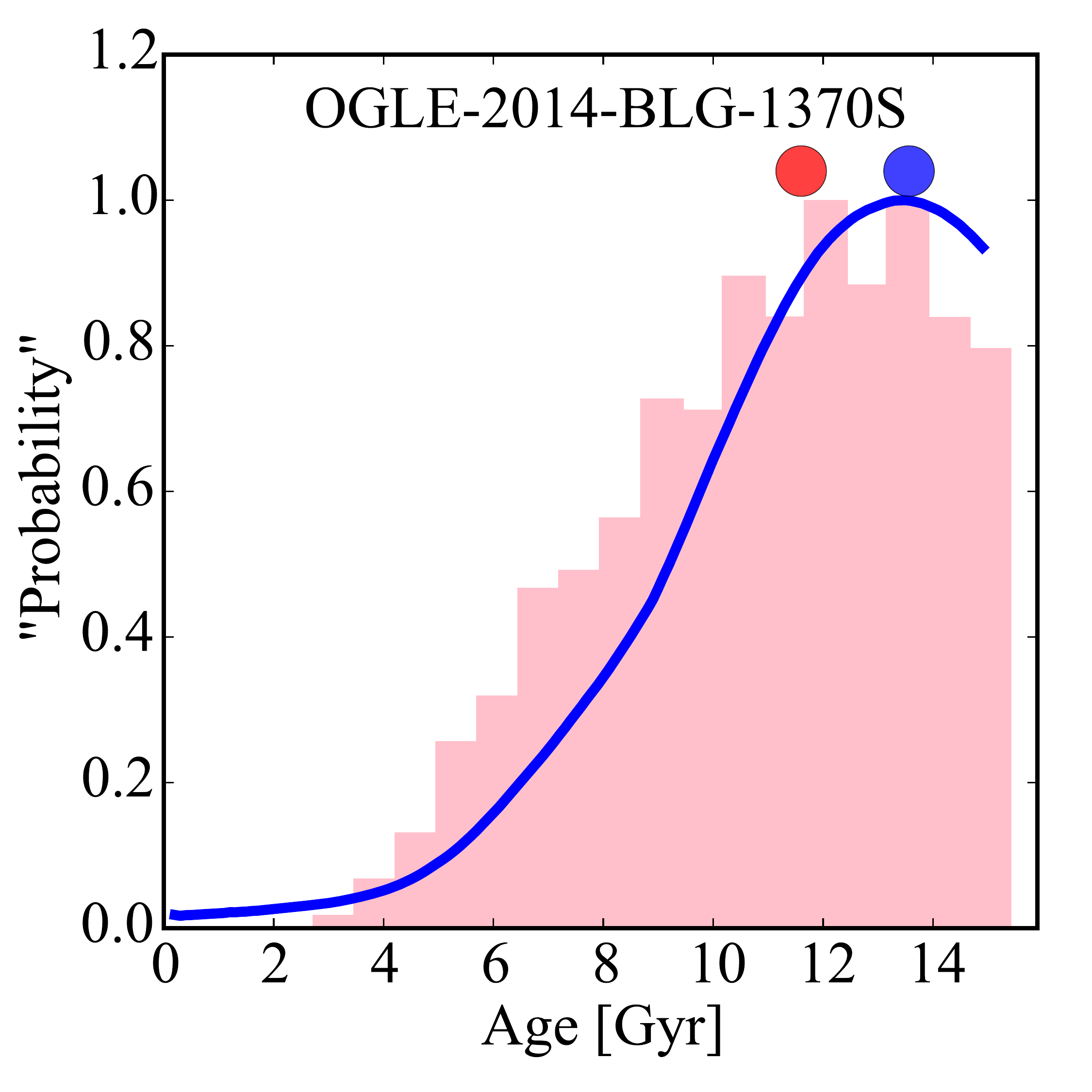}
\includegraphics[viewport= 93 0 648 648,clip]{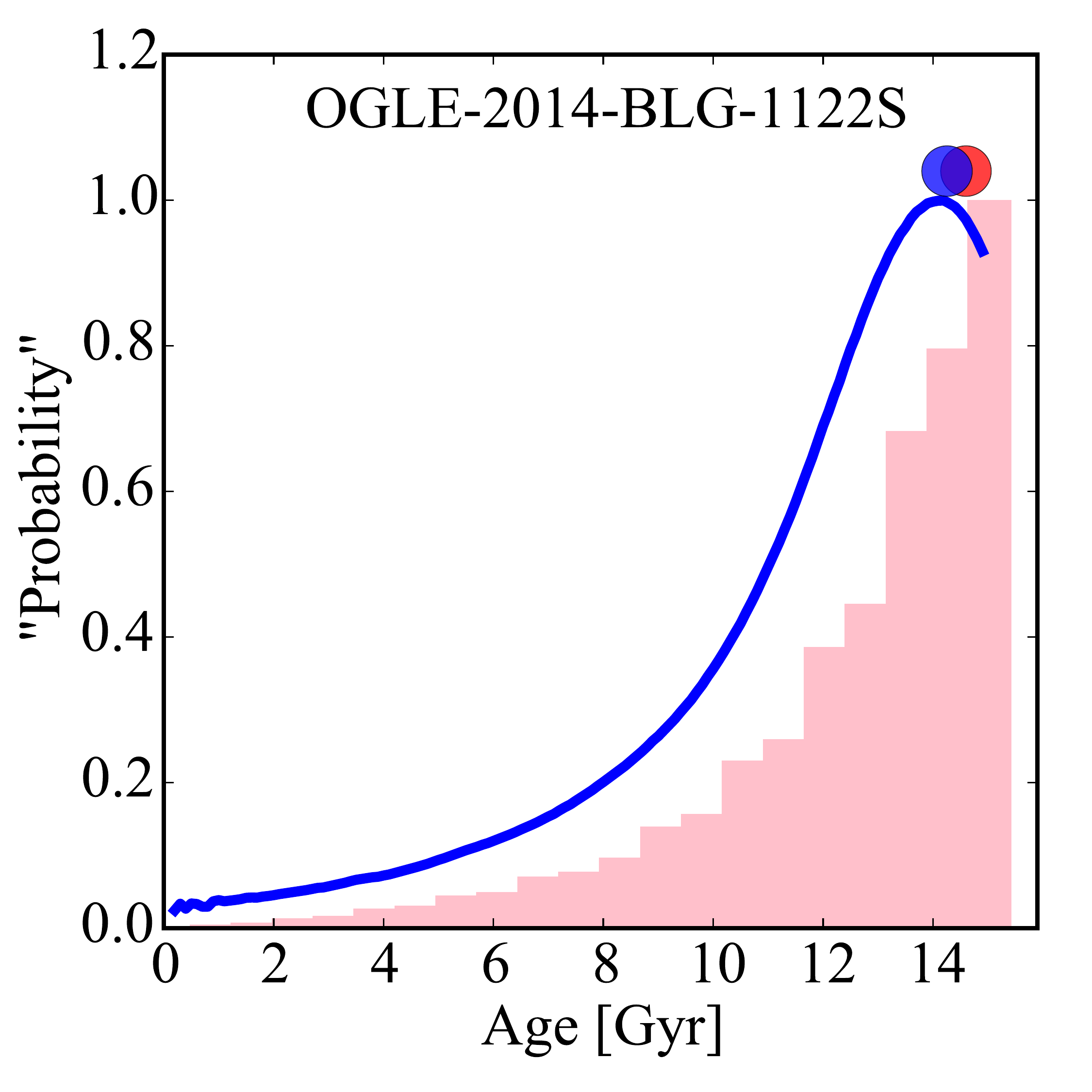}
\includegraphics[viewport= 93 0 648 648,clip]{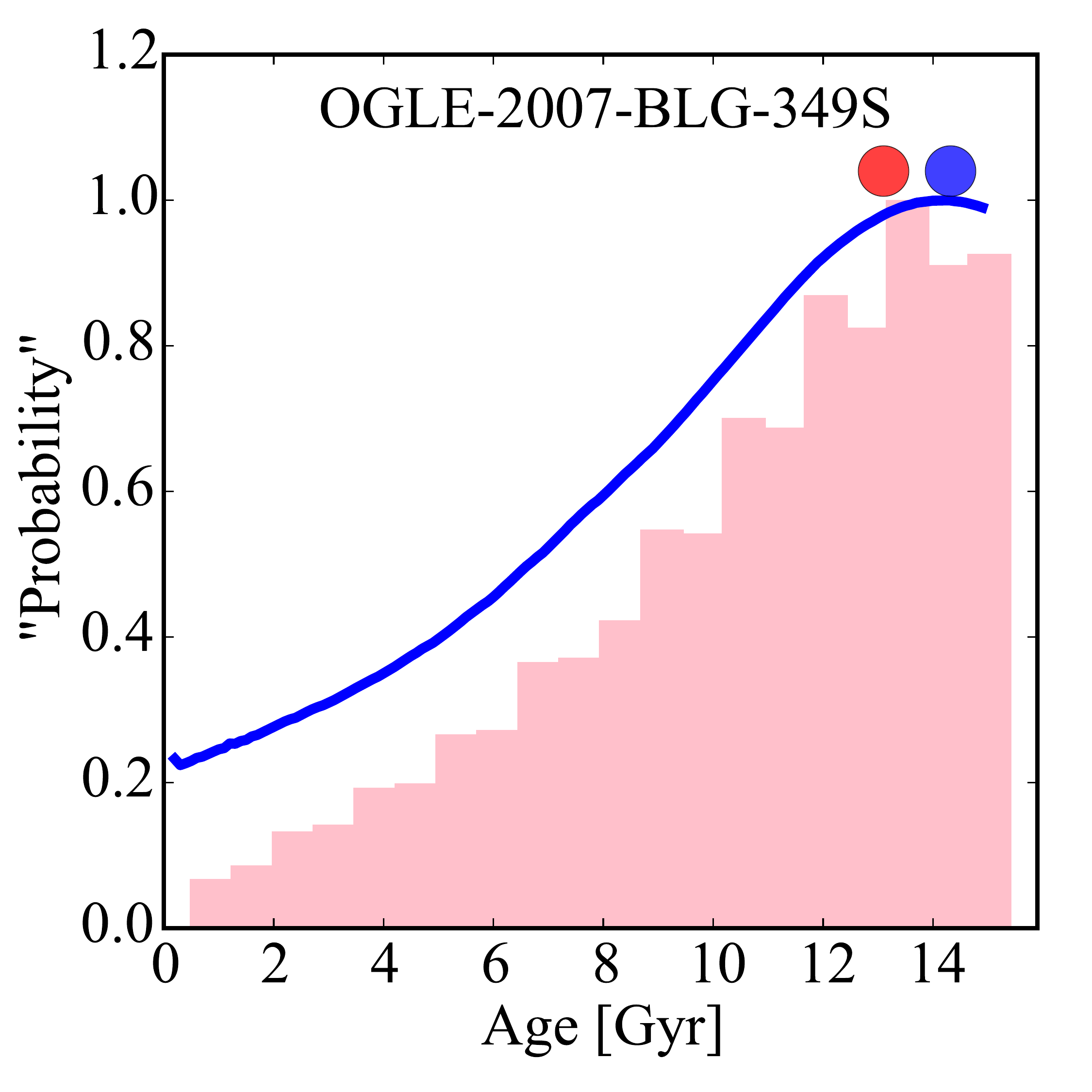}
\includegraphics[viewport= 93 0 648 648,clip]{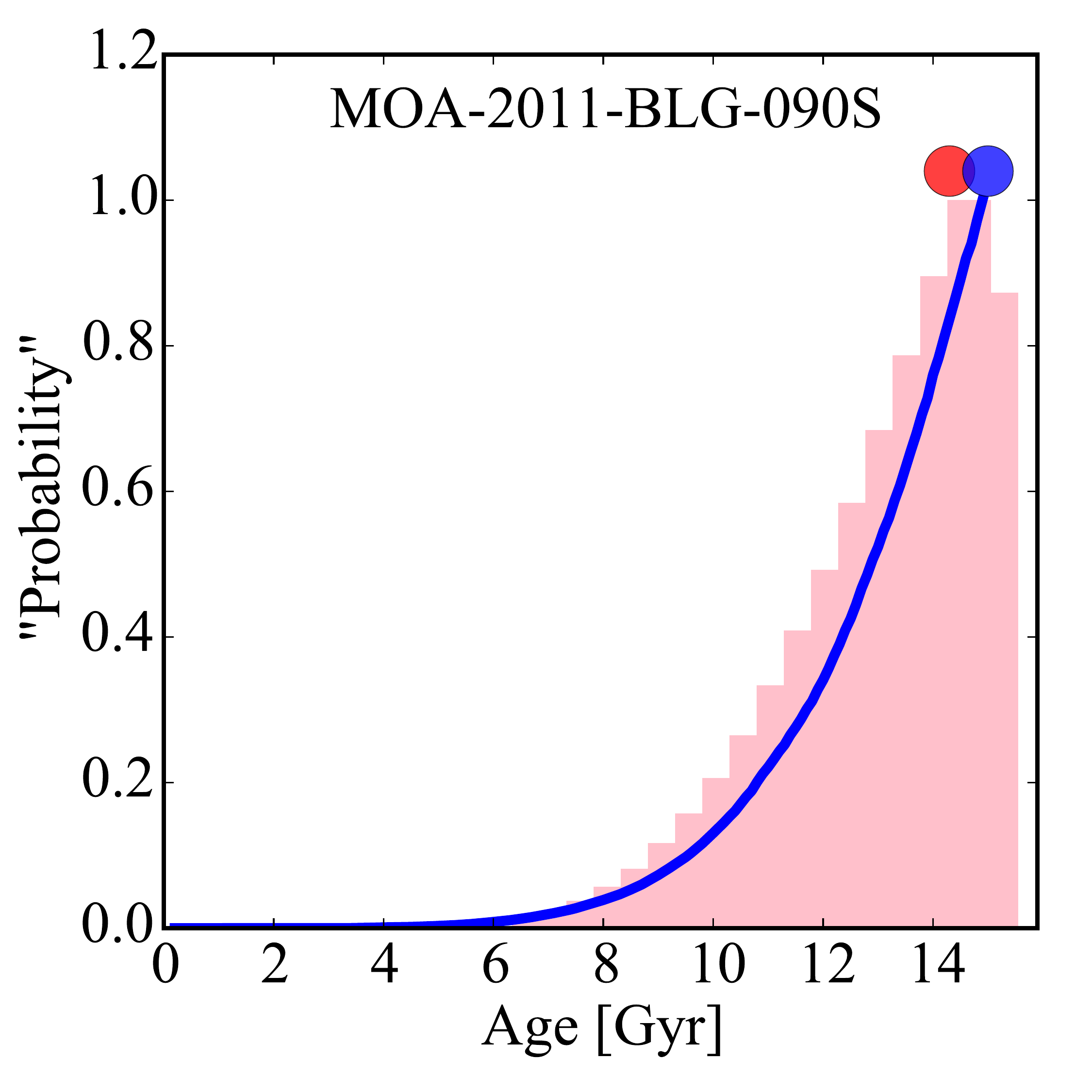}}
\resizebox{\hsize}{!}{
\includegraphics[viewport= 0 0 648 648,clip]{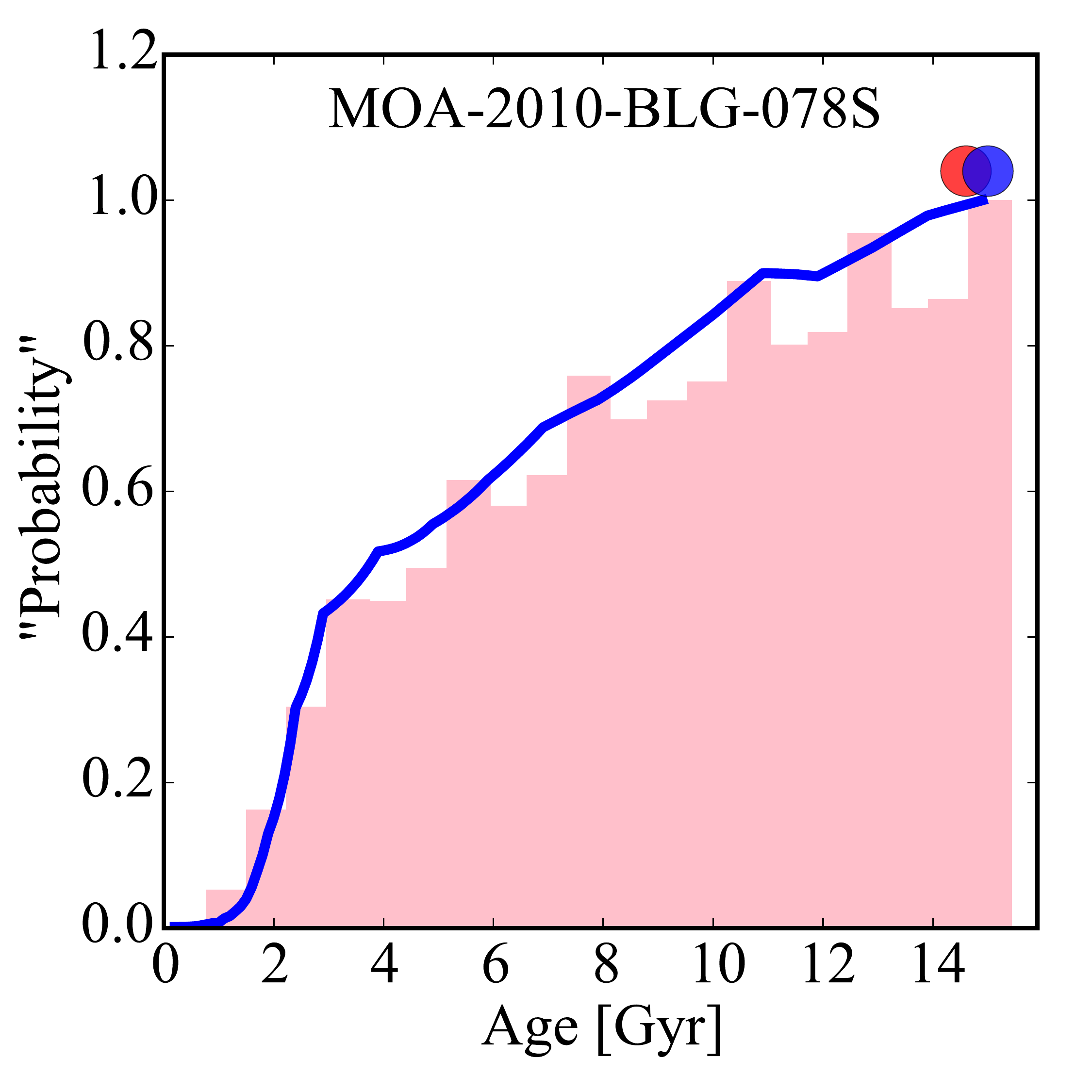}
\includegraphics[viewport= 93 0 648 648,clip]{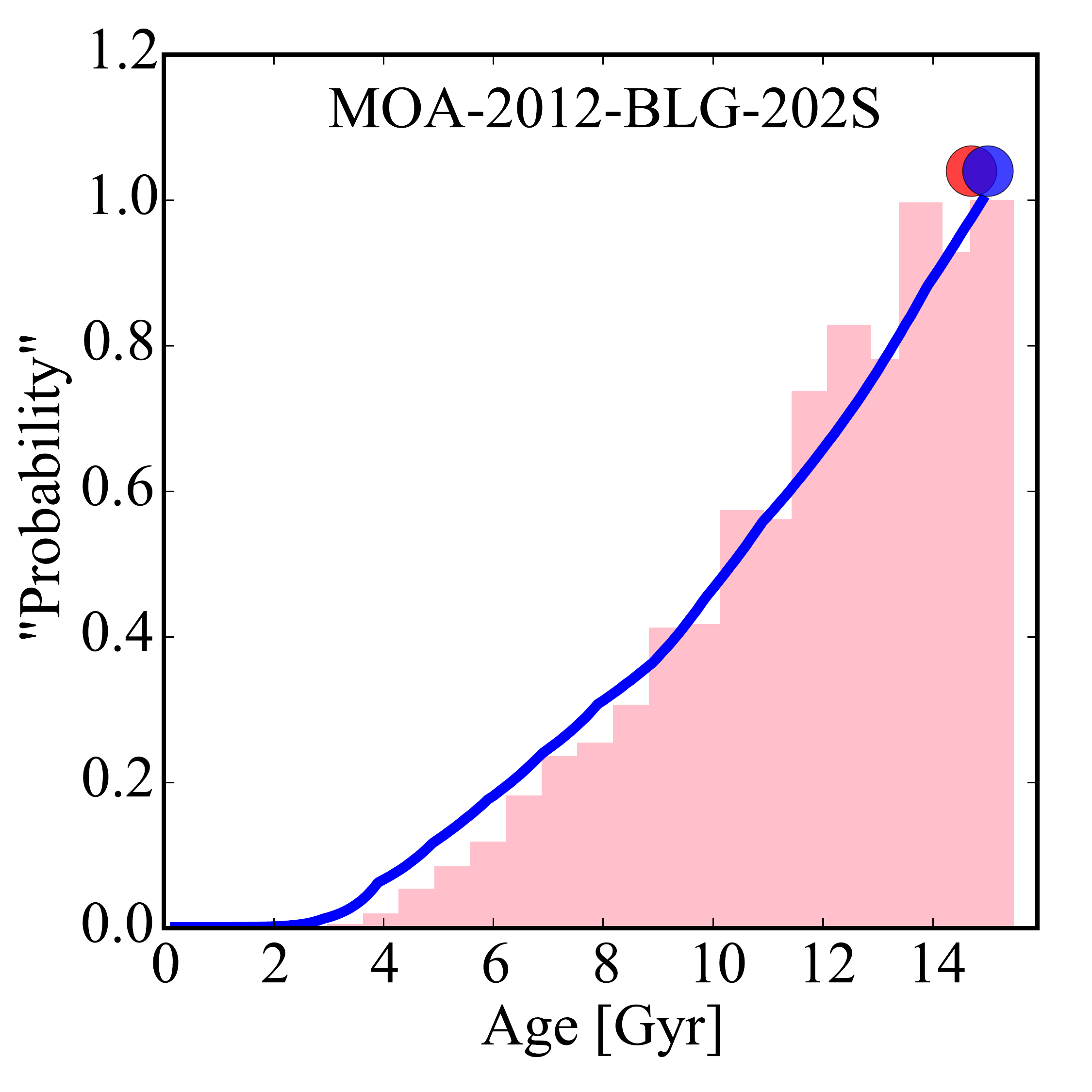}
\includegraphics[viewport= 93 0 648 648,clip]{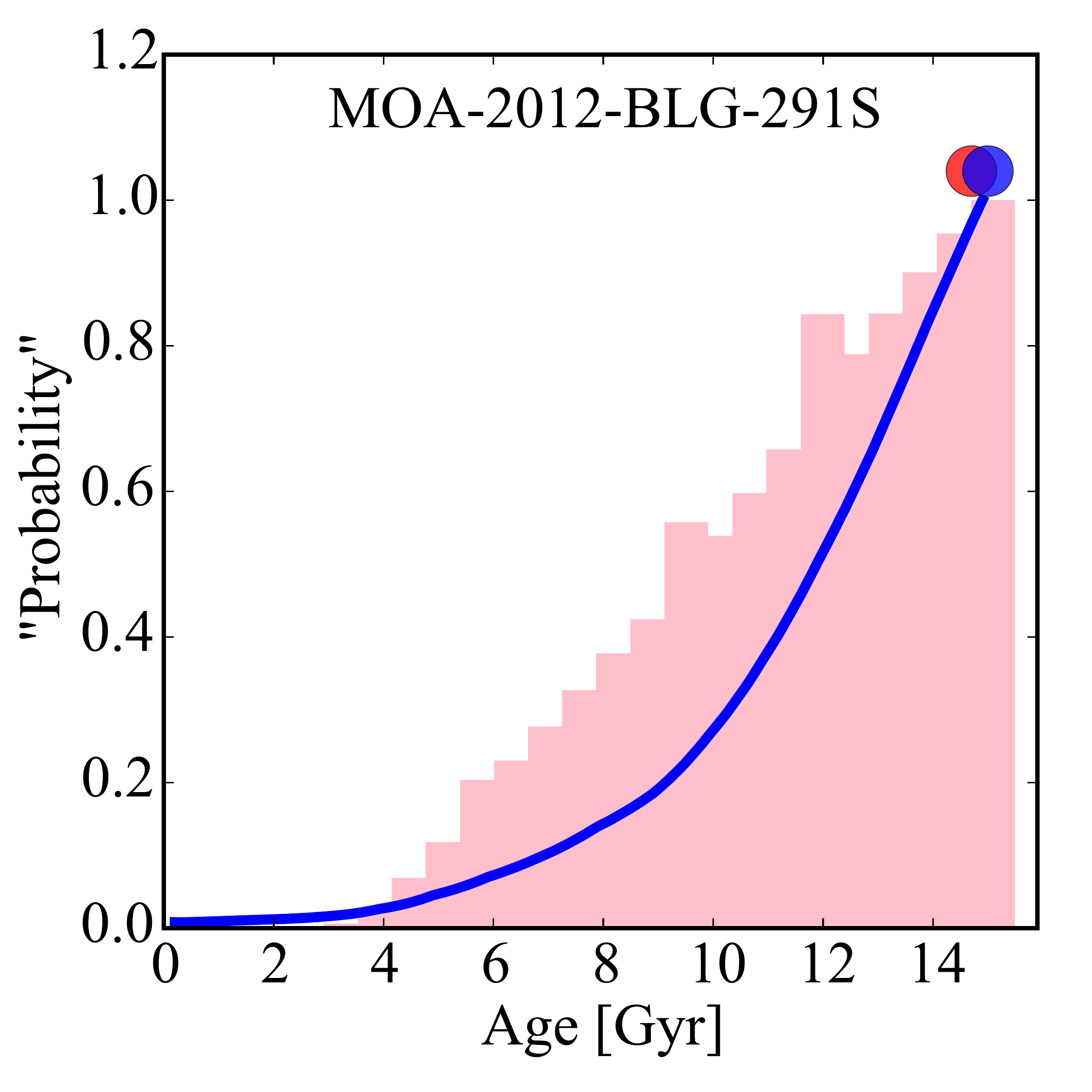}
\includegraphics[viewport= 93 0 648 648,clip]{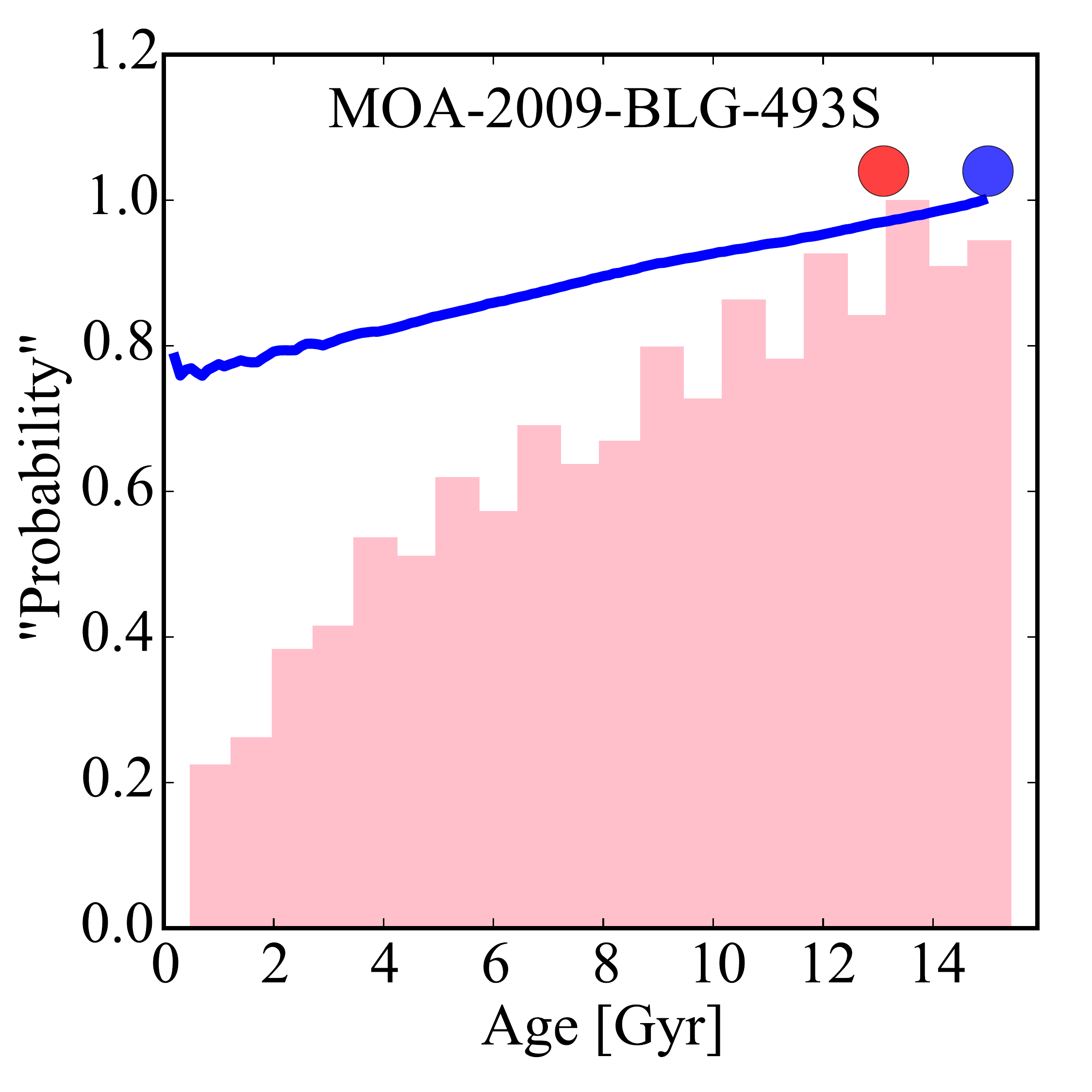}
\includegraphics[viewport= 93 0 648 648,clip]{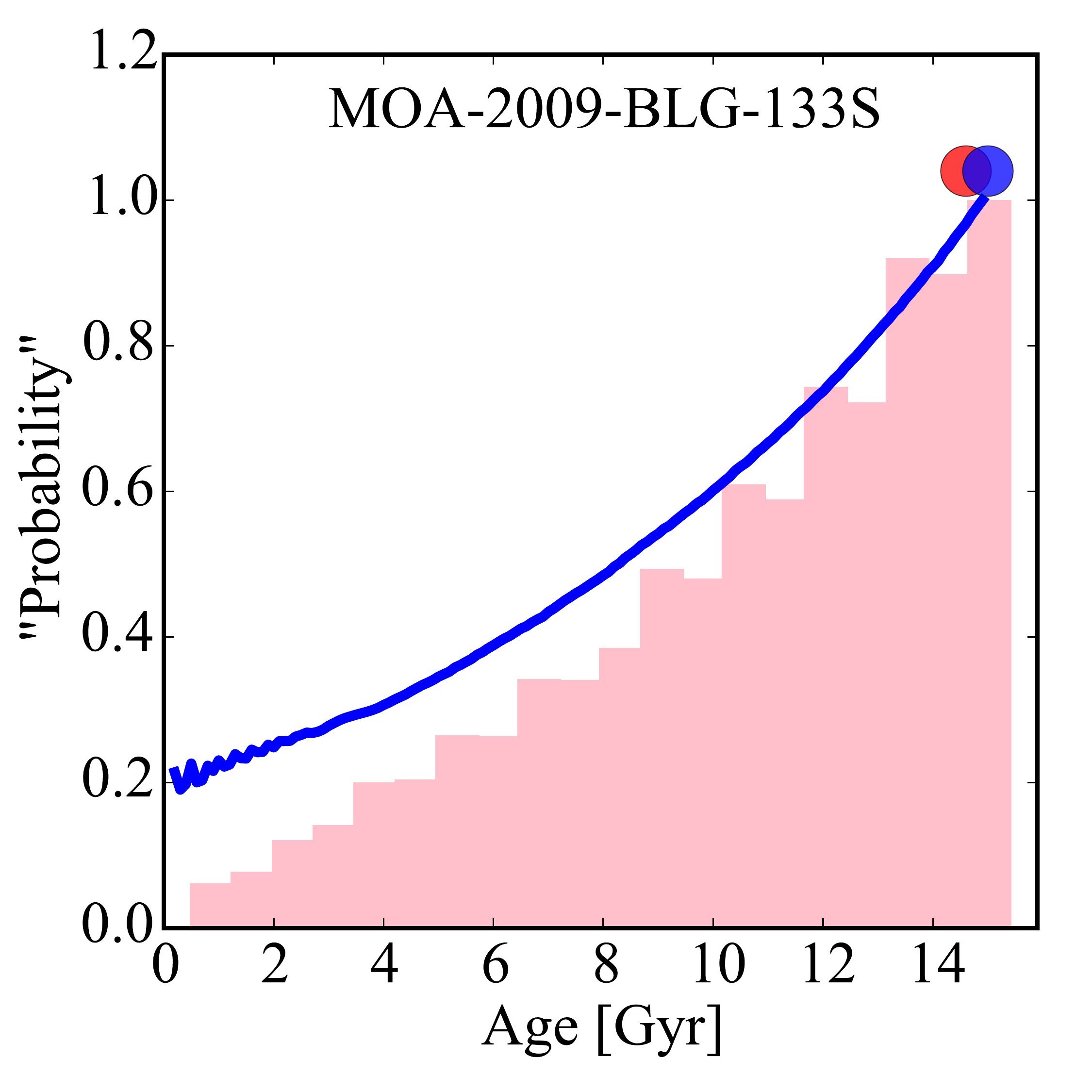}}
\resizebox{\hsize}{!}{
\includegraphics[viewport= 0 0 648 648,clip]{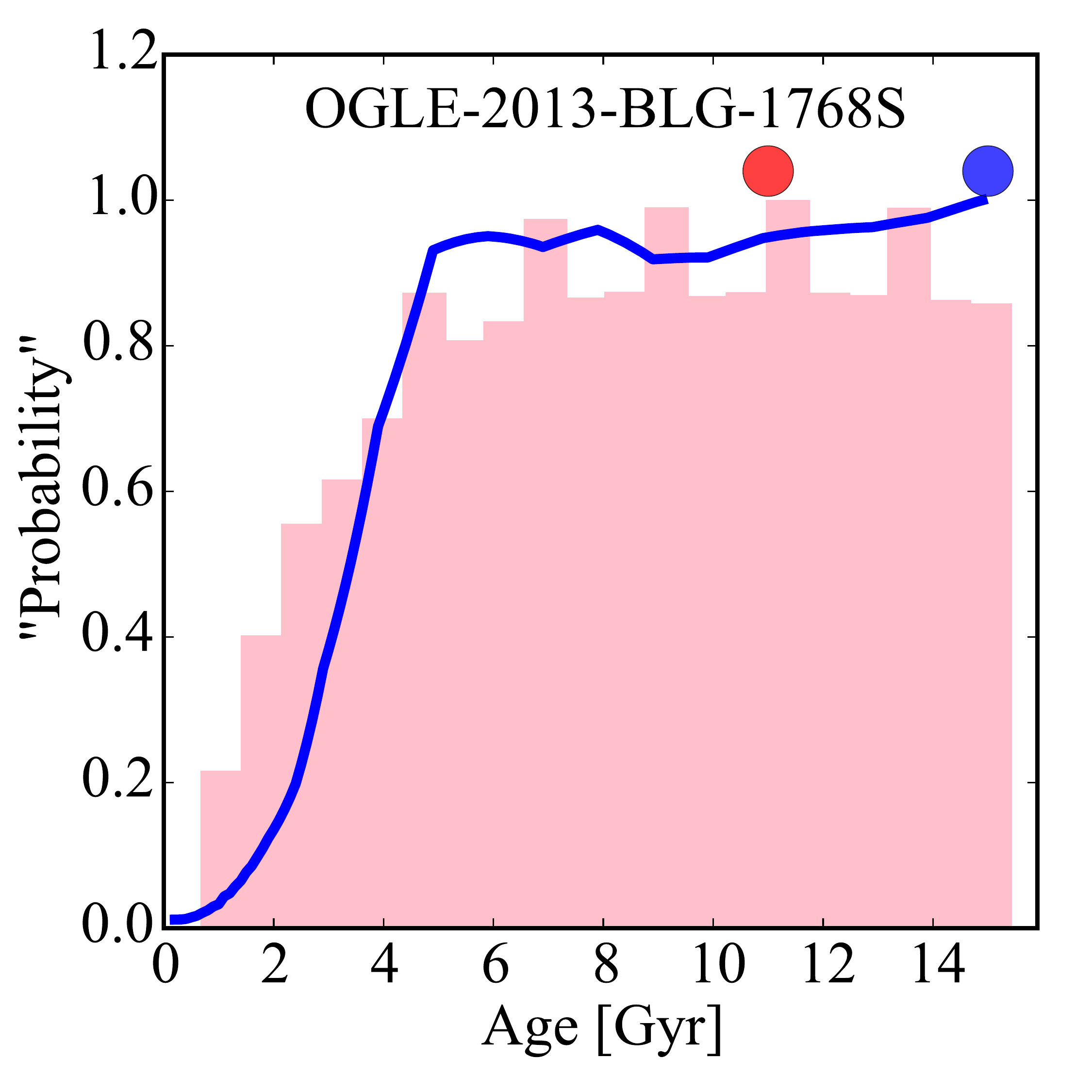}
\includegraphics[viewport= 93 0 648 0,clip]{agefunctions_931768.pdf}
\includegraphics[viewport= 93 0 648 0,clip]{agefunctions_931768.pdf}
\includegraphics[viewport= 93 0 648 0,clip]{agefunctions_931768.pdf}
\includegraphics[viewport= 93 0 648 0,clip]{agefunctions_931768.pdf}}
\caption{\sl continued
}
\end{figure*}

\begin{figure*}[h]
\centering
\resizebox{0.85\hsize}{!}{
\includegraphics[viewport=0 56 510 225,clip]{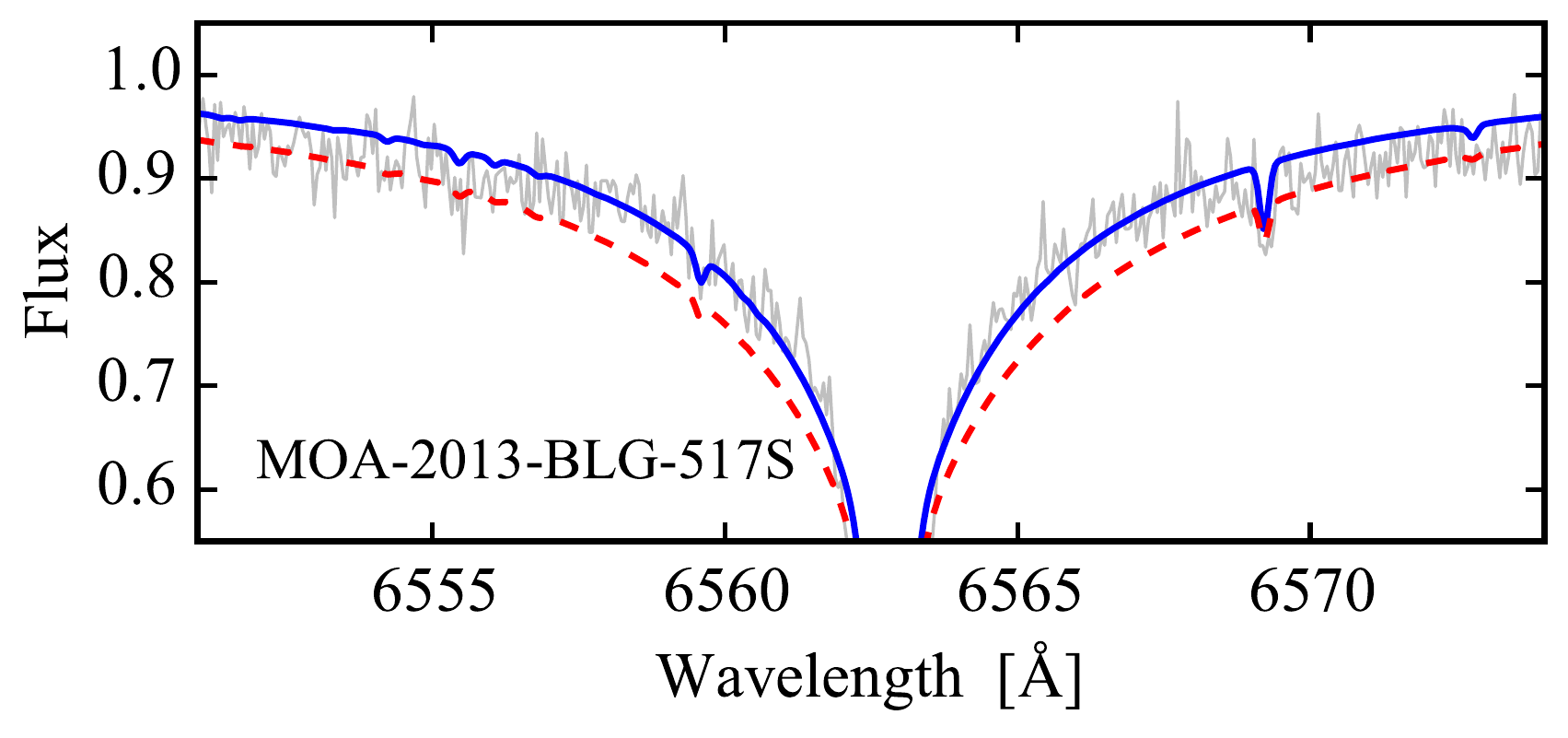}
\includegraphics[viewport=56 56 510 225,clip]{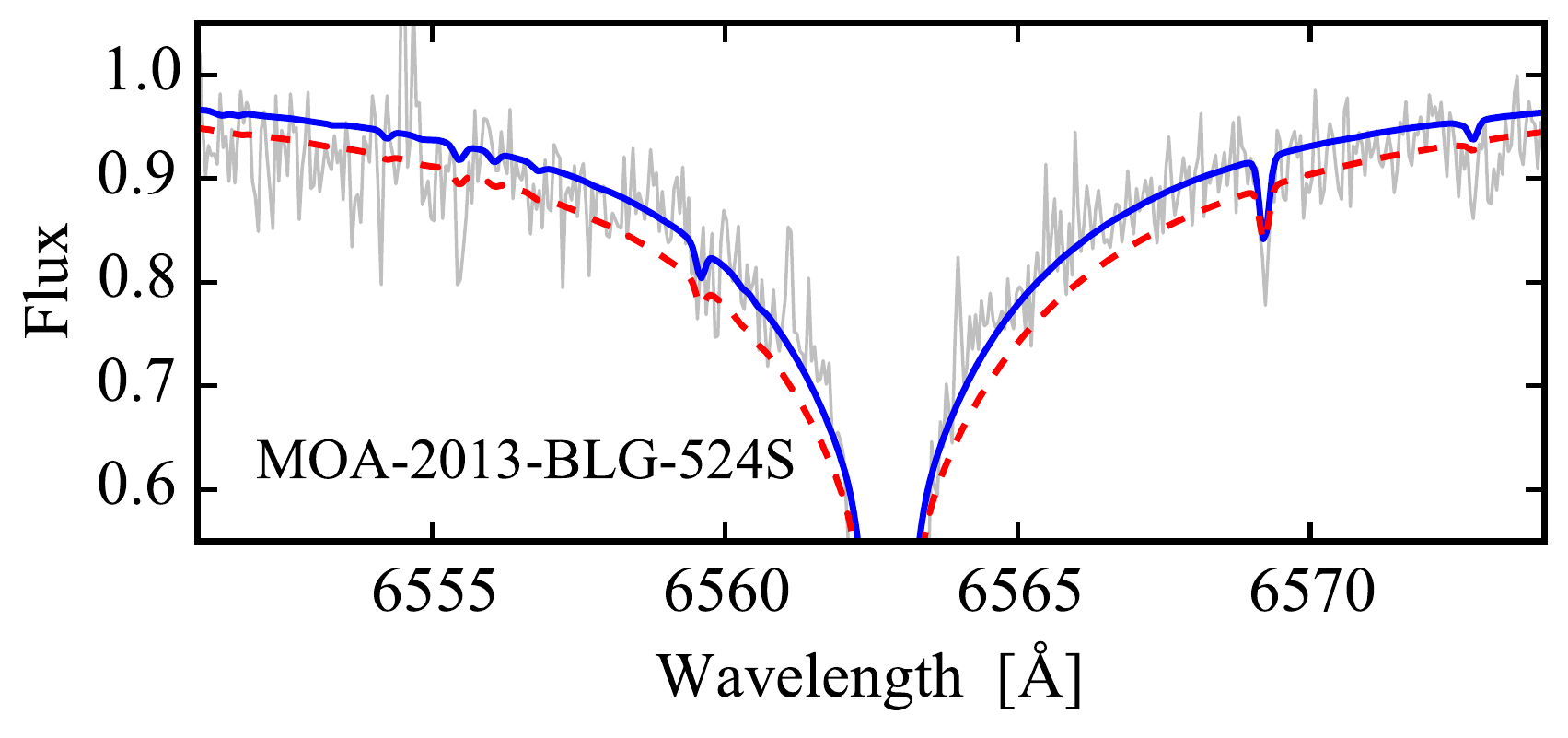}
\includegraphics[viewport=56 56 485 225,clip]{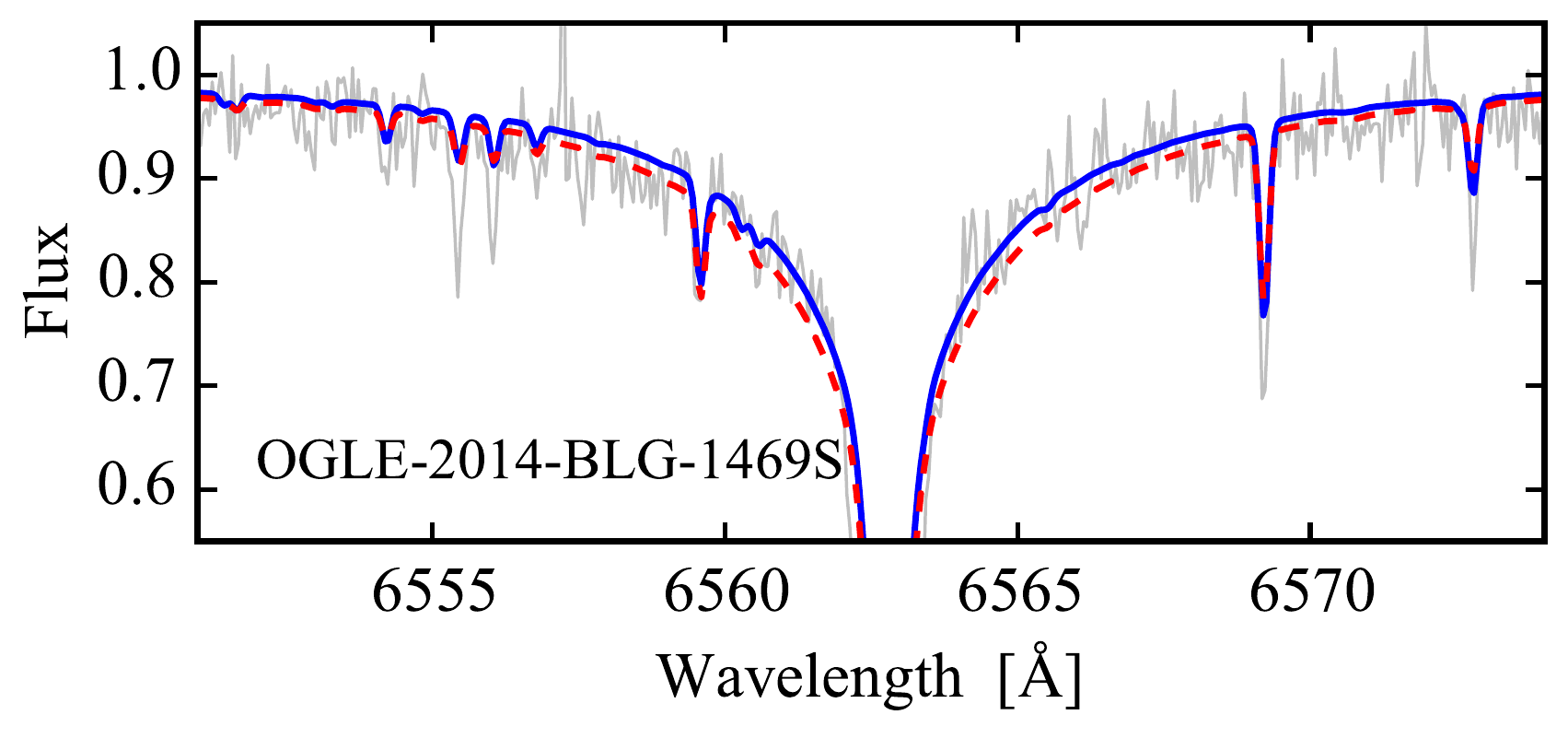}
}
\resizebox{0.85\hsize}{!}{
\includegraphics[viewport=0 56 510 225,clip]{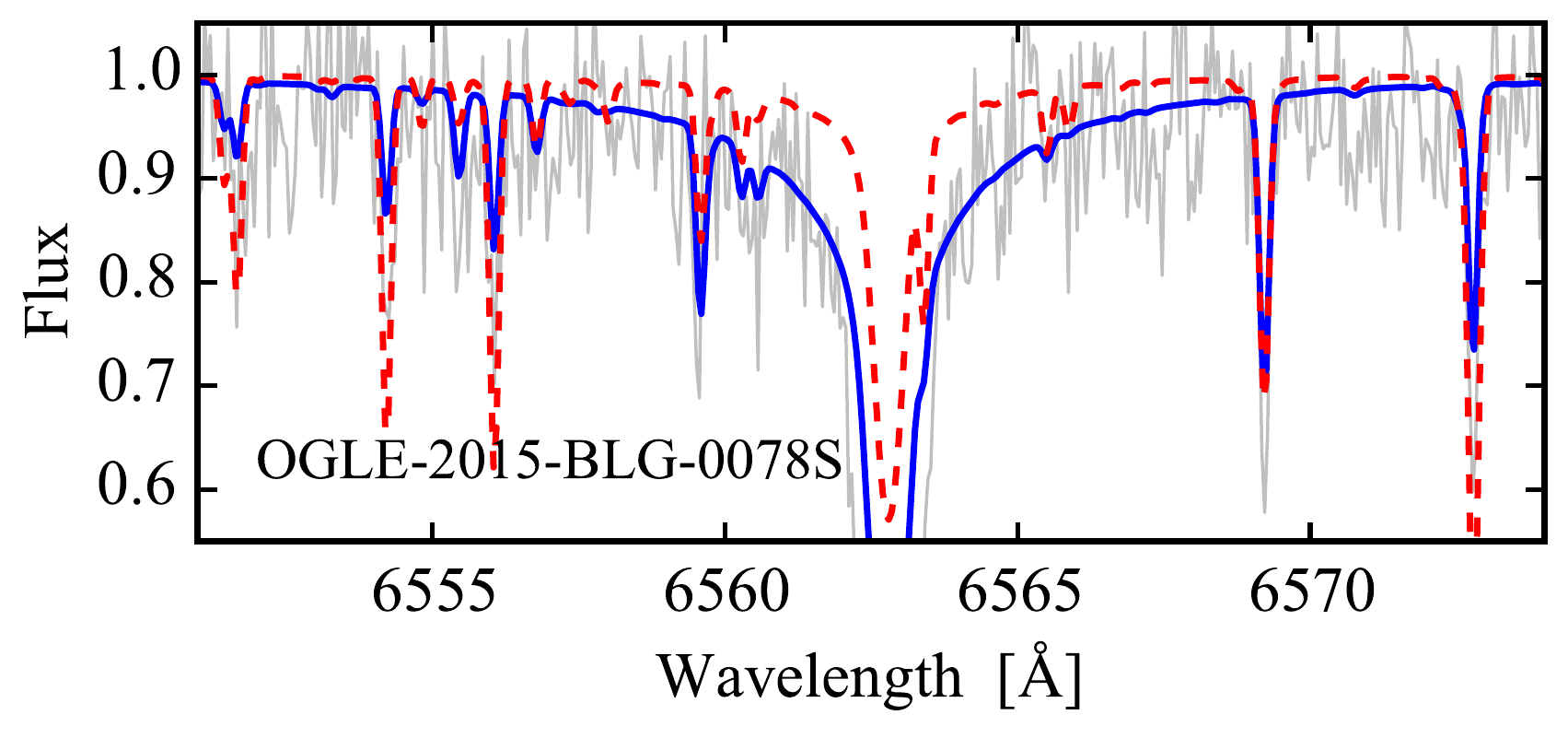}
\includegraphics[viewport=56 56 510 225,clip]{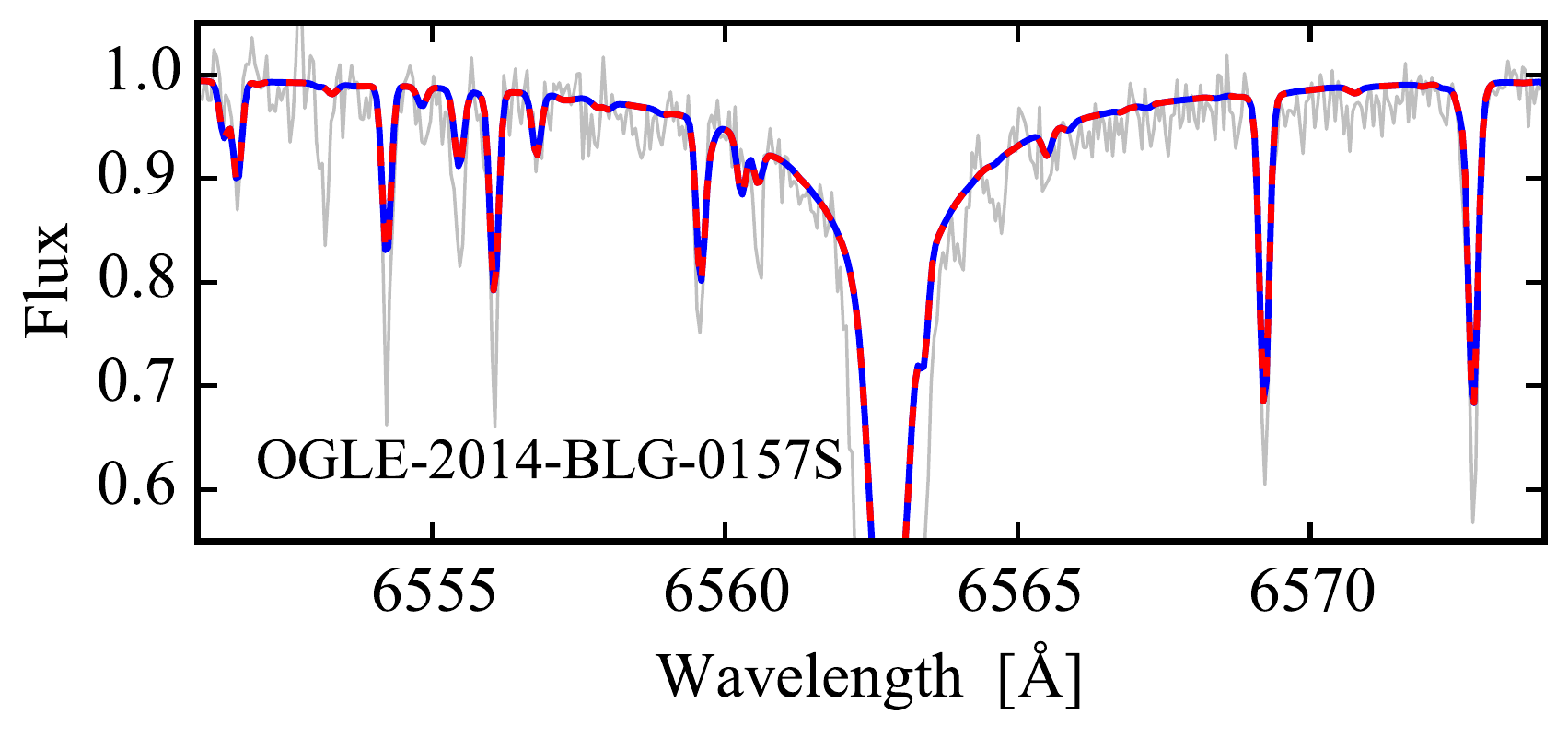}
\includegraphics[viewport=56 56 485 225,clip]{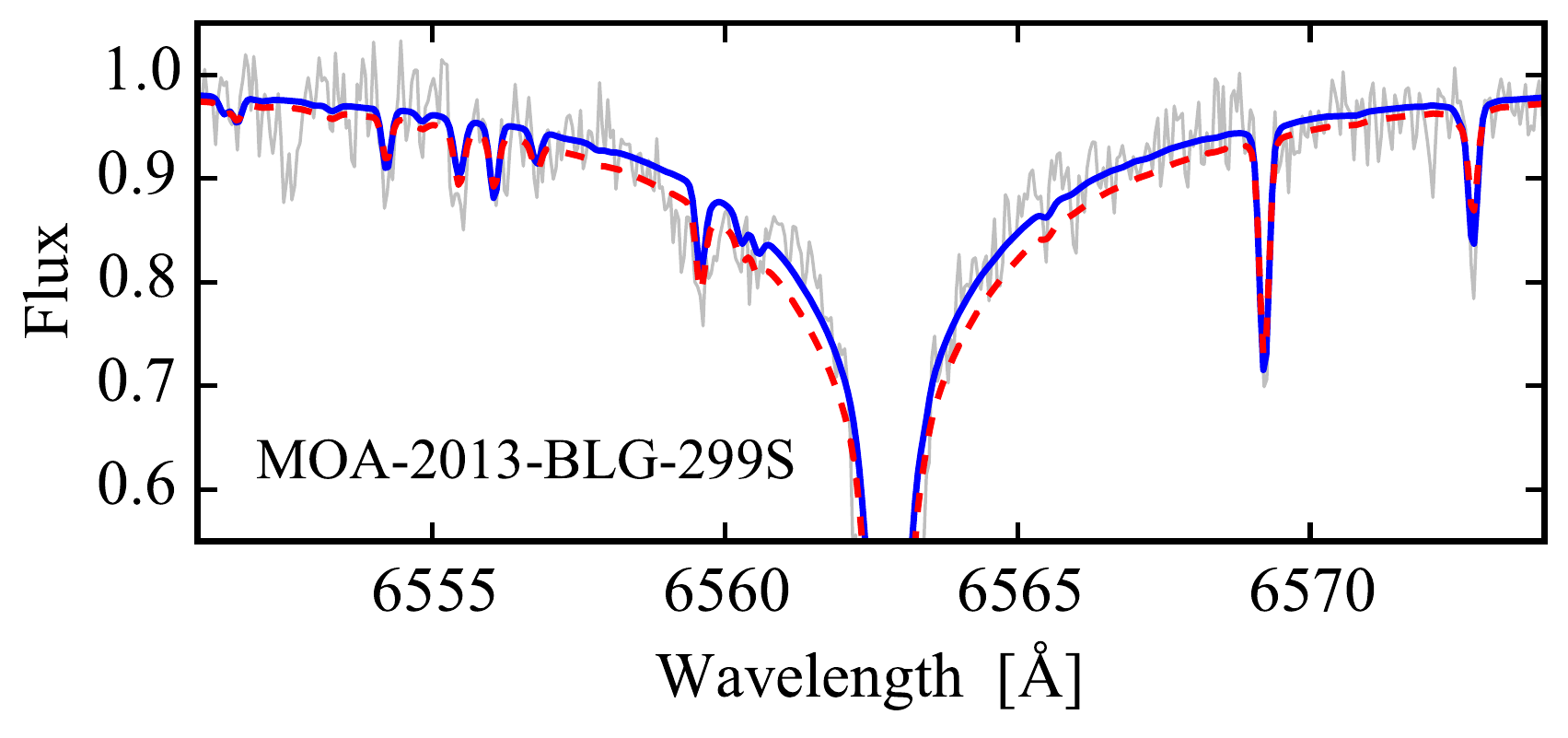}
}
\resizebox{0.85\hsize}{!}{
\includegraphics[viewport=0 56 510 225,clip]{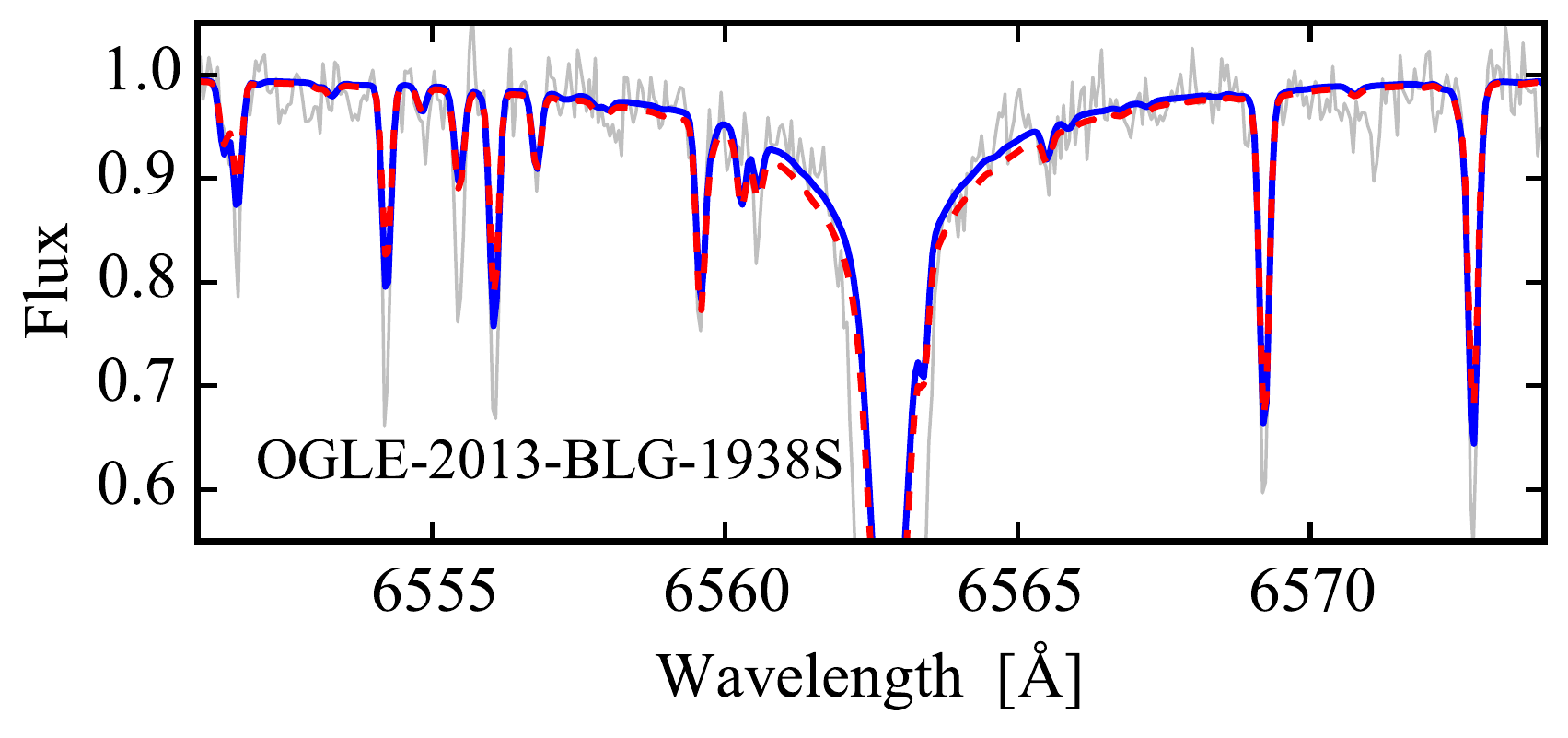}
\includegraphics[viewport=56 56 510 225,clip]{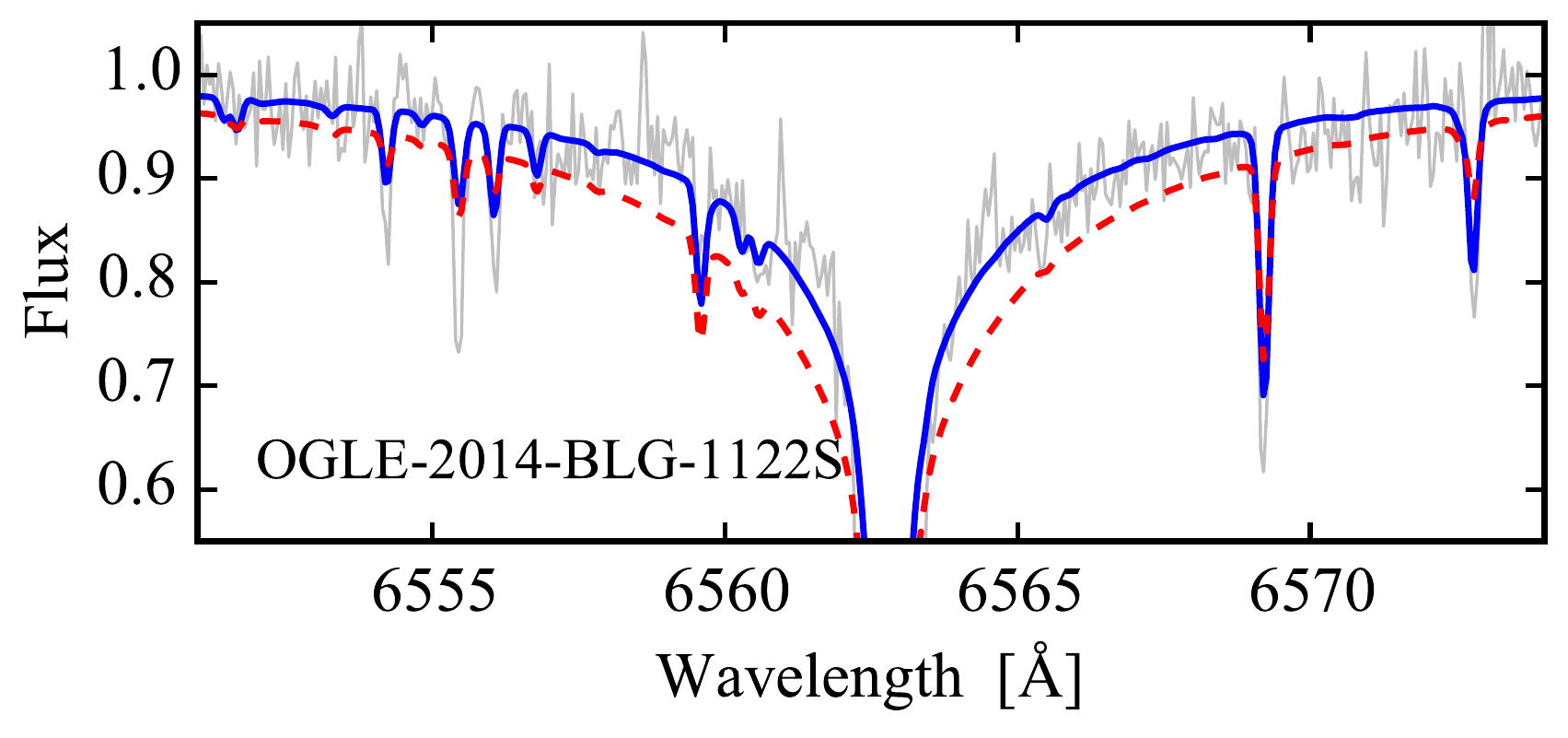}
\includegraphics[viewport=56 56 485 225,clip]{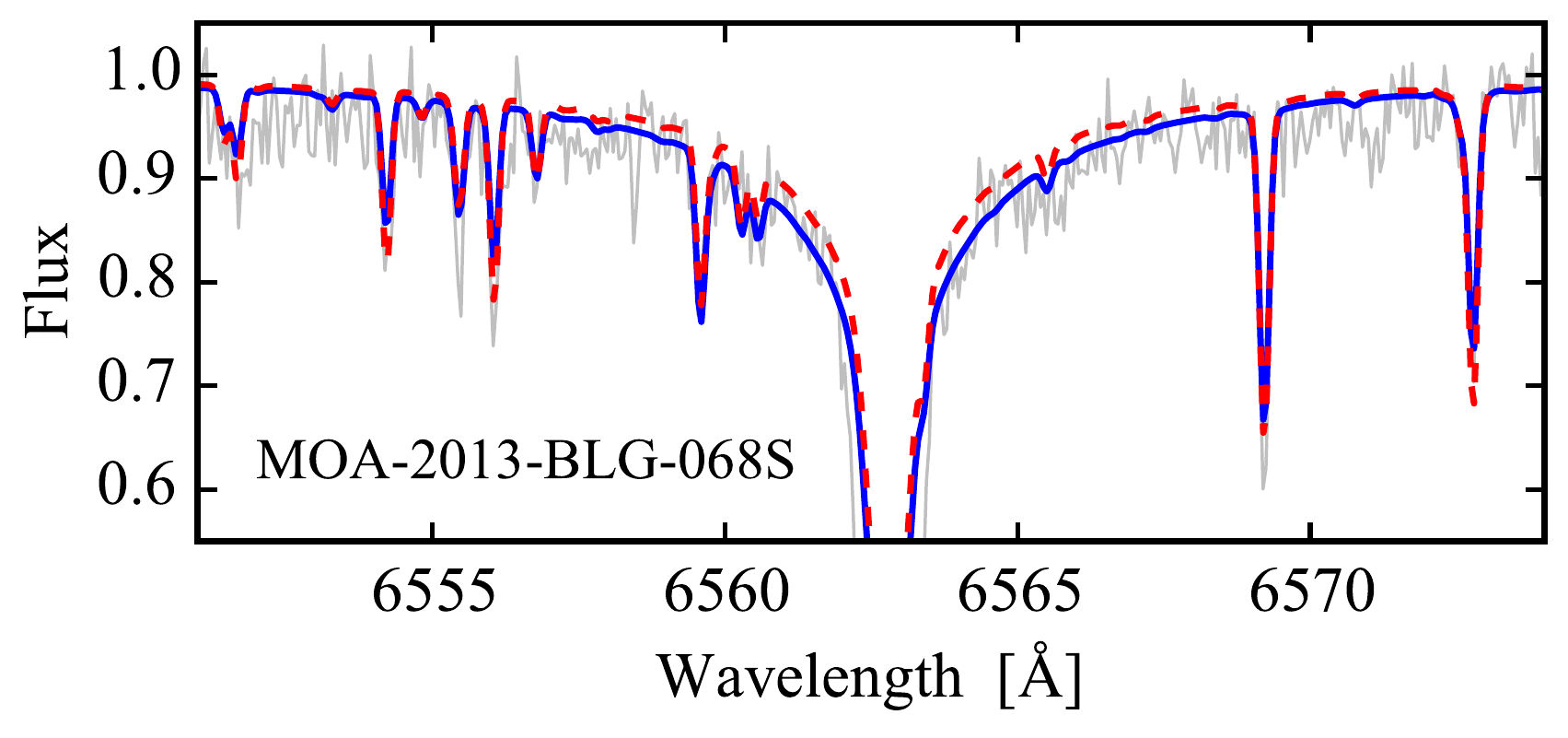}
}
\resizebox{0.85\hsize}{!}{
\includegraphics[viewport=0 56 510 225,clip]{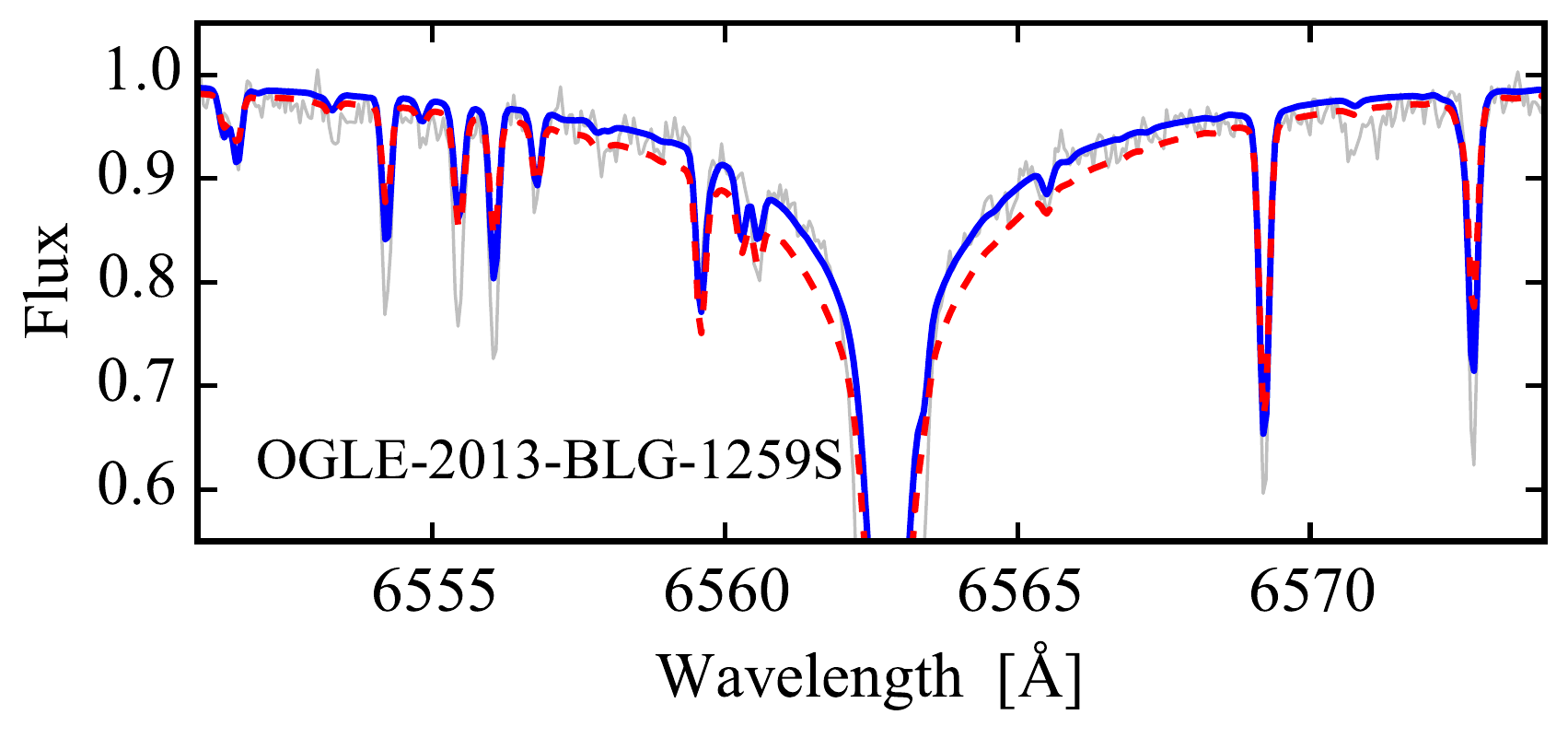}
\includegraphics[viewport=56 56 510 225,clip]{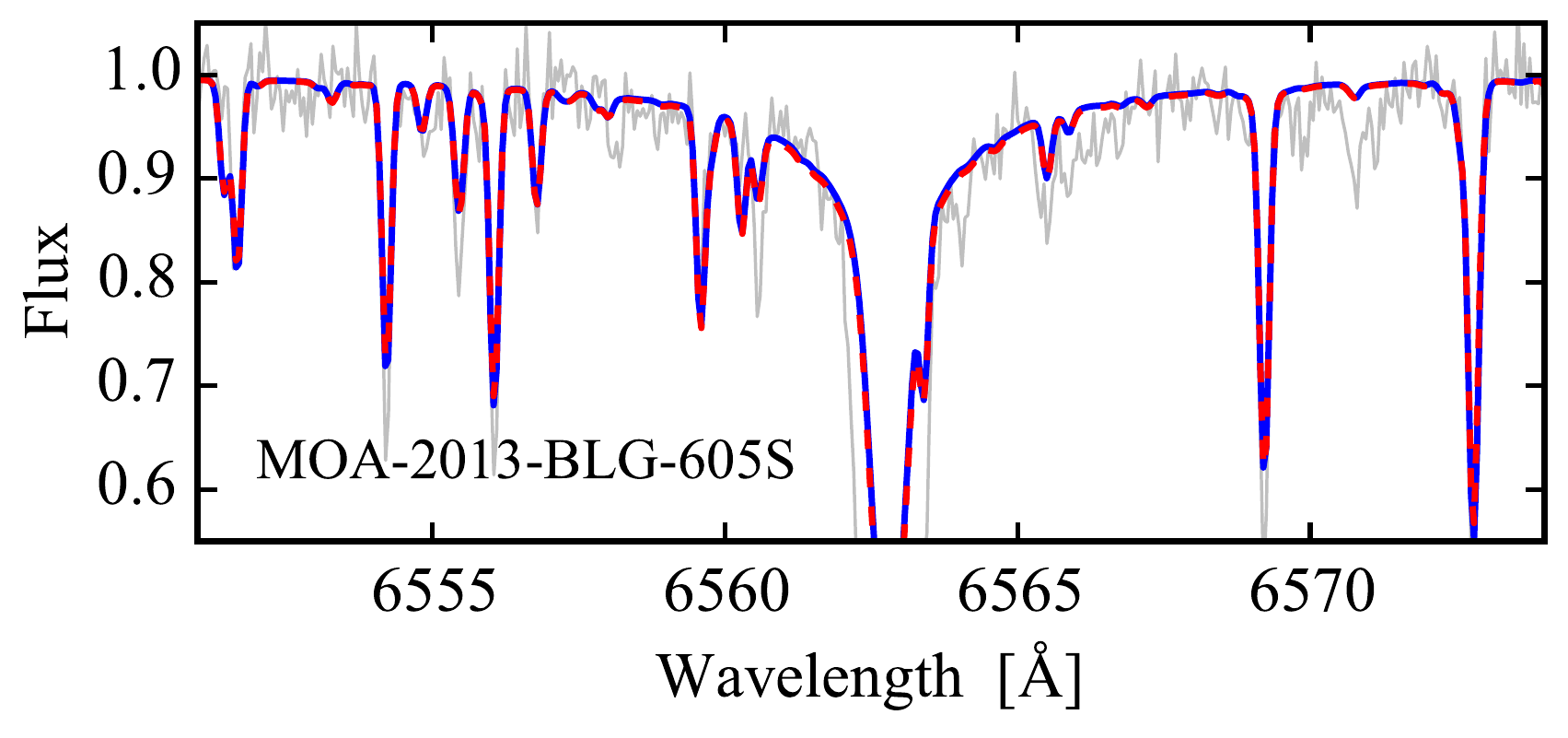}
\includegraphics[viewport=56 56 485 225,clip]{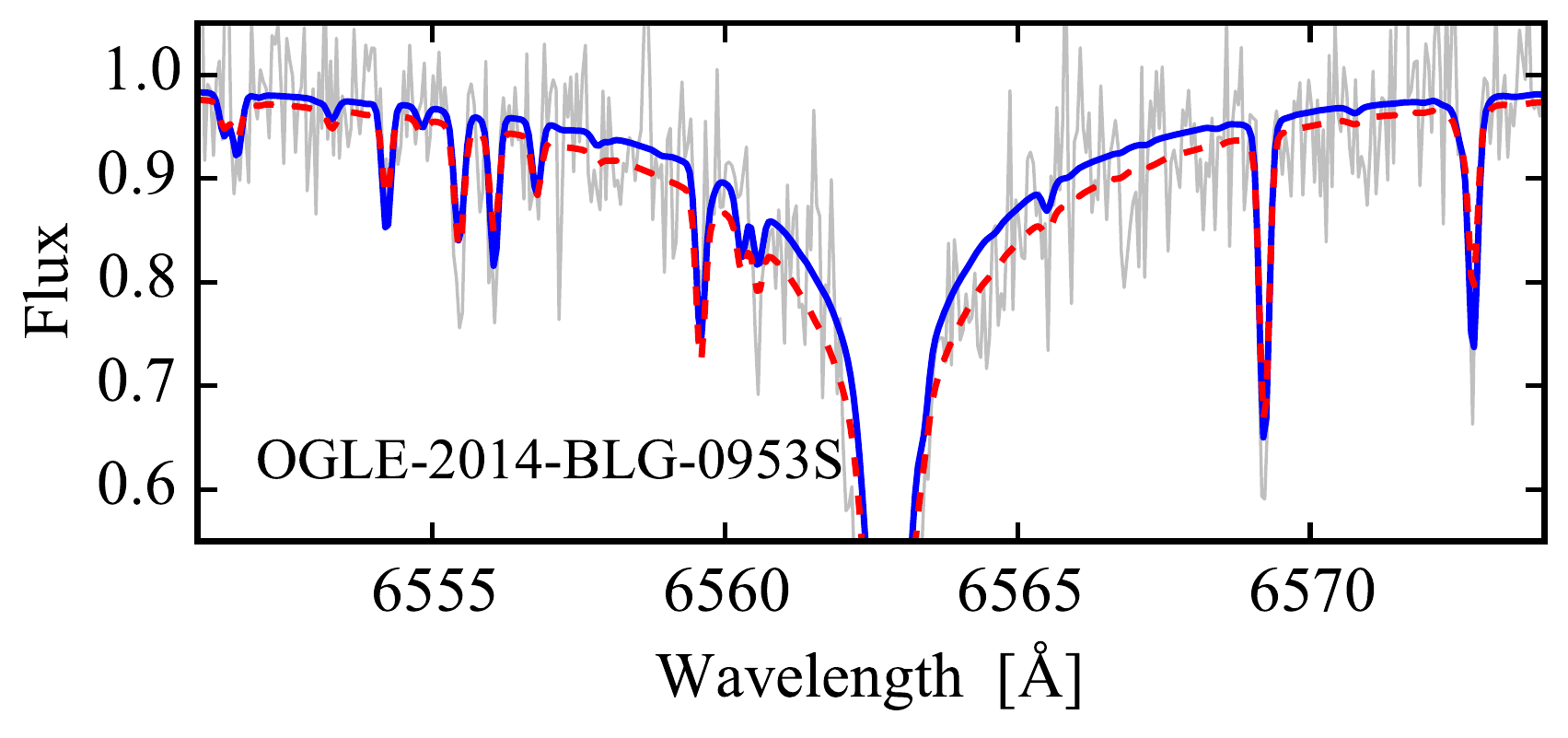}
}
\resizebox{0.85\hsize}{!}{
\includegraphics[viewport=0 56 510 225,clip]{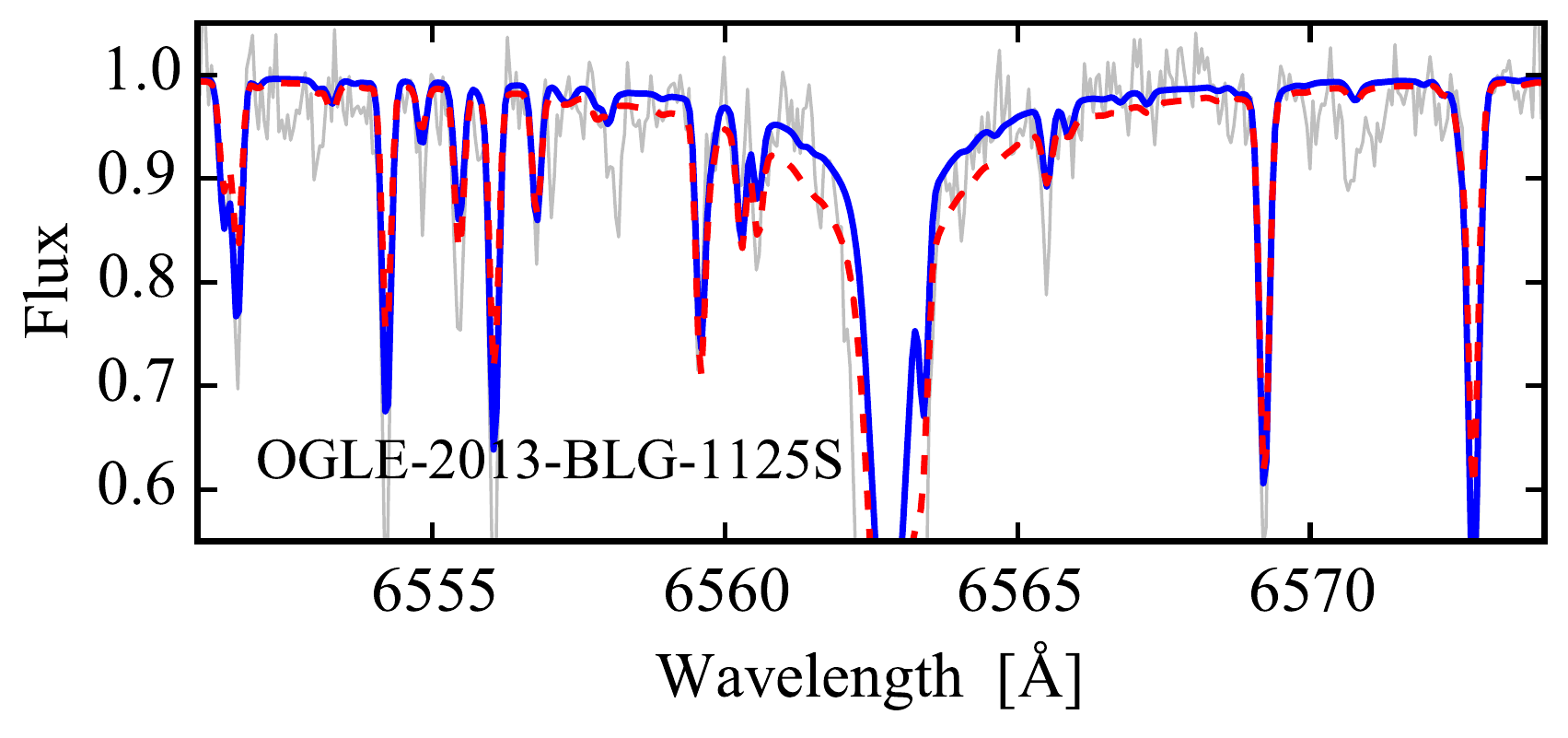}
\includegraphics[viewport=56 56 510 225,clip]{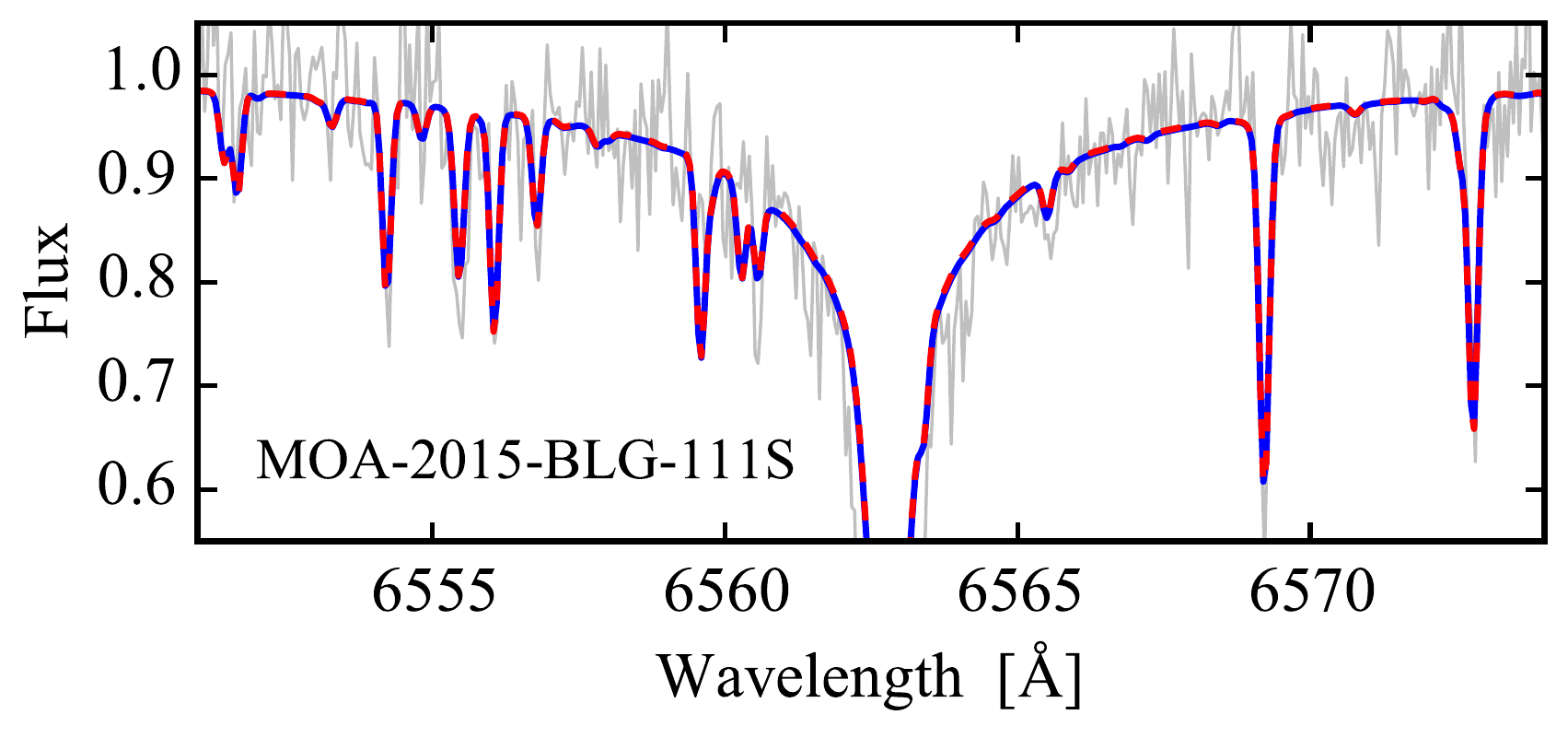}
\includegraphics[viewport=56 56 485 225,clip]{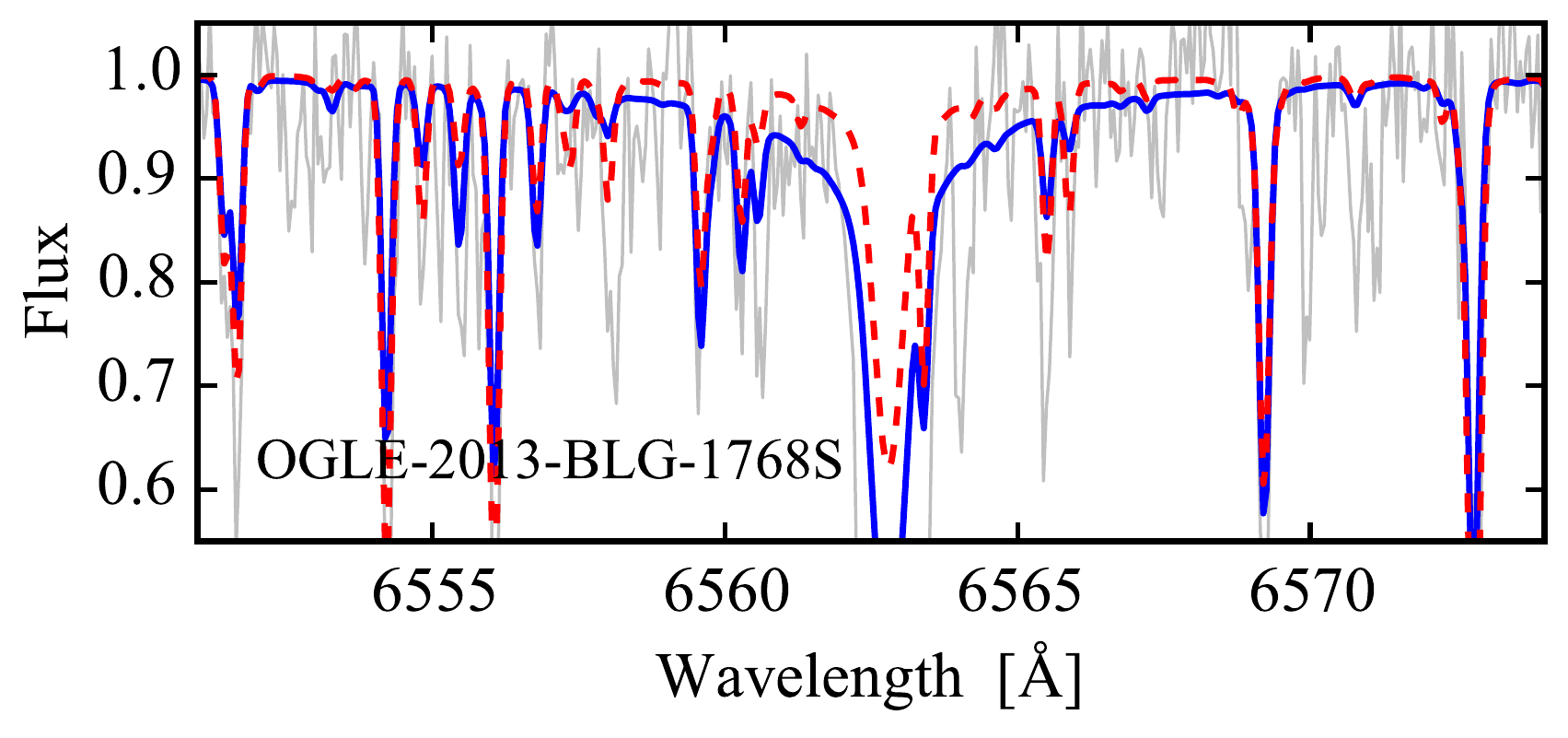}
}
\resizebox{0.85\hsize}{!}{
\includegraphics[viewport=0 56 510 225,clip]{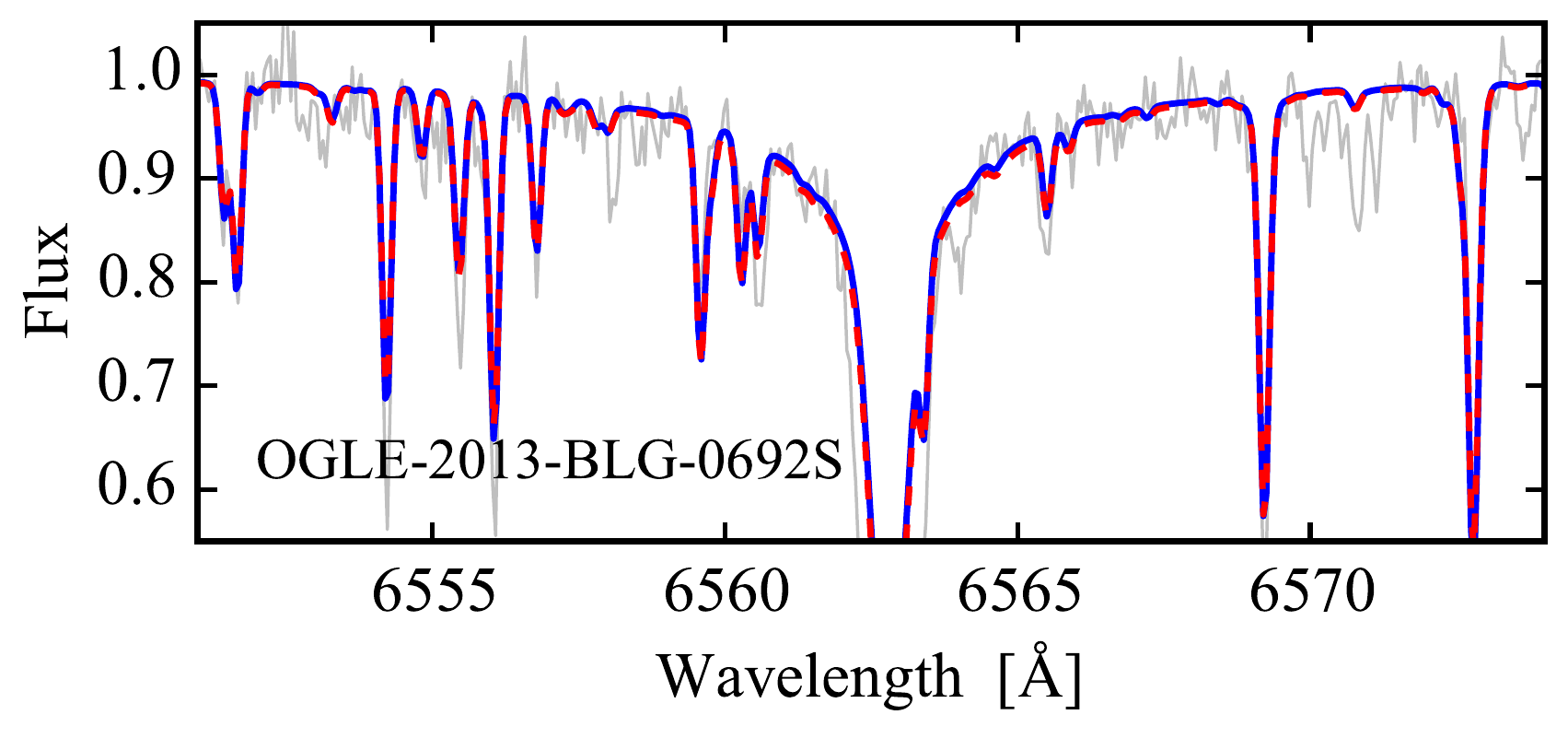}
\includegraphics[viewport=56 56 510 225,clip]{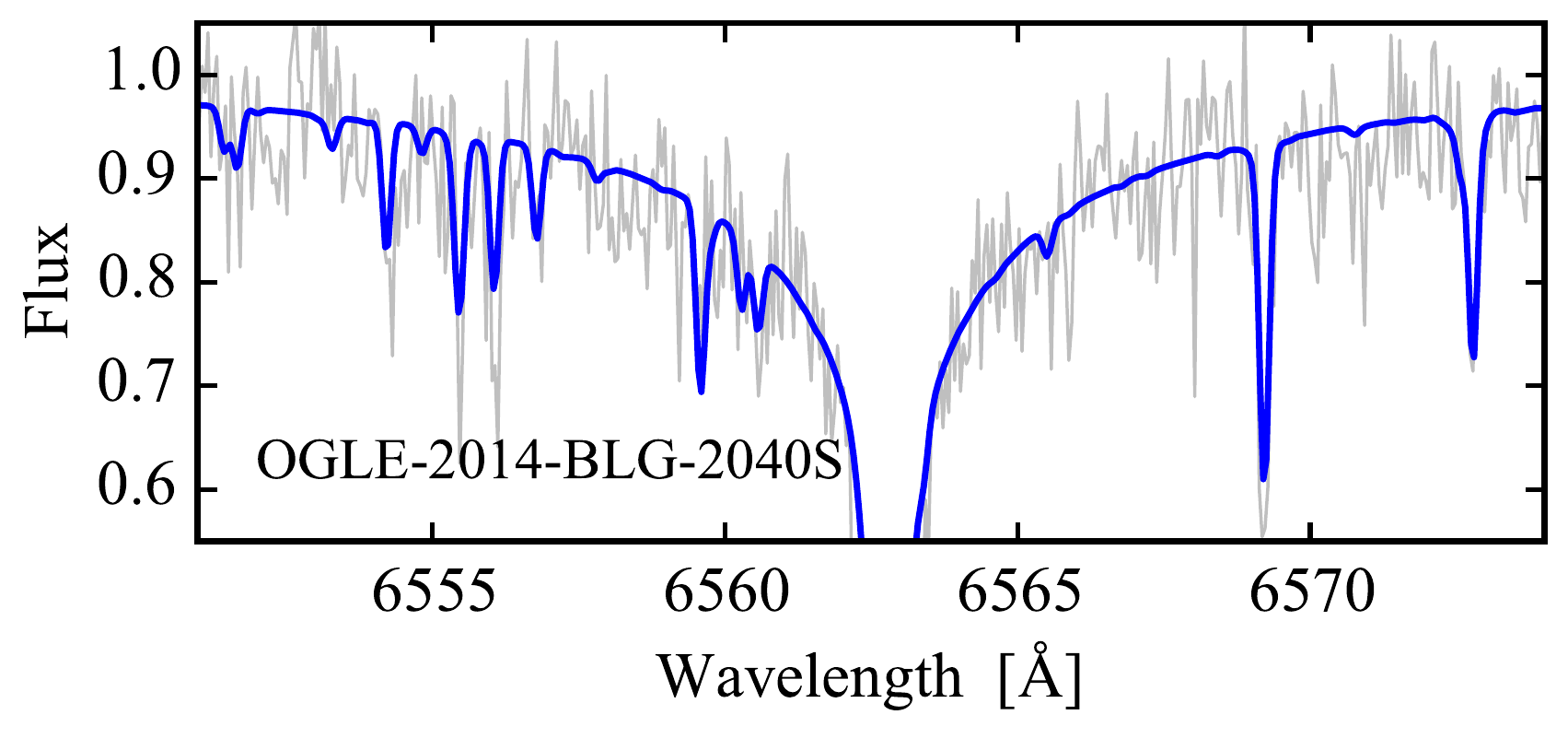}
\includegraphics[viewport=56 56 485 225,clip]{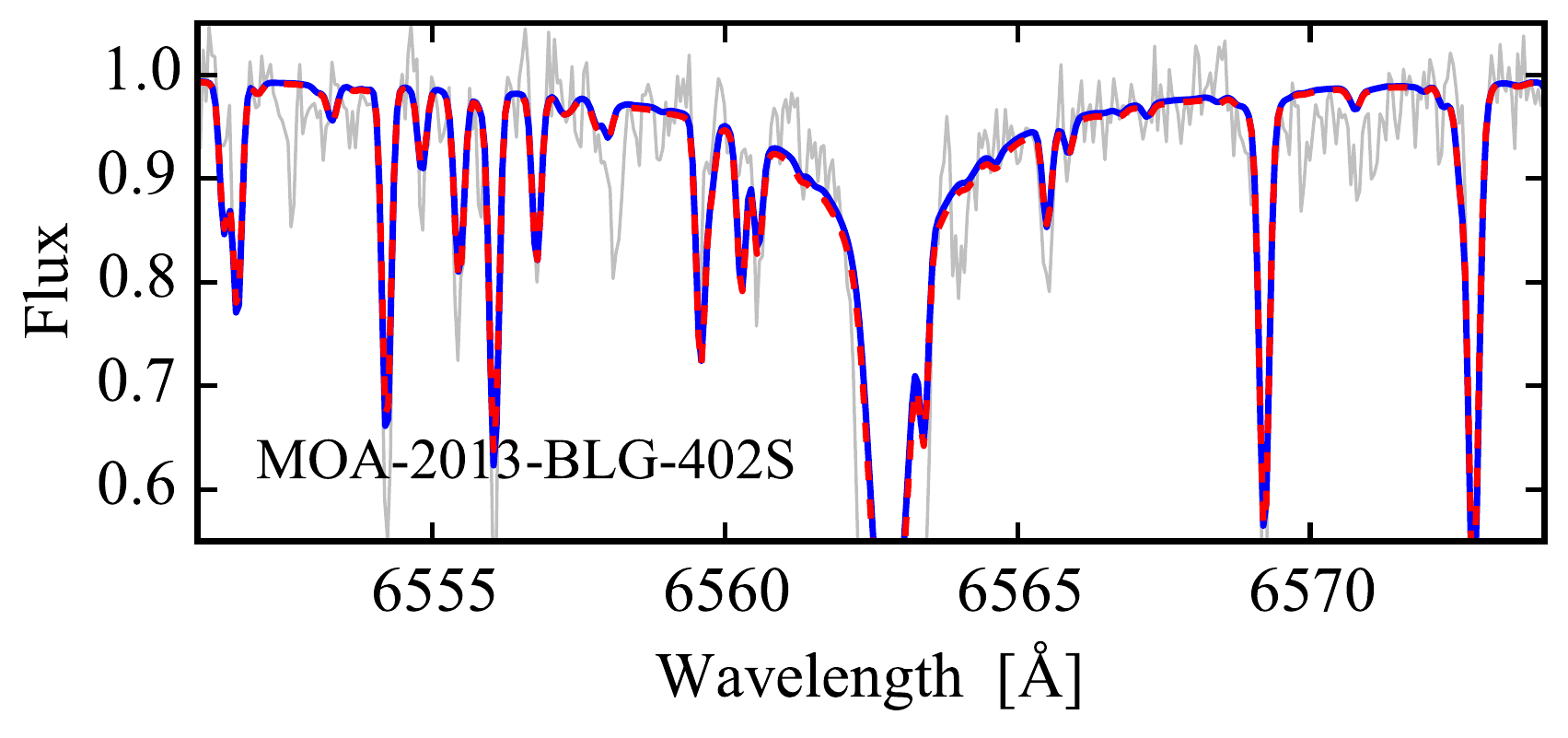}
}
\resizebox{0.85\hsize}{!}{
\includegraphics[viewport=0 56 510 225,clip]{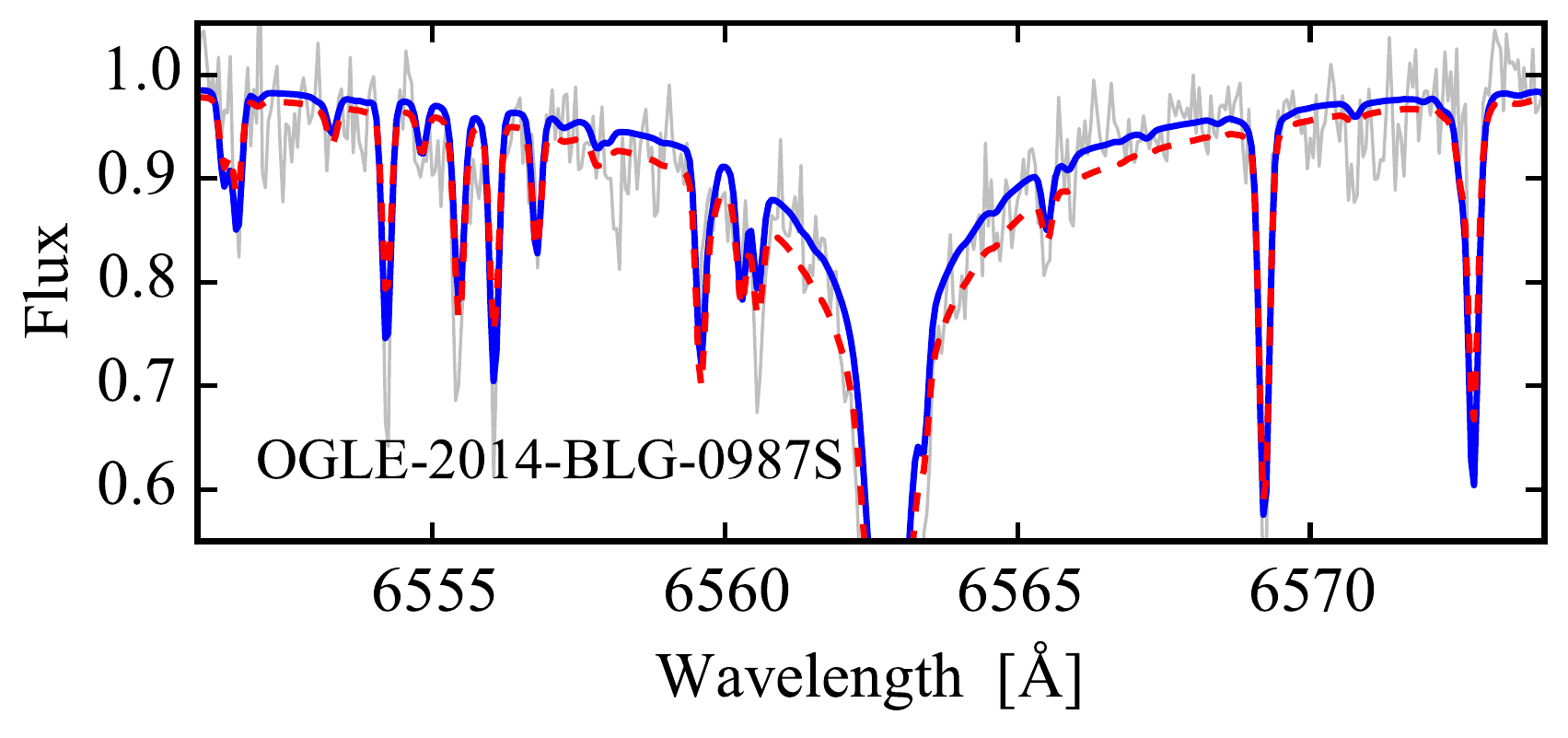}
\includegraphics[viewport=56 56 510 225,clip]{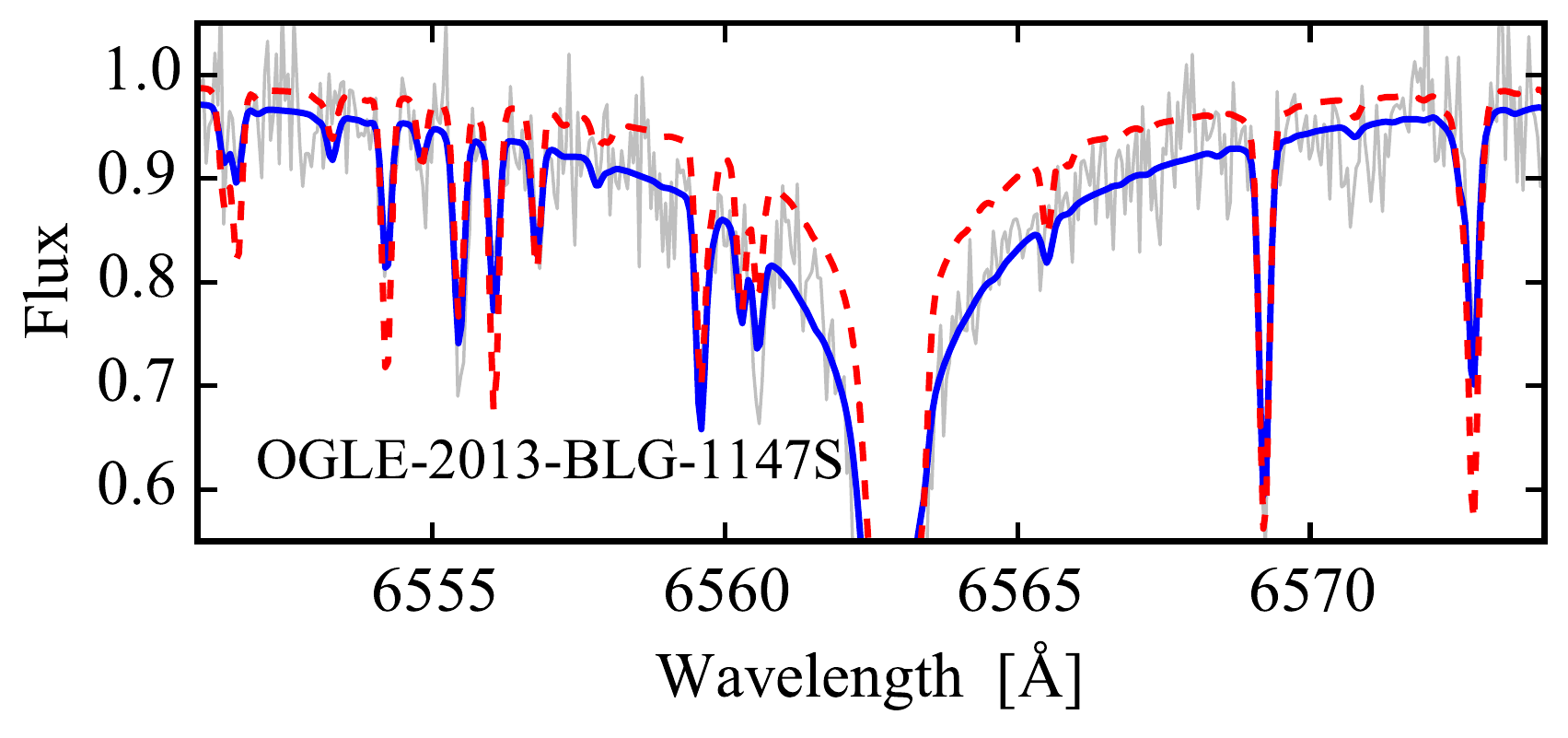}
\includegraphics[viewport=56 56 485 225,clip]{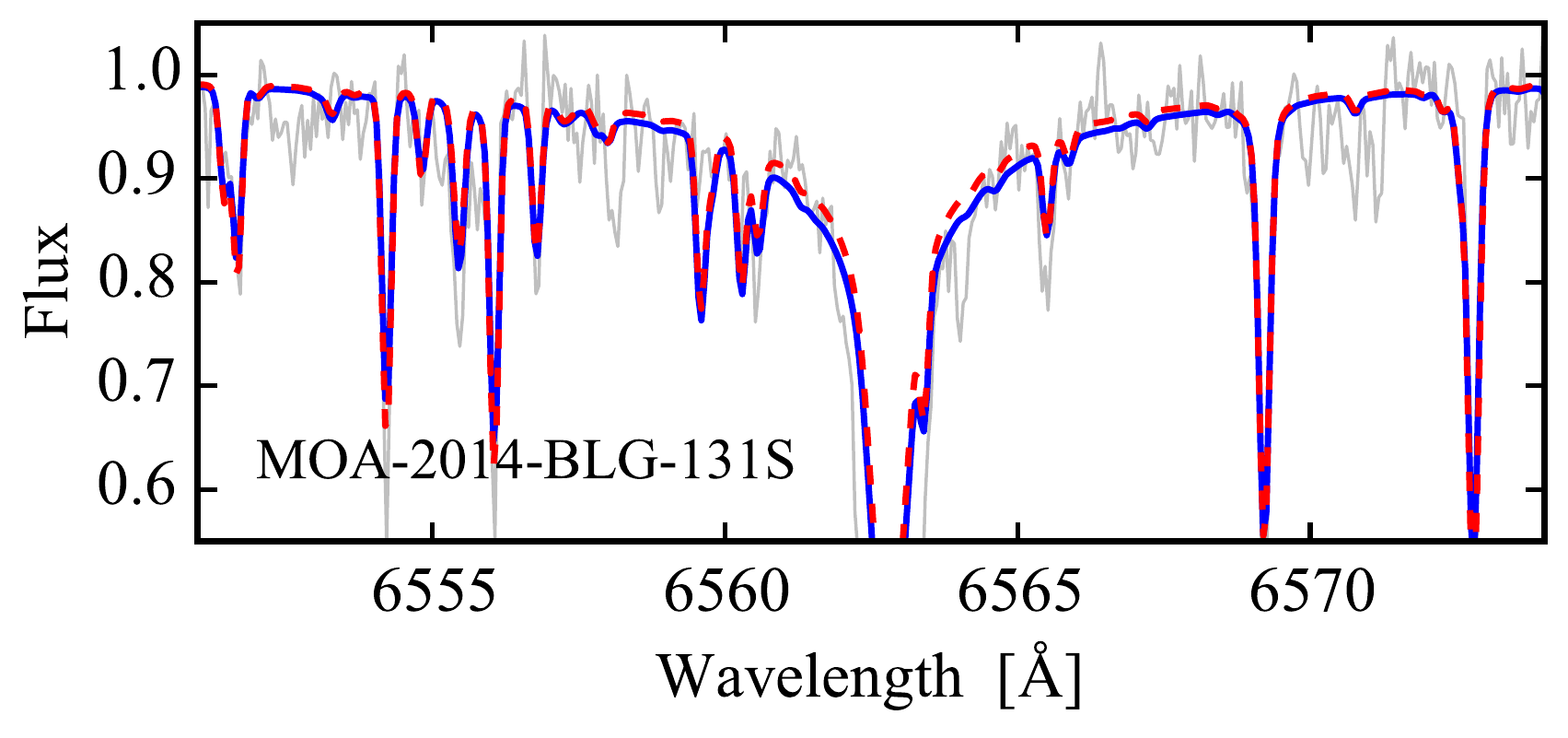}
}
\resizebox{0.85\hsize}{!}{
\includegraphics[viewport=0 56 510 225,clip]{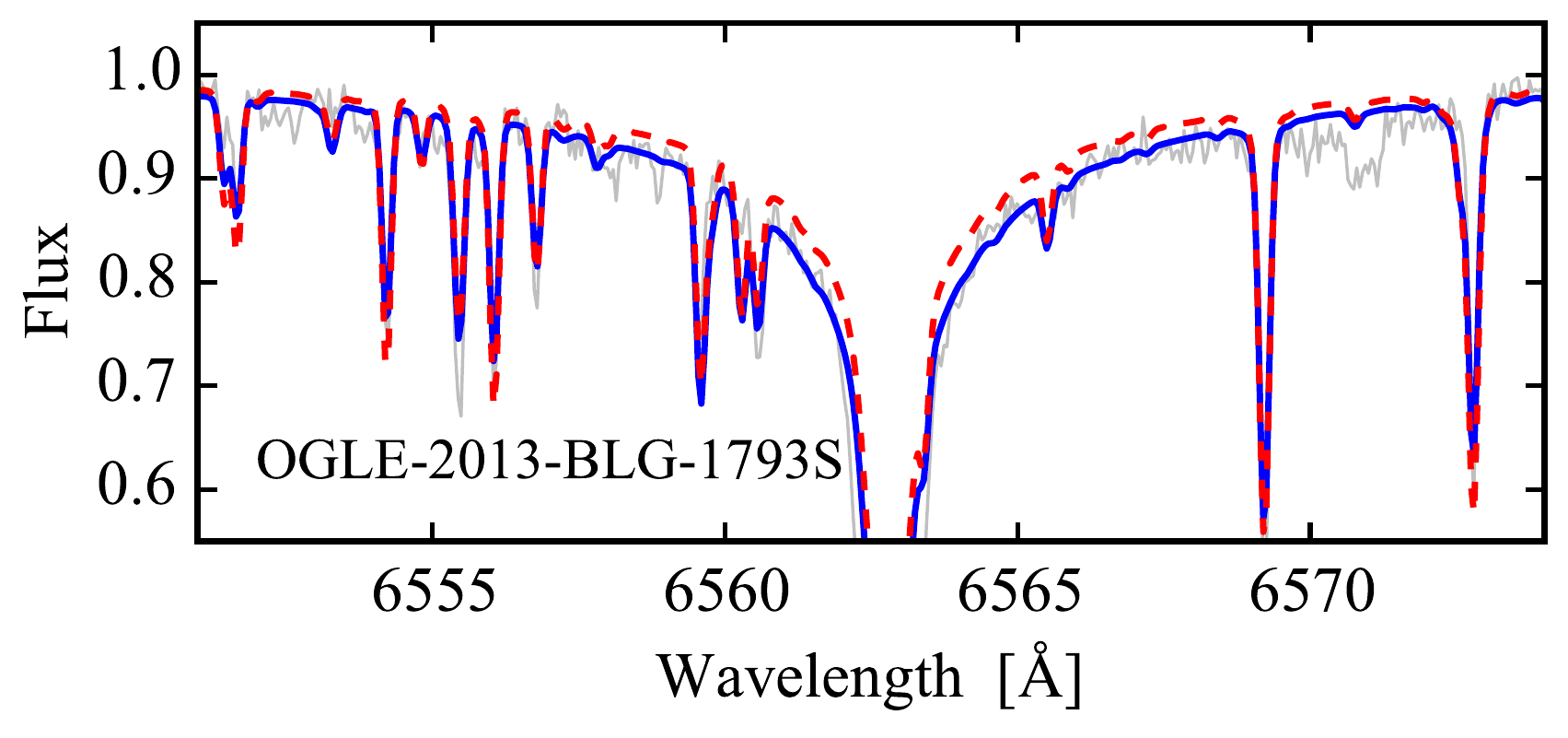}
\includegraphics[viewport=56 56 510 225,clip]{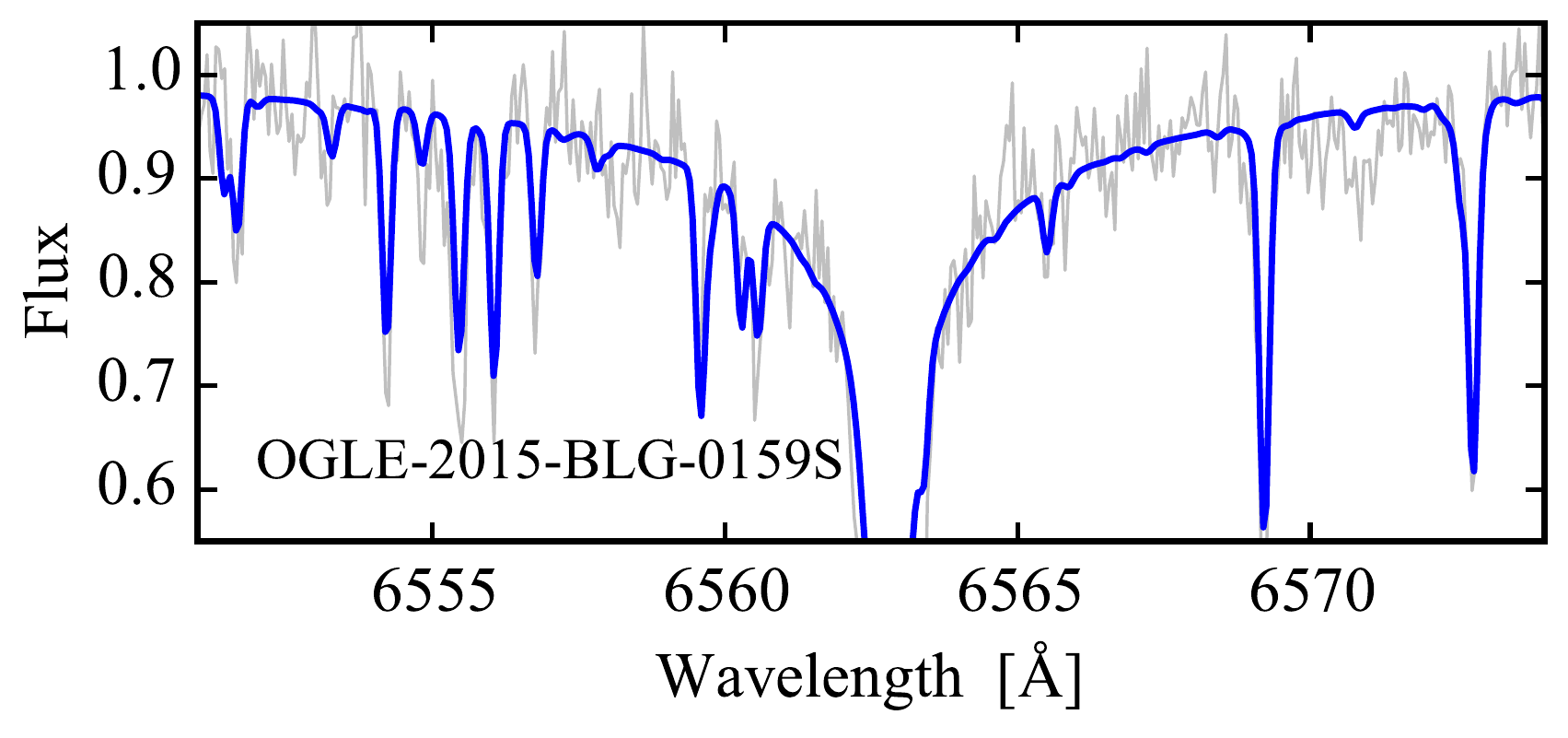}
\includegraphics[viewport=56 56 485 225,clip]{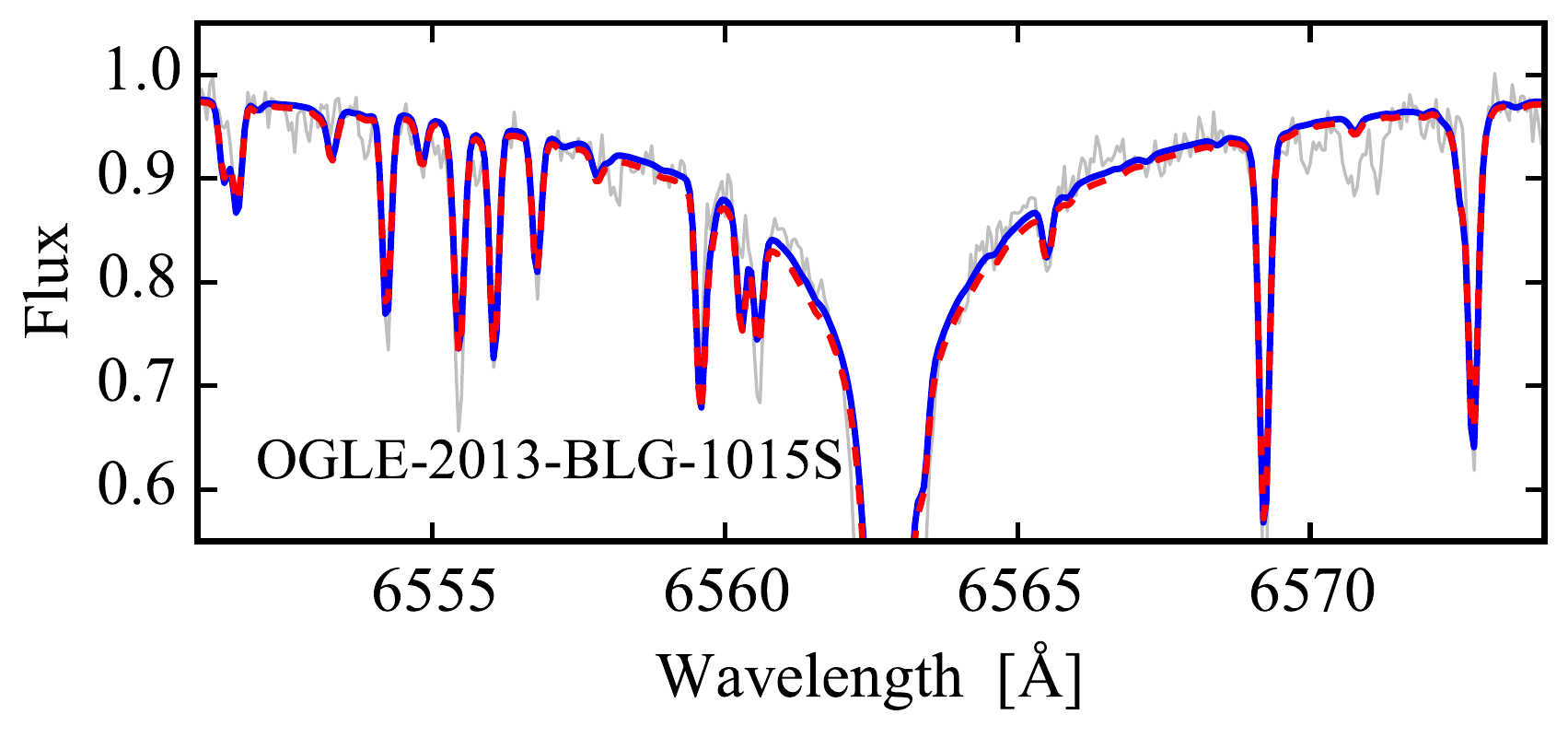}
}
\resizebox{0.85\hsize}{!}{
\includegraphics[viewport=0 56 510 225,clip]{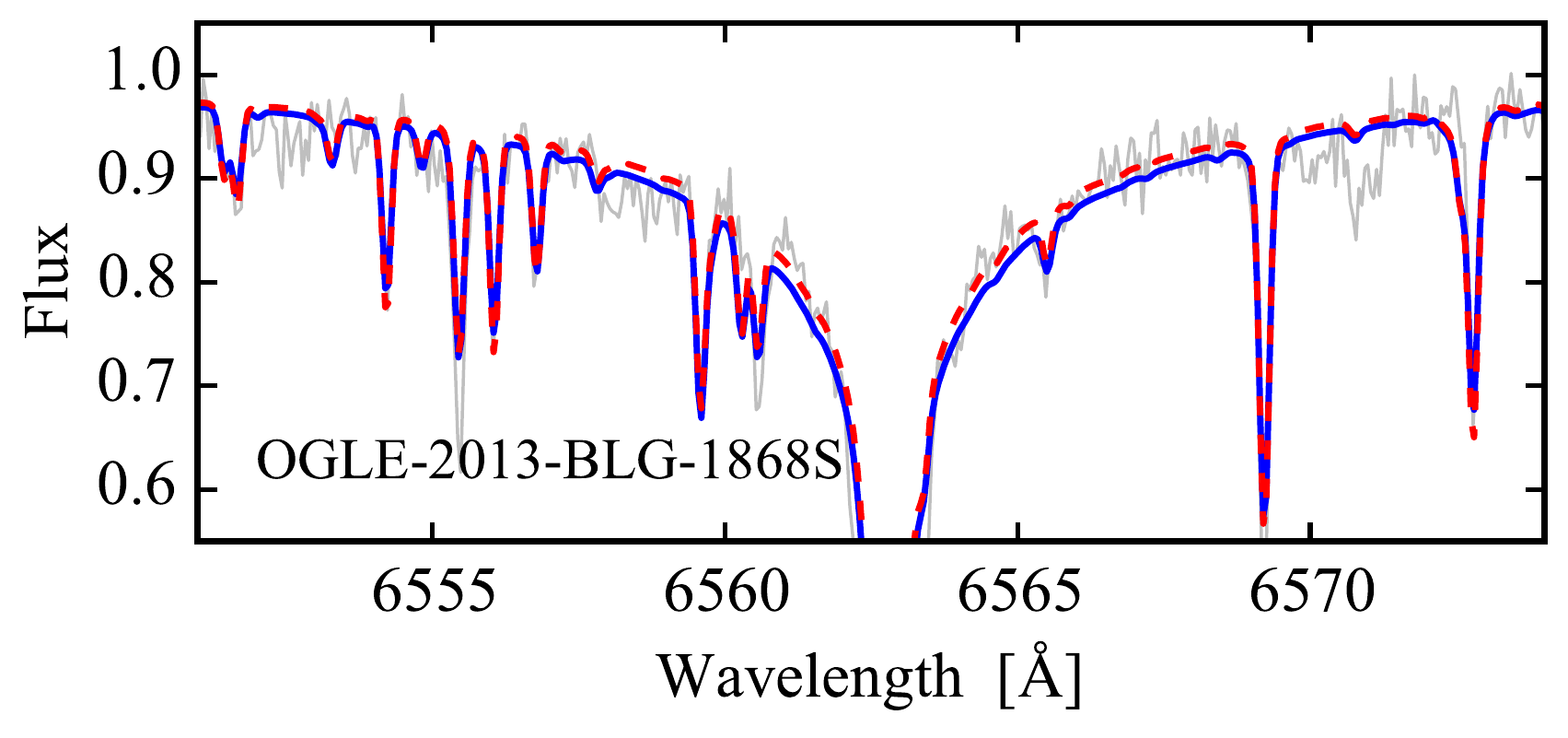}
\includegraphics[viewport=56 56 510 225,clip]{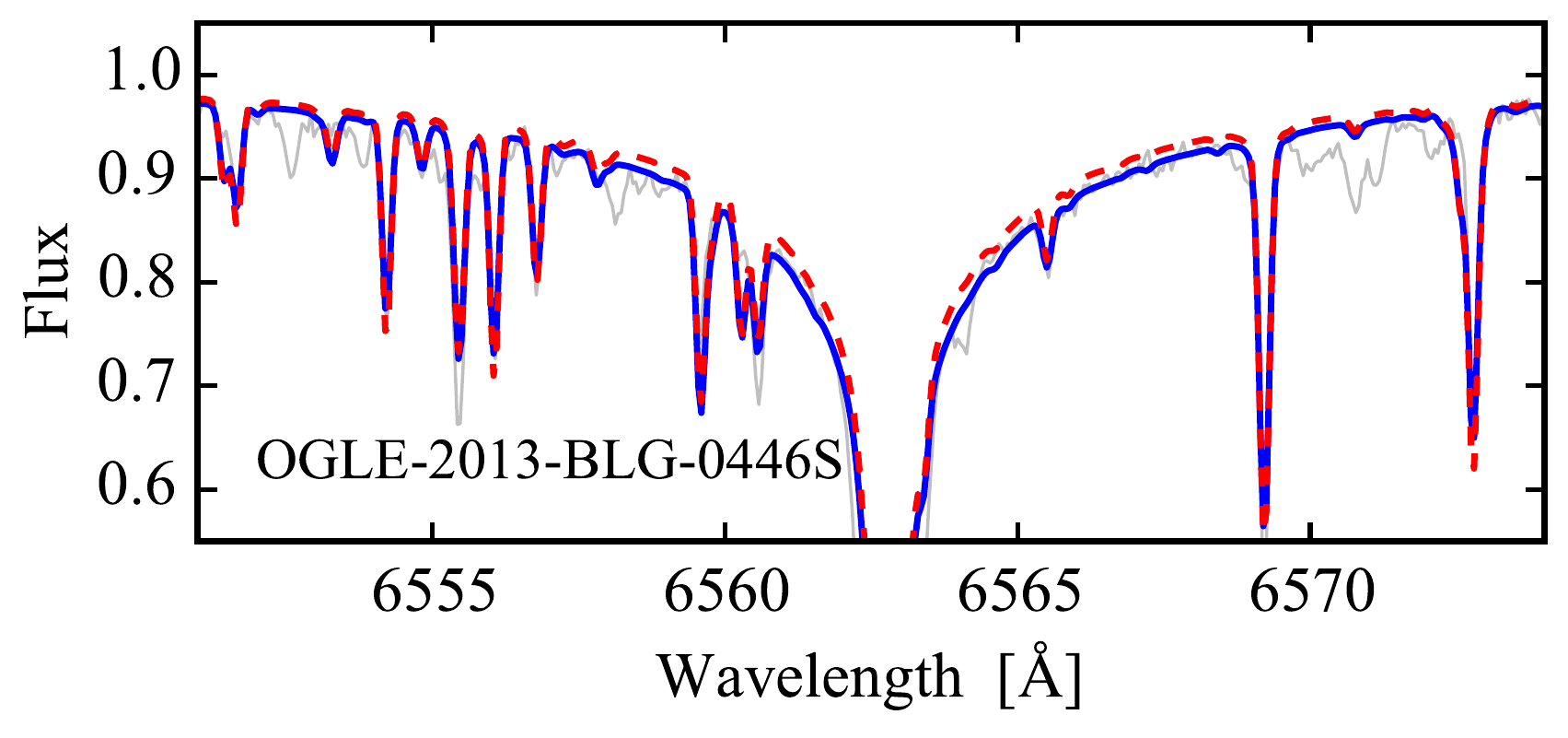}
\includegraphics[viewport=56 56 485 225,clip]{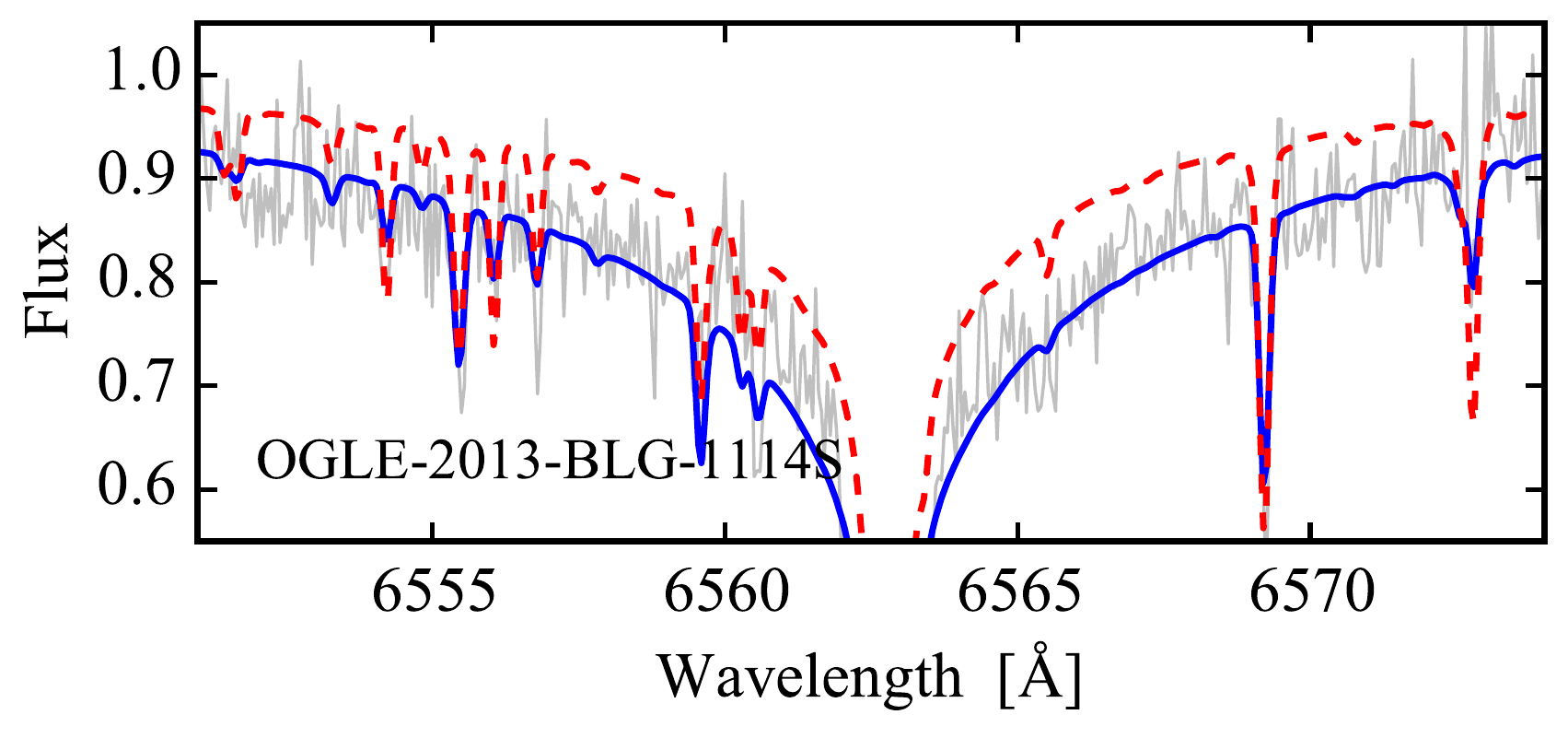}
}
\resizebox{0.85\hsize}{!}{
\includegraphics[viewport=0 56 510 225,clip]{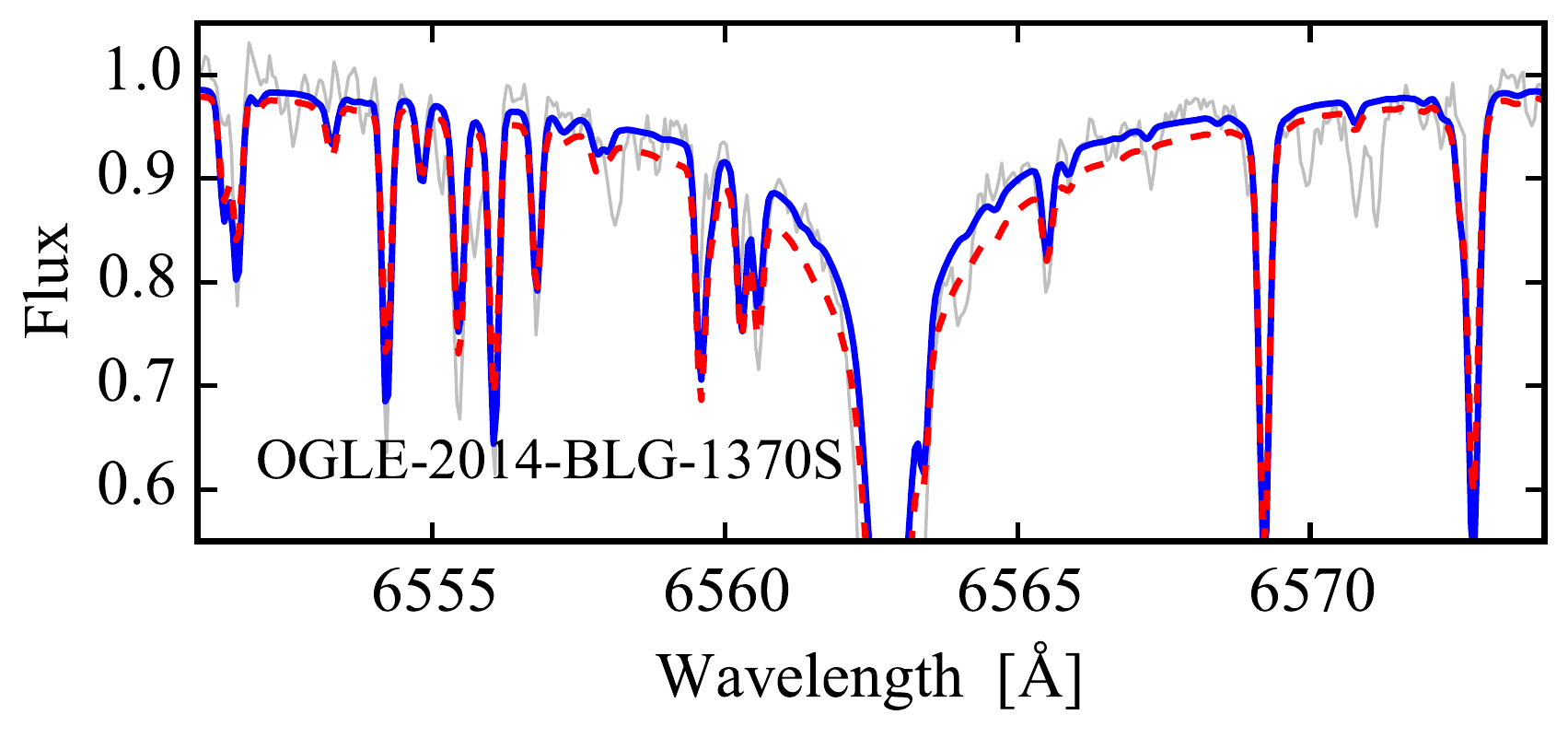}
\includegraphics[viewport=56 56 510 225,clip]{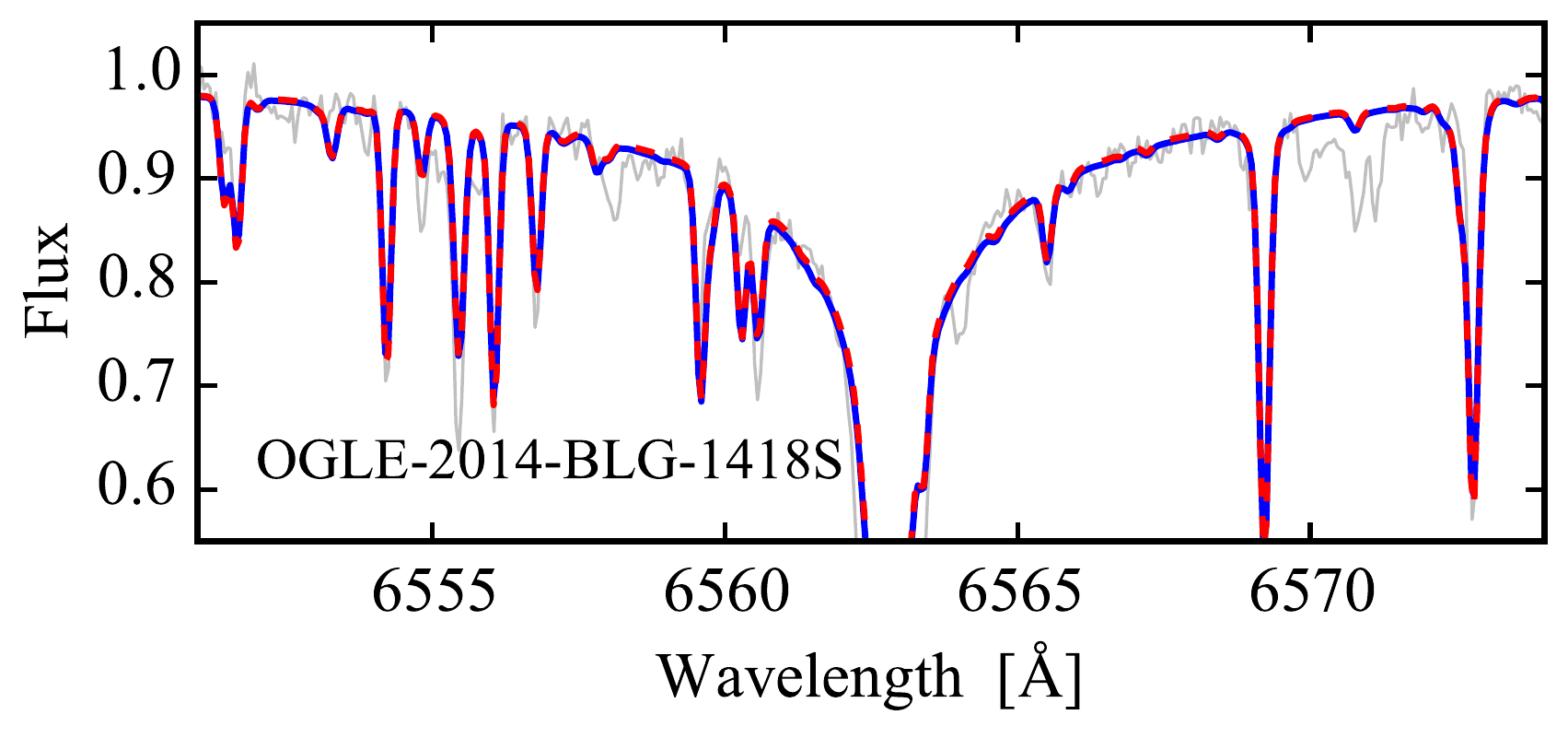}
\includegraphics[viewport=56 56 485 225,clip]{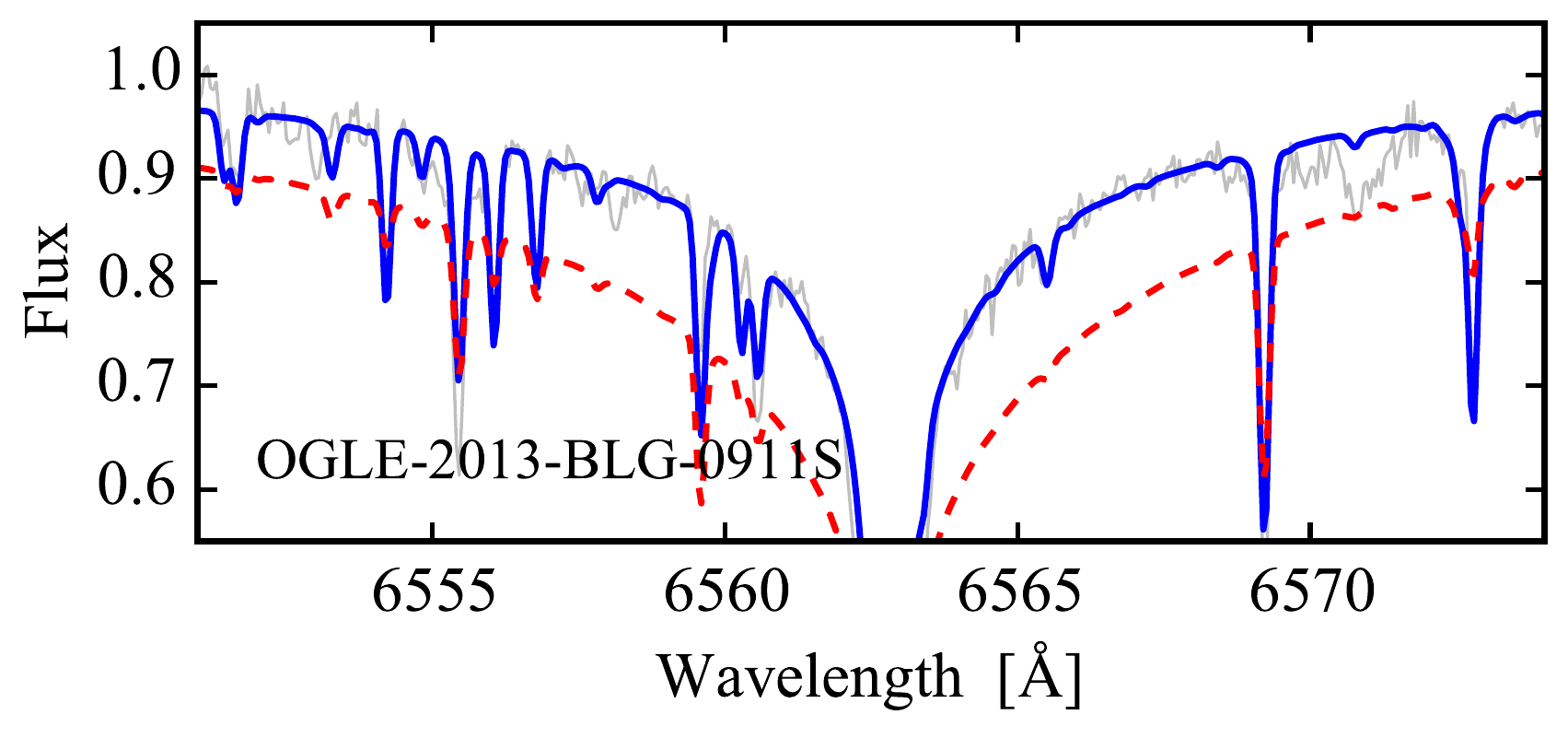}
}
\resizebox{0.85\hsize}{!}{
\includegraphics[viewport=0 0 510 225,clip]{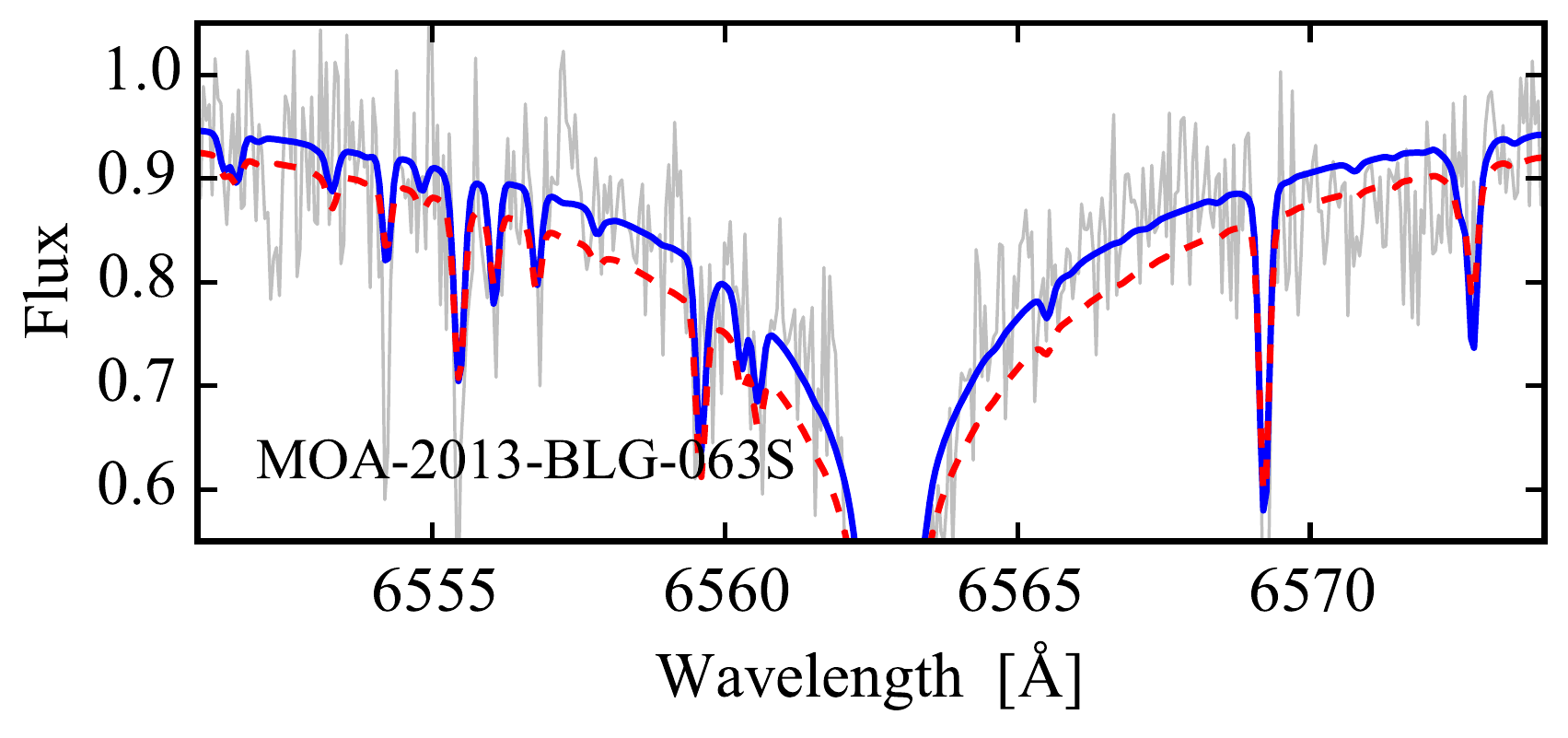}
\includegraphics[viewport=56 0 510 225,clip]{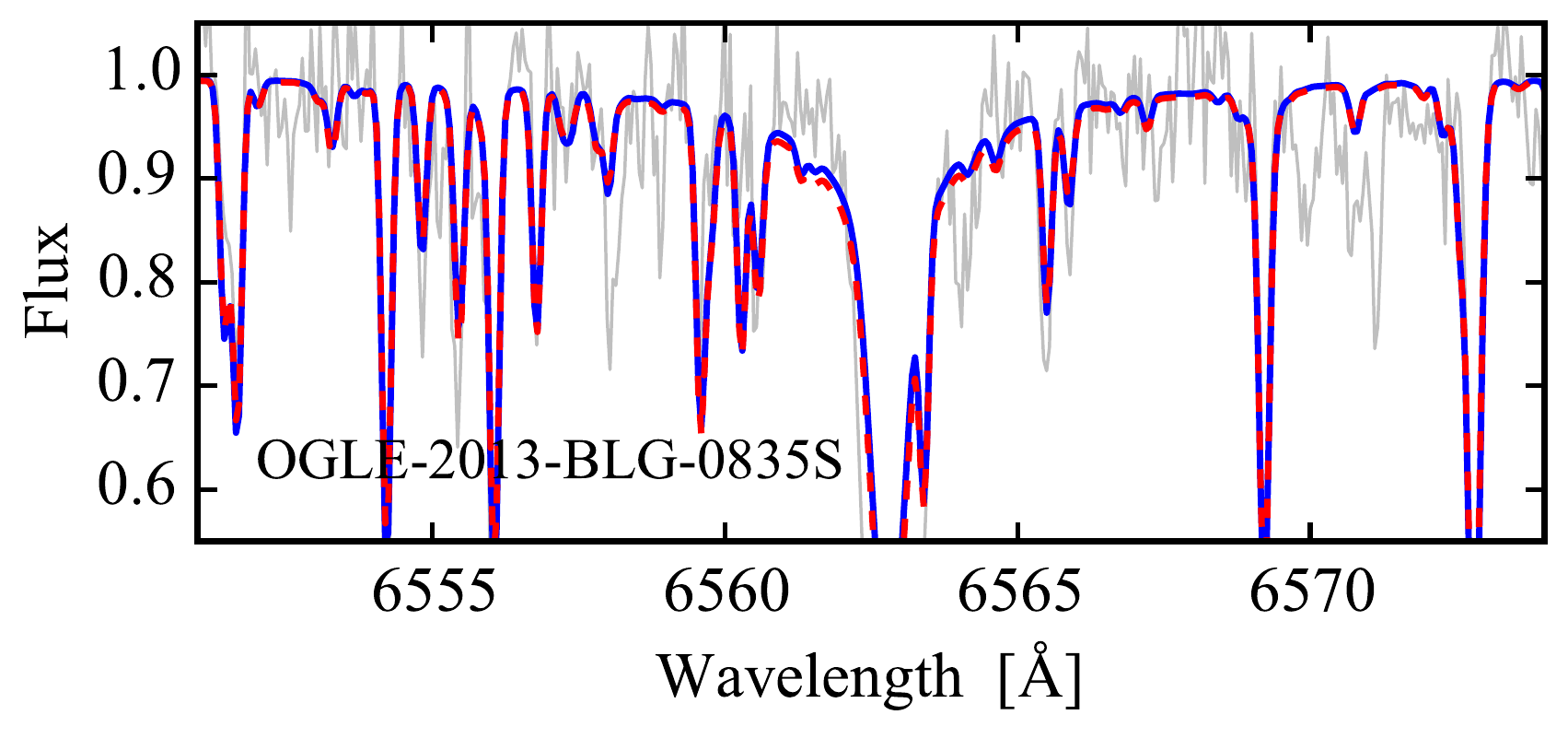}
\includegraphics[viewport=56 0 485 225,clip]{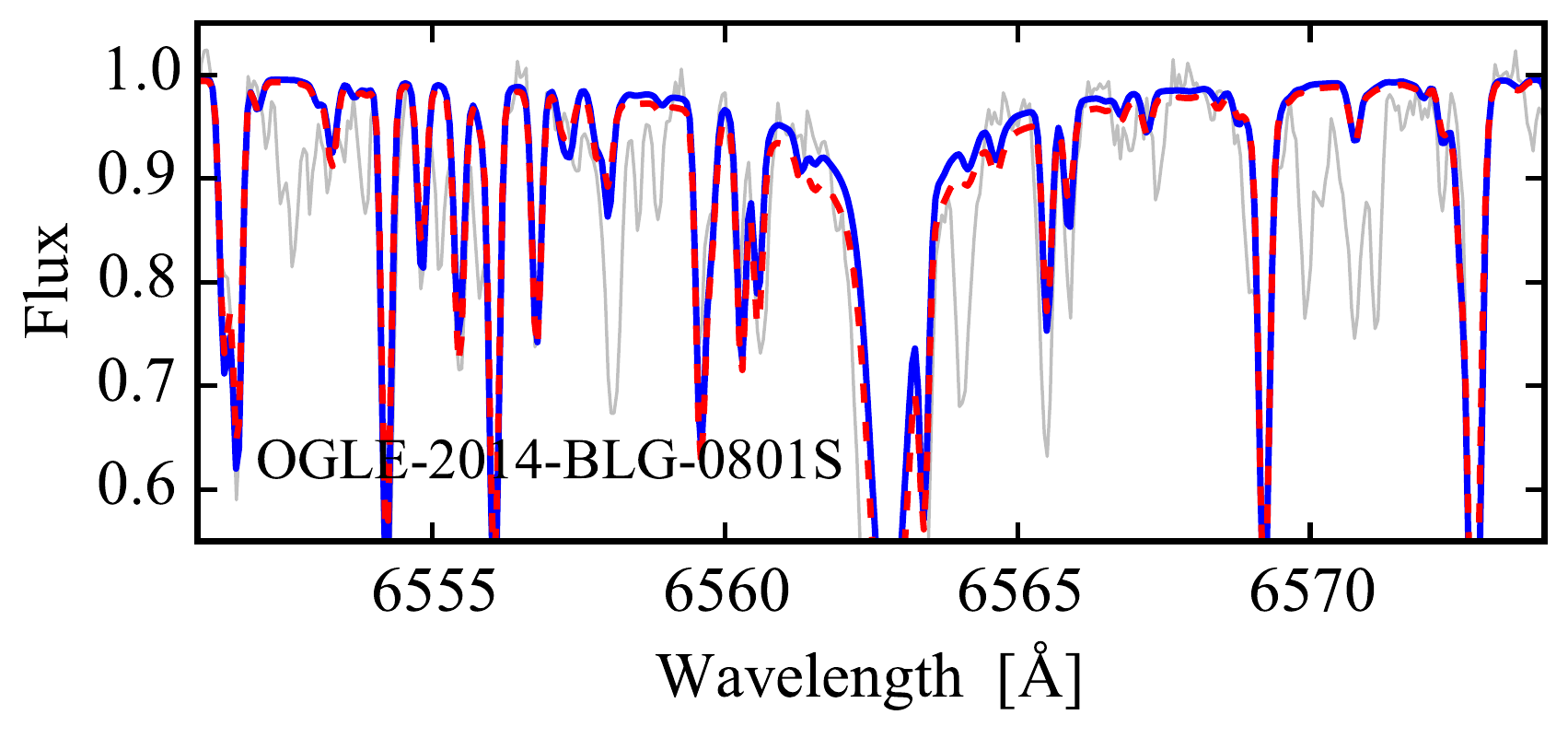}
}
\caption{Comparison between synthetic spectra based on the spectroscopic temperatures (full blue lines) and temperatures from microlensing techniques (dashed red lines) to the observed H$\alpha$ line profiles at 6563\,{\AA} for the 33 new stars.  The stars are sorted by metallicity. Similar plots can be found for the other 58 stars in \cite{bensby2011} and \cite{bensby2013}.
\label{fig:balmer}
}
\end{figure*}

\clearpage
\section{Balmer line wing profiles (Figure~\ref{fig:balmer})}

\end{appendix}

\end{document}